\pdfoutput=1

\documentclass[epj]{svjour}
\usepackage{amssymb}
\usepackage{amsmath}
\usepackage{citesort}
\usepackage{latexsym}
\usepackage{graphics}
\usepackage{epsfig}
\newcommand{\mc}{\multicolumn}
\newcommand{\lsim}{\mathrel{\mathop{\kern 0pt \rlap
  {\raise.2ex\hbox{$<$}}}
  \lower.9ex\hbox{\kern-.190em $\sim$}}}
\newcommand{\gsim}{\mathrel{\mathop{\kern 0pt \rlap
  {\raise.2ex\hbox{$>$}}}
  \lower.9ex\hbox{\kern-.190em $\sim$}}}

\begin{document}

\hyphenation{stran-ge}
\hyphenation{me-so-nic}
\hyphenation{pa-ra-me-tri-zed}
 
\hyphenation{FI-NU-DA}
\hyphenation{DA-PHNE}
\hyphenation{me-so-nic}
\hyphenation{pa-ra-me-tri-zed}
\hyphenation{star-ted}
\hyphenation{phe-no-me-non}
\hyphenation{phy-sics}
\hyphenation{di-ba-ryon}
\hyphenation{ba-sed}

\title{Strangeness Nuclear Physics: a critical review on selected topics}
\author{Elena Botta, Tullio Bressani \and  Gianni Garbarino 
}                     
%
%
\institute{INFN - Sezione di Torino and Dipartimento di Fisica, Universit\`a di Torino 
}
%
%
\abstract{
Selected topics in strangeness nuclear physics are critically reviewed. This
includes production, structure and weak decay of $\Lambda$--Hypernuclei, the $\bar K$ nuclear 
interaction and the possible existence of $\bar K$ bound states in nuclei.
Perspectives for future studies on these issues are also outlined.
\PACS{
      {21.80.+a}{Hypernuclei}   \and
      {25.80.Pw}{Hyperon--induced reactions}  \and 
      {21.60.-n}{Nuclear structure models and methods} \and
      {21.45.+v}{Few--body Systems} \and
      {25.80.Nv}{Kaon--induced reactions}  \and
      {13.75.Jz} {Kaon--baryon interactions}
     } 
} 
\maketitle

\tableofcontents


\section{Introduction}
\label{intro}

The interaction between hadrons is  one of the most interesting topics among the basic 
subjects since the very beginning of the discoveries of hadrons. Due to its relatively easy access, the 
nucleon--nucleon ($NN$) interaction played historically the key r$\hat{\mathrm{o}}$le.
The theoretical works at low energies constructed either purely phenomenological interaction 
potentials or potentials based on the
meson--exchange picture initiated by Yukawa \cite{yukawa}. A good description of the $NN$ interaction 
is of paramount importance since
it is the basic ingredient for the understanding of the nuclear world. There has been a recent 
theoretical progress to derive the nuclear force
potential from quantum chromodynamics (QCD) using Lattice QCD techniques. It successfully 
reproduced long--range attraction and short--range repulsion at the same time.

Effective field theories (EFTs) provide a crucial test for analyzing the properties and 
interactions of hadrons and nuclei. The theories that describe
low--energy hadron physics embody the symmetries of the underlying theory, QCD, in effective 
Lagragians for the relevant degrees of freedom,
such as pions, photons and nucleons: they lead to systematic expansions of observables in ratios of 
low--energy to high--energy scales, making it
possible to estimate the uncertainties in their predictions and to systematically improve the 
accuracy of the pertinent calculations.

Chiral perturbation theory (ChPT) is the most important of these for low--energy hadronic and 
nuclear physics. It embodies the approximate
$SU(2)_{R} \times SU(2)_{L}$ symmetry of QCD, which is a consequence of the very small current 
masses of the $u$ and $d$ quarks.
Pions are, to a good approximation, the Goldstone bosons that arise from the spontaneous breaking 
of this symmetry by the condensation of
quark--antiquark pairs in the QCD vacuum. The small mass of the pions and the weakness of their 
interactions at low--energy are all consequences
of this symmetry and are embodied in the effective Lagrangian of ChPT. Extensions of the theory can 
describe heavy particles, such as nucleons or
$D$ or $B$ mesons interacting with pions. Enlarging the symmetry to $SU(3)_{R} \times SU(3)_{L}$ leads 
to EFTs that include strange particles.

EFTs can be used to connect different experimental observables within the symmetries of QCD.
Increasingly, they are also being used to relate experimental observables to quantities that can be 
calculated using the methods of Lattice QCD.

Considerable progress has been made in the derivation of nuclear forces from chiral EFT. Although 
there is still an on--going debate about the  
renormalisation of these forces and the resulting expansion scheme, in practice very successful 
potentials have been constructed using
Weinberg's original scheme at order--Q4 or N3LO.  They provide an accurate representation of 
two--body scattering data and are currently
being used as input to ab initio nuclear structure calculations, using, for example, the 
no--core shell model or the coupled cluster method.
In addition, EFT provides a consistent framework for the construction of three-- and four--body 
interactions and also effective electroweak current   
operators. Calculations of light nuclei show that these three--body forces are essential for an 
accurate description of their binding energies.

These techniques are now being extended to hyperon--nucleon ($YN$) and hyperon--hyperon ($YY$) forces. 
The longest--range parts of these interactions arise from pion--exchange and so they can be calculated using 
the same chiral EFT. Here the smaller scattering lengths and weaker pion--hyperon coupling mean that
the Weinberg's expansion scheme is appropriate.

A relevant agreement from theoreticians is the strong request of more experimental data on the $YN$ and $YY$ interactions.
At present, the experimental data on the $\Lambda N$ and $\Sigma N$ interactions consist of not more than 
850 spin--averaged scattering events, in the momentum region from 200 to 1500 MeV/c, while no data is available for $YY$ scattering. 
The low--energy data, in particular, fail to adequately define even the sizes of the dominant $s$--wave spin--triplet and 
spin--singlet scattering lengths and effective ranges.

Due to the inherent difficulty in obtaining precise direct experimental data with beams of low momentum 
(200-400 MeV/c) hyperons 
due to their short life, the only way to obtain such data is that of performing
experiments on nuclear systems which bind one or more hyperons, i.e. Hypernuclei. The possibility 
of exploiting in different ways the Pauli principle for
complex systems composed by many nucleons and one or more hyperons is the key issue for obtaining 
reliable and precise data.

In the last decade, spectroscopy of $\Lambda$--Hypernuclei sho-wed an impressive step forward in 
determining the energy levels of nearly all $p$--shell Hypernuclei 
both by $\gamma$--ray spectroscopy from the decay of low--lying excited states and by missing 
mass measurements with magnetic spectrometers at KEK (SKS), JLab (Hall A and Hall C) and LNF--INFN (FINUDA). 
A $\Lambda N$ effective interaction which is expressed by 
five strength parameters was introduced and determined for the first time from a fit of $\gamma$--ray data on $p$--shell Hypernuclei:
of particular relevance, we mention the extraction of the spin--spin term in the $\Lambda N$ interaction. In 
the spectroscopy of $\Lambda$--Hypernuclei the Pauli principle plays the
crucial r$\hat{\mathrm{o}}$le that all the possible states of the $\Lambda$ in the many--body system 
are allowed, contrarily to nucleons. For this reason, the calculation of the 
quantum numbers of Hypernuclei is fully reliable.

In addition to information on the strong $YN$ interaction, Hypernuclei may give access to 
experimental information, not otherwise accessible, on the weak $YN$
interaction by their decays, in particular the non--mesonic decay. A free $\Lambda$ decays almost at 
100$\%$ into a pion and a nucleon, with a release of
kinetic energy of about 5 MeV to the nucleon, corresponding to a final momentum of about 100 MeV/c. 
The situation changes dramatically when the
$\Lambda$ is embedded in a nucleus, since the nucleon of the mesonic decay is emitted with a 
momentum which is much smaller than the nucleon Fermi
momentum $k_{F}\sim280$ MeV/c. The Pauli principle acts now in the opposite way and the pionic 
decay modes are strongly hindered in all but the lightest
Hypernuclei. However, new non--mesonic weak decay (NMWD) modes are switched on and are due to the weak 
interaction of the $\Lambda$ with one nucleon of the medium
$(\Lambda n \rightarrow nn,\, \Lambda p \rightarrow n p)$. The above weak decay reactions are 
possible only in Hypernuclei, thanks to the stability 
of the $\Lambda$ against the mesonic weak decay due to the Pauli blocking and to the impossibility 
of strong interaction processes involving the hyperon due to Strangeness conservation.

The study of the NMWD is of fundamental importance, since it provides primary means of exploring 
the four--baryon, strangeness changing, weak interactions.
The non--mesonic process resembles the weak $\Delta S=0$ $NN\to NN$ 
interaction, which was studied experimentally in
pari-ty--violating $NN$ scattering. However, the $\Lambda N \rightarrow nN$ two--body 
interaction mode offers more information, since it can
explore both the parity--conserving and the parity--violating sectors of the $\Delta S=1$ weak 
four--baryon interaction.
In the $NN$ system, the strong force masks the signal of the weak parity--conserving part of the 
interaction.

In addition, the large momentum transfer in NMWD processes implies that they probe short distances 
and might therefore explore the r$\hat{\mathrm{o}}$le of explicit
quark/gluon substructures of the baryons.
It is also possible to envisage in a nuclear medium a weak interaction of a $\Lambda$ with two 
correlated nucleons, $\Lambda NN \rightarrow nNN$.
It appears then that an exhaustive study of the weak interaction in Hypernuclei needs a big 
amount of correlated information from  a very systematic and coordinated series of measurements.
After several decades during which the experimental information was scarce, due to the hardness of 
the measurements, in the last years some very interesting new measurements were carried out. 
For NMWD, coincidence measurements of nucleon pairs emitted in quasi back--to--back
kinematics performed at KEK provided values of the observables in agreement with the most updated 
theoretical calculations.

The spectroscopic study of the proton distributions from NMWD of all the $p$--shell Hypernuclei carried out 
at DA$\Phi$NE--LNF allowed to ascertain experimentally
the existence of the $\Lambda NN\to NNN$ weak interaction, whose contribution to the total NMWD width
turned out to be non--negligible.
The NMWD of Hypernuclei may be considered, literally, as the first and more spectacular example of 
nuclear medium modifications of the properties of
an hadron. However with this definition or more concisely with nuclear medium 
modifications we intend today the changes of the mass
and width of a strong interacting particle when produced inside a nucleus.

Theory predicts strong modifications of the kaon and antikaon properties in a dense hadronic 
environment. While a repulsive kaon--nucleus potential of a few tens
of MeV is predicted for the $K^{+}$, the potential could be attractive up to 100 MeV or more for the $K^{-}$, 
depending on the model. A broadening is also predicted for the antikaon.
Note that in the case of the strange mesons, no clean decay channel can be used to access directly 
their in--medium mass and width. The data taken so far, mostly based on the study of the yield under 
different kinematics and geometrical conditions (flow pattern) are well explained by
transport models who consistently lead to a repulsive $K^{+}$N potential, as predicted by QCD based 
calculations.
The case of the $K^{-}$ meson is especially difficult because strangeness exchange reactions 
considerably disturb the already weak signals.
More theoretical work, taking into account off--shell dynamics, is severely needed here.

Recently, a quite unexpected observation related to the production of Hypernuclei by $K^{-}$ at rest 
on nuclei at LNF--DA$\Phi$NE was reported. The mesonless
absorption of $K^{-}$ at rest by 2, 3, 4 nucleons in light nuclei occurs with similar and quite high 
rates ($10^{-3}$/stopped $K^{-}$). Rates of a similar order of
magnitude were found only for the production of some Hypernuclear states 
($^{12}_{\Lambda}\mathrm{C}$). This fact suggests that the absorption of $K^{-}$
at rest should lead to the formation of some intermediate state(s), decaying at rest into 
back--to--back correlated pairs, $\Lambda p$, $\Lambda d$ and
$\Lambda t$ for 2, 3 and 4 nucleon absorption, respectively. 
These states could be the so--called AntiKaon Nuclear Clusters (AKNC), 
i.e., states in which a $\overline{K}$ ($K^{-}$) is strongly bound to
few nucleons. A phenomenological approach to this problem predicted for these systems a large 
binding energy (more than 100 MeV) with narrow width
(20-30 MeV) and a density that should be several times larger than the normal nuclear density.
The determination of the invariant mass of the $\Lambda p$ back--to--back correlated events seems 
to confirm the existence of the AKNCs, but this conclusion 
must be strengthened by other experiments with better statistics.

The study of nuclear and hadronic systems with strange-ness
--including hyperonic atoms, Hypernuclei, kaonic atoms and nuclei,
exotic hadronic states such as strangelets, $H$-dibaryons
and pentaquark baryons, compact stars and the phenomenon
of strangeness production in heavy--ion collisions-- has attracted great interest and
experienced major advances in the last years \cite{SI08,SI08b}.
Nowadays, one can safely say that
strangeness nuclear physics plays a major r$\hat{\mathrm{o}}$le in modern nuclear and hadronic
physics and involves important connections with astrophysical processes and
observables as well as with QCD. 

In the present paper we critically review some selected topics in strangeness
nuclear physics, including production, structure and weak decays of
$\Lambda$--Hypernuclei and the controversial issue of kaonic nuclear states. 
We discuss the present status of these issues from both the experimental and theoretical viewpoints.
We shall point out the most recent achievements but also some flaws in the present understanding.
The key experimental aspects and the calculations
that must be considered in the near future will be discussed.

The paper is organized as follows.
In Section \ref{basics} we summarize the most important issues concerning general properties of Hypernuclei: 
the production reactions and the evolution of the experimental facilities for $\Lambda$--Hypernuclei are described in 
details. 
In Section \ref{spect} we discuss the recent results on Hypernuclear
spectroscopy and their relevance on the study of the baryon--baryon strong interaction.
In Section \ref{sec:decay} we discuss the recent results on Hypernuclear weak decays 
and their relevance on the study of the baryon--baryon weak interaction.
In Section \ref{kns} we describe recent experimental results on an unexpected 
high rate of mesonless multinucleon absorption of stopped K$^{-}$, with hints
on the possible existence of exotic Antikaon Nuclear aggregates. Finally, 
in Section \ref{concl} we conclude with a critical summary of the last achievements of strangeness nuclear physics, 
with hints on the key measurements and calculations that must be carried out in order to
achieve the final results so far anticipated.

\section{Basic Issues of Hypernuclear Physics}
\label{basics}

In the present Section we discuss general properties of 
strangeness $-1$ and $-2$ Hypernuclei with emphasis on the production reactions and the development
of the facilities in the experimental study of $\Lambda$--Hypernuclei. 

\subsection{General properties of Hypernuclei and relevance for other fields}
\label{basics-Hypernuclei}

Hypernuclei are bound nuclear systems of non--strange and strange baryons.
On the one hand, studies on the production
mechanism and the structure of Hypernuclei \cite{tamu,vare07} are of interest
since they provide indications on the
Y--N and Y--Y strong interactions which cannot be
determined from the difficult scattering experiments. Hypernuclear (strangeness--violating)
weak decay is on the other hand the only available tool to acquire
knowledge on strangeness--changing weak baryon interactions \cite{AG02}.
The mentioned strong and weak interactions,
whose determination requires the solution of complex many--body problems,
have a direct connection with astrophysics: they
are important inputs when investigating the composition and
macroscopic properties (masses and radii) of compact stars, their
thermal evolution and stability \cite{Sch10}. Moreover, interactions involving 
hyperons are also relevant in the physics of heavy--ion collisions, 
whose main purpose is the study of the nuclear equation of state, the
possible phase transition from hadronic matter to a quark--gluon plasma
and the modification of hadron properties in dense strong interacting matter.

Among Hypernuclei, the most known and long studied are $\Lambda$--Hypernuclei, in which a
$\Lambda$ hyperon replaces a nucleon of the nucleus; they are indicated by the
suffix $\Lambda$ preceding the usual notation of nuclei: $^{A}_{\Lambda}Z$.
$\Lambda$--Hypernuclei are stable at the nuclear time scale (10$^{-23}$ s) since the
$\Lambda$ particle, the lightest of the hyperons, maintains its identity even if embedded
in a system of  other nucleons, the only strong interaction which conserves strangeness
and is allowed being, in fact, $\Lambda N$ scattering.
Similarly, two $\Lambda$'s may stick to a nuclear core, forming the so--called
double $\Lambda$--Hypernuclei, indicated by the symbol $^{A}_{\Lambda \Lambda}Z$.
$\Sigma$--Hypernuclei do not exist, at least as nuclear
systems surviving for times longer than 10$^{-23}$ s. The reason is that the
strong $\Sigma N \rightarrow \Lambda N$ conversion reaction occurring in nuclei prevents the
observation of narrow $\Sigma$ nuclear states as for $\Lambda$--Hypernuclei.
A similar situation occurs for $\Xi^{-}$--Hypernuclei, 
whose importance lies in the fact that the strong conversion 
reaction $\Xi^{-}\ p \rightarrow \Lambda \Lambda$ in nuclei may be used as a 
breeder for the formation of $\Lambda \Lambda$--Hypernuclei. 
We may also think of $\Omega^{-}$--Hypernuclei, possibly
breeders of $3\Lambda$--Hypernuclei through a sequence of the two strong conversion reactions:
$\Omega^{-} n \rightarrow \Lambda \Xi^{-}$,  $\Xi^{-} p \rightarrow \Lambda \Lambda$
occurring in the same nucleus.

A series of experiments (BNL, KEK, JLab, FINUDA)  and calculations
\cite{Mi07} clearly showed the hyperon single--particle structure
of light to heavy $\Lambda$--Hypernuclei (from $^5_\Lambda$He to $^{208}_\Lambda$Pb)
and a substantial part of the energy level structure (spin--doublets, spin--orbit separations,
etc) of $s$-- and $p$--shell $\Lambda$--Hypernuclei. This enabled us to obtain
important information on the $\Lambda$--nucleus mean potential and
on the spin--dependence of the $\Lambda N$
interaction. The next generation of high--precision spectroscopy experiments
of $\Lambda$--Hypernuclei is under preparation at J--PARC and JLab.

Analyses \cite{Ko04,Da08} of $\Sigma$ formation spectra in the $(K^-,\pi^\pm)$ \cite{Ba99}
and $(\pi^+,K^+)$ reactions \cite{No02,Sa04} showed that the $\Sigma$--nucleus
potential has a substantial isospin--dependence and, with the exception of very light
systems, is repulsive: ${\rm Re}\, V_\Sigma(\rho_{0})$ $\sim +(10$-$50$) MeV, $\rho_0=0.17$ fm$^{-3}$ 
being the normal density of nuclear matter.
The $\Sigma N\to \Lambda N$ conversion width is theoretically evaluated to be of 
about 30 MeV for a $\Sigma$ in nuclear matter at normal density 
(see for instance the $\Sigma N$ G--matrix calculation of Ref.~\cite{Ya10}).
We note that repulsive values for the $\Sigma$--nucleus potential can be obtained 
theoretically \cite{Ya10} only in calculations with the most recent Nijmegen $YN$ potentials, ESC08, 
\cite{Ri10b} or with quark--cluster models \cite{Fu07}, thanks to the 
particular modeling of the short--range part of the $\Lambda N$ interaction provided 
by these approaches. We note that the construction of the repulsive core part of the
ESC08 potentials was guided by quark--cluster model studies. 
The only quasibound state of a $\Sigma$ in a nucleus has been
observed in $^4_\Sigma$He. By using an isospin decomposition of the
hyperon--nucleus potential, one has
$V_\Sigma=V^0_\Sigma+V^1_\Sigma {\vec T_A}\cdot {\vec t_\Sigma}/A$, 
$V^0_\Sigma$ and $V^1_\Sigma$ being the isoscalar and isovector potentials,
respectively. Phenomenologically, one has positive values for both
${\rm Re}\, V^0_\Sigma$ and ${\rm Re}\, V^1_\Sigma$, which produces the
above--mentioned repulsion.
However, for a nuclear core with $N<Z$ and small $A$, the isovector term
produces a significant attraction and can bind the $\Sigma$ in $^4_\Sigma$He.

Despite their implications on the possible existence of dibaryon states and
multi--strangeness Hypernuclei and on the study of compact stars, 
much less is known for strangeness $-2$ Hypernuclei
\cite{SI08,SI08b}. We have very little phenomenological information on the $\Xi$--nucleus potential
$V_\Xi$: fits performed on an inclusive $(K ^-,K^+)$ production experiment \cite{Fu98} provided
a shallow potential, ${\rm Re}\,V_\Xi(\rho_0) \sim -14$ MeV. This experiment is interpreted as the 
formation of a $\Xi^-$ bound state in $^{12}_{\Xi^-}$Be. Coulomb--assisted $\Xi^-$ bound states
are expected for medium and heavy Hypernuclei. In calculations based on
$\Xi N$ NSC89 and NSC97 Nijmegen potentials, the $\Xi$--nucleus potential turns out
to be repulsive; only the recent ESC04 and ESC08 models were constructed to provide
an attractive $\Xi$--nucleus potential \cite{Ri10b}. 
The existence of the strong $\Xi^-p \to \Lambda \Lambda$ reaction
makes $\Xi$--Hypernuclei unstable with respect to the strong interaction.
According to $\Xi N$ G--matrix calculations in nuclear matter at normal density \cite{Ya10}, 
the width of the $\Xi^-p \to \Lambda \Lambda$ conversion reaction is of about
3-8 MeV for the NSC08 interaction model but can be as large as 15 MeV for the NSC04 
model, thus making it difficult the experimental investigation of the level structure
of $\Xi$--Hypernuclei in the latter case.  
Recently, an experiment to study the spectroscopy of $^{12}_{\Xi^-}$Be via the $(K ^-,K^+)$ 
reaction on $^{12}$C target has been approved at J--PARC \cite{E05} as a first priority. 
Valuable information on the S=-2 baryon-baryon interaction will be obtained with a high 
resolution magnetic spectrometer; the first experimental results will also stimulate further 
theoretical work.  

Only a few $\Lambda \Lambda$--Hypernuclei events have been studied experimentally up
to date. In recent KEK experiments, $^4_{\Lambda \Lambda}$H, $^6_{\Lambda \Lambda}$He
and $^{10}_{\Lambda \Lambda}$Be have been identified, while less unambiguous 
events were recorded for $^6_{\Lambda \Lambda}$He and $^{10}_{\Lambda \Lambda}$Be
in the 60's and for $^{13}_{\Lambda \Lambda}$B in the early 90's \cite{Nak10}. 
The recent observations imply a weak and attractive $\Lambda \Lambda$ interaction, 
i.e., a bond energy $\Delta B_{\Lambda \Lambda}(^6_{\Lambda \Lambda}{\rm He})\equiv
B_{\Lambda \Lambda}(^6_{\Lambda \Lambda}{\rm He})
-2B_\Lambda(^5_{\Lambda}{\rm He})= (0.67 \pm 0.16)$ MeV for the so--called 
NAGARA event \cite{Tak01}. However, this indication may not be conclusive, 
as it has not yet been possible to interpret with certainty all the old events 
consistently with the NAGARA event.
Theoretical approaches to the structure of strangeness $-2$ nuclear systems have been
proposed using various models. Small and attractive $\Lambda \Lambda$ bond energies 
can be obtained by the NSC89, NSC97 and ESC04 and ESC08 Nijmegen $YY$ models 
in three--body $2\Lambda+{\rm core}$ calculations based on a $\Lambda \Lambda$ 
and $\Lambda\ N$ 
G--matrix interaction \cite{Ri10b,Ya10}. However, the microscopic calculation of 
Ref.~\cite{Sc10}, based on realistic NSC97 $\Lambda N$ and $\Lambda \Lambda$ interactions,
provided values of the bond energy which are smaller than the experimental value for
the NAGARA event.
%
Future experiments on strangeness $-2$ Hypernuclei will be carried out
at J--PARC and FAIR (PANDA Collaboration).

We note that information derived from Hypernuclear structure have important implication in
the study of compact astrophysical objects. The composition and equation of state
of supernovae and neutron star cores sensitively depend on the hyperon content and
are poorly known at present: these properties are mainly controlled by the depths 
of the hyperon--nucleus mean field potentials $V_Y(\rho)$ at high densities $\rho$. 
Various scenarios can be envisaged which correspond to different hyperon compositions and 
influence observable properties such as the mass--radius relation, the maximum mass 
and the temperature of neutron stars. Hyperons and hyperon interactions are also important in the 
study of failed supernovae. These stars evolve by first collapsing into proto--neutrons 
stars, which have very high density and temperature. The appearance of hyperons in failed 
supernovae makes the equation of state of the proto--neutron star softer and triggers a 
second collapse which finally ends with the formation of a black hole \cite{Su10}. The 
black hole formation can be studied by detecting neutrino bursts on Earth
originated in the failed supernova collapse. The duration of neutrino bursts becomes shorter in 
the presence of hyperons: this could allow one to probe the hyperon content in failed 
supernovae.

Concerning the weak decay, for $\Lambda$--Hypernuclei one has the mesonic mode,
$\Lambda\to \pi N$, which resembles what happens to the $\Lambda$ in the free space,
and the so--called non--mesonic mode, with channels $\Lambda N\to nN$, $\Lambda NN\to nNN$,
etc, which can only occur in nuclear systems. Nowadays, a good agreement has been
reached between the theoretical description and the experimental data on the mesonic
decay of $\Lambda$--Hypernuclei.
Because of the Pauli blocking on the final nucleon,
the mesonic decay is disfavored for medium and (especially) heavy Hypernuclei
(say from $^{12}_\Lambda$C).
Innovative experiments (KEK, FINUDA) on non--mesonic
decay of $\Lambda$--Hypernuclei, which dominates for medium and heavy
Hypernuclei, have been performed in recent times \cite{kang,mjkim,mkim,fnd_nmwd,nmwd_n}.
These advances were accompanied by the advent of elaborated theoretical models
(a variety of both finite nucleus and nuclear matter approaches)
and allowed us to reach a reasonable agreement between data and predictions
for the non--mesonic rate $\Gamma_{\rm NM}$, the ratio $\Gamma_n/\Gamma_p\equiv
\Gamma(\Lambda n\to nn)/\Gamma(\Lambda p\to np)$ between the neutron and the proton--induced 
decay rates and the intrinsic
asymmetry parameter $a_\Lambda$ for the decay of polarized Hypernuclei.
However, discrepancies between theory and experiment
(and in some case even between different experiments) are still
present for the emission spectra involving protons \cite{BGPR,BG-PRC}.
This could signal an imperfect implementation of final state interaction effects
in the theoretical and/or experimental analyses.
Furthermore, it is still unclear the r$\hat{\mathrm{o}}$le played in the non--mesonic decay
by the $\Delta$--baryon resonance and
by possible violations of the $\Delta I=1/2$ rule on the isospin change.
New experiments on non--mesonic decay will be carried out at J--PARC, GSI and FAIR (HypHI 
Collaboration), while new analyses are expected also from FINUDA.

No data is available on the $\Lambda$--induced $\Lambda$ decays 
which are possible in $\Lambda \Lambda$--Hypernuclei:
$\Lambda \Lambda\to \Lambda n$, $\Lambda \Lambda\to \Sigma^- p$, $\Lambda \Lambda\to \Sigma^0 n$ 
($\Delta S=1$) and $\Lambda \Lambda\to n n$ ($\Delta S=2$),
apart from the claim for the observation of a single event at KEK~\cite{Wa07}.
Only a couple of predictions (in disagreement with each other) are available  
for such interesting strangeness--chan-ging processes \cite{It01,Pa02}.
Realistic calculation of these decays could also provide important
hints on the possible existence of the $H$--dibaryon.
Moreover, a reliable calculation will be important in 
the design of future experiments at J--PARC and FAIR (PANDA Collaboration), 
where the corresponding processes could be observed for the first time.

Finally, we note that also Hypernuclear non--mesonic weak decay has a relevant
impact on the physics of dense stars. The equilibrium between $NN \to \Lambda N$ and $\Lambda N \to NN$
non--mesonic weak processes is expected
to be extremely important for explaining the stability of rotating neutron stars 
with respect to the emission of gravitational waves \cite{Sch10}. 
The fundamental property which controls the stability of a neutron star is the 
viscosity, which dominantly depends on non--mesonic weak processes involving hyperons.
Other related hyperon--induced weak interactions (the so--called Urca processes) 
provide a relevant contribution to the cooling mechanism of neutron stars.
The occurrence of hyperon superfluidity is indispensable for this hyperon cooling scenario 
to be consistent with the observations of cold neutron stars.
In particular, the weakly attractive $\Lambda \Lambda$ interaction derived from the
mentioned NAGARA event turned out to affect very much the hyperon cooling 
scenario \cite{Ta06}.

\subsection{Production of Hypernuclei and Experimental Facilities}
\label{prod}
\subsubsection{The most used reactions for production of Hypernuclei}
\label{basics-prod}

The first observation of an Hypernucleus (or more precisely an Hyperfragment) is due to Danysz and Pniewski in 1953 \cite{danysz}, by analyzing the events 
recorded by a stack of photographic emulsions exposed to the cosmic radiation at about 26 km from the Earth surface with a balloon. 
Up to early '70 visualizing techniques (photographic emulsions, but also bubble chambers filled with $^{4}$He or heavy liquids) were the experimental tool 
used to study the properties of Hypernuclei.
Cosmic rays and afterward K$^{-}$ beams were employed for producing Hypernuclei. Notwithstanding the inherent difficulty of this technique (low statistics), 
the basic features of Hypernuclear Physics and more generally of Strangeness Nuclear Physics (production, decay modes, inclusive spectra of charged particles) 
were assessed. 20 Hypernuclear species were unambiguously identified by the kinematic analysis of the disintegration star. 
In particular measurements were performed of the $\Lambda$ binding energy, B$_{\Lambda}$, of the $^{A}_{\Lambda}Z$ Hypernuclear ground state, defined as:
\begin{equation}
B_{\Lambda}  =  M_{core} +  M_{\Lambda} -  M_{hyp}
\label{bl}
\end{equation}                                                          
where $M_{core}$ is the mass of the $^{(A-1)}$Z nucleus well known from Nuclear Physics, M$_{\Lambda}$ is the $\Lambda$ Hyperon mass and M$_{hyp}$ 
is the measured $^{A}_{\Lambda}Z$ Hypernuclear mass. It was found that B$_{\Lambda}$ varies linearly with A, with a slope of around 1 MeV/ (unit of A), 
saturating at about 27 MeV for heavy Hypernuclei. This result is the most important from the emulsion era and even today, in spite of the tremendous experimental 
effort on the magnetic spectrometers that we will discuss later, the B$_{\Lambda}$ from the emulsion technique are the most precise and serve as a reference, 
due to their extraordinary precision.
 
\begin{figure*}
\begin{center}
\resizebox{0.65\textwidth}{!}{%
  \includegraphics{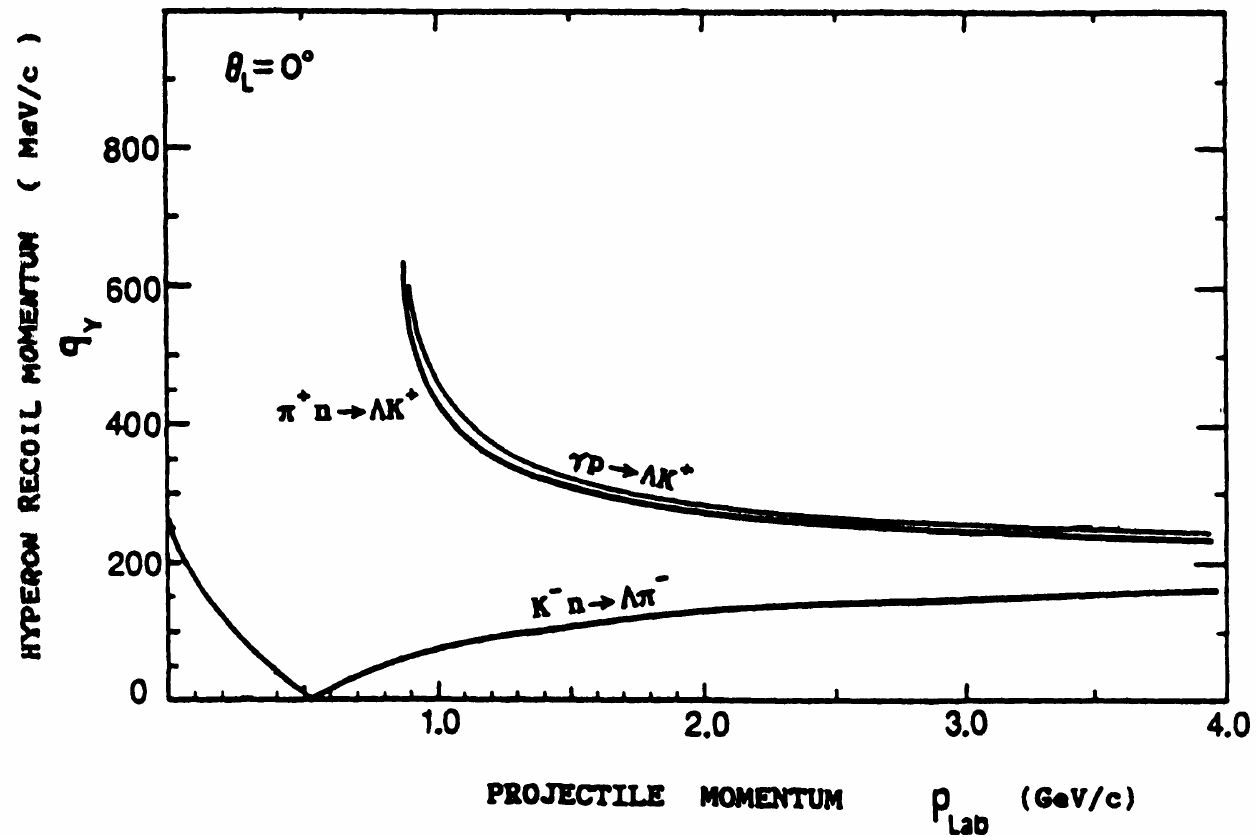}
}
\end{center}
\caption{Kinematics of the reactions (\protect{\ref{eq:eq1}}),
(\protect{\ref{eq:eq2}}) and photoproduction (\protect{\ref{eq:eq4}}) for forward $\Lambda$ emission ($\theta_{L}$=0$^{\circ}$).
The momentum transferred to the hyperon is plotted as a function of the incoming particle ($K^{-}$,$\pi^{+}$ or $\gamma$) in the Laboratory system. 
From Ref.~\cite{vare07}.}
\label{fig:fig1}
\end{figure*}
        
From 1972 two--body reactions producing $\Lambda$ on a nuclear target were studied. The two--body reactions that led to practically all the present 
bulk of experimental information are the following. 
\begin{enumerate}
\item{} The  ``Strangeness Exchange'' reaction:
\begin{equation}
K^{-} + {\cal{N}} \rightarrow \Lambda +\pi
\label{eq:eq1}
\end{equation}
exploited mainly in the $K^{-} + n \rightarrow \Lambda +\pi^{-}$ charge state, for evident reasons of easiness in spectroscopizing the $\pi$ final state. 
The reaction can be seen as a transfer of the $s$--quark from the incident meson to the struck baryon.
\item{}  The ``Associated Production'' reaction:
\begin{equation}
\pi^{+} + n \rightarrow \Lambda +K^{+}.
\label{eq:eq2}
\end{equation}
This reaction proceeds by the creation of a  $(s\bar{s})$ pair by the incident meson.
\item{} The electroproduction of strangeness on protons in the very forward direction:
\begin{equation}
e   +   p \rightarrow e' + \Lambda +K^{+}                                                                       
\label{eq:eq3}
\end{equation}
exploited quite recently. The virtual photons associated to the reaction (\ref{eq:eq3}) can be 
regarded as quasi--real and reaction (\ref{eq:eq3}) is often rewritten 
as a two--body photoproduction reaction:
\begin{equation}
\gamma + p \rightarrow \Lambda +K^{+}                                                                                
\label{eq:eq4}
\end{equation}
\end{enumerate}

It must be noted that whereas with reactions (\ref{eq:eq1}) and (\ref{eq:eq2}) it is possible to replace a neutron in the target nucleus by a $\Lambda$, 
with reaction (\ref{eq:eq3}) a proton is replaced by a $\Lambda$, and the neutron--rich mirror Hypernucleus is obtained out from the same nuclear 
target. Just for example, with reactions (\ref{eq:eq1}) and (\ref{eq:eq2}) on a $^{12}$C target 
$^{12}_{\Lambda}$C is obtained, while with reaction (\ref{eq:eq3}) $^{12}_{\Lambda}$B is produced. 

Each one of the aforesaid reactions has its own characteristics in the elementary cross section, internal quantum numbers transfer, momentum 
transfer, 
absorption of incoming and outgoing particles in nuclear matter. The most important parameter in determining the selectivity of the different reactions is 
the momentum transfer. Figure \ref{fig:fig1} shows the momentum transferred to the $\Lambda$ hyperon, $q_{Y}$, as a 
function of the momentum of the projectile in the laboratory frame, $p_{Lab}$, for reactions 
(\ref{eq:eq1}), (\ref{eq:eq2}) and (\ref{eq:eq3}) at $\theta_{Lab}=0^{\circ}$. 
A striking kinematics difference appears: for
(\ref{eq:eq1}), which is exoenergetic with a Q value of $\sim$178 MeV,  there is a value of $p_{Lab}$ (505 MeV/$c$), usually referred to as 
``magic momentum'' \cite{fesh} for which  $q_{\Lambda}$ vanishes and recoiless production takes place; for (\ref{eq:eq2}), which is endoenergetic 
with a Q value of $\sim -$ 530 MeV, $q_{\Lambda}$ decreases monotonically with $p_{Lab}$, staying 
always at values exceeding 200 MeV/$c$. 
A further degree of freedom in selecting  $q_{\Lambda}$  is given by the detection angle  of the produced meson. It is important to notice that
$q_{\Lambda}$  for (\ref{eq:eq1}) with $K^{-}$ at rest is not very different from that for (\ref{eq:eq2}).

\hspace{3mm} In order to get insight on how the features of the elementary  reaction affect the production of Hypernuclei in well defined states the impulse 
approximation  may be used. The two--body reaction $t$--matrix in the nuclear medi-um is replaced by the free space $t$--matrix at the same 
incident  momentum and the differential cross section  for the reaction:
\begin{equation}
\left.
\begin{array}{rr}
K^- \\
\pi^+
\end{array} \right\}
 + ^A{\mathrm Z} \rightarrow ^A_{\Lambda}
{\mathrm Z} + \left\{
\begin{array}{ll}
\pi^- \\
K^+
\end{array}
\right.
\label{eq:eq5}
\end{equation}
may be written as

\begin{equation}
d\sigma (\theta)/d\Omega_{L} = \xi \left[ d\sigma
(\theta)/d\Omega_{L} \right]_{\mathrm{free}} N_{\mathrm{eff}} (i \rightarrow
f, \, \theta).
\label{eq:eq6}
\end{equation}

In (\ref{eq:eq6}) $d\sigma (\theta)/d\Omega_{L}$ is the Lab. cross section for the production of the Hypernucleus
$^{A}_{\Lambda}{\mathrm Z}$ in a given final state $f$, $\xi$ a kinematic factor arising from two--body to many--body 
frame transformation, $\left[ d\sigma (\theta)/d\Omega_{L} \right]_{\mathrm{free}}$ the cross section for the
elementary (or free) reactions (\ref{eq:eq1}) and (\ref{eq:eq2}) and $N_{\mathrm {eff}}(i\rightarrow f,\, \theta)$
the so--called ``effective nucleon number''.

Eq.~(\ref{eq:eq6}), simple and easy to understand, was used by Bonazzola {\it et al.}  \cite{bona}
for the first time to describe the production of Hypernuclei by the $(K^{-}, \, \pi^{-})$ reaction at 390 MeV/c. 
All the complications and difficulties related to the many--body strong interaction system are obviously  contained in 
the term $N_{\mathrm{eff}}(i \rightarrow f, \, \theta)$ (the Nuclear Physics ``black box'').
In the simplest plane wave approximation (PWA) $N_{\mathrm{eff}}(i \rightarrow f, \, \theta)$ can be written as
\begin{equation}
N_{\mathrm{eff}}(i \rightarrow f, \, \theta)=(
{\mathrm{Clebsh-Gordan\ coefficients}}) \times F(q),
\label{eq:eq5a}
\end{equation}
in which $F(q)$ is a form factor, related to q by simple relationships, at least in the frame of the independent particle nuclear model. 

PWA is only a rough approximation for describing Hypernuclei production reactions and is unable to provide reliable values of
the cross sections. A better insight is given by the Distorted Wave Impulse Approximation (DWIA). The $K$ and $\pi$ distorted waves are calculated
separately by solving the Klein--Gordon equation with the use of appropriate optical potentials, conventionally taken proportional to the nucleon 
density. 
DWIA calculations are quite successful in describing the features of the production of Hypernuclei by meson beams. A comprehensive and updated 
account on this subject can be found in Ref.~\cite{moto_ito_yama}. 

Similar theoretical considerations may be applied also for describing the electroproduction of Hypernuclei by me-ans of the elementary reaction (\ref{eq:eq3}), 
even though the relevant expressions are somehow complicated by the three--body final state kinematics and by the dynamics described by the 
electromagnetic interaction. The relevant relationships can be found, e.g., in Ref.~\cite{tamu} and in the recent paper of Ref.~\cite{moto_bydz}.

From the above considerations and from figure \ref{fig:fig1} it appears that, with the 
(K$^{-}$, $\pi^{-}$) in--flight reaction, substitutional states of the 
Hypernucleus, obtained by converting a neutron of the target nucleus into a $\Lambda$ hyperon in the same orbit and spin state with no orbital angular 
momentum transfer, are preferentially populated. On the other hand, the ($\pi^{+}$, K$^{+}$) and (e, e' K$^{+}$) reactions and also the (K$^{-}$, $\pi^{-}$) 
reaction at rest transfer a significant recoil momentum (200$\div$300 MeV/c) to the Hypernucleus, and then several high--spin Hypernuclear states are 
preferentially populated. 

This feature is of paramount importance for Hypernuclear Spectroscopy studies (see Sec.~\ref{spect}) and also for investigations on the Weak Decay, which occurs 
mainly for Hypernuclei in the ground state (see Sec.~\ref{sec:decay}). 
          Furthermore, since the spin--flip amplitudes in reactions (\ref{eq:eq1}) and (\ref{eq:eq2}) are generally small, they populate mainly non 
spin-flip states of Hypernuclei. On the contrary,  since the amplitude for reaction (\ref{eq:eq3}) has a sizable spin--flip component even at zero 
degrees due to the spin 1 of the photon, both spin--flip and spin--non--flip final Hypernuclear states may be populated in electroproduction. 
           From the above considerations it appears that in principle we have at disposal all the tools needed for a complete study of all the Hypernuclei 
we wish to investigate. In practice the experimenters have to face many other constraints in designing their detectors. They are:\\
a) the values of the  cross sections for producing Hypernuclei with reactions (\ref{eq:eq1}), (\ref{eq:eq2}) and (\ref{eq:eq3});\\
b) the presence of physical backgrounds in which the signals due to Hypernuclei formation could be blurred;\\
c) the intensity and energy resolution of the beams of projectiles (K$^{-}$,$\pi^{+}$, e);\\
d) the availability of the above beams at the different Laboratories, in competition with other experiments in Nuclear and Particle Physics.\\
A simple answer to question (a) is given by Fig.~2. 
\begin{figure}
\begin{center}
\resizebox{0.45\textwidth}{!}{%
  \includegraphics{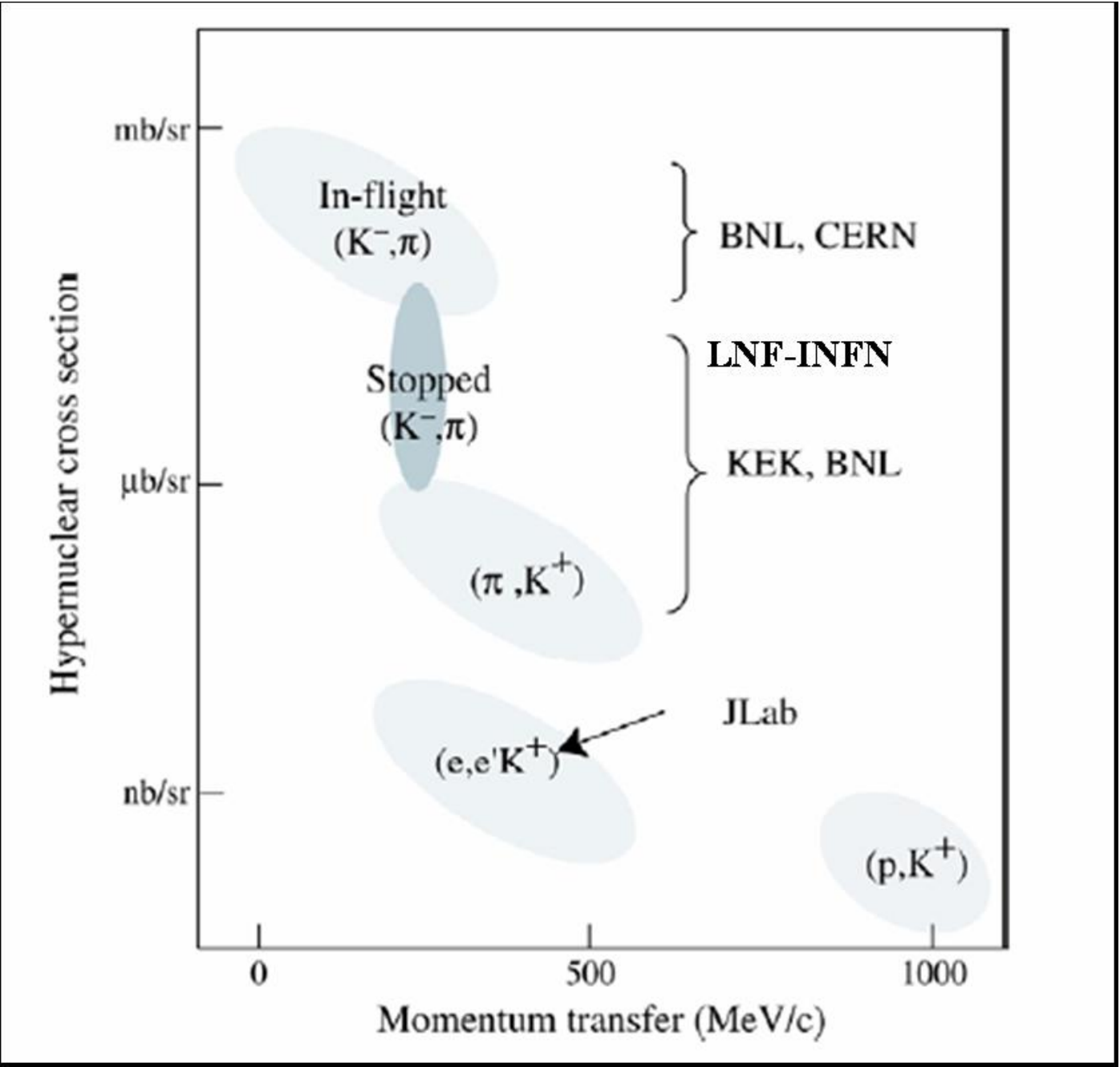}
}
\end{center}
\caption{Hypernuclear production cross section for typical reactions versus momentum transfer. From Ref.~\cite{vare07}.}
\label{fig:fig2}
\end{figure}
It must be noted that the production of Hypernuclei by means of the ($K^{-}$,$\pi$) reaction with $K^{-}$ at rest is reported too on a cross section 
scale, which is formally wrong since the production by particles at rest is defined by a capture rate and not by a cross section \cite{vare07}. 
Unfortunately the intensities and the qualities (point (c)) of the beams scale inversely to the values of the production cross sections. 
The electron beams were always excellent concerning the intensity and the energy resolution. Unfortunately in the past years the duty cycle of the machines 
was very low, and the smallness of the cross section joined to the need of measuring the e' and $K^{+}$ from (\ref{eq:eq3}) in coincidence did not allow to study the 
production of Hypernuclei. 
Only the advent of CEBAF, a machine with an excellent duty cycle, allowed the start up of experiments on Hypernuclei.  
The best projectiles for producing Hypernuclei, and in general Strange Nuclei, are $K^{-}$. Unfortunately the intensity of such secondary beams is quite low, 
and even worse is the quality (contamination by $\pi^{-}$, energy spread). For these reasons the magnetic spectrometers that were designed were quite 
complicated and the running times needed to obtain adequate statistics long. The powerful complex of machines J--PARC, that was recently put into 
operation and for which Strange Nuclei and Atoms Physics will be among the top priorities, will allow a real breakthrough in this field. 
Production of Hypernuclei by pion beams is midway concerning both the cross sections and the quality of the beams. For these reasons experiments 
using reaction (\ref{eq:eq2}) performed mostly at KEK with the SKS spectrometer produced a great part of experimental information in the last ten years by 
means of advanced state--of--art technologies at an old machine, the KEK 12 GeV ProtonSynchrotron. An overview of  the measurements done with the SKS 
can be found in Refs.~\cite{tamu}.

A special discussion must be given in relation to the use of the ($K^{-}$, $\pi^{-}$) reaction at rest.  Looking at Figures \ref{fig:fig1} and \ref{fig:fig2} this method seems 
easier to exploit, since no experimental information on the momentum of the incoming particle is needed. The main reason it has not been so widely 
used stems in point (b). Taking again as an example the production of $^{12}_\Lambda$C from a $^{12}$C target through reaction (\ref{eq:eq1}) with $K^{-}$ at rest:
\begin{equation}
K^{-}_{\rm stop} + ^{12}C \rightarrow ^{12}_\Lambda C  + \pi^{-}                                         
\label{eq:eq6a}
\end{equation}
at first sight one could think that the simple spectroscopy of the $\pi^{-}$ from (\ref{eq:eq6a}) should be enough to get information on the excitation spectrum 
of $^{12}_\Lambda$C. The fastest $\pi^{-}$ should be those related to the formation of $^{12}_\Lambda$C in the ground state. Unfortunately other processes 
following the capture of $K^{-}$ at rest produce $\pi^{-}$ with momenta even larger than those from reactions like (\ref{eq:eq6a}). The most dangerous for 
Hypernuclear spectroscopy is the capture of a $K^{-}$ by a correlated (np) pair in the nucleus:
\begin{equation}
K^{-} + (np)\rightarrow \Sigma^{-} +  p    
\label{eq:eq7}
\end{equation}                                                                         
followed by the decay in flight of the $\Sigma^{-}$  into $\pi^{-}$+n. This process leads to the production of $\pi^{-}$ with a flat momentum spectrum that extends  
beyond the kinematics limit due to the two--body production of Hypernuclei. This drawback was clearly understood in the first series of experiments 
on Hypernuclei with $K^{-}$ at rest, done at KEK ~\cite{tamura}, and the background reasonably modeled by accurate simulations \cite{haya}. 
This circumstance generated ambiguity on the subsequent experiments with $K^{-}$ at rest, described in the following.

\par
After the visualizing techniques era, in the early '70s the study of Hypernuclei was started by means of electronic techniques (magnetic spectrometers), following 
the progress in the design of new intense $K^{-}$ beams and pioneering the use of the newly developed multiwire proportional chambers \cite{charpa}.
However, the magnetic spectrometers that were used by the different Groups were not designed specifically and optimized for Hypernuclear Physics. 
The first measurement of the $\pi^{-}$ spectra emitted from $K^{-}$ capture at rest by $^{12}$C was performed by means of an apparatus dedicated to the 
measurement of the $K_{e2}$ decay, with some modifications of the target assembly and the trigger \cite{faess}. 
The energy resolution of the Hypernuclear energy levels was 6 MeV FWHM. A similar resolution was attained in the first experiment on the production of 
Hypernuclei with the reaction (\ref{eq:eq1}) in flight, at 390 MeV/c. The apparatus was realized \cite{bonanim} specifically for the study of the production 
of Hypernuclei, but using pre--existing general purpose magnets. The energy resolution on the Hypernuclear energy levels was about 6 MeV FWHM.  

$\gamma$--rays spectra emitted from low--lying excited Hypernuclear states  were too measured from $A=4$ systems in a pilot experiment performed with 
NaI detectors \cite{bambe} at CERN.
A double spectrometer for the study of the ($K^{-}$, $\pi^{-}$) reaction on some nuclear targets at 900 MeV/c was assembled soon afterwards, always at CERN, 
again using pre--existing beam elements and magnets, providing a better resolution (5 MeV) and cleaner spectra with higher statistics \cite{bruck}.

The first magnetic spectrometer which was specifically designed for Hypernuclear Physics, SPES 2, had a quite funny story. It had a quite sophisticated and nice magnetic optics, and was assembled in 1976 and afterwards used at the proton Synchrotron SATURNE in Saclay in an attempt to produce Hypernuclei by  the
 (p,p' K$^{+}$) three body reaction in the forward direction on a nuclear target.
Unfortunately the experiment failed, due to the overwhelming background, but the spectrometer was so 
properly designed that, transported at CERN and installed at 
a K$^{-}$ beam in 1978, produced an excellent amount of data on Hypernuclear Physics by the ($K^{-}$, $\pi^{-}$) reaction at 900 MeV/c \cite{bruck78}. 

A great step of quality in Hypernuclear Physics was made possible by these experiments. Production of excited states was observed for several Hypernuclei, mostly 
in the $p$--shell, and their quantum numbers determined in some cases.
These data constituted the input for several theoretical investigations on the description of the Hypernuclear excited states in terms of microscopic models starting 
from the $\Lambda N$ potential.
The most important result from this series of measurements is the observation that the spin--orbit term 
in the $\Lambda$--nucleus optical potential is consistent with zero. 
This conclusion was recently questioned by recent experiments on heavier Hypernuclei, but the original hypothesis of a very small value seems still to be the more valid. 

Looking at the very promising results obtained at CERN, the BNL management decided to start a strong effort on Hypernuclear Physics. The first step 
was to set--up 
a focusing spectrometer system placed on a rotatable platform, called Moby--Dick. A suitable beam--line was designed to be operated 
in connection with the spectrometer.

From the late '70 BNL produced an impressive set of excellent data, practically on all the experimental aspects of Hypernuclear Physics. The greater amount of 
data was obtained by means of the  $(K^{-},\pi^{-})$ reaction, thanks to the fact that BNL had the better K beams in the world. However, BNL pioneered also to 
exploit the associated production reaction (\ref{eq:eq2}) on nuclei in a very efficient and controlled way. As a matter of fact, the spectra obtained on heavy targets
like $^{89}Y$ showed for the first time, in a very impressive way, the validity of the single particle 
model even for the inner shells \cite{millener}. 
It is difficult to enumerate here all the contributions to Hypernuclear Physics provided by the BNL experiments, without omissions. They can be found, e.g., in the 
review by Chrien \cite{chrien}. 
BNL pioneered practically all the experimental techniques that are the basis for the present and future installations, but did not perform a systematic study.

At the 12 GeV Proton Synchrotron of KEK the experimental effort started with a series of experiments in which the $\pi$ spectra emitted following the capture 
of $K^{-}$ at rest by several nuclei were measured with a magnetic spectrometer \cite{tamura}. 
At the beginning the apparatus was the one used for the study of rare and 
exotic decays of $K^{+}$. In some sense the situation was similar to that one occurred in 1973 for the first CERN experiment, but the results, 
immediate and also far reaching, were completely different. 
Exploiting the better beam quality, the use of dedicated detectors for coincidence measurements and 
also the physics information gained by more than ten years of experimental results with the in--flight reactions, the KEK results showed that  reaction (\ref{eq:eq1}) 
with $K^{-}$ at rest could provide results on Hypernuclear Physics of very high quality, in contrast to the initial perception. 
\begin{figure}[h]
\begin{center}
\resizebox{0.5\textwidth}{!}{%
  \includegraphics{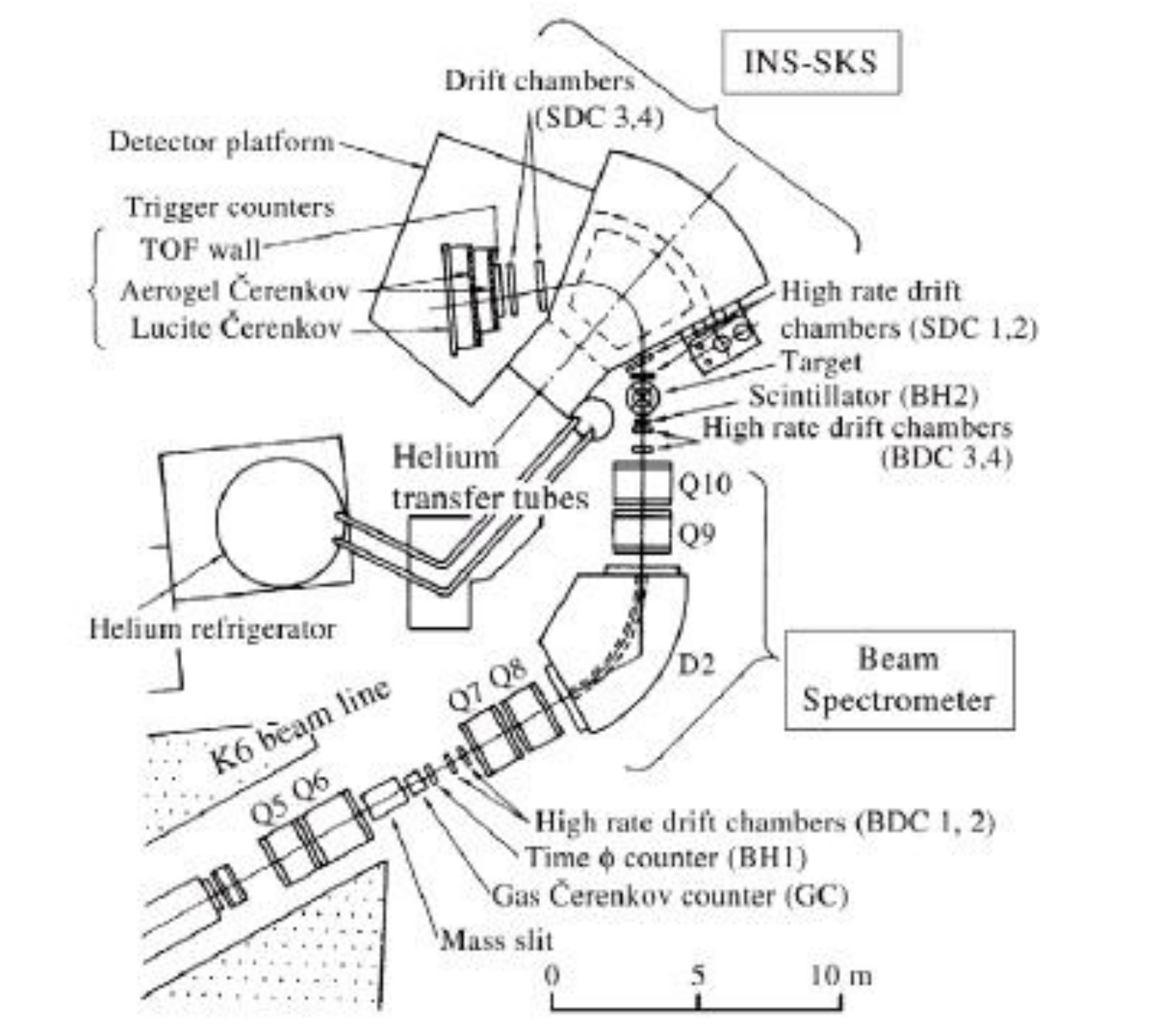}
}
\end{center}
\caption{Schematic drawing of the SKS spectrometer system with the beam line spectrometer. From Ref.~\cite{tamu}.}
\label{fig:fig7}
\end{figure}
The main results from the KEK experiment are summarized in Ref.~\cite{tamura}. 
          
\begin{figure}
\begin{center}
\resizebox{0.5\textwidth}{!}{%
  \includegraphics{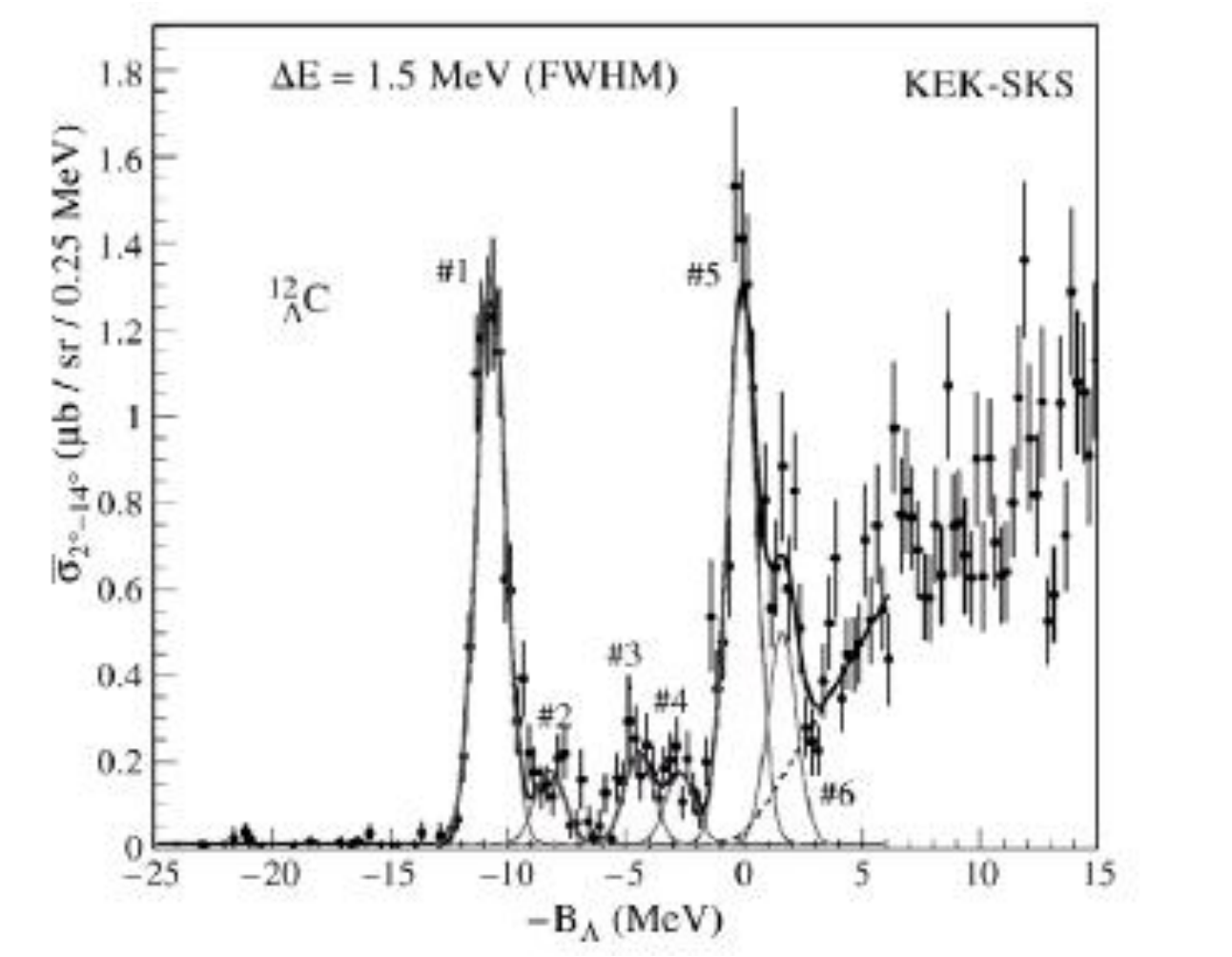}
}
\end{center}
\caption{Missing mass spectrum of the $^{12}$C($\pi^{+}$, K$^{+}$)$^{12}_{\Lambda}$C reaction at the KEK 12 GeV PS using the SKS spectrometer. 
From Ref.~\cite{hot}.}
\label{fig:fig8}
\end{figure}

At the end of the 90's, a powerful detector dedicated to Hypernuclear Physics was approved for installation at KEK. It was based on a large superconducting 
dipole magnet with a pole gap of 50 cm and a maximum field of 3T, optimized to provide a large solid 
angle (100 msr) for the reaction $(\pi^{+},K^{+})$ on nuclei.
A dedicated new beam line, providing $\sim 2 \times 10^{6} \pi^{+}/spill$ at about 1 GeV/c was put into operation. The spectrometer, named SKS (acronym for 
Superconducting Kaon Spectrometer) was equipped with state--of--art drift chambers and detectors for particle identification. Figure \ref{fig:fig7} 
shows a sketch of the detector. The final resolution of the device was about 1.5 MeV on the Hypernucleus energy spectrum as shown in 
Figure \ref{fig:fig8} \cite{hot}.

As a first step, excitation energy spectra were measured for $^{7}_{\Lambda}$Li,  $^{9}_{\Lambda}$Be,  
$^{10}_{\Lambda}$B, $^{12}_{\Lambda}$C, $^{13}_{\Lambda}$C, $^{16}_{\Lambda}$O, $^{28}_{\Lambda}$Si, 
$^{51}_{\Lambda}$V, $^{89}_{\Lambda}$Y, $^{139}_{\Lambda}$La, $^{208}_{\Lambda}$Pb. 
It is the more systematic study ever done, and confirmed the previous result on the physical reality of 
single particle states deeply located in nuclei as heavy as  A=208 systems. This bank of data allowed a 
reliable determination of the parameters of the potential well seen by the $\Lambda$ in a Hypernucleus. 
Furthermore, thanks to the improved energy resolution, a splitting of the peaks corresponding to the  
$\Lambda$ in one shell was observed. The splitting seemingly increased with the angular momentum of the 
observed shell model orbit, asking then for a non--zero value of the spin--orbit term. 

After this series of systematic measurements, the SKS was used to ``prepare'' Hypernuclei in well defined states (ground or excited) and study their decay by 
other arrays of complex detectors. This was possible thanks to the space left free around the targets used in the  $(\pi^{+},K^{+})$ reaction.
The most important of these additional complex detectors was certainly the so--called Hyperball, since it pioneered a new field in Hypernuclear 
Physics, the $\gamma$--rays spectroscopy of excited states, allowing the achievement of an energy resolution  (of the order of keV) better by about 
three orders of magnitude than that reached with the magnetic spectrometers. 
It is important to notice that the device had in any case to be coupled to a magnetic spectrometer in order to prepare, with a resolution of  some MeV, the 
Hypernuclear energy excitation spectrum to be analyzed by means of a further coincidence with the Hyperball Detector. 
Details on the Physics results achieved with this technique will be given in Section \ref{spect}. 
The first Hyperball consisted of fourteen n--type coaxial Ge detectors, arranged in a compact geometry. The Ge detectors were surrounded by BGO 
counters 
for  background suppression (Compton scattering, high energy electromagnetic showers from $\pi^{0}$ and penetration of high energy pions and muons). 
The energy resolution of the detector was $\sim$3 keV FWHM and the total efficiency was 0.38$\%$ for $\gamma$--rays of 1 MeV. 
The system was cooled at LN$_{2}$ temperature and equipped with fast--readout electronics, necessary to operate at high rates. 
The very successful results provided by the first Hyperball led to a continuous improvement of the detector. 
The last one, named Hyperball--J (J stands for J--PARC) is equipped with a mechanical cooling with a refrigerator instead of LN$_{2}$ cooling, 
PWO counters instead than BGO for background suppression (they have much shorter time for light emission) and new readout electronics.

The second, complex detector fully dedicated to Hypernuclear physics, built taking into account the very interesting results obtained by using $K^{-}$ at rest, 
is FINUDA.
FINUDA (acronym for FIsica NUcleare a DA$\Phi$NE) may be considered the more complete detector built up to now and is completely 
different from all the other detectors and spectrometers described up to now. It was installed at an ($e^{+}$,$e^{-}$) collider, DA$\Phi$NE, at Laboratori 
Nazionali di Frascati. Thanks to the large angular coverage for detection of charged 
and neutral particles from the formation and decay of Hypernuclei, many observables (excitation energy spectra, partial decay widths for mesonic, 
non--mesonic decay and multi--nucleon absorption), were measured simultaneously and provided a bulk of data of unprecedented completeness, precision 
and cleanliness. Furthermore, spectra corresponding to different targets were measured simultaneously, avoiding possible systematic errors in 
comparing the properties of different Hypernuclei. 
Under this point of view, FINUDA is the modern electronic experiment approaching more closely the original 
visualizing technique by Danysz and Pniewski.

The idea of performing an experiment on the production and decay of Hypernuclei, which is inherently a fixed target physics, at a 
collider seems not good at first sight. The main decay channel of the $\Phi$--meson, produced nearly at rest in the collision of 510 MeV 
electrons and positrons at a rate of 
$\sim 4.4 \times 10^{2}\ \mathrm{s}^{-1}$ at 
the luminosity ${\cal{L}} = 10^{32}\ \mathrm{cm}^{-2} \mathrm{s}^{-1}$, is $(K^{+}, K^{-})$, BR $\sim 49 \%$. Since the $\Phi$--meson is produced at rest, 
DA$\Phi$NE at LNF is a source of $\sim 2.2 \times 10^{2}\ (K^{+}, K^{-})$ pairs/s, which are collinear, background free, and, very important, 
of very low energy ($\sim 16$ MeV). The low energy of the produced charged kaons is the key advantage for an experiment 
on Hypernuclei production and decay by means of the strangeness--exchange reaction with $K^{-}$ at rest:

\begin{figure}
\begin{center}
\resizebox{0.5\textwidth}{!}{%
  \includegraphics{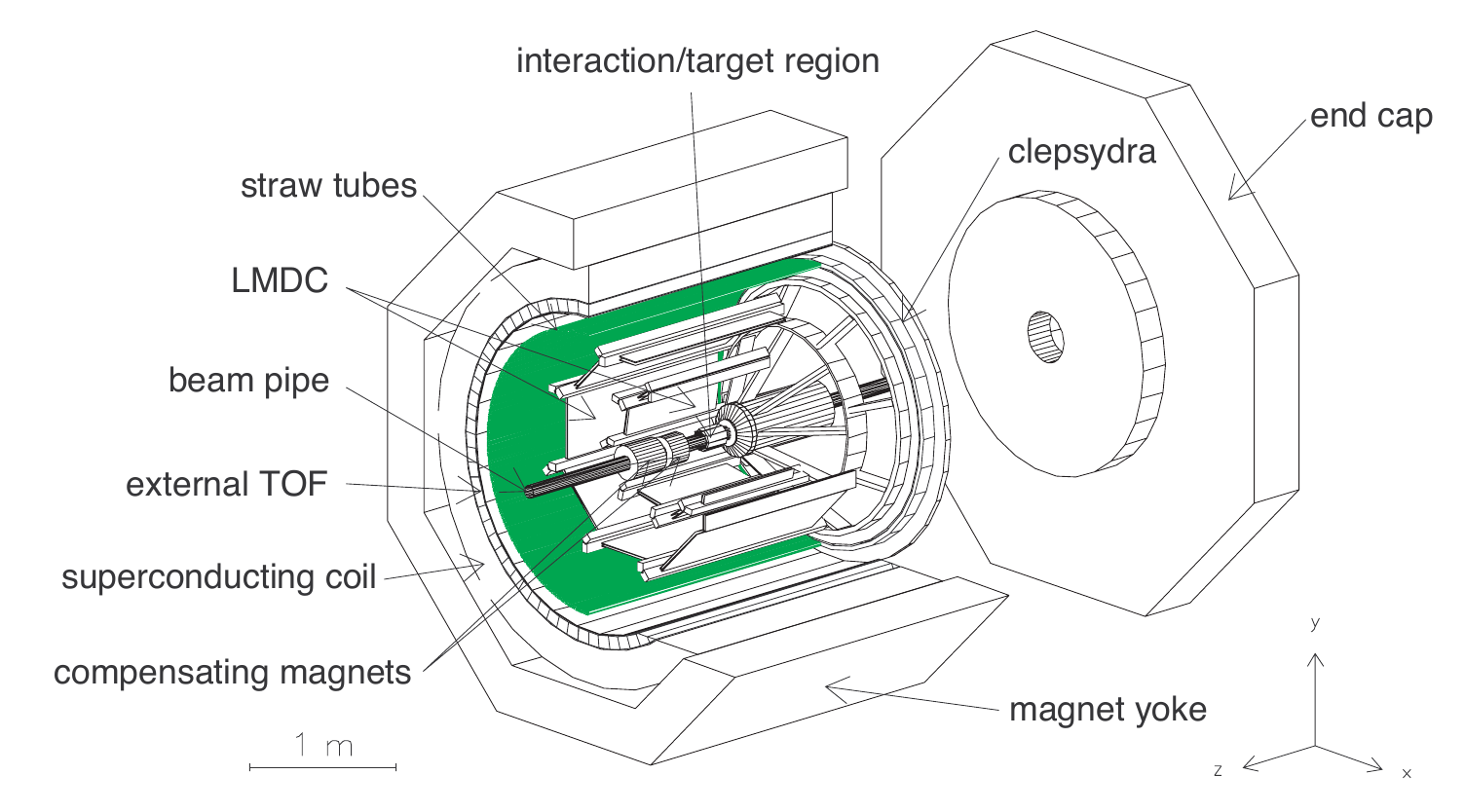}
}
\end{center}
\caption{Global view of the FINUDA detector with the superconducting solenoid and the iron yoke.}
\label{fig:fig9}
\end{figure}

\begin{equation}
K^{-}_{\rm stop} + ^{A}Z \rightarrow ^{A}_{\Lambda}Z + \pi^{-}
\label{eq_uno}
\end{equation}
in which $^{A}Z$ indicates the nuclear target and $^{A}_{\Lambda}Z$ the produced Hypernucleus. The $K^{-}$ can be stopped in very thin 
targets (a few hundreds of mg/cm$^{2}$), in contrast to what happens in experiments with stopped $K^{-}$ at hadron machines, where from 80\% up 
to 90\% of the incident $K^{-}$ beam is lost in the degraders facing the stopping target. Furthermore, the cylindrical geometry of the 
interaction region at a collider allows the construction of cylindrical high resolution spectrometers with solid angles for detecting 
the $\pi^{-}$ from (\ref{eq:eq1}) very much bigger $(> 2\pi\ \mathrm{sr})$ than those at fixed target machines ($\sim 100$ msr). Last but not least, 
the use of thin targets introduces lower cuts on the measurement of the energy spectra of charged particles ($\pi^{-},\ p,\ d$) from the weak decay of 
the Hypernuclei produced by (\ref{eq_uno}) and they are detected by the same arrays of detectors with large solid angles $(> 2\pi\ \mathrm{sr})$. 
These considerations were put forward in 1991~\cite{Lnf} and a proposal was afterwards elaborated~\cite{finrep} and soon accepted 
by the Scientific Committee of LNF. 
Fig.~\ref{fig:fig9} shows a sketch of the detector, immersed in a superconducting solenoid which provides a homogeneous magnetic field of 
1.0 T inside a cylindrical volume of 146 cm radius and 211 cm length.  
We emphasize that different experimental features relevant to different measurements (e.g. momentum resolution for the different particles versus 
the statistical significance of the spectra) may be optimized out of the same data bank by applying with software different selection criteria. 
The approach is then quite different from that followed by SKS, in which different hardware layouts were used in subsequent runs in order to obtain 
excellent results, some of which we will present in the following Sections. 

Let us now emphasize a peculiar feature of FINUDA, not adopted in the detectors so far mentioned. It is the presence of a powerful vertex detector, 
composed by two barrel--shaped arrays of Si Microstrip detectors that allow the separation of the primary interaction vertex of reaction 
(\ref{eq:eq1}) from the secondary vertexes from the decay of the produced Hyperons,
$\Lambda$ and $\Sigma$, to better than 1 mm. For studies of Hypernuclear spectroscopy it is of 
paramount importance to measure the Hypernuclear energy 
spectra minimizing the background, that should forbid the observation of  states produced with a low strength.
As discussed before, the more significant contribution in the bound Hypernuclear states region comes from the absorption of the $K^{-}$ on a $NN$ pair, 
producing a $\Sigma^{\pm}$ and a nucleon. The $\Sigma$ decays in flight into N+$\pi$, and the $\pi$ from this process has a broad flat spectrum  
extending beyond the kinematics limit corresponding to the formation of the Hypernucleus in the ground state ($^{12}_\Lambda$C in the example).  

\begin{figure}
\begin{center}
\resizebox{0.5\textwidth}{!}{%
  \includegraphics{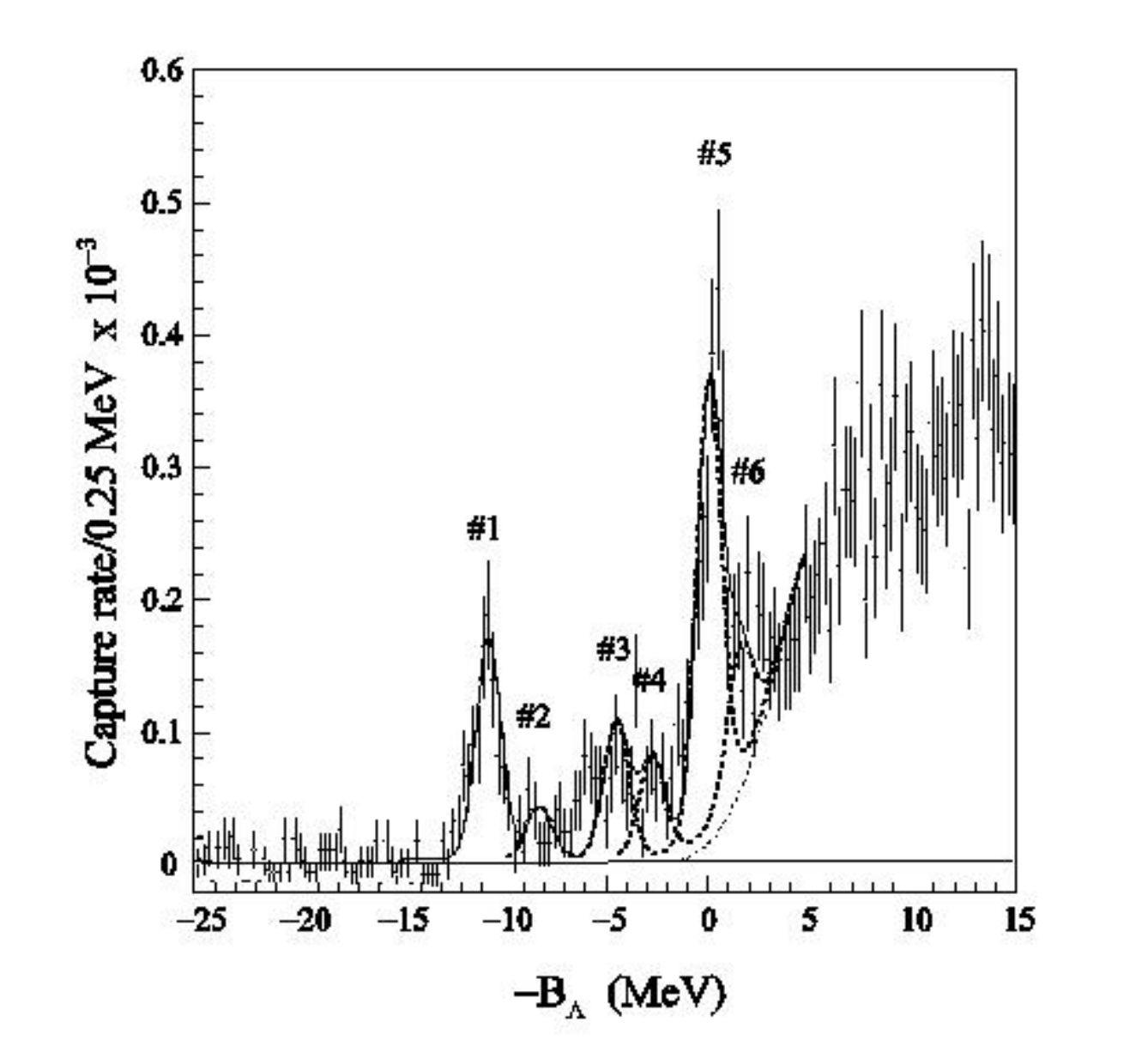}
}
\end{center}
\caption{$\Lambda$--binding energy spectrum of $^{12}_\Lambda$C measured by FINUDA. The solid line represents the best fit obtained with seven Gaussian 
functions. From Ref.~\cite{agnello}.}
\label{fig:fig13}
\end{figure}

Thanks to these features FINUDA provided nice results on  Hypernuclear Spectroscopy featuring the best energy resolution on the Hypernuclear energy 
levels (1.29 MeV FWHM) so far reached with magnetic spectrometers for meson--induced reactions like (\ref{eq:eq1}) and (\ref{eq:eq2}). 
Fig.~\ref{fig:fig13} shows the excitation energy spectrum measured for $^{12}_\Lambda$C ~\cite{agnello}. 

\begin{figure}
\begin{center}
\resizebox{0.5\textwidth}{!}{%
  \includegraphics{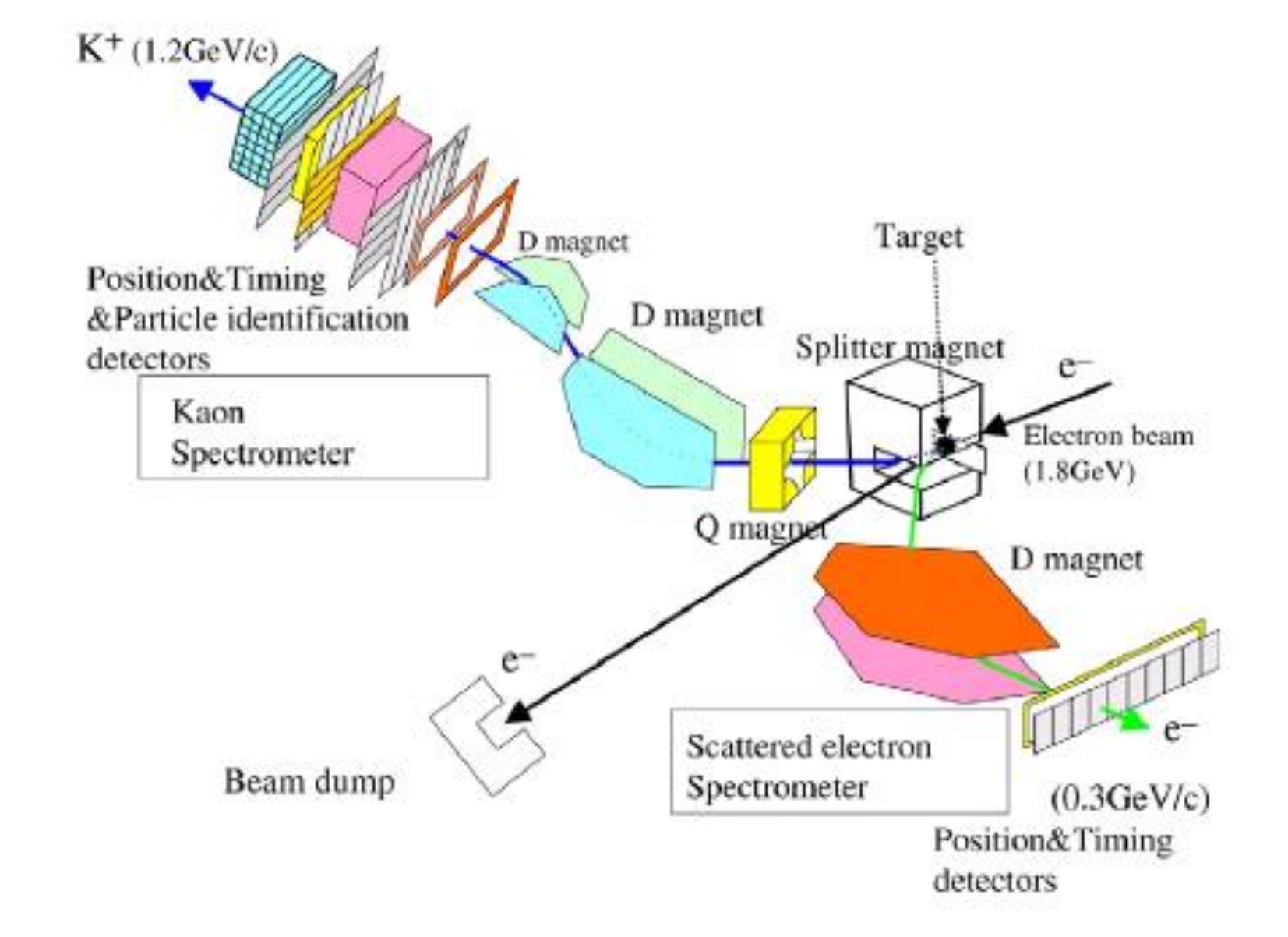}
}
\end{center}
\caption{Hypernuclear spectrometer system in JLab Hall C for E89--009. From Ref.~\cite{tamu}.}
\label{fig:fig14}
\end{figure}

\begin{figure}
\begin{center}
\resizebox{0.5\textwidth}{!}{%
  \includegraphics{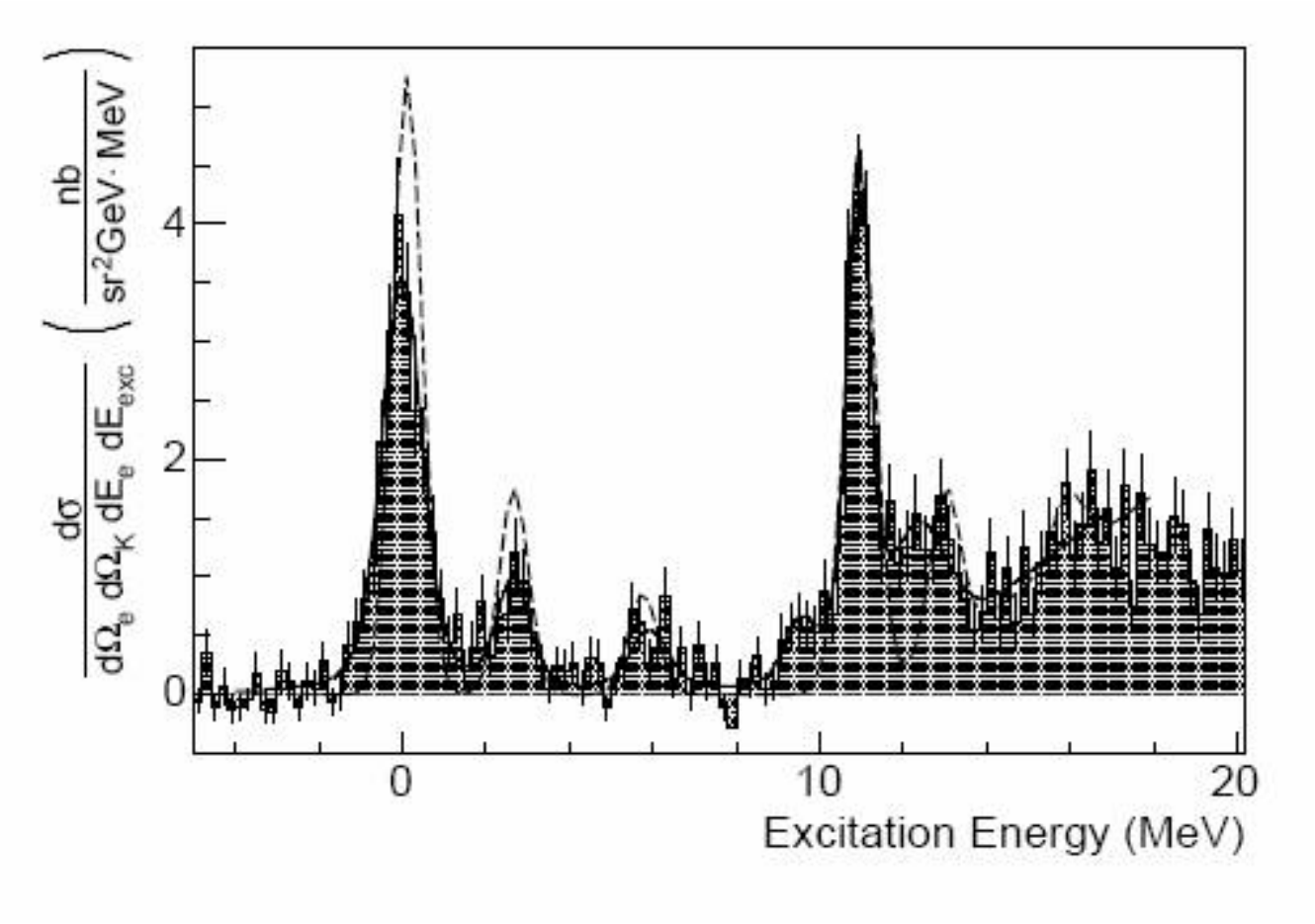}
}
\end{center}
\caption{The $^{12}_\Lambda$B excitation--energy spectrum. The best fit (solid curve) and a theoretical prediction (dashed curve) are superimposed on 
the data. From Ref.~\cite{iodice}}
\label{fig:fig16}
\end{figure}

As mentioned before, only recently reaction (\ref{eq:eq3}) has been exploited at JLab for the production of Hypernuclei. Two performing complexes 
of spectrometers were installed in Hall A and Hall C respectively. Since reaction (\ref{eq:eq3}) leads to three bodies in the final state, two magnetic 
spectrometers with large acceptance had to be installed for the event--by--event reconstruction. A further difficulty, for the Hall C spectrometer, was represented by the necessity 
of using a splitter magnet after the target, keeping at minimum the disturbance introduced by such an element in the magnetic optics of the spectrometers. 
Fig.~\ref{fig:fig14} shows a sketch of the spectrometer complex installed in Hall C. 
The difficulty of spectroscopizing two charged particles in the final state is overcompensated by the fact that spectrometers for electrons offer the best momentum 
resolution and then the final energy resolution on the Hypernuclear levels is considerably improved. As a matter of fact,  a resolution of 0.67 MeV FWHM was very recently 
attained \cite{iodice} by using the spectrometry's complex installed in Hall A, in a measurement on a $^{12}$C target, 
with good statistics. Fig.~\ref{fig:fig16} shows the obtained $^{12}_\Lambda$B excitation--energy spectrum, with peaks corresponding to the hypernucleus ground state and 
to both $\Lambda$ and nuclear core excited states \cite{iodice}.

\subsubsection{Other reactions for the production of Hypernuclei}
\label{other_prod}

Several projectiles other than mesons and electrons were used in attempts to produce Hypernuclei  and measure some observable with a precision better than that reached 
with the standard reactions of Sec.~\ref{basics-prod}.
We recall that low momentum $\bar{\mathrm{p}}$ from LEAR were used to produce Heavy Hypernuclei from Bi and U targets and measure their lifetime by 
the delayed fission using the recoil shadow method \cite{bocquet,armstrong}. The same method was used some years later \cite{cassing} using the p of 3 
GeV from COSY to produce fissionable heavy $\Lambda$--Hypernuclei. Both experiments provided values for the Heavy Hypernuclei with errors  which at 
the end were at the same percentage level of those measured in experiments using mesons as projectiles. 
However, the reported values of the lifetimes for heavy Hypernuclei look too low when compared to those obtained for light--medium A Hypernuclei.
We notice that, otherwise than other experiments, the lifetime determination was not direct, but needed some additional hypotheses; a detailed 
description will be given in Sec.~\ref{subsec:tau_meas}.

With the advent of the relativistic ion beams,  some measurements were performed on Hypernuclear observables. The first attempt was the measurement of the 
lifetime for 
$^{16}_{\Lambda}$O performed at the LBL Bevalac with a 2.1 GeV/nucleon $^{16}_{\Lambda}$O beam \cite{nield}. However, the obtained value was about 3 
times lower than the 
lifetime for neighboring Hypernuclei, indicating that some serious drawback was affecting the measurement. The method was improved at the JINR 
Synchrophasatron, providing reliable results for the lifetime of light Hypernuclei \cite{avra}. More details will be given in 
Sec.~\ref{subsec:tau_meas}.

Heavy--ion collisions experiments are of extreme importance for exploring the
phases of strongly interacting matter as a function of temperature and net baryon
density. In heavy--ion collisions at moderate energies, one has
access to the high net baryon density region of the QCD phase diagram. Experiments
of this kind, where particles with moderate strangeness quantum number are produced,
will be performed at FAIR by the Compressed Baryonic Matter (CBM) 
Collaboration: their main aim is to study the nature of the
transition between hadronic and quark matter, the search for a QCD critical endpoint
and for a chiral phase transition.
Instead, in nuclear collisions at high--energy, hot and dense matter with almost vanishing 
net baryonic density is produced in the early stage of the fireball evolution \cite{Ab09}. 
High--energy heavy--ion collisions experiments have been carried out at RHIC and, at higher energies, are 
currently running at LHC. The matter created in such experiments
contains an almost equal number of quarks and anti--quarks; it has been interpreted as the signature 
of the formation of a quark--gluon plasma (QGP) \cite{Ad05} and is suited for the production of 
Hypernuclei and anti--Hypernuclei in addition to many other particles and their anti--particles
with large strangeness content.

The production of $^3_\Lambda$H Hypernuclei and $^3_{\bar \Lambda} \bar {\rm H}$ 
antimatter Hypernuclei was observed by the STAR Collaboration at RHIC in Au$+$Au collisions 
at a center--of--mass energy per nucleon--nucleon collision $\sqrt{s_{NN}}=200$ GeV 
\cite{Ab10,Ch10}. This is the first observation of an antimatter Hypernucleus.

The lifetime of hypertriton and anti--hypertriton were measured by observing
$^3_{\Lambda} {\rm H} \to {^3{{\rm He}}}+ \pi^-$ and
$^3_{\bar \Lambda} \overline {\rm H} \to {^3{\overline{\rm He}}}+ \pi^+$ mesonic decays,
with a value $\tau=(182^{+89}_{-45}\pm 27)$ ps determined from the combined
set of hypertriton and anti--hypertriton data (the separate hypertriton and 
anti--hypertriton showed no difference within errors, as predicted by the
matter--antimatter symmetry). The determined masses,
$m(^3_\Lambda {\rm H})=(2.989\pm 0.001\pm 0.002)$ GeV and
$m(^3_{\bar \Lambda} \bar {\rm H})=(2.991\pm 0.001\pm 0.002)$ GeV,
are consistent with the best value for hypertriton from the literature.

The experiment determined a $^3_{\bar \Lambda} \bar {\rm H}/^3_\Lambda {\rm H}$ ratio 
between the $^3_{\bar \Lambda}\bar {\rm H}$ and $^3_\Lambda {\rm H}$ production yields 
of about 0.5. This result is consistent with $^3_{\bar \Lambda}\bar {\rm H}$ and 
$^3_\Lambda {\rm H}$ production via a hadron
coalescence mechanism by the overlapping of the $\Lambda$, 
$p$, $n$ baryon wave functions for $^3_\Lambda {\rm H}$ and $\bar \Lambda$, $\bar p$, $\bar n$ 
anti--baryon wave functions for $^3_{\bar \Lambda} \bar {\rm H}$ in the final stage of
the collision: in a coalescence picture one has
$^3_{\bar \Lambda} \bar {\rm H}/^3_\Lambda {\rm H}\sim ({\bar \Lambda}/\Lambda)\times
(\bar p/p)\times (\bar n/n)$, where with $\bar B/B$ ($B=\Lambda,\,p\,,n$) we indicate
the baryon yield ratios. 
Relativistic heavy--ion collisions are expected to produce large
amounts of hyperons and anti--hyperons with two or three units of strangeness,
thus may be regarded as interesting sources of multi--strangeness Hypernuclei
and anti--Hypernuclei. 

Equilibration between strange and light quarks has been considered
as a signature of the formation of a QGP. According to recent studies \cite{Ko05},
the evolution of the correlation between the baryon and the strangeness quantum numbers
as a function of the collision energy provides information on the onset of deconfinement
which separates the QGP and the hadron gas phases.
These baryon--strangeness correlations can be measured from the Hypernuclear production 
yields by introducing the so--called strangeness population factor 
$S_3={^{3}_\Lambda{\rm H}}/({^3{\rm He}}\times (\Lambda/p))$. This observable is
expected to increase as a function of the collision energy for a deconfined partonic 
phase while for a systems which remains in a hadronic phase is almost
independent of energy \cite{Zh10}.
Heavy--ion collision measurements at high energy can thus distinguish
between the two phases of matter in the high temperature regime.
Although with large errors, $S_3$ was measured to be about $1/3$
for AGS@BNL collisions at $\sqrt{s_{NN}}=5$ GeV, and about $1$
for RHIC collisions at $\sqrt{s_{NN}}=200$ GeV \cite{Ab10}. A value of $S_3$ close 
to 1 indicates that the phase space populations for strange and light quarks
are similar to each other and favours the formation of a high temperature matter made 
up of deconfined partons. New experiments at RHIC and LHC could elucidate this important 
issue.


\section{Hypernuclear Structure and the $\Lambda N$ Interaction}
\label{spect}


As said in Sec. \ref{intro}, the spectroscopic study of $\Lambda$--Hypernu-clei has become an indispensable
tool for the determination of the $\Lambda N$ interaction.
One can distinguish between reaction spectroscopy and $\gamma$--ray
spectroscopy, which are complementary to each other.
Reaction spectroscopy provides the gross features of the $\Lambda N$ interaction,
i.e., the central, spin--independent part of this interaction,
through the determination of the Hypernuclear masses, the hyperon binding
energies and the production reaction cross sections.
With $\gamma$--ray spectroscopy one has access to high--resolution
measurements of the Hypernuclear excitation energies
(the energy resolution is about three orders of magnitude better than with reaction 
spectroscopy), thus opening the possibility to study the fine structure and the 
hyperon spin--orbit splittings of Hypernuclei 
and therefore to determine the spin--dependent part of the $\Lambda N$ interaction.

\subsection{Theoretical Models of the Bare $\Lambda N$ Interaction}
\label{th-LN}

Before discussing the empirical results obtained in reaction and $\gamma$--ray 
spectroscopy experiments and their implications on Hypernuclear structure, we wish to 
briefly summarize the theoretical approaches used to determine the free--space
$YN$ and $YY$ interactions which are usually adopted in those Hypernuclear
calculations which are compared with the aforementioned experiments.

Various free--space $YN$ and $YY$ meson--exchange
potentials have been constructed by the Nijmegen group 
by extending the $NN$ interaction models using (broken)
flavour $SU(3)$ symmetry together with a fitting procedure of the scarce 
$\Lambda N$ and $\Sigma N$ scattering data and the rich body of $NN$ data.
The one--boson--exchange picture led to the so--called soft--core interaction models
NSC89 \cite{Ri89} and NSC97 \cite{Ri99}, while the
extended soft--core models ESC04 \cite{Ri06} and ESC08 \cite{Ri10b,Ri10}
consist of one--boson--exchange, two--pseudoscalar--exchange, 
meson--pair--exchange and mul-tiple--gluon--exchange. 
The Juelich group also proposed meson--exchange $YN$ and $YY$ 
potentials \cite{Ju89,Ju94,Ju05} and other potentials based on leading--order chiral 
effective field theory \cite{Po06,Po07}.
In Ref.~\cite{Sa06} a chiral unitary approach is adopt-ed to study the central
part of the $\Lambda N$ and $\Lambda \Lambda$ interactions.
Models of the baryon--baryon interactions also including quarks degrees of freedom have been
developed, for instance in Refs.~\cite{Fu96,Fu96b,Fu07} within a spin--flavour SU(6) 
quark--model supplemented by an effective--meson exchan-ge potential.
Finally, we mention that Lattice QCD calculations of the baryon--baryon interactions
are being developed in recent years \cite{Be07,Ne09,In10}, providing results
for the scattering parameters in the strangeness $-1$ and $-2$ worlds and the
first simulations for the $YN$ potentials. Especially the meson--exchange 
models differ from each other concerning their spin--isospin structure,
which cannot be fixed by the few existing baryon--baryon scattering data.
Therefore, often these models provide results in disagreement with each other
concerning Hypernuclear observables such as the hyperon--nucleus potential depths,
the Hypernuclear energy level structure, etc, as we discuss in Sections
\ref{s-1} and \ref{theor-LN}.

\subsection{Structure of $\Lambda$--Hypernuclei from Reaction Spectroscopy Experiments}
\label{s-1}

\begin{figure} [h]
\begin{center}
\resizebox{0.5\textwidth}{!}{%
  \includegraphics{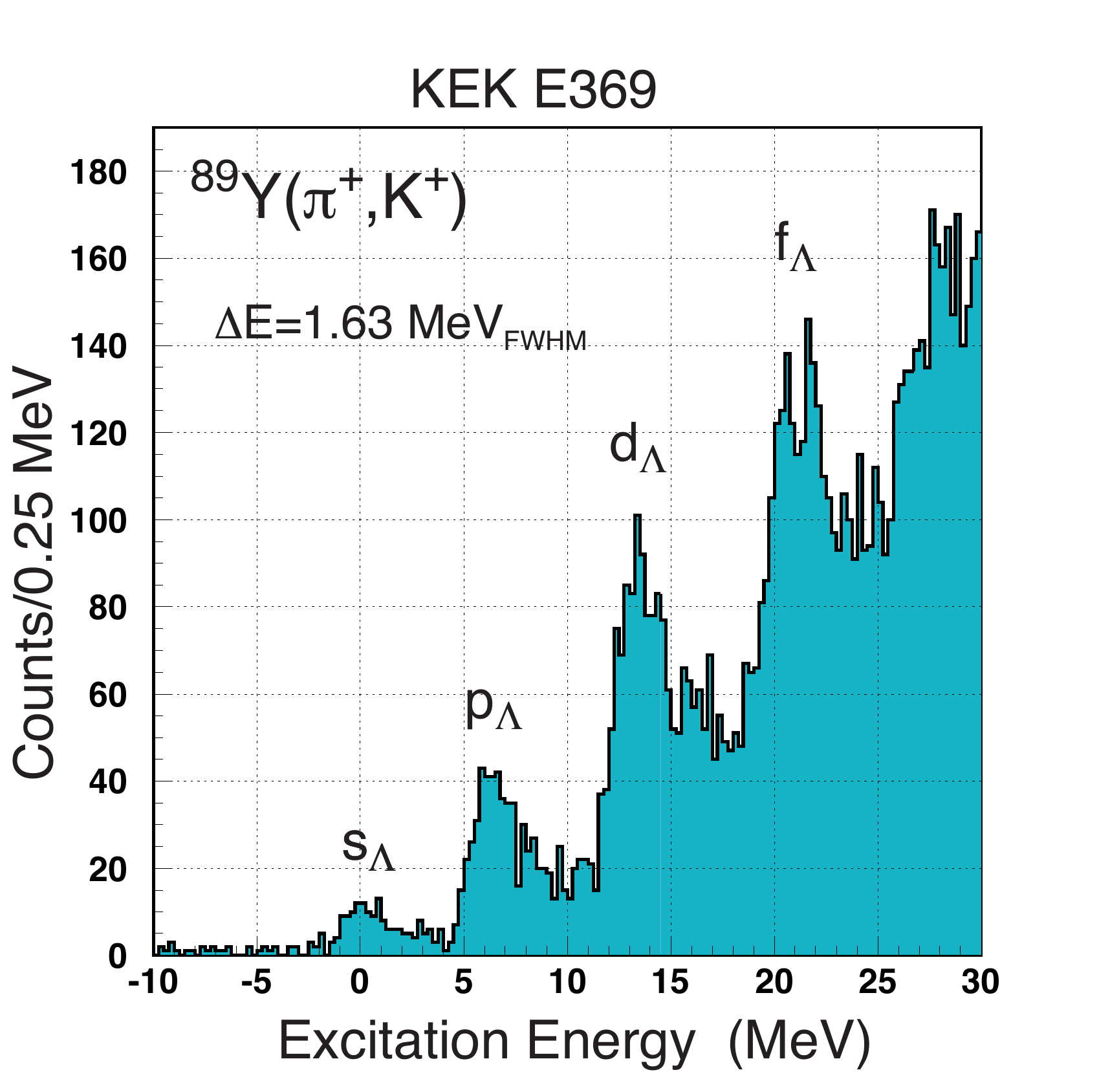}
}
\caption{Excitation energy spectrum of $^{89}_\Lambda$Y obtained with the 
$(\pi^+,K^+)$ reaction at KEK--E369 \protect\cite{hot}. The peaks
shows the $s_\Lambda$, $p_\Lambda$, etc, $\Lambda$ single--particle states.}
\label{spel}       
\end{center}
\vspace{-4mm}
\end{figure}

Since the 70's up to the beginning of the new millennium, 
many reaction spectroscopy experiments have been performed at CERN,
BNL, KEK and LNF with the $(K^-,\pi^-)$ and $(\pi^\pm,K^+)$ production reactions 
\cite{tamu} and more recently at JLab with the $(e,e' K^+)$ reaction \cite{Ha10}. 
They achieved good energy resolution 
and provided good quality excitation spectra and the binding energy (\ref{bl}) 
of the hyperon in Hypernuclei in wide excitation energy and
mass ranges, from $^3_\Lambda$H to $^{208}_\Lambda$Pb. 
In such experiments, the energies of the various $s_\Lambda$, $p_\Lambda$, etc, 
$\Lambda$ levels are directly obtained from the production reaction. 
Fig.~\ref{spel} clearly shows, through the well separated peaks, the $\Lambda$ 
single--particle levels structure of $^{89}_\Lambda$Y .
As a matter of fact, since there is only one $\Lambda$, all the single particle states are allowed. 
This is not the case for nucleons in a nucleus, due to the Pauli principle. 
One often refers to this case as a textbook example of the $\Lambda$ independent particle 
behavior in Hypernuclei.

The observed excitation spectra were rather well explained by DWIA calculations 
based on the shell model \cite{Mo94}. The observed Hypernuclear states were
instead interpreted with shell model \cite{Mi01} and cluster model calculations \cite{Hi00}.

With the exception of $s$-- and $p$--shell Hypernuclei, mean field models with Wood--Saxon 
$\Lambda$--nucleus potentials reproduce rather well the Hypernuclear mass dependence of 
the hyperon binding energies of the various orbits determined by the reaction spectroscopy 
experiments by using a radius $R=1.1(A-1)^{1/3}$ fm and a depth of 28-30 MeV \cite{Mi88}: 
${\rm Re}\, V_\Lambda(\rho_0) \sim -(28$-$30)$ MeV. 
The fact that a simple mean field picture turns 
out to be a fairly accurate description of the bulk Hypernuclear properties 
means that the hyperon, being a distinguishable particle, maintains its single 
particle behavior in the medium; this occurs even for states well below the 
Fermi surface, a property which is not observed for the nucleons.
Due to its position in the inner part of the nucleons' core, on a single particle 
level which is forbidden, by the Pauli principle, to the nucleons, the $\Lambda$
plays the r$\hat{\mathrm{o}}$le of stabilizer of the Hypernucleus.

The spectroscopy by the (K$^{-}_{\rm stop}$, $\pi^{-}$) reaction performed with the best achieved resolution 
on $^{12}$C (Fig.~\ref{fig:fig13}) \cite{agnello} showed new features of relevance for the description of 
Hypernuclear states. Besides the two wide peaks corresponding to the formation of $^{12}_{\Lambda}$C with 
the $\Lambda$ in the $s$-- and $p$--states, there is a considerable strength for the production of excited states 
lying between these two. It is of the same order (10$^{-3}$/stopped K$^{-}$) of that for the production of the ground 
state of $^{12}_{\Lambda}$C. These states may be described by mixed L--excited nuclear core configurations 
and are predicted by some theoretical calculations concerning their energies, not at all, to our knowledge, their 
production rates. The importance of the excited nuclear core states was confirmed, with a better resolution, by an 
experiment with the (e, e' K$^{+}$) reaction on a $^{12}$C target \cite{iodice} (see Figure \ref{fig:fig16}).

\begin{figure} [h]
\begin{center}
\resizebox{0.5\textwidth}{!}{%
  \includegraphics{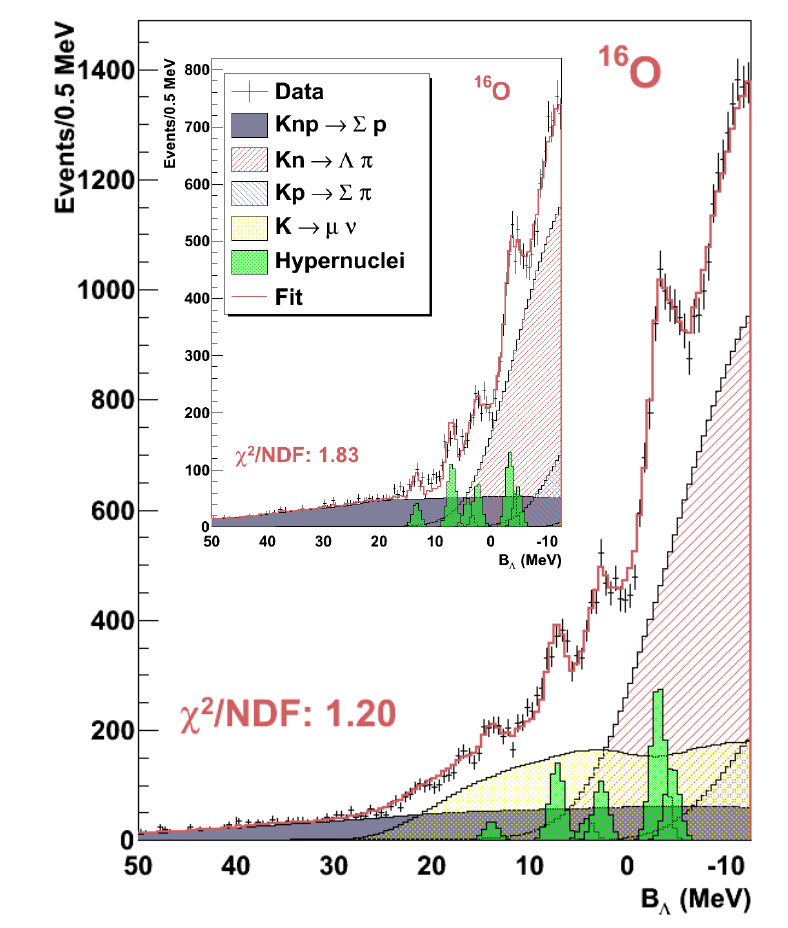}
}
\caption{Excitation energy spectrum of $^{16}_\Lambda$O obtained with the 
(K$^{-}_{\rm stop}$, $\pi^{-}$) reaction by FINUDA \protect\cite{fnd_spec}. In the inset the binding energy distribution for a sample free from the K$^{-}$ in flight 
decay background is shown.}
\label{fig:fig8b}       
\end{center}
\vspace{-4mm}
\end{figure}

Stimulated by this observation, the FINUDA Collaboration published recently \cite{fnd_spec} new data on the spectroscopy 
of $^{7}_{\Lambda}$Li, $^{9}_{\Lambda}$Be, $^{13}_{\Lambda}$C and $^{16}_{\Lambda}$O produced by K$^{-}$ at rest. 
The measured pattern of excited states is similar to that obtained with the ($\pi^{+}$, K$^{+}$) reaction at 1.05 GeV/c  \cite{tamu}; 
this is expected  since the two experiment feature a comparable energy resolution and a comparable momentum transfer 
(300 MeV/c). As an example, the spectrum obtained with the $^{16}$O target is shown in Fig.~\ref{fig:fig8b}. 
The only disagreement was found for the binding energy of the ground state of $^{16}_{\Lambda}$O, 13.4$\pm$0.4 MeV, 
to be compared with 12.42$\pm$0.05 MeV of Ref.~\cite{tamu}.  
We believe that, in spite of the larger error, the value from FINUDA is more reliable since it agrees with a previous measurement 
\cite{tamura} with stopped K$^{-}$ (12.9$\pm$0.4 MeV) and with a recent result \cite{cusanno} on the ($e,e'K^{+}$) electroproduction 
reaction on $^{16}$O leading to the formation of $^{16}_{\Lambda}$N (13.76$\pm$0.16 MeV). Furthermore, FINUDA had the 
advantage of a continuous self--calibrating control on the stability on the absolute value of the momenta of the spectroscopized pions 
by means of the monochromatic $\mu^{+}$ from the K$_{\mu2}$ decay. 

\begin{figure} [h]
\begin{center}
\resizebox{0.5\textwidth}{!}{%
  \includegraphics{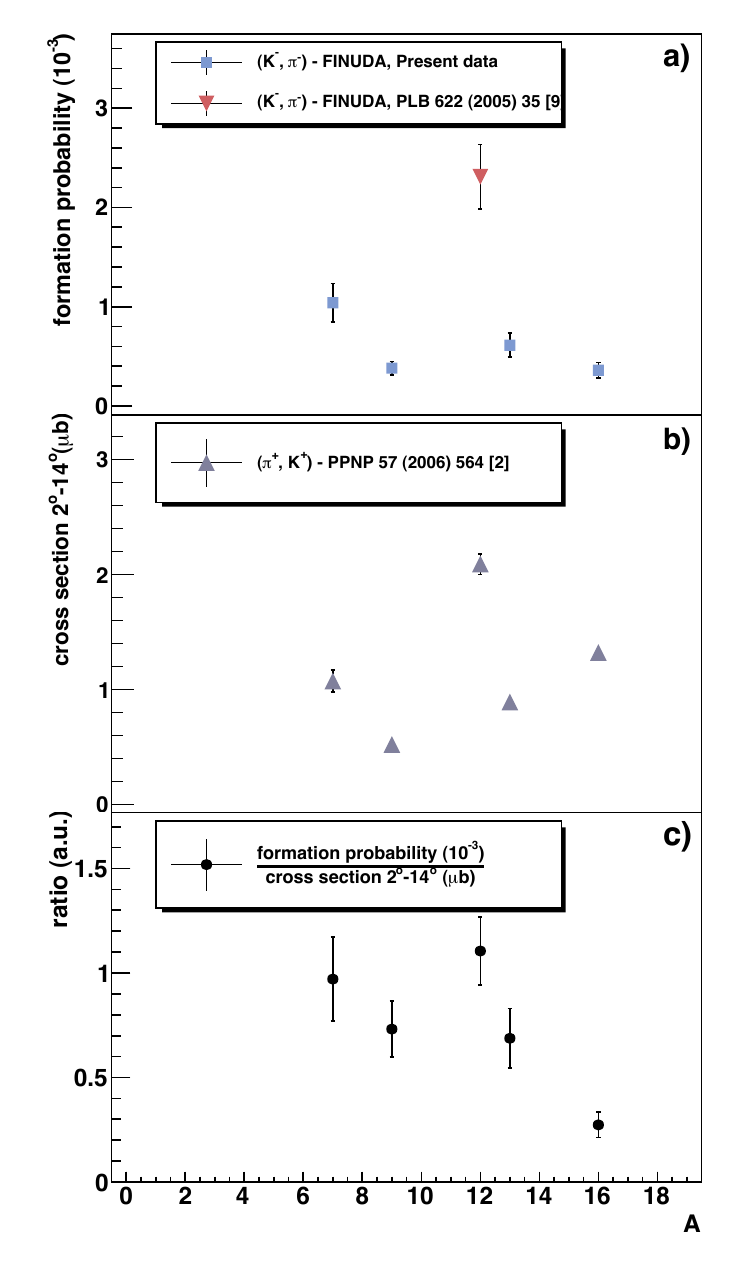}
}
\caption{Formation probabilities from FINUDA (a) and cross section from E336 \cite{tamu} (b) for bound states. In (c) the ratio between the two is shown.
From Ref.~\protect\cite{fnd_spec}. }
\label{fig:fig_sp_fnd}       
\end{center}
\vspace{-4mm}
\end{figure}
\begin{figure} [h]
\begin{center}
\resizebox{0.5\textwidth}{!}{%
  \includegraphics{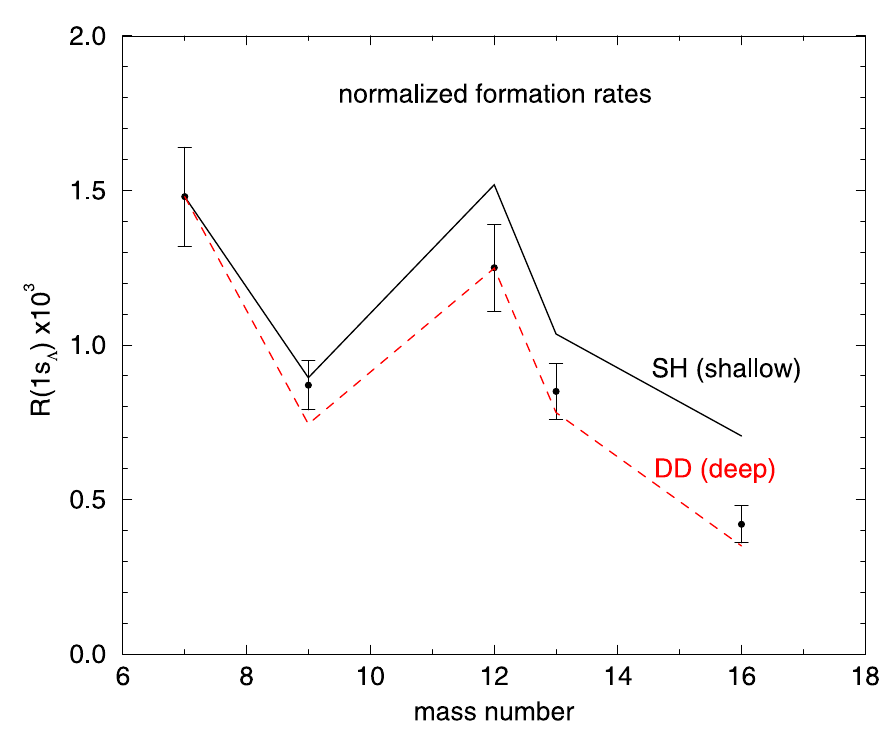}
}
\caption{Comparison between 1s$_{\Lambda}$ formation capture rates from FINUDA \protect\cite{fnd_spec} and DWIA calculations normalized to the 
1s$_{\Lambda}$ formation rate of $^{7}_{\Lambda}$Li for shallow (SH) and deep (DD) K$^{-}$ nuclear potentials. 
From Ref.~\protect\cite{cieply}. }
\label{fig:fig_cieply}       
\end{center}
\vspace{-4mm}
\end{figure}

It was not easy to draw from the data on K$^{-}$ capture probabilities of formation of the different states a simple $A$--dependence for the 
examined $p$--shell Hypernuclei. As a matter of fact, if we consider only their ground states, there are strong differences in the description 
of the nuclear configurations, and in some cases a not completely clean separation, from an experimental point of view, of the ground 
state from low--lying excited states (e.g. $^{7}_{\Lambda}$Li or $^{12}_{\Lambda}$C, in which the doublet of states, one of which is 
the ground state, has a spacing of some hundreds keV). 
On the other hand, considering all states with binding energy greater than zero, we know that in some cases the states at higher excitation 
are described by configurations containing an Hypernucleus with a lower A. For these reasons, in order to consider only well defined 
Hypernuclides, only excited states with energy below the threshold for the decay by nucleon emission were selected.  
Fig.~\ref{fig:fig_sp_fnd}(a) shows a plot of the capture rates chosen following the above criterion as a function of A. 
A smoothly decreasing behavior appears, with the exception of a strong enhancement corresponding to the formation of $^{12}_{\Lambda}$C 
bound states. 
Fig.~\ref{fig:fig_sp_fnd}(b) shows for comparison the differential cross section integrated in the forward direction (2$^{\circ}$--14$^{\circ}$) 
for the production of the same states observed with the ($\pi^{+}$, K$^{+}$) reaction \cite{tamu} and Fig.~\ref{fig:fig_sp_fnd}(c) the ratio between 
the two values. This ratio ranges from a large value for $^{7}_{\Lambda}$Li and 
$^{12}_{\Lambda}$C to a small value for $^{16}_{\Lambda}$O, showing a distinct $A$--dependence for the two reactions, K$^{-}$ capture at rest and 
in--flight ($\pi^{+}$, K$^{+}$).
The strong $A$--dependence of the (K$^{-}_{\rm stop}$, $\pi^{-}$)  rates with respect to the weak dependence of the ($\pi^{+}$, K$^{+}$) differential 
cross sections reflects the sizable difference between the strongly attractive K$^{-}$--nuclear interaction at threshold and the weakly repulsive 
K$^{+}$--nuclear interaction. 
This remark was the starting point for an attempt \cite{cieply} to determine the value of Re V$_{K^{-}}$, quite important for several items of K--nuclear 
Physics, as it will be discussed with more details in Sec.~\ref{saga}. 
Theoretical approaches varying from shallow potentials (40-60 MeV) at threshold to very deep density dependent potentials (150-200) MeV 
were in fact proposed. The authors considered only 1s$_{\Lambda}$ capture rates and for this reason they scaled the experimental capture rates 
given in Ref.~\cite{fnd_spec} by appropriate structure functions derived from neutron pick--up spectroscopic fractions in the target nuclei. 
The comparison of these normalized experimental formation rates with the calculations performed with a shallow or deep potential  slightly favored 
the second one, as shown in fig.~\ref{fig:fig_cieply}. 

\subsubsection{Theoretical Predictions based on $\Lambda N$ Effective Interactions}
\label{tplni}

In various calculations, the $\Lambda$ single--particle potential $V_\Lambda$ is obtained via a
folding procedure from G--matrix $\Lambda N$ interactions based on free--space
$YN$ potential. G--matrix calculations in nuclear matter at normal densities based
on the Nijmegen soft--core (NSC97, ESC04 and ESC08) $YN$ potentials \cite{Ri10b} can reasonably
reproduce the depth of the $\Lambda$--nucleus potential, 
${\rm Re}\, V_\Lambda(\rho_0)\sim -30$ MeV,
determined phenomenologically and (especially for the 'a' version of the ESC08 potential)
the $\Lambda$ binding energies for the $s_\Lambda$, $p_\Lambda$, $d_\Lambda$, ect,
single--particle states in medium and heavy Hypernuclei \cite{Ya10} (this is the case,
for instance, of $^{89}_\Lambda$Y, see Fig.~\ref{spel}). 
In Ref.~\cite{Hi10b}, a $\Lambda N$ G--matrix interaction is used in a
$\alpha$-$\alpha$-$\Lambda$ three--body calculation of $^{9}_\Lambda$Be.
The $\Lambda N$--$\Sigma N$ coupling was introduced effectively. 
The main purpose was to study the central, spin--independent $\Lambda N$ interaction
by the $B_\Lambda(^{9}_\Lambda{\rm Be})$ binding energy. The experimental value
of $B_\Lambda(^{9}_\Lambda{\rm Be})$ could be reproduced for different versions
of the bare potentials (NSC97, ESC04 and ESC08) by fixing the nuclear Fermi
momentum of the $\Lambda N$ G--matrix interaction to give the experimental $\Lambda$ binding 
energy in $^{5}_\Lambda$He, without the need of any additional $\Lambda NN$ three--body force.
Calculations of the $\Lambda$ single--particle potential
\cite{Ko00} have also been performed via nuclear matter $\Lambda N$ G--matrices derived from  
a $SU(6)$ quark model \cite{Fu96}: these calculations tend to overestimate the 
potential depth at normal density.

Other many--body calculations based on Nijmegen $YN$
potentials were carried out within the extended Skyrme--Hartree--Fock model 
(a review of these calculations is given in Ref.~\cite{Sc10}). 
The adopted $YN$ G--matrix interactions were obtained in a Bruekner--Hartree--Fock approach.
The calculations with the NSC89 and NSC97f potentials provided 
$s$, $p$, $d$, etc, $\Lambda$ single--particle energies which compare well with data. 

In addition, low--momentum $\Lambda N$ effective interactions $V^\Lambda_{\rm low\, k}$ have been
constructed in Ref.~\cite{Sc06} starting from Nijmegen bare potentials. In Ref.~\cite{Dap08}
$V^\Lambda_{\rm low\, k}$ effective interactions have been applied for predicting the momentum
and density dependence of the Hypernuclear $\Lambda$--nucleus single--particle potential
in Hartree--Fock approximation. Reasonable results for ${\rm Re}\, V_\Lambda(\rho_0)$
have been obtained for bare NSC97 and chiral effective field theory \cite{Po06} $\Lambda N$
interactions, while the use of NSC89 potentials (Juelich potentials of
Ref.~\cite{Ju05}) has led to a strong underestimation (overestimation) of 
${\rm Re}\, V_\Lambda(\rho_0)$.

Effects beyond the mean field such as three--body interactions and two--body 
interactions with strangeness exchange are also relevant in Hypernuclei. 
The $\Lambda NN$ three--body force 
is an important ingredient to investigate the structure of $\Lambda$--Hypernuclei, especially 
in light systems. This is due to the $\Lambda N$--$\Sigma N$ strong coupling, which is
sizable in the nuclear medium \cite{Ak00}, and, on the other hand,
leads to a non--negligible second order tensor force in the $\Lambda N$ interaction.
Especially because of the relatively small $\Lambda$--$\Sigma$ mass difference, 
$m_\Sigma-m_\Lambda\sim 78$ MeV, in Hypernuclei the 
$\Lambda N$--$\Sigma N$ coupling is more important than the $NN$--$\Delta N$ coupling 
in conventional nuclei, where it plays a very small r$\hat{\mathrm{o}}$le in binding few--nucleon 
systems since $m_{\Delta}-m_N\sim 293$ MeV. Four--body cluster model calculations 
based on $YN$ NSC97f interactions have demonstrated the relevance of the $\Lambda N$--$\Sigma N$ 
coupling in $A=4$ Hypernuclei ($^4_\Lambda$H and $^4_\Lambda$He) structure calculations \cite{Hi02}, 
although the data for the $1^+$ excited states binding energies were underestimated.
New calculations with the most recent Nijmegen potentials, ESC08, are in progress \cite{Hi10b}. 

Another signal of the $\Lambda N$--$\Sigma N$ coupling probably comes from the observation that 
in $S$--wave relative states the $\Lambda p$ interaction is more attractive than the $\Lambda n$
interaction. This follows from a comparison of the experimental $\Lambda$ binding 
energies for the ground states and the first excited states in the $A=4$, $I=1/2$ doublet,
formed by $^4_{\Lambda}{\rm He}$ and $^4_{\Lambda}{\rm H}$
(with separations of $0.35$ MeV between the ground states and of 
$0.24$ MeV for the excited states, the latter being determined from the $\gamma$--ray
spectroscopy experiments discussed in Sec.~\ref{spect-lambda}), which implies an important charge 
symmetry breaking (CSB) for the $\Lambda N$ interaction. The origin of this CSB is still not known. 
Also the relevant experimental data, from old measurements, must be taken with some care. 
A cluster model study on the CSB was performed for $A=7$ and $A=8$ iso--multiplets 
on the basis of the phenomenological information available for $A=4$ Hypernuclei \cite{Hi10b}. 
The calculation showed that the few existing emulsion data on $A=7$ and $A=8$ Hypernuclei may 
be inconsistent with the large CSB indicated by the $A=4$ data.
New measurements for $A=4,\,7,\,8$ Hypernuclei are thus important.

Another important aspect of the $\Lambda N$ interaction is its spin--dependence.
A qualitative indication of the difference between the singlet ($J=0$) and
triplet ($J=1$) $\Lambda N$ interactions comes from the comparison of the $\Lambda$
binding energy in isobar nuclei not related by charge symmetry. For example,
$B_{\Lambda}(^7_{\Lambda}{\rm Li})$ is larger than
$B_{\Lambda}(^7_{\Lambda}{\rm Be})$ by about $0.42$ MeV.
The greater $B_{\Lambda}$ value corresponds to the Hypernucleus whose 
nucleons' core has non--zero spin; the above difference
can be explained by the effect of the spin--dependent $\Lambda$ interaction with
the unpaired nucleons, a proton and a neutron, in $^7_{\Lambda}$Li.
Moreover, reaction spectroscopy experiments showed for the first time
that the spin--orbit splittings of the $\Lambda$ in Hypernuclei are small
(the evidence came from the splitting between the $p_{1/2}$ and the $p_{3/2}$ 
$\Lambda$ states in $^{16}_\Lambda$O \cite{bruck78}).
However, as we will discuss in Sec.~\ref{spect-lambda}, a complete study
of the spin--dependence of the $\Lambda N$ interaction requires
precision experiments which measure the detailed Hypernuclear excitation spectra.

\subsection{Structure of $p$--shell $\Lambda$--Hypernuclei from Gamma--Ray Spectroscopy Experiments}
\label{spect-lambda}
To have access to the complete information on the spin--dependence
of the $\Lambda N$ interaction, one has to perform experiments with high
energy resolution. One needs to observe the de--excitation,
caused by the electromagnetic interaction, of the Hypernucleus
produced in various excited states and study the small (sub--MeV) energy separation 
of the spin--doublet and the $\Lambda$ spin--orbit splittings together with the larger 
non--doublet separations.

$\gamma$--ray spectroscopy with Ge detectors is currently the only 
method to access few tens keV Hypernuclear splittings. Experiments on Hypernuclear $\gamma$--ray 
spectroscopy started in 1998 at KEK by using the Hyperball Ge detector and continued within 
systematic programs at KEK and BNL with Hyperball and Hyperball2 
\cite{tamu,Ta00,Aj01,Ak02,Uk04,Ta10}: 
they allowed to measure with a few keV precision various electromagnetic transitions
and in some cases their lifetimes and relative intensities.

Studies of the level structure of various $p$--shell Hypernuclei 
($^7_\Lambda$Li, $^9_\Lambda$Be, $^{10}_\Lambda$B, $^{11}_\Lambda$B,
$^{12}_\Lambda$C, $^{15}_\Lambda$N and $^{16}_\Lambda$O) with the reactions $(\pi^+,K^+)$ 
and $(K^-,\pi^-)$ have provided us with reliable information on the spin--dependence 
of the $\Lambda N$ interaction and some indication on the $\Lambda N$--$\Sigma N$ strong coupling 
\cite{Mi07,Mi10}. This has been possible by identifying 22 $\gamma$--ray
transitions for a total of 7 Hypernuclei  \cite{tamu,boh4}, which led to the measure 
of 9 spin--doublet spacings. 


\subsubsection{Spin--Dependent $\Lambda N$ Effective Interaction}   
\label{spin-dep}
\begin{figure} [h]
\begin{center}
\resizebox{0.55\textwidth}{!}{%
  \includegraphics{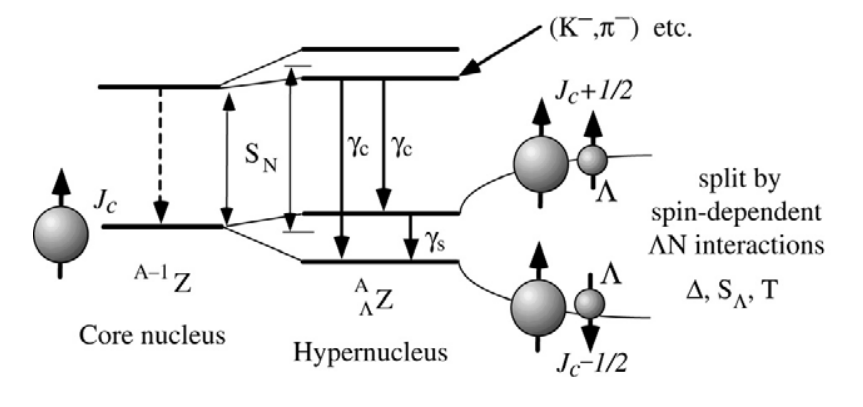}
}
\caption{Schematic low--lying level structure and $\gamma$--ray transitions
for $\Lambda$--Hypernuclei in the weak--coupling limit.
With $\gamma_c$ and $\gamma_s$ we indicate the
core transitions and the $\Lambda$ spin--flip transitions, respectively. From Ref.\cite{tamu}.
}
\label{scheme-schematic}       
\end{center}
\vspace{-4mm} 
\end{figure}

In the weak--coupling picture, information on the spin--dependence of the 
$\Lambda N$ interaction can be obtained from the energy spacings of the $J=J_c \pm 1/2$ 
spin--doublets in $\Lambda$--Hypernuclei, where the hyperon 
in the $s$--shell is coupled to a (ground state) nuclear core with non--zero spin $J_c$. 
A schematic representation of the low--lying Hypernuclear levels together with
some $\gamma$ transition is shown in Fig.~\ref{scheme-schematic}.
The weak--coupling scheme turns out to work well for $\Lambda$--Hypernuclei with the
hyperon in the $s$ state due to the weakness of the $\Lambda N$ interaction. 

Shell model calculations have been performed to interpret the $\gamma$--ray 
experiments. As input, they use a $\Lambda N$ interaction parametrized as follows 
\cite{Ga71,Da78}:
\begin{eqnarray}
\label{pot-eff}
V_{\Lambda N}(r)&=& V_0(r)+V_\sigma(r){\vec s_N}\cdot {\vec s_\Lambda}
+V_\Lambda(r){\vec l_{\Lambda N}}\cdot {\vec s_\Lambda} \\
&&+V_N(r){\vec l_{\Lambda N}}\cdot {\vec s_N}
+V_T(r)S_{12}~, \nonumber
\end{eqnarray}
where $S_{12}=3({\vec \sigma_\Lambda}\cdot {\vec r})
({\vec \sigma_N}\cdot {\vec r})/r^2-{\vec \sigma_\Lambda}\cdot {\vec \sigma_N}$
is the tensor operator,
${\vec r}={\vec r_N}-{\vec r_\Lambda}$ is the relative $\Lambda N$ coordinate,
${\vec s_N}$ (${\vec s_\Lambda}$) is the nucleon (hyperon) spin operator
and ${\vec l_{\Lambda N}}$ is the $\Lambda N$ relative angular momentum operator.
Moreover, $V_0=(V_1+3V_3)/4$ is the spin--averaged central interaction, 
$V_\sigma=V_3-V_1$ is the difference between the triplet and the singlet central
interactions, $V_\Lambda=V^{\rm LS}_\Lambda+V^{\rm ALS}_\Lambda$ 
($V_N=V^{\rm LS}_\Lambda-V^{\rm ALS}_\Lambda$) is the sum (difference) between the
symmetric and antisymmetric spin--orbit interactions and
$V_T$ is the tensor interaction. Once $\Sigma$ degrees of freedoms
are introduced, one has similar expressions to Eq.~(\ref{pot-eff}) for
the $\Lambda N\to \Sigma N$ and $\Sigma N\to \Sigma N$ effective interactions.

For the radial form of the effective interaction (\ref{pot-eff}),
$\Lambda N$ multirange Gaussian potentials, i.e., expansions in terms of
a number of Gaussian functions, are often used \cite{Mi07}.
From these radial potentials, in a shell model approach one defines 
two--body matrix elements with respect to a set 
of $\Lambda N$ relative wave functions. 
For Hypernuclei with an $s$--level $\Lambda$ coupled
with a $p$--shell nuclear core there are 5 $p_Ns_\Lambda$ two--body matrix elements.
These matrix elements are determined by 5 radial integrals \cite{Mi07}, 
usually denoted as $\bar V$ (spin--independent interaction), 
$\Delta$ (spin--spin interaction), $S_\Lambda$ ($\Lambda$--spin--dependent spin--orbit
interaction), $S_N$ (nucleon--spin--dependent spin--orbit interaction) and 
$T$ (tensor interaction), associated to the $V_0$, $V_\sigma$, $V_\Lambda$, $V_N$ and $V_T$
terms of Eq.~(\ref{pot-eff}), respectively. 
In a schematic way, for the spin--spin matrix element one has:
\begin{equation}
\label{radial-int}
\Delta \sim \int d {\vec r}\, \phi_{\Lambda N}^\star({\vec r}) 
V_\sigma(r)\phi_{\Lambda N}({\vec r})\, ,
\end{equation}
where $\phi_{\Lambda N}$ is the ${\Lambda N}$ radial relative wave function in the 
Hypernucleus. 
Similar equations may be written for the other terms in (\ref{pot-eff}). 
From Eqs.~(\ref{pot-eff}) and (\ref{radial-int}) one thus obtain 
the following $\Lambda N$ effective interaction in the $p$--shell:
\begin{eqnarray}
\label{pot-eff2}
V_{\Lambda N}
&=& {\bar V} +\Delta\, {\vec s_N}\cdot {\vec s_\Lambda}
+S_\Lambda\, {\vec l_{N}}\cdot {\vec s_\Lambda} \\
&&+S_N\, {\vec l_{N}}\cdot {\vec s_N} +T\, S_{12}~. \nonumber
\end{eqnarray}
Note that, since the $\Lambda$ is in the $s$ level,
the angular momentum $\vec l_{\Lambda N}$ in Eq.~(\ref{pot-eff}) is proportional 
to the angular momentum ${\vec l_{N}}$ in Eq.~(\ref{pot-eff2}).

%
The spin--independent matrix element $\bar V$ contributes to the $\Lambda$ binding energy $B_\Lambda$. 
It is well known already from emulsion data that $B_\Lambda$ increases linearly with $A$
with a slope of about 1 MeV/(unit of $A$) for $6 \lsim A\lsim 20$ and then saturates at about 27 MeV 
for heavy Hypernuclei. In first approximation, neglecting three--body $\Lambda NN$ interaction,
for $6 \lsim A\lsim 20$ 
the simple relation $B_\Lambda \sim \bar V (A-1)$ (based on a shell model picture) turns out to 
fit the data quite well: a value of $\bar V$ of about 1 MeV is thus expected.
In the weak--coupling limit of the shell model, the
energy separations of the $p$--shell Hypernuclear spin--doublets for nuclear
core states with non--zero spin are linear combinations of the $\Lambda N$ spin--dependent
parameters $\Delta$, $S_\Lambda$ and $T$ and depend on the $\Lambda N$--$\Sigma N$ coupling
(the shell model fixes the coefficients of the different parameters).
In the same limit, the parameter $S_N$ does not affect these doublet separations but
contributes to the nuclear spin--orbit interaction and is derived from
the splittings between Hypernuclear states based on different nuclear core states.

\begin{figure*}
\begin{center}
\resizebox{0.9\textwidth}{!}{%
  \includegraphics{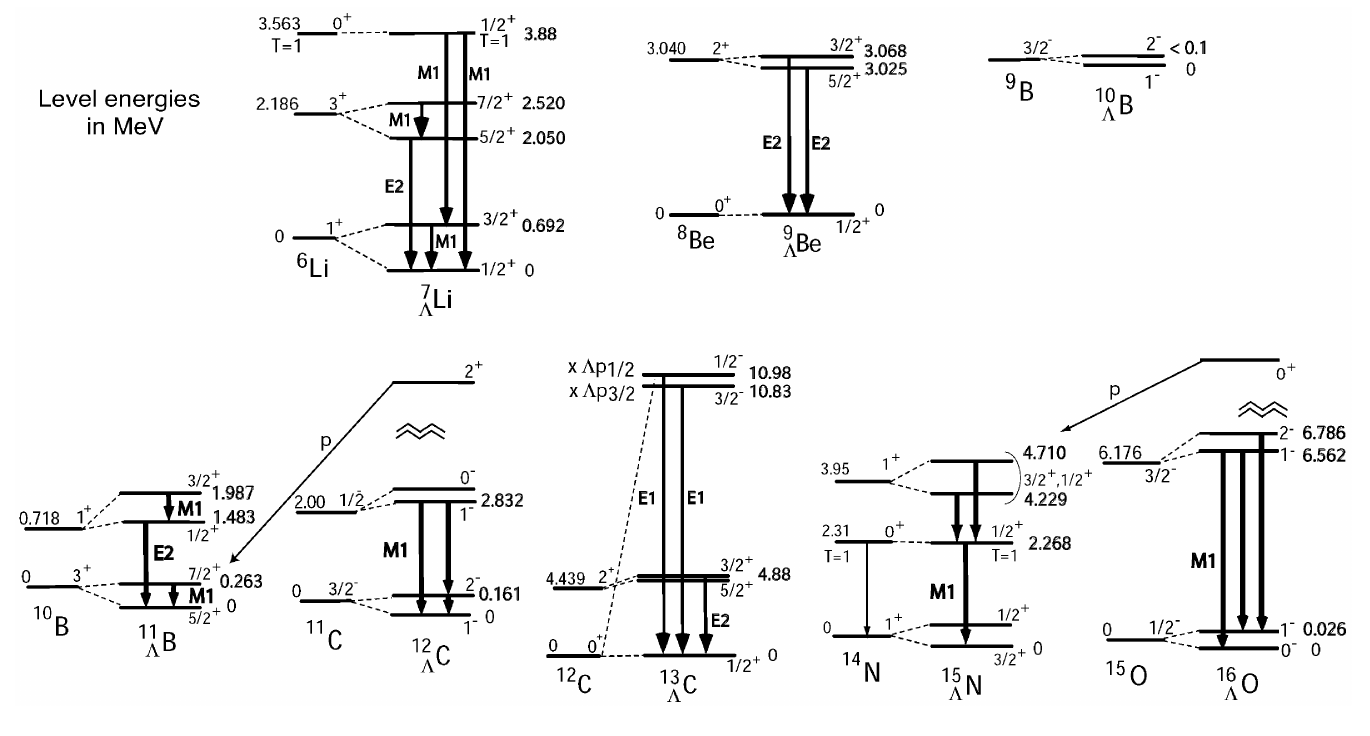}
}
\caption{Energy level scheme and $\gamma$--ray transitions observed for 
$p$--shell Hypernuclei in $\gamma$--ray spectroscopy experiments 
with Hyperball and Hyperball2 \protect\cite{tamu}
(the scheme for $^{13}_\Lambda$C was measured with a NaI
detector \protect\cite{Ko02}). Energies are in MeV. 
From Ref. \cite{tamu_hypx}.} 
\label{scheme-gamma}       
\end{center}
\vspace{-4mm}
\end{figure*}

To give an example, the $(3/2^+,1/2^+)$ and $(7/2^+,5/2^+)$ spin--doublet splittings 
in $^7_\Lambda$Li (see Fig.~\ref{scheme-gamma}) are given by \cite{Da78}:
\begin{equation}
E(3/2^+)-E(1/2^+) = \frac{3}{2}\Delta\, ,
\end{equation}
\begin{equation}
E(7/2^+)-E(5/2^+) = \frac{7}{6}\Delta + \frac{7}{3}S_\Lambda - \frac{14}{5} T\, ,
\end{equation}
respectively, in the limit of pure nuclear core configurations, i.e., in the 
LS coupling limit, which is a fairly good approximation for light systems. 
However, configuration mixing in the nuclear core modifies
these simple relations, and for the above doublet separations Ref.~\cite{Mi10} found:
\begin{eqnarray}
\label{3/21/2}
E(3/2^+)-E(1/2^+) &=& 1.461\, \Delta + 0.038\, S_\Lambda  \\
&&+ 0.011\, S_N - 0.285\, T\, , \nonumber \\
E(7/2^+)-E(5/2^+) &=& 1.294\, \Delta + 2.166\, S_\Lambda  \\
&&+ 0.020\, S_N - 2.380\, T\, . \nonumber
\end{eqnarray}

A fit of the experimental determinations of 
the Hypernuclear excitation spectra thus allows one to determine
the values of the $\Lambda N$ interaction parameters.
For details on the shell model and the fitting procedure we refer to 
Refs.~\cite{Mi07,Da78,Mi85,Au83}. In the following we only discuss the final 
results for the fitting parameters and the energy level separations. 
Fig.~\ref{scheme-gamma} reports the Hypernuclear level schemes and the
transitions determined since 1998 in $\gamma$--ray spectroscopy experiments. 
Three types of $\gamma$ transitions are possible in the weak--coupling limit:
core transitions, $\Lambda$ spin--flip transitions and $\Lambda$ inter--shell
transitions. Core transitions, denoted by $\gamma_c$ in 
Fig.~\ref{scheme-schematic}, connect different doublets and are 
due to transitions in the nuclear core: it is the case, for instance, of
the $E2(5/2^+\to 1/2^+)$ transition in $^7_\Lambda$Li of Fig.~\ref{scheme-gamma}.
$\Lambda$ spin--flip transitions, denoted by $M1$ in Fig.~\ref{scheme-gamma} 
and by $\gamma_s$ in Fig.~\ref{scheme-schematic}, connect the upper and
lower members of the same doublet and are due to a spin--flip of the hyperon
spin: it is the case, for instance, of the $M1(3/2^+\to 1/2^+)$ transition again
in $^7_\Lambda$Li. Spin--flip $M1$ transitions are also possible between 
Hypernuclear doublet members: two of these transitions are shown
in Fig.~\ref{scheme-gamma} for $^7_\Lambda$Li.
Finally, in a $\Lambda$ inter--shell transition, denoted
by $E1$ in Fig.~\ref{scheme-gamma}, the hyperon changes the shell orbit from $p_\Lambda$ 
to $s_\Lambda$: it is the case of the $\Lambda\, p_{1/2}\to \Lambda\, s_{1/2}$ and
$\Lambda\, p_{3/2}\to \Lambda\, s_{1/2}$ transitions in $^{13}_\Lambda$C.

The first experiment with Hyperball, KEK--E419, was performed in 1998 and studied 
$^7_\Lambda$Li, produced with the $(\pi^+,K^+)$
reaction. Fig.~\ref{scheme-gamma} shows that $^7_\Lambda$Li is the best 
studied Hypernucleus: 5 $\gamma$--rays were measured for it, which led to
the determination of 2 doublet splittings and 2 core transition separations. 
Note that here and in the following the considered number of core transition separations does not take
into account those ones which are not independent of the other (core or 
doublet) separations.
The $^9_\Lambda $Be Hypernucleus was then studied, again in 1998 and with Hyperball,
at BNL--E930 by using the $(K^-,\pi^-)$ reaction, with the observation of 2 $\gamma$--rays,
corresponding to 1 doublet splitting and 1 core transition measurement. 
After including the $\Lambda N$--$\Sigma N$ coupling, a fit of the 
experimental $^7_\Lambda$Li and $^9_\Lambda $Be energy separations 
(the 3 doublets and the 3 independent core transitions of Fig.~\ref{scheme-gamma})
has led to the following determination of the radial matrix elements \cite{Mi10}:
\begin{eqnarray}
\label{7-9}
&&A=7\div 9: \\
&&\Delta=0.430,\, S_\Lambda=-0.015,\, S_N=-0.390,\, T=0.030\, , \nonumber
\end{eqnarray}    
where all numbers are in MeV.

$^{13}_\Lambda$C, $\gamma$--ray measurements were performed in 1998 at BNL--E929 
with the $(K^-,\pi^-)$ reaction and a NaI detector \cite{Ko02}, with the determination of 
2 $\Lambda$ inter--shell transitions and 1 core transition.
For $^{11}_\Lambda$B, $^{12}_\Lambda$C, $^{15}_\Lambda$N and $^{16}_\Lambda$O 
Hypernuclei, a total of 6 spin--doublet splittings and 5 core transitions were determined 
experimentally with the Hyperball and Hyperball2 detectors. 
$^{15}_\Lambda$N and $^{16}_\Lambda$O were studied at BNL--E930 in 2001 with the
$(K^-,\pi^-)$ reaction and the Hyperball detector, while $^{11}_\Lambda$B
and $^{12}_\Lambda$C were investigated in 2005 at KEK--E518 and KEK--E566, respectively,
with the $(\pi^+,K^+)$ reaction and the Hyperball2 detector.
The data for Hypernuclei from $^{11}_\Lambda$B to $^{16}_\Lambda$O
(the 6 doublets and the $1^-_2\to 0^-$ core transition for $^{16}_\Lambda$O 
of Fig.~\ref{scheme-gamma})
led to a determination of smaller values for the $\Lambda$ spin--dependent
parameters $\Delta$ and $T$ and for the parameter $S_N$ \cite{Mi10}:
\begin{eqnarray}
\label{11-16}
&&A=11\div 16: \\
&&\Delta=0.330,\, S_\Lambda=-0.015,\, S_N=-0.350,\, T=0.024\, ,\nonumber
\end{eqnarray}
where again the values are in MeV. 

For all the $p$--shell Hypernuclei, the shell model approach~\cite{Mi10} adopted
the following predictions of the $\Lambda N$--$\Sigma N$ interaction parameters:
\begin{equation}
\label{par-sigma}
\bar V'=1.45,\, \Delta'=3.04,\, S'_\Lambda=S'_N=-0.09,\, T'=0.16\,
\end{equation}
(values in MeV), derived from a multirange Gaussian potential fitted to
a G--matrix calculation \cite{Ak00} performed with the NSC97a,f potentials.

The $\gamma$--ray measurements of the Hypernuclear energy level separations
do not allow one to determine the $\Lambda N$ spin--independent parameter, $\bar V$,
which is related to the $\Lambda$ binding energy $B_\Lambda$. 
To have an idea of the magnitude of this parameter, we quote the predictions of the shell model 
calculation of Ref.~\cite{Mi10c} derived from the experimental $B_\Lambda$ values: 
$\bar V=-0.84$ MeV for $^9_\Lambda$Be and $\bar V=-1.06\pm 0.03$ MeV for $^{10}_\Lambda$B, 
$^{11}_\Lambda$B and $^{12}_\Lambda$C.

The values obtained for the spin--spin parameter $\Delta$ inform us that the $\Lambda N$
singlet central interaction is more attractive than the $\Lambda N$ triplet central interaction.
Some $\gamma$ transitions turned out to be particularly important in the determination
of the various parameters. The $M1(3/2^+\to 1/2^+)$ transition in $^7_\Lambda$Li
mainly contributes to the determination of the spin--spin
strength $\Delta$ (see Eq.~(\ref{3/21/2})).
The small value of the $\Lambda$--spin--dependent spin--orbit parameter
$S_\Lambda$ and the relatively large value of the nucleon--spin--dependent
spin--orbit parameter $S_N$ mean that the symmetric ($S^+$) and antisymmetric 
($S^-$) spin--orbit interactions have almost the same strength and opposite sign:
$S^+=(S_\Lambda+S_N)/2 \sim -0.19$ MeV and $S^-=(S_\Lambda-S_N)/2 \sim 0.18$ MeV.
The $3^+\to 1^+$ transition for the nuclear
core in $^7_\Lambda$Li is particularly important to determine $S_N$. 
The value of $S_\Lambda$ mainly originates from the smallness 
of the $3/2^+$--$5/2^+$ doublet separation, $(43\pm 5)$ keV, in $^9_\Lambda$Be. 
Information on $S_\Lambda$ is in principle also given by the 
spin--orbit splittings of $\Lambda$ single--particle states, for instance the 
$(152\pm 54 \pm 36)$ keV separation between the $(\Lambda\, p_{1/2})$ and the 
$(\Lambda\, p_{3/2})$ states in $^{13}_\Lambda$C. However, a 
sizable effect of the tensor interaction and a loosely bound $p_\Lambda$ orbit make it 
difficult to draw definite conclusions on $S_\Lambda$ and $T$ starting from the 
$(\Lambda\, p_{1/2})$--$(\Lambda\, p_{3/2})$ spin--orbit splitting 
\cite{Mi07}. The small tensor parameter $T$ is mainly determined by the small 
$1_1^-$--$0^-$ doublet splitting for $^{16}_\Lambda$O, once $\Delta$ is fixed 
from $^7_\Lambda$Li.


The assumption of a mass--independent $\Lambda N$ interaction is justified
for $p$--shell Hypernuclei if one considers that the rms charge radii of $p$--shell
nuclei is rather constant through the shell. Nevertheless,
the spin--dependent parameters of Eqs.~(\ref{7-9}) and (\ref{11-16}) exhibit some
dependence on the Hypernuclear mass number. However, this is really evident only
for the spin--spin parameter: the value of $\Delta$ obtained from $^7_\Lambda$Li is considerably
larger than the determinations from heavier Hypernuclei. This $A$--dependence is
however not confirmed by calculations of the $\Lambda N$ interaction parameters based on
realistic free--space $\Lambda N$ potentials \cite{Mi01}. For the future it is thus necessary to clarify the 
reason of the $A$--dependence of the shell model fit.

In Table \ref{exp-sm} we compare the Hyperball and Hyperball2 experimental data on $p$--shell 
level separations (both doublet spacings and core transition separations)
with the shell model predictions of Refs.~\cite{Mi07,Mi10,Mi10c} obtained with the values of
Eqs.~(\ref{7-9}), (\ref{11-16}) and (\ref{par-sigma}) of the $\Lambda N$ and $\Lambda N$--$\Sigma N$ 
interaction strengths. 
We see that the $\Lambda N$--$\Sigma N$ coupling affects in a
non--negligible way the $p$--shell Hypernuclear spacings, especially the doublet ones.
However, we remind the reader that the strongest effect of this coupling has been
obtained from the analysis of $s$--shell Hypernuclei, where its contribution
to the $0^+$--$1^+$ doublet splittings in $^4_\Lambda$H and $^4_\Lambda$He
is of the same order of the one coming from the spin--spin $\Lambda N$ interaction
\cite{Ak00} and is essential to explain the level structure of all $s$--shell Hypernuclei.
The shell model calculation well reproduces the data, even for the energy separations not 
included in the fit, except for the underestimation of a few core transitions 
separations: the $1/2^+$--$5/2^+$ separation in $^{11}_\Lambda$B and, to a lesser
extent, the $1_2^-$--$1_1^-$ separation in $^{12}_\Lambda$C and the 
$3/2^+$--$1/2^+$ separation in $^{13}_\Lambda$C. 
Apart from these inconsistencies, which deserve further 
investigation, the spin--dependence of the $\Lambda N$ effective interaction is rather 
well understood at present. The interpretation in terms of the shell model calculation 
could thus serve to improve the theoretical models of the free-space $\Lambda N$ interaction.
\begin{table*}[t]
\begin{center}
\caption{Energy contributions (in keV) to the spacings for the spin--doublet  
and the core transition separations obtained with the shell model fitting procedure
of Refs.~\cite{Mi07,Mi10,Mi10c}. The transitions are indicated with $J^P_{\rm upper}\to J^P_{\rm 
lower}$, $J^P_{\rm upper}$ ($J^P_{\rm lower}$) indicating the spin--parity of the upper (lower)
energy level. The contributions to the energy separation originating from the 
$\Lambda N$--$\Sigma N$ coupling are indicated
in the $\Lambda \Sigma$ column. The theoretical energy separations 
$\Delta E^{\rm th}$ are compared with the experimental determinations $\Delta E^{\rm exp}$.
Small contributions from the energies of admixed core states are not shown.
$\Delta E_c$ is the contribution of the nuclear core level spacing; the values 
of $\Delta E_c$ are taken from experiment.}
\label{exp-sm}       
\begin{tabular}{ccrrrrrrrr}
\hline\noalign{\smallskip}
& Transition & $\Delta$ & $S_\Lambda$ & $S_N$ & $T$ & $\Lambda\Sigma$ &
 $\Delta E_c$ & $\Delta E^{\rm th}$ & $\Delta E^{\rm exp}$  \\
\noalign{\smallskip}\hline\noalign{\smallskip}
$^7_\Lambda$Li & $3/2^+\to 1/2^+$ & $628$ & $-1$ & $-4$ & $-9$ & $72$ & $0$ & $693$ & $692$ \\
$^7_\Lambda$Li & $7/2^+\to 5/2^+$ & $557$ & $-32$ & $-8$ & $-71$ & $74$ & $0$ & $494$ & $471$ \\
$^7_\Lambda$Li & $5/2^+\to 1/2^+$ & $77$ & $17$ & $-288$ & $33$ & $4$ & $2186$ & $2047$ & $2050$ \\
$^7_\Lambda$Li & $1/2^+_1\to 1/2^+_0$ & $418$ & $0$ & $-82$ & $-3$ & $-23$ & $3565$ & $3883$ & $3877$ \\

$^9_\Lambda$Be & $3/2^+\to 5/2^+$ & $-14$ & $37$ & $0$ & $28$ & $-8$ & $0$ & $44$ & $43$ \\

$^{10}_\Lambda$B & $2^-\to 1^-$ & $188$ & $-21$ & $-3$ & $-26$ & $-15$ & $0$ & $120$ & $<100$ \\

$^{11}_\Lambda$B & $7/2^+\to 5/2^+$ & $339$ & $-37$ & $-10$ & $-80$ & $56$ & $0$ & $267$ & $264$ \\  
$^{11}_\Lambda$B & $3/2^+\to 1/2^+$ & $424$ & $-3$ & $-44$ & $-10$ & $61$ & $0$ & $475$ & $505$ \\
$^{11}_\Lambda$B & $1/2^+\to 5/2^+$ & $-88$ & $-19$ & $391$ & $-38$ & $5$ & $718$ & $968$ & $1483$ \\

$^{12}_\Lambda$C & $2^-\to 1^-$ & $175$ & $-12$ & $-13$ & $-42$ & $61$ & $0$ & $153$ & $162$ \\
$^{12}_\Lambda$C & $1_2^-\to 1_1^-$ & $117$ & $17$ & $309$ & $20$ & $49$ & $2000$ & $2430$ & $2834$ \\

$^{13}_\Lambda$C & $3/2^+\to 1/2^+$ & $-11$ & $22$ & $203$ & $-22$ & $1$ & $4439$ & $4630$ & $4880$ \\ 

$^{15}_\Lambda$N & $1/2_1^+\to 3/2_1^+$ & $244$ & $34$ & $-8$ & $-214$ & $44$ & $0$ & $99$ & $> -100$ \\
$^{15}_\Lambda$N & $3/2_2^+\to 1/2_2^+$ & $451$ & $-2$ & $-16$ & $-10$ & $65$ & $0$ & $507$ & $481$ \\
$^{15}_\Lambda$N & $1/2^+;T=1\to 3/2_1^+$ & $86$ & $11$ & $-6$ & $-71$ & $-57$ & $2313$ & $2274$ & $2268$ \\
$^{15}_\Lambda$N & $1/2_2^+\to 3/2_1^+$ & $-208$ & $13$ & $473$ & $-67$ & $-16$ & $3948$ & $4120$ & $4229$ \\

$^{16}_\Lambda$O & $1_1^-\to 0^-$ & $-123$ & $-20$ & $1$ & $188$ & $-33$ & $0$ & $23$ & $26$ \\
$^{16}_\Lambda$O & $2^-\to 1_2^-$ & $207$ & $-21$ & $1$ & $-41$ & $92$ & $0$ & $248$ & $224$ \\
$^{16}_\Lambda$O & $1_2^-\to 0^-$ & $-207$ & $-2$ & $524$ & $170$ & $-70$ & $6176$ & $6582$ & $6562$ \\
\noalign{\smallskip}\hline
\end{tabular}
\end{center}
\end{table*}



We conclude this subsection with more detailed comments on the consistency 
problems arising from the comparison of the shell model
fit results with the $\gamma$--ray data. The large value of $\Delta$ required in the 
beginning of the $p$--shell is unlikely to be simply the effect of a smaller rms radius 
of the $^7_\Lambda$Li nuclear core size with respect to the other $p$--shell cores.
Given its important r$\hat{\mathrm{o}}$le, alternative and improved descriptions of the 
$\Lambda N$--$\Sigma N$ coupling should be considered in new fits of data. 
For instance, a contribution of the $\Lambda N$--$\Sigma N$ coupling in the
end of the $p$--shell smaller than the one in the beginning of the $p$--shell could 
led to a determination of a value of $\Delta$ for A$=$15, 16
in agreement with the determination for A$=$7, 9 \cite{boh4}.
In addition, new shell model calculations 
should explore the effect of an expansion of the nuclear core basis states beyond the
$p$--shell configurations \cite{Mi10c}. A second major problem consists in the fact that for
Hypernuclei in the middle of the $p$--shell
(specifically, $^{11}_\Lambda$B, $^{12}_\Lambda$C and $^{13}_\Lambda$C)
the parameter $S_N$ does not supply a sufficiently attractive contribution to 
reproduce the observed splittings for some core transitions (see Table~\ref{exp-sm}),
which indeed would require more negative values of $S_N$ than in Eq.~(\ref{11-16}).
The explicit inclusion of three--body $\Lambda NN$ interactions,
induced by the $\Lambda N$--$\Sigma N$ coupling,  could also help
in solving this second problem \cite{Mi10c}, thanks to an antisymmetric 
spin--orbit $\Lambda NN$ interaction contribution that has the same effect of the 
parameter $S_N$ on the core transition splittings. We note that double one--pion--exchange
$\Lambda NN$ interactions were already included long ago \cite{Ga71}, in an approximated way, in 
shell model analyses of the $\Lambda$ binding energies of $p$--shell Hypernuclei.
Additional experimental $\gamma$--ray data for $p$--shell Hypernuclei will also play
an essential r$\hat{\mathrm{o}}$le for a deeper understanding of the $\Lambda N$ interaction.


\subsubsection{Theoretical Predictions based on $\Lambda N$ Interactions}
\label{theor-LN}


Various theoretical approaches, mainly based on the $YN$ Nijmegen 
meson--exchange potentials, have been used for modeling the in--medium $\Lambda N$ 
interaction and thus study the structure of Hypernuclei. 

In Table~\ref{freeYY} we summarize the results obtained with some of these approaches 
for the parametrization of Eq.~(\ref{pot-eff2}) of the $\Lambda N$ effective interaction
in the case of $^7_\Lambda$Li. 
The reported results are obtained from fixed radial forms of the potentials using 
Woods--Saxon wave functions and Gaussian \cite{Ri99} or Yukawa \cite{Ha08} 
representations of $\Lambda N$ G--matrix elements. The $\Lambda \Sigma$ contributions adopted 
in the calculations, not shown in the Table,
are dominated by the central matrix elements $\bar V'$ and $\Delta'$.
For comparison, in the Table we also report the phenomenological values of the parameters
determined with the shell model fit (\ref{7-9}) of the $\gamma$--ray data.
Despite the 6 versions of the NSC97 potentials predict values of the $\Delta$ parameter
in a rather wide range, only NSC97f provides a
prediction for $\Delta$ in agreement with the shell model fit. Unlike the
ESC04a interaction, the ESC08a one strongly underestimates the phenomenological $\Delta$ 
value. All the Nijmegen interaction models are inconsistent with the phenomenological 
determination of the $S_\Lambda$ and $S_N$ parameters: this indicates that these models 
need to be revised. Finally, the Nijmegen potentials can approximately explain the small
parameter $T$ of the phenomenological determination.
The smallness of $T$ is expected on the basis of the absence of $\pi$-- and 
$\rho$--exchange at tree--level in the $\Lambda N$ interaction.
\begin{table}[h]
\begin{center}
\caption{Predictions for the spin--dependent parameters of the $\Lambda N$ effective 
interaction (\protect\ref{pot-eff2}), in MeV, for $^7_\Lambda$Li obtained from free--space
baryon--baryon interactions. The NSC97f prediction is from
Ref.~\protect\cite{Ri99}, the ESC04a and ESC08a are from Ref.~\protect\cite{Ha08},
while the phenomenological (Phen.) values are from Eq.~(\protect\ref{7-9}).}
\label{freeYY}       
\begin{tabular}{lcccc}
\hline\noalign{\smallskip}
Model & $\Delta$ & $S_\Lambda$ & $S_N$ & $T$  \\
\noalign{\smallskip}\hline\noalign{\smallskip}
NSC97f & $0.421$ & $-0.149$ & $-0.238$ & $0.055$  \\
ESC04a & $0.381$ & $-0.108$ & $-0.236$ & $0.013$ \\
ESC08a & $0.146$ & $-0.074$ & $-0.241$ & $0.055$ \\
Phen.  & $0.430$ & $-0.015$ & $-0.390$ & $0.030$ \\
\noalign{\smallskip}\hline
\end{tabular}
\end{center}  
\end{table}

We now comment on the general results obtained with the approaches based 
on free--space baryon--baryon interactions for the energy level splittings in various 
Hypernuclei. In the next paragraphs we describe the individual frameworks.
For $^4_\Lambda$H, $^4_\Lambda$He and $^7_\Lambda$Li, some calculations reproduce
the central parameters $\bar V$ and $\Delta$ as well as the $\Lambda N$--$\Sigma N$ 
coupling adopted in the shell model fits. For some of the $YN$ interaction models, also the 
predictions for the tensor parameter $T$ agree fairly well with the phenomenological 
determinations of Eqs.~(\ref{7-9}) and (\ref{11-16}). However, although these models provide 
a wide range of results for the $\Lambda N$ spin--dependent channels, for most of 
them the $(43 \pm 5)$ keV separation for the
$(3/2^+,5/2^+)$ spin--doublet in $^9_\Lambda$Be and/or the
$(152\pm 54 \pm 36)$ keV splitting between the 
$(\Lambda\, p_{1/2})$ and $(\Lambda\, p_{3/2})$ states in $^{13}_\Lambda$C are overestimated: 
this implies the incapacity to reproduce the phenomenological value of the parameter $S_\Lambda$. 
This is well established for approaches based on the NSC97 and ESC04 potentials 
\cite{Mi10,Mi10b}, which underestimate the antisymmetric spin--orbit
parameter $S^-=(S_\Lambda-S_N)/2\sim 0.18$ MeV of the shell model fit;
only with the use of the most recent extended soft--core potential, ESC08, which was designed 
to explain a large amount of Hypernuclear data in addition to $NN$ and $YN$ scattering data,
one can obtain an improvement of the predictions for the small $\Lambda$--nucleus spin--orbit splittings 
observed experimentally \cite{Ya10,Ri10b}. Nevertheless, the ESC08 potentials
predict an incorrect ordering for the ground state spin--doublets in $^{11}_\Lambda$B and 
$^{12}_\Lambda$C.


Cluster models based on one--range Gaussian $\Lambda N$ effective interactions 
have been used since the 80's \cite{Ba85}.
Other cluster models were based on G--matrix effective interactions \cite{Ya90} 
derived from free--space $YN$ meson--exchange potentials
(NSC89 \cite{Ri89} and the JA and JB 
versions of the Juelich potential \cite{Ju89}): none of these approaches could
reproduce the ($1^+,0^+$) splitting observed for the ground state spin--doublet in $^4_\Lambda$H 
and $^4_\Lambda$He. Only with the introduction of the NSC97 Nijmegen potentials 
\cite{Ri99}, cluster model calculations based on $YN$ G--matrix interactions 
\cite{Ya94} could reproduce the above spin--doublet separation: six versions, a to f, were 
introduced for the NSC97, each one of them predicting a different $\Lambda N$ spin--spin 
interaction, with the NSC97a and NSC97f versions giving the best description of the spin--doublet 
separations in $^4_\Lambda$H and $^4_\Lambda$He. 

After the advent of $\gamma$--ray spectroscopy, studies of the spin--dependence of 
the effective $\Lambda N$ interaction were performed for $p$--shell Hypernuclei 
by comparing the obtained data with three-- and four--body calculations based on 
the Gaussian expansion method, by adopting NSC97, ESC04 and ESC08 $YN$ 
interactions \cite{hiyama,Hi10b,nemura,Hi99,Hi06}.

In Ref.~\cite{Hi06} the $\Lambda N$ spin--spin interaction was initially set to reproduce 
the 1.04 MeV ($1^+,0^+$) doublet spacing observed in $^4_\Lambda$H, while the
$\Lambda$--spin--dependent spin--orbit interaction was fixed to reproduce the
43 keV $(3/2^+,5/2^+)$ doublet spacing observed in $^9_\Lambda$Be.
Then, by an $\alpha+p+n+\Lambda$ four--body cluster model based on a $\Lambda N$ G--matrix 
obtained from the NSC97f potential, for the $(3/2^+,1/2^+)$ and $(7/2^+,5/2^+)$ 
spin--doublets in $^7_\Lambda$Li the authors predicted 690 keV and 460 keV separation energies, 
respectively, in good agreement with the $\gamma$--ray data reported in Table \ref{exp-sm}.
In Ref.~\cite{Hi06} it was thus demonstrated that it is possible to explain consistently
the ground and excited state spin--doublet splittings in $^7_\Lambda$Li
and the $(3/2^+,5/2^+)$ spin--doublet splitting in $^9_\Lambda$Be.
However, a similar $3\alpha+\Lambda$ four--body calculation
strongly overestimated the spin--orbit splittings of the $\Lambda$ 
single--particle states observed in $^{13}_\Lambda$C \cite{Hi00}.

In Ref.~\cite{Hi10b} the spin--orbit part of the $\Lambda N$ interaction was studied with a
few--body cluster model for $^9_\Lambda$Be ($2\alpha+\Lambda$) and $^{13}_\Lambda$C
($3\alpha+\Lambda$) based on a $\Lambda N$ G--matrix interaction derived form
ESC08 potentials. The G--matrix interaction was fixed to reproduce the empirical 
values of the $\Lambda$ binding energies in $^5_\Lambda$He, $^{9}_\Lambda$Be
and $^{13}_\Lambda$C; this required the consideration of density--dependent
effects for the interacting hyperon, which were introduced by a phenomenological $\Lambda NN$ 
three--body interaction. The best result for the $(3/2^+,5/2^+)$ spin--doublet separation 
in $^{9}_\Lambda$Be was predicted to be of 80 keV with the ESC08a potential
(instead of the experimental $(43\pm 5)$ keV), largely 
improving the predictions based on NSC97 potentials thanks to a less attractive
symmetric spin--orbit interaction and a more repulsive antisymmetric spin--orbit
interaction. Unfortunately, the best prediction for the splitting between the
$(\Lambda\, p_{1/2})$ and $(\Lambda\, p_{3/2})$ states in $^{13}_\Lambda$C
was obtained as 400 keV (instead of the experimental $(152\pm 54 \pm 36)$ keV), 
again with the ESC08a potential.

We can thus summarize the preformed G--matrix calculations as follows.
The approaches based on the NSC97 and ESC04 interactions 
can reproduce the spin--spin strength of the $\Lambda N$ interaction derived
with the shell model analysis of $\gamma$--ray spectroscopy experiments, but strongly 
overestimate the $\Lambda$ single--particle spin--orbit splittings. Only with the most 
recent version of the Nijmegen extended soft--core potentials, ESC08 \cite{Ri10b}, this 
overestimation problem can be partially solved \cite{Ya10}.

Quark--based models of the baryon--baryon interaction were considered in Hypernuclear
structure calculations. Some of them can improve in a natural way the prediction of the 
(empirically large) antisymmetric spin--orbit strength $S^-=(S_\Lambda-S_N)/2$ of the
$\Lambda N$ effective interaction (they reproduce the small $\Lambda$ spin--orbit separations) 
\cite{Ts98,Gu08,Sh02}, but others still have problems in reproducing the phenomenological value of 
$S^-$ \cite{Fu07,Fu04}. 

The Kyoto-Niigata group \cite{Fu07,Fu04} describes the free--space $YN$ interaction within 
the QCD inspired spin--flavour SU(6) quark model supplemented by an effective--meson exchange 
potential (containing scalar, pseudoscalar and vector mesons) acting between quarks. An 
$\alpha+\alpha+\Lambda$ three--cluster Faddeev calculation was performed in Ref.~\cite{Fu04}
for $^9_\Lambda$Be. While the energies of the $1/2^+$ ground state and the $3/2^+$ excited state 
could be reproduced, the obtained $(3/2^+,5/2^+)$ splitting overestimated the $\gamma$--ray 
measurement by a factor ranging from 3 to 5. The authors attribute this result to an inappropriate
description of the $P$--wave $\Lambda N$--$\Sigma N$ coupling.
The $\Lambda$ spin--orbit interaction strengths were previously examined in Ref.~\cite{Ko00},
by a $\Lambda N$ G--matrix interaction derived from the SU(6) quark model, 
in a calculation for symmetric nuclear matter at normal density.
The results imply a ratio $S_\Lambda /S_N$ between the $\Lambda$--spin--dependent spin--orbit
and the nucleon--spin--dependent spin--orbit effective parameters ranging from 0.08 and 0.25,
which surely improves the predictions of meson--exchange approaches 
but overestimates the phenomenological ratio, $S_\Lambda /S_N \sim 0.04$.

Hypernuclear structure studies were also performed with the quark--meson coupling (QMC) model. 
Within this framework, which is similar to the self--consistent quantum hadrodynamics,
the baryon--baryon interactions are described by the exchange of scalar and vector 
mesons, while the SU(6) quark model is adopted for the bound nucleons and hyperon,
with a baryon internal structure modeled by the MIT bag. 
The first application of the QMC model was done in Ref.~\cite{Ts98}, where the exchange of $\sigma$, 
$\omega$ and $\rho$ mesons and a phenomenological $\Lambda N$--$\Sigma N$ coupling were taken into 
account. Hypernuclei with closed--shell nuclear core were considered.
This approach naturally incorporates a weak spin--orbit interaction for the 
$\Lambda$ in a Hypernucleus, due to the explicit quark structure of the $\Lambda$ given by the SU(6) 
quark model (where non--strange mesons only couple to light $u$ and $d$ quarks of baryons and the 
$\Lambda$ spin is carried by the $s$ quark), but the predicted $\Lambda$ single--particle energies 
overestimates the experimental data.
More recently \cite{Gu08}, the QMC model was improved by including the self--consistent 
effect of the scalar field on the one--gluon--exchange hyperfine interaction between quarks,
that in free space leads to the $\Delta$--$N$ and $\Sigma$--$\Lambda$ mass splittings. 
With the hyperfine interaction, the MIT bag model can reproduce the $\Delta$--$N$ mass splitting. 
A single--particle shell model was used and Hypernuclei with closed--shell nuclear cores
were considered. Again, the QMC calculation leads to a natural explanation of the small $\Lambda$ 
spin--orbit splittings. Moreover, improved predictions are obtained for the $\Lambda$ 
single--particle energies, although the calculation  tends to overbind the less bound $\Lambda$ 
states.

A quark mean field (QMF) model, in which baryons are described by the constituent quark model,
was instead considered in Ref.~\cite{Sh02}. Again, baryon--baryon interactions were
given by $\sigma$, $\omega$ and $\rho$ meson exchange, where the mesons coupled exclusively 
to light quarks and Hypernuclei consisting of a closed--shell nuclear core were studied.
Small $\Lambda$ spin--orbit splittings were obtained, while the
experimental $\Lambda$ single--particle energies were underestimated.
The QMF and QMC models are rather similar to each other.	
The underbinding in the QMF model and the overbinding in the QMC model can be explained by
the different $\Lambda$--$\sigma$ coupling adopted in the two calculations.



Only a recent calculation, with a relativistic energy density functional based
on SU(3) chiral effective field theory, predicted both $\Lambda$ single--particle energies
and $\Lambda$--nucleus spin--orbit splittings which compare well with data \cite{Fi09}.
In SU(3) chiral effective field theory, the $\Lambda N$ interaction is described as follows:
the longest range interactions are described by two--pion exchange diagrams including
an intermediate $\Sigma N$ state and by one--kaon--exchange, while the short--range
interactions are described by contact terms (scalar and vector mean fields contributions).
The mechanism responsible for the improvement of Ref.~\cite{Fi09} is an almost complete 
cancellation between the short--range scalar and vector mean fields contributions and the
intermediate--range terms originated by the two--pion exchange.

Finally, we note that studies on the spin--dependence of the $\Lambda N$ interaction
can also be performed by mesonic weak decay spectra measurements,
which allows one to assign the spin--parity to the decaying Hypernucleus
(spin--parity of Hypernuclear states are normally assigned by angular correlations and 
polarizations of the emitted $\gamma$--rays). This will be discussed in 
Sec.~\ref{subsubsec:mwd_p}.

\subsubsection{New Generation $\gamma$--ray Experiments at J--PARC}
\label{new-gamma}   

The next generation of $\gamma$--ray experiments is planned at J--PARC with the
Hyperball--J detector, by using intense and pure $K^-$ beams from the 50 GeV
proton synchrotron \cite{tamu,Ta10}.
Improved data for $s$-- and $p$--shell Hypernuclei ($^4_\Lambda$He, $^{10}_\Lambda$B,
and $^{11}_\Lambda$B) are expected from the E13 experiment. They will help in clarifying 
the present inconsistencies of the shell model approach. The experiment on 
the $1^+\to 0^+$ transition energy in $^4_\Lambda$He is 
designed to determine the real effect of the charge symmetry breaking in the $\Lambda N$ 
interaction. Then, $sd$--shell Hypernuclei will be considered by E13, starting from 
$^{19}_\Lambda$N, which requires a shell model approach with 8 $(sd)s_\Lambda$ additional 
matrix elements, up to $A=30$, by using the ($K^-,\pi^-$) production reaction. 
The experiments will allow one to determine the spin--parity of the observed 
states and the size of $sd$--shell Hypernuclei, thus also sheding light on the $A$--dependence 
of the spin--spin parameter $\Delta$. At a later stage of E13, studies on medium and 
heavy Hypernuclei (up to $A=208$), again with the ($K^-,\pi^-$) reaction, could provide 
information on the $P$--wave $\Lambda N$ 
interaction, through the observation of inter--shell $E1(p_\Lambda\to s_\Lambda)$ transitions.


\subsection{Impurity Nuclear Physics}
\label{impurity}
   
In a $\Lambda$--Hypernucleus the Pauli principle allows the hyperon
to occupy the $s$--level of the $\Lambda$--nucleus mean field potential, thus making
the hyperon a good probe of the inner part of the Hypernucleus.
A $\Lambda$ hyperon bound in a Hypernucleus can thus be considered as an impurity
for the nuclear core. Some properties of the nuclear core, for instance the size,
shape, shell structure and collective motion, are modified by the presence of the
strange baryon.

\subsubsection{The $\Lambda$ Glue--Like R$\hat{\mathrm{o}}$le}
\label{glue-like}

Theoretical studies predicted a contraction of the nuclear core caused by 
the presence of the $\Lambda$ \cite{Mo83,Hi99}: the hyperon plays a 
glue--like r$\hat{\mathrm{o}}$le in Hypernuclei.

This shrinking effect was then confirmed experimentally when the KEK--E419 
$\gamma$ spectroscopy experiment sho-wed that the $\Lambda$ makes smaller 
a loosely--bound nuclear core such as $^6$Li \cite{Ta01}. This was possible
by the observation of the $E2(5/2^+\to 1/2^+)$ core 
transition in $^7_\Lambda$Li (see Fig.~\ref{scheme-gamma}). The corresponding reduced 
transition probability was derived from the lifetime (measured with the Doppler shift 
attenuation method) of the $5/2^+$ Hypernuclear state and the branching
ratio of the $E2(5/2^+\to 1/2^+)$ transition, with the result
$B(E2; ^7_\Lambda{\rm Li},5/2^+\to 1/2^+)=3.6\pm 0.5^{+0.5}_{-0.4}$ e$^2$ fm$^4$.

$B(E2)$ turns out to be approximately proportional to the fourth power of the nuclear core 
radius. The $E2(5/2^+\to 1/2^+)$ Hypernuclear transition is induced by the $E2(3^+\to 1^+)$ 
transition of the $^6$Li core. In the weak--coupling limit, Ref.~\cite{Hi99} introduced 
a size factor
\begin{equation}
S=\left[\frac{9}{7}\frac{B(E2; ^7_\Lambda{\rm Li},5/2^+\to 1/2^+)}
{B(E2; ^6{\rm Li},3^+\to 1^+)}\right]^{1/4}\, 
\end{equation}
which provides the change in the $^6$Li core radius due to the addition
of the hyperon. From a previous measurement of $B(E2; ^6{\rm Li},3^+\to 1^+)$
a sizable shrinkage of the $^6$Li core was obtained:
$S=0.81\pm 0.04$. This result agrees with the cluster model calculation of Ref.~\cite{Mo83}. 
By a more recent $\alpha+n+p+\Lambda$ cluster model \cite{Hi99} one can interpret the 
above size factor as a reduction of the distance between the $\alpha$ cluster and the center of 
mass of the $np$ pair by $(19\pm 4)$\%, while the shrinkage of the distance between the proton 
and the neutron of the $np$ pair is negligible.


The nuclear core compression effect is expected also in heavier Hypernuclei,
although generally at a lower level \cite{Ta02,Ya11}.
In Ref.~\cite{Hi97} a $3\alpha+\Lambda$ four body model was applied to the
study of the shrinkage in the nuclear core of $^{13}_\Lambda$C. 
This Hypernucleus has a $^{12}$C core with peculiar characteristics:
the $0_1^+$ ground state has a shell--model--like compact structure, while
the $2_2^+$ excited state at $E_x=7.65$ MeV has a loosely bound $3\alpha$ structure
which can hardly be explained in a shell model approach. Because of this behaviour,
for the $^{13}_\Lambda$C$(1/2_1^+)$ and $^{12}$C$(0_1^+)$ ground states the presence of the 
hyperon implies a few \% shrinkage for the distance between two $\alpha$ clusters, but
for $^{13}_\Lambda$C$(5/2_2^+)$ a significant compression of the $\alpha$-$\alpha$ 
distance, of about 28\%, is predicted with respect to the same distance in the 
$^{12}$C$(2_2^+)$ core \cite{Hi97}.


\subsubsection{Deformation of $\Lambda$--Hypernuclei}
\label{deformation}

Various open--shell nuclei are deformed in the ground state. This deformation
results in the existence of quadrupole transitions and rotational spectra which can
be observed experimentally.

Relativistic mean field models \cite{Wi08}, non--relativistic Skyrme--Hartree--Fock
approaches \cite{Zh07,Sc10b,Wi11} and the framework of non--relativistic nuclear
energy density functionals \cite{Ya11} were applied to the study of
the collective motion and the deformation properties of Hypernuclei.

The relativistic mean field model of Ref.~\cite{Wi08} predicted a strong reduction of the
deformation for $^{12}$C and some of the $sd$--shell nuclear cores in the presence of a 
$\Lambda$ hyperon in the $s$ orbit: in some cases, the change is from an oblate shape (this is 
the
case of $^{28}$Si) to a spherical shape (the $^{28}$Si core in the $^{29}_\Lambda$Si
Hypernucleus). This result does not
seem to be corroborated by other studies, for instance by the Skyrme--Hartree--Fock approach
of Ref.~\cite{Wi11}, which found that the gross features of the potential energy surfaces
in the $(\beta\,,\gamma)$ plane ($\beta$ and $\gamma$ being the Quadrupole deformation 
parameters) remain basically unaltered by the addition of the $\Lambda$. 

These predictions need empirical confirmation. Future $\gamma$--ray spectroscopy experiments at 
J--PARC could help in clarifying the deformation properties and the collective motions in 
Hypernuclei by measuring the $B(E2)$ values and possibly the excitation energies in the rotational (and 
vibrational) bands.

\subsubsection{Neutron--Rich $\Lambda$--Hypernuclei}
\label{neutron-rich}

It is well known from experiments with radioactive nuclear beams
that nuclei produced near the neutron (proton) drip line exhibit a
neutron (proton) halo, i.e., a density of weakly bound valence neutron(s) (proton(s)) 
which extend well beyond the radius of the nuclear core. Neutron--rich nuclei
are particularly important in nuclear astrophysics, as they play a r$\hat{\mathrm{o}}$le
in neutron capture nucleosynthesis.

\begin{figure} [h]
\begin{center}
\resizebox{0.4\textwidth}{!}{%
  \includegraphics{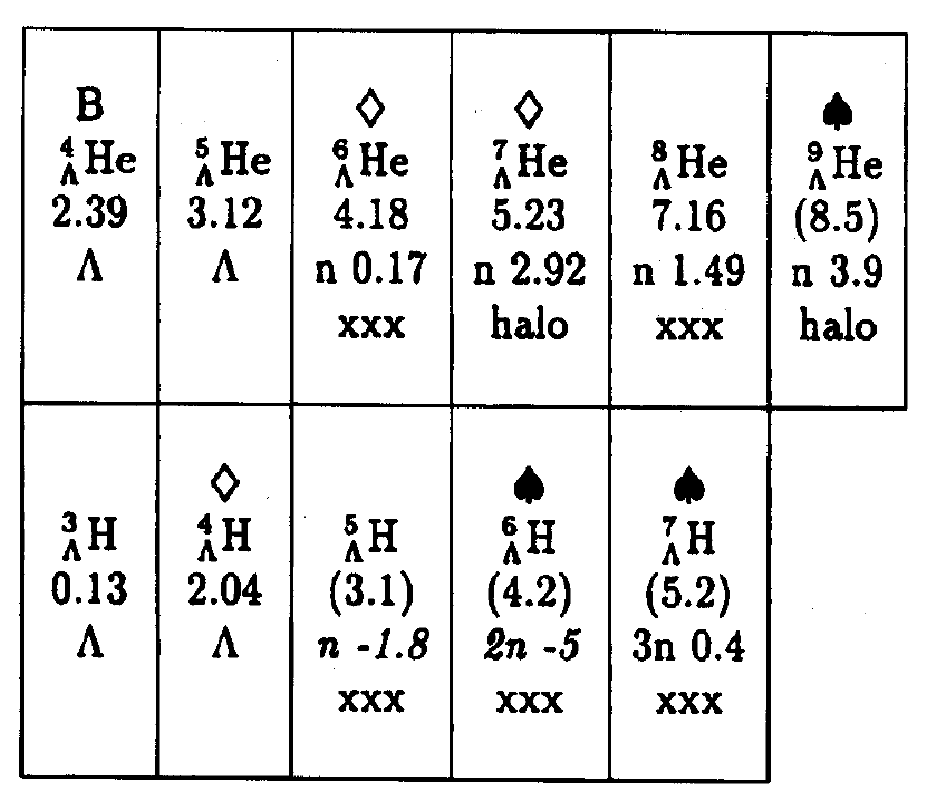}
}
\caption{Part of the light $\Lambda$--Hypernuclei chart with binding energy and particle 
instability threshold in MeV. Unstable cores are indicated by xxx; $\Lambda$--Hypernuclei 
produced by the (K$^{-}$, $\pi^{0}$) reaction, the so--called mirror Hypernuclei, are 
indicated by $\diamondsuit$; $\Lambda$--Hypernuclei produced by the (K$^{-}$, $\pi^{+}$) reaction, 
neutron--rich Hypernuclei, by $\spadesuit$. From Ref.~\protect{\cite{Ma95}}.}
\label{fig:maj}       
\end{center}
\vspace{-4mm}
\end{figure}

Since most of the Hypernuclear experiments performed to date with kaon, pion and electron 
beams used targets of stable nuclei, spectroscopy study of Hypernuclei has been 
mainly restricted to systems in the $\beta$--stability valley. 
However, as a consequence of the modification of nuclear structure introduced by the 
presence of the hyperon, one can expect that particle bound neutron--rich (proton--rich) 
Hypernuclei exist even for large (small) values of the neutron to proton ratio
\cite{Ma95}. Note that we do not consider as ``neutron--rich'' Hypernuclei obtained by the 
reaction (\ref{eq:eq3}), that are better defined as ``mirror'' Hypernuclei of those obtained by 
reactions (\ref{eq:eq1}) and (\ref{eq:eq2}).
Fig.~\ref{fig:maj} shows some of the predicted neutron--rich Hypernuclei.

Various neutron--rich Hypernuclei are in principle accessible to the $(K^-, \pi^+)$ and 
$(\pi^-, K^+)$ double charge--exchange reactions using stable targets.
In particular, thanks to the glue--like r$\hat{\mathrm{o}}$le of the hyperon, one could produce 
a Hypernucleus with a neutron halo even by starting from a weakly unbound nuclear core
(see the cases of Fig.~\ref{fig:maj} indicated by xxx), thus 
possibly extending the neutron drip line beyond the standard limits of neutron--rich nuclei 
\cite{Vr98,Zh08}. We remind that this effect was predicted already in the early days of Hypernuclear 
Physics \cite{dalitz_levi}.

Hypernuclei with a neutron or proton excess are suitable systems for studying  
the $\Lambda N$ effective interaction, the $\Lambda$--nucleus mean field potential
and the behaviour of the $\Lambda$ hyperon in a low density neutron or proton halo matter. 

Relevant contributions of the coherent $\Lambda N$--$\Sigma N$ coupling and the associated
$\Lambda NN$ three--body interaction are predicted for Hypernuclei with a neutron excess
\cite{Um09,Um11}. Studies of neutron--rich Hypernuclei could thus put constraints on 
the $\Lambda NN$ interaction. In a shell model approach within a perturbation 
theory framework, the second order $\Lambda N\to \Sigma N \to \Lambda N$ 
interaction is found to provide a contribution to the spin--doublet separations in neutron--rich 
Hypernuclei which is of the same order of magnitude of the contribution from the direct 
$\Lambda N\to \Lambda N$ interaction \cite{Um11}. Moreover, the
$\Sigma$ hyperon admixture in neutron--rich Hypernuclei is found to increase
with the isospin, i.e. with the neutron excess number $N-Z$ \cite{Um09,Ak00,Ak08}, 
thus it could reveal particularly important also in studies of the composition and 
equation of state of neutron stars; in various hadronic models, these compact 
stars indeed have a substantial content of neutrons as well as hyperons,
but particularly the effect of $\Sigma$ hyperons is not yet known.

Since the valence nucleons in Hypernuclei with neutron and proton halos are
weakly bound, these systems can be conveniently described by resorting to cluster models.
Halo and skin structures were studied theoretically in Ref.~\cite{Hi96,Hi09}
with a three--body cluster model for $A=6$ ($^6_\Lambda$He and $^6_\Lambda$Li) and $A=7$ 
($^7_\Lambda$He and $^7_\Lambda$Li and $^7_\Lambda$Be) Hypernuclei.
The $\alpha+\Lambda +N$ three--body ($\alpha+\Lambda +N+N$ four--body) model 
reproduces the binding energies of the ground states observed in $A=6$ ($A=7$) Hypernuclei.
The case of $^6_\Lambda$He is particularly interesting, as its ground state
is only 0.17 MeV below the $^5_\Lambda$He$+n$ threshold and the $^5$He core
is unbound.
For $^6_\Lambda$He the calculation indeed predicted a structure consisting of three 
distinct layers of matter: an $\alpha$ nuclear core, a $\Lambda$ skin and a neutron halo. 
The halo of the valence neutron is more extended than in the neutron halo nucleus
$^6$He, the $\Lambda$ density distribution clearly extends beyond the $\alpha$
core but is well confined within the volume of the neutron halo.
Analogous three layer structure of matter are exhibited for $A=7$ Hypernuclei.

Two--body reactions in which neutron--rich Hypernuclei could be produced are:
\begin{equation}
K^{-} + {^{A}Z}  \rightarrow {^{A}_{\Lambda}(Z-2)} + \pi^{+}\, ,
\label{prodK}
\end{equation}  
induced on nuclear targets by stopped or in flight K$^{-}$, and
\begin{equation}
\pi^{-} + {^{A}Z}  \rightarrow {^{A}_{\Lambda}(Z-2)} + K^{+}\, .
\label{prodpi}
\end{equation}  
by in flight $\pi^{-}$. 
In these double charge--exchange reactions, two protons of the target nucleus are replaced by a $\Lambda$ and a neutron.

The simplest description of the above reactions is a two--step process on two different protons of the same nucleus, converting them into a neutron 
and a $\Lambda$,  with the additional condition that the final nuclear system is bound. For (\ref{prodK}) it should mean a $K^{-} p \rightarrow \Lambda \pi^{0}$ 
reaction followed by $\pi^{0} p \rightarrow n \pi^{+}$ , for (\ref{prodpi}) a  $\pi^{-} p \rightarrow n \pi^{0}$ reaction followed by $\pi^{0} p \rightarrow K^{+} \Lambda$ 
or $\pi^{-} p \rightarrow K^{0} \Lambda$ followed by $K^{0} p \rightarrow K^{+} n$. 
Another mechanism which was proposed \cite{Ak00,Ak08} is a single--step process such as $\pi^-pp\to K^+\Sigma^- p\to K^+ \Lambda n$
via the $\Sigma$--admixture in the $\Lambda$--Hypernuclear state due to the coherent $\Sigma^{-} p \rightarrow \Lambda n$ coupling. 
Owing to these features, both single--step and two--step processes are expected to occur at a much lower rate than 
the production of normal $\Lambda$--Hypernuclei by means of the usual two--body reactions.
According to the study of Ref.~\cite{Yu03}, the two--step production
reactions have a larger cross section than the one--step mechanisms even when considering their
lar-gest theoretical prediction for the $\Sigma^-$ admixture probability.
Instead, the analysis of Ref.~\cite{Ha09} favoured one--step reactions over two--step reactions
and larger $\Sigma^-$ admixture probabilities.

The first experimental attempt to produce neutron--rich Hypernuclei by the reaction (\ref{prodK}) was carried out at KEK \cite{Ku96}. 
An upper limit (per stopped K$^{-}$) was obtained for the production of $^{9}_{\Lambda}$He, $^{12}_{\Lambda}$Be and $^{16}_{\Lambda}$C Hypernuclei: 
the results are in the range (0.6$\div$2.0) $\cdot$ 10$^{-4}$/K$^{-}_{\rm stop}$, while the theoretical predictions \cite{tretyak01} for 
$^{12}_{\Lambda}$Be and $^{16}_{\Lambda}$C lie in 
the interval (10$^{-6}\div$10$^{-7}$)/K$^{-}_{\rm stop}$, that is at least one order of magnitude lower than the experimental results and three orders 
of magnitude smaller than the usual (K$^{-}_{\rm stop}$, $\pi^{-}$) one--step reaction rates on the same targets (10$^{-3}$/K$^{-}_{\rm stop}$).

\begin{figure} [h]
\begin{center}
\resizebox{0.5\textwidth}{!}{%
  \includegraphics{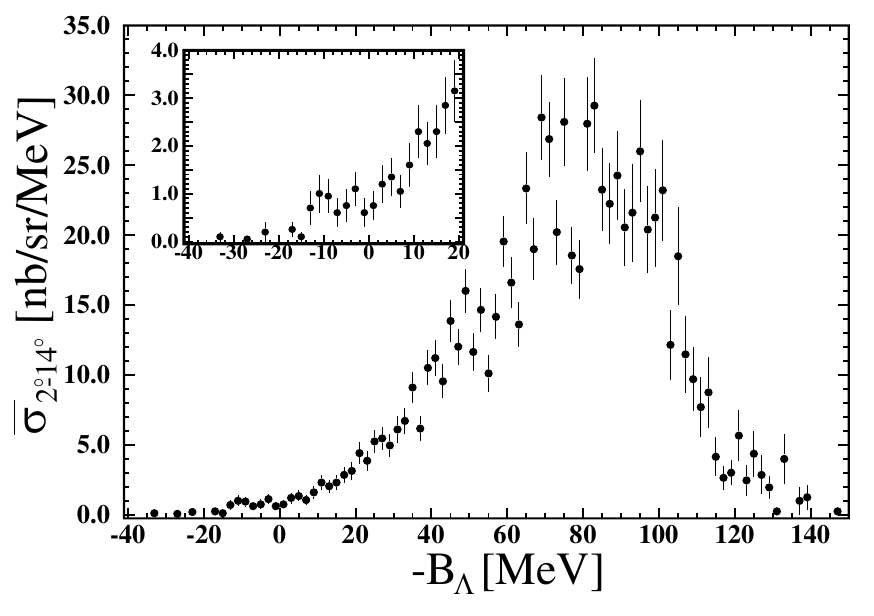}
}
\caption{Binding energy spectrum in the ($\pi^{-}$, K$^{+}$) reaction on a $^{10}$B target at 1.2 GeV/c. In the inset an expanded view containing the 
$\Lambda$ bound region is shown. 
From Ref.~\protect{\cite{Sa05}}.}
\label{fig:saha}       
\end{center}
\vspace{-4mm}
\end{figure}

Another KEK experiment \cite{Sa05} reported the observation of $^{10}_{\Lambda}$Li with the ($\pi^{-}$, K$^{+}$) double charge--exchange reaction on a 
$^{10}$B target. This experiment was affected by much less background than in the previous one with the $(K_{\rm stop}^-,\pi^+)$ reaction.
Figure \ref{fig:saha} shows the binding energy spectrum obtained by Ref.~\cite{Sa05}: the production cross section (in nb/sr/MeV) is drawn versus the 
$\Lambda$ binding energy (in MeV). 
The results are not directly comparable with theoretical calculations \cite{Yu03} since no discrete structure was observed and the 
production cross section was integrated over the whole bound region (0 MeV $<$ B$_{\Lambda}$ $<$ 20 MeV). 
Significant production yields were observed in the $\Lambda$ bound state region,
with an integrated cross section of $11.3\pm 1.9$ nb/sr ($5.8\pm 2.2$ nb/sr) for a 1.2 GeV/c
(1.05 GeV/c) incident momentum and a scattering angle in the laboratory system
ranging from $2^\circ$ to $14^\circ$. Unfortunately, no clear peak was observed in the $\Lambda$ bound region, possibly due to the low statistics and 
resolution. The above results were obtained by assuming that all counts recorded in the $\Lambda$ bound state
region correspond to the formation of the $^{10}_\Lambda$Li Hypernucleus.
The measured cross sections are in disagreement with the theoretical study of Ref.~\cite{Yu03}, which favoured a two--step production mechanism,
but agree with the prediction of Ref.~\cite{Ha09} in the case of a 1.2 GeV/c incoming momentum.
By assuming a dominant one--step production mechanism and a $\Sigma$ admixture probability of the
order of 0.1\%, Ref.~\cite{Ha09} could reproduce the magnitude of the $^{10}_\Lambda$Li production
spectrum obtained in Ref.~\cite{Sa05}.

A further attempt to observe neutron rich Hypernuclei by means of the reaction (\ref{prodK}), with K$^{-}$ at rest, was done at the DA$\Phi$NE collider at LNF by the FINUDA 
experiment \cite{nrich1}, on $^{6}$Li and $^{7}$Li targets. 
The limited data sample collected during the first run period of the experiment was used to look for the production rates per 
stopped K$^{-}$ for $^{6}_{\Lambda}$H and $^{7}_{\Lambda}$H; the inclusive $\pi^{+}$ spectra from $^{6}$Li and $^{7}$Li targets were analyzed in momentum regions 
corresponding, through momentum and energy conservation, to the B$_{\Lambda}$ values predicted by Ref.~\cite{Ma95,Ak08}. Due to the strong 
contribution of the background reactions:
\begin{eqnarray}
K^{-} + p & \rightarrow & \Sigma^{+} + \pi^{-} \nonumber \\
                &                      & \downarrow \nonumber \\
                &                      &  n + \pi^{+}
\label{Kp}
\end{eqnarray}
and
\begin{eqnarray}
K^{-} + p p & \rightarrow & \Sigma^{+} + n \nonumber \\
                &                      & \downarrow \nonumber \\
                &                      &  n + \pi^{+}
\label{Kpp}
\end{eqnarray}
which give the main contributions of the inclusive $\pi^{+}$ spectra for absorption of stopped K$^{-}$  on nuclei, and to the limited statistics, only upper 
limits could be evaluated, leading to the values ($2.5 \pm 0.4_{\rm stat}$ $^{+0.4}_{-0.1\, \rm syst}) \cdot10^{-5}$/K$^{-}_{\rm stop}$ and 
($4.5\pm0.9_{\rm stat}$$^{+0.4}_{-0.1\, \rm syst}) \cdot$10$^{-5}$/K$^{-}_{\rm stop}$ 
for the production of, respectively, $^{6}_{\Lambda}$H and $^{7}_{\Lambda}$H; also the production rate of $^{12}_{\Lambda}$Be was obtained, 
($2.0 \pm 0.4_{\rm stat}$ $^{+0.3}_{-0.1\, \rm syst}) \cdot10^{-5}$/K$^{-}_{\rm stop}$, lowering by a factor $\sim$3 the previous determination of 
Ref.~\cite{Ku96}. 

\begin{figure} [h]
\begin{center}
\vspace{-6mm}
\resizebox{0.5\textwidth}{!}{%
  \includegraphics{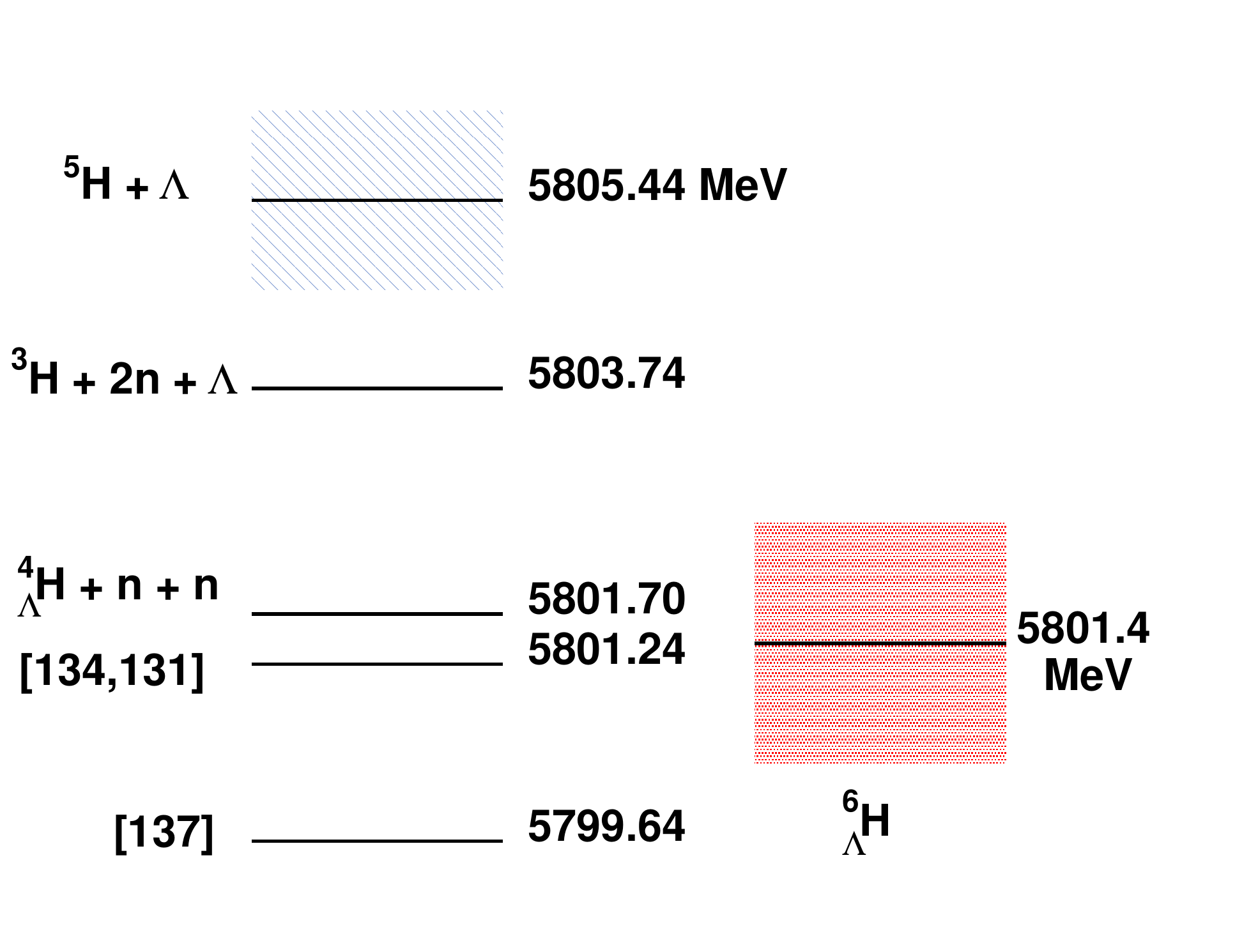}
}
\caption{Energy level scheme for $^{6}_{\Lambda}$H ground state. The red box represents the error on the mass mean value obtained by FINUDA
\protect{\cite{hadron11}}, the blue box indicates the width of $^{5}$H.}
\label{fig:L6H}       
\end{center}
\vspace{-4mm}
\end{figure}

Very recently, the full statistics collected by FINUDA on $^{6}$Li was analyzed by requiring a coincidence between the $\pi^{+}$ signaling the formation of 
$^{6}_{\Lambda}$H and its subsequent decay into $^{6}$He + $\pi^{-}$ within a very tight energy window \cite{hadron11}. 
This approach was possible thanks to the very good energy resolution and instrumental stability of the detectors. Three events were found, corresponding 
unambiguously to the formation and subsequent decay of  $^{6}_{\Lambda}$H. 
A very careful analysis showed that they could not be ascribed to possible instrumental or physical backgrounds, mainly coming from reaction (\ref{Kp}). 
The mass of  $^{6}_{\Lambda}$H deduced from the above events was found to be 5801.4$\pm$1.1 MeV, corresponding to a $\Lambda$ binding energy 
$B_{\Lambda}$=4.0$\pm$1.1 MeV with respect to the mass of the unbound $^{5}$H$+\Lambda$. 
Fig.~\ref{fig:L6H} shows the energy level scheme for the ground state of $^{6}_{\Lambda}$H deduced by FINUDA.
We remark that the measured value of $B_{\Lambda}$ compares well with the theoretical evaluation of Ref.~\cite{Ma95}
but is lower by about 1.8 MeV than the theoretical prediction $B_{\Lambda}$=5.8 MeV  
for the $^{5}$H$+\Lambda$ choice of the zero energy level obtained in Refs.~\cite{Ak00,Ak08}, which also included the
effect of the $\Lambda N$--$\Sigma N$ coupling. Note that this 1.8 MeV disagreement is close to the value 
attributed by Refs.~\cite{Ak00,Ak08} to the contribution to $B_\Lambda(^6_\Lambda {\rm H})$ arising from
the $\Lambda NN$ three--body interaction resulting from the coherent $\Lambda$--$\Sigma$ mixing.
From this experiment, which provides the first observation of the hyperheavy hydrogen $^{6}_{\Lambda}$H, we may not
conclude that the $\Lambda NN$ force is negligible, but only that its influence looks lower than predicted. The production rate of $^{6}_{\Lambda}$H,
assuming a 50$\%$ probability for the $^{6}_{\Lambda}$H$\to \pi^{-} + ^{6}$He decay, fully justified by the measured decay rates of neighbour or 
similar Hypernuclei ($^{4}_{\Lambda}$H, $^{7}_{\Lambda}$Li), is (5.9$\pm$4.0) $\cdot$10$^{-6}$/ stopped  K$^{-}$, about three orders of magnitude 
lower than the production rates for bound $\Lambda$--Hypernuclei, as expected. 

An experiment to produce $^{6}_{\Lambda}$H via the ($\pi^{-}$, K$^{+}$) 
reaction on $^{6}$Li at 1.2 GeV was recently approved at J--PARC \cite{P10} and should run as quickly as the Laboratory will be restored to operation. 
The expected resolution is 2.5 MeV, on a sample of about 100 events. Further results on neutron-- and proton--rich Hypernuclei are expected from the
HypHI experiment at GSI and FAIR by using stable heavy--ion beams and rare isotope beams \cite{hyphi}.

\subsection{In--Medium Hyperon Properties} 
\label{in-medium}

Hyperon properties in Hypernuclei are influenced by the nuclear medium. A partial 
restoration of chiral symmetry in the nuclear medium or even a partial 
quark deconfinement may affect $\Lambda$ properties such as the mass, size and 
magnetic moment for an hyperon located in the inner core of the Hypernucleus.

\subsubsection{The $\Lambda$ Magnetic Moment}
\label{magnetic}

The magnetic moment of hadrons provides interesting information on the internal structure of 
these particles. The magnetic moment of an Hypernucleus is an important observable which is
sensitive to the spin and angular momentum structure of this many--body system,
to the spin--dependent part of the $\Lambda N$ effective interaction
and to the $\Sigma$ admixture in the Hypernucleus.

The magnetic moment operator of a Hypernucleus in the weak--coupling limit is given by:
\begin{equation}
\vec \mu = \frac{1}{2} \int d^3 \vec r\, \vec r \times
[{\vec j}^{\rm em}_c(\vec r)+{\vec j}^{\rm em}_{\Lambda}(\vec r)] \, ,
\end{equation}
where ${\vec j}^{\rm em}_c$ and ${\vec j}^{\rm em}_\Lambda$ are the electromagnetic currents of
the nuclear core and of the hyperon, respectively. One can rewrite this definition as:
\begin{eqnarray}
\vec \mu &\equiv & \vec \mu_c + \vec \mu_\Lambda = (g_c \vec J_c +g_\Lambda \vec J_\Lambda)\mu_N \\
&=& [g_c \vec J + (g_\Lambda-g_c)\vec J_{\Lambda}]\mu_N\, , \nonumber
\end{eqnarray}
$\vec J_c$ ($\vec J_\Lambda$) being the total spin of the nuclear core (hyperon),
$\vec J=\vec J_c +\vec J_\Lambda$ the total spin of the Hypernucleus,
$g_c$ ($g_\Lambda$) the effective $g$--factor of the nuclear core (hyperon)
and $\mu_N$ the nuclear magneton.


Already the simple quark model reproduces quite well the magnetic moments of the stable (non--strange and 
strange) baryons by fixing the magnetic moments of the $u$, $d$ and $s$ quarks to reproduce
the magnetic moments of neutron, proton and $\Lambda$. 
For free $\Lambda$ and $\Sigma$ hyperons the measured values of the $g$--factors are:
$g^{\rm free}_\Lambda=-1.226\pm0.008$, $g^{\rm free}_{\Sigma^+}=4.916 \pm 0.020$, 
$g^{\rm free}_{\Sigma^-}=-2.320 \pm 0.050$ and $g^{\rm free}_{\Sigma \Lambda}=3.22 \pm 0.16$ 
(the last being the $\Sigma^0 \to \Lambda$ transition $g$--factor
for the $\Sigma^0\to \gamma \Lambda$ electromagnetic transition) \cite{Na10}.
The change in the quark structure of a $\Lambda$ bound in the inner core of a Hypernucleus is 
expected to introduce a variation in the $\Lambda$ magnetic moment compared to the above free--space value.

Before discussing the important question of a possible deviation
of $g_\Lambda$ in Hypernuclei from the $g^{\rm free}_\Lambda$ value,
a related issue has to be clarified: it concerns the r$\hat{\mathrm{o}}$le played by the nuclear 
core in the Hypernuclear magnetic moments.
Relativistic mean field models neglecting the $\Lambda N$--$\Sigma N$ coupling
predicted Hypernuclear magnetic moments which are very close to the 
(non--relativistic) Schmidt values \cite{Ga91}
(the Schmidt value is $\mu=\mu_\Lambda=g_\Lambda\, \mu_N/2=-0.613\, \mu_N$ for a Hypernucleus 
with a closed shell nuclear core and an $s$--level $\Lambda$).
Ref.~\cite{Co92} showed that the global core contribution to the Hypernuclear magnetic
moment is vanishingly small ($\mu_c\sim 0$) due to the large contribution of the
$\omega \Lambda \Lambda$ tensor coupling induced by the $\omega$ vector field,
which almost cancels the $\omega \Lambda \Lambda$ vector coupling contribution.

However, in the presence of an important $\Sigma$ admixture in Hypernuclei
and due to large differences among the above free $\Lambda$ and $\Sigma$ hyperons $g$--factors,
the value of the Hypernuclear magnetic moments are expected to deviate from the Schmidt values.
This is especially true for $p$--shell and heavier odd--$A$ Hypernuclei with isospin 1 nuclear 
cores, as demonstrated in the qualitative treatment of Ref.~\cite{Do95} based on perturbation
theory. For instance, $\mu(^{15}_\Lambda{\rm C})\sim \mu_\Lambda+\sqrt{2}\, \beta\, 
\mu_{\Sigma \Lambda}$, where $\beta$ is evaluated in the range from $+0.03$ to $+0.07$ in 
Ref.~\cite{Do95}. According to this result, 
the deviation of the Hypernuclear magnetic moments from the Schmidt values can be interpreted as a 
signal of the relevance of the $\Lambda N$--$\Sigma N$ coupling. More quantitative studies 
within the shell model are needed to settle this important issue.

The effect of the meson exchange currents on the magnetic moment of light Hypernuclei
is studied in Ref.~\cite{Sa97} with an effective Lagrangian method in the 
harmonic oscillator shell model. The calculation takes into account the $NN$
and $\Lambda N$ $\pi$-- and $K$--exchange currents and the nuclear core polarization.
For $^5_\Lambda$He the magnetic moment is reduced in absolute value by 9\% from the Schmidt 
value $\mu_\Lambda=-0.613\, \mu_N$: $\mu(^5_\Lambda {\rm He})=\mu_\Lambda+0.054 \mu_N=-0.559\, \mu_N$.
This change is originated only by the isoscalar $\Lambda N$ $K$--exchange currents, as both the
$\Lambda$ and the $^4$He nuclear core have isospin 0. Smaller variations from the Schmidt
values are instead obtained for the ($^6_\Lambda$He, $^6_\Lambda$Li) isodoublet,
despite in these cases the magnetic moment also has an isovector part.


We come now to the question concerning the in--medium $\Lambda$ $g$--factor, $g_\Lambda$.
Since it is impossible to directly measure $g_\Lambda$, due to the very short lifetime for spin 
precession in a magnetic field, an indirect method consists in
the measurement of the reduced transition probability $B(M1)$ for the strengths of 
$\Lambda$ spin--flip transitions $M1$, which in the weak--coupling limit is 
related to $g_\Lambda$ by \cite{Da78}:
\begin{equation}
B(M1)= \frac{3}{8\pi}\frac{2J_{\rm down}+1}{2J_c+1}(g_{\Lambda}-g_{\rm c})^2\mu^2_N\, ,
\end{equation}
where $J_{\rm down}$ is the spin of the lower state of the doublet.

$B(M1)$ can be obtained experimentally by measuring the $\gamma$--ray energy 
for the $\Lambda$ spin--flip transition $\Delta E$ and the lifetime $\tau$
of the excited state (determined with the Doppler shift attenuation method): 
$B(M1)\propto 1/(\tau\, \Delta E^3)$. 
A few measurements of $B(M1)$ have already been performed, at BNL and KEK,
but with insufficient accuracy to establish the modification of the $\Lambda$
$g$--factor in a Hypernucleus: the most precise determination is indeed
$g_\Lambda=-1.04\pm 0.41$ \cite{Ta10}.

A future measurement of $B(M1)$ for the $M1(3/2^+$ $\to 1/2^+)$ transition in 
$^7_\Lambda$Li will be carried out at J--PARC E13 with a 5\% level accuracy \cite{Ta10}.
In order to study the density-- and isospin--dependence of $g_\Lambda$, experiments on
the $B(M1)$ values will be performed at J--PARC for heavier Hypernuclei. 
In experiments with heavy--ion beams such as HypHI \cite{hyphi} Hypernuclei are produced 
with large velocities and fly in free--space for tens of centimeters before decaying,
thus enabling direct measurements of the $\Lambda$ magnetic moment
by observing the spin precession in strong magnetic fields.

\section{Weak Decay of Hypernuclei}
\label{sec:decay}
$\Lambda$--Hypernuclei are produced in the ground state or in an excited state of the $\Lambda$--particle 
neutron--hole configuration. When a $\Lambda$--Hypernucleus is excited above the particle emission threshold 
it decays dominantly by the strong interaction, through nucleon or cluster emission; the remaining strange nuclear 
system then de--excites to its ground state via electromagnetic transitions.    

A $\Lambda$--Hypernucleus in the ground state decays to non--strange nuclear 
systems through the mesonic (MWD) or non--mesonic (NMWD) weak decay 
mechanisms.
 
In MWD the $\Lambda$ hyperon decays into a nucleon and a pion 
in the nuclear medium, similarly to the weak decay mode in free space: 
\begin{equation} 
\Lambda_{\rm free}  \rightarrow  p + \pi^{-} + \mathrm{37.8~MeV}\ \ \ (B.R. = 64.2 \%) 
\label{lfreep} 
\end{equation} 
\begin{equation} 
\Lambda_{\rm free}  \rightarrow  n + \pi^{0} + \mathrm{41.1~MeV}\ \ \ (B.R. = 35.8 \%) 
\label{lfreen} 
\end{equation} 
in which the emitted nucleon carries a momentum $q \approx 100$ MeV/c, for a decay at rest, 
corresponding to a Q--value of about 40 MeV.  The branching ratios of the channels (\ref{lfreep}) and (\ref{lfreen}) 
are consistent with the empirical $\Delta$I = 1/2 rule, valid for all non--leptonic strangeness--changing 
processes, like the $\Sigma$ hyperon decay and pionic $K$ decays.  

Inside a Hypernucleus the binding energy of the $\Lambda$ ($\sim 3$ MeV for $^{5}_{\Lambda}{\rm He}$, 
$\sim 11$ MeV for $^{12}_{\Lambda}{\rm C}$, 
$\sim 27$ MeV for $^{208}_{\Lambda}{\rm U}$) further reduces the energy available to 
the final state particles; 
MWD is thus suppressed in Hypernuclei with respect to the free--space decay due 
to the Pauli principle, since the momentum of the emitted nucleon is by far 
smaller than the nuclear Fermi momentum ($k_F \sim 270$ MeV/c) in all 
nuclei except for the lightest, $s$--shell ones. 

In NMWD the $\Lambda$--Hypernucleus decays through processes which involve a weak 
interaction of the constituent $\Lambda$ with one or more core nucleons. If the pion emitted in the hadronic 
vertex $\Lambda \rightarrow \pi N$ is virtual, then it can be absorbed by the nuclear 
medium, resulting in a non--mesonic decay of the following types:
\begin{equation}
\Lambda n \rightarrow nn \quad(\Gamma_{n})~,
\label{gamman} 
\end{equation}
\begin{equation}
\Lambda p \rightarrow np \quad(\Gamma_{p})~,
 \label{gammap} 
\end{equation}
\begin{equation}
\Lambda NN \rightarrow nNN \quad(\Gamma_{2})~,
\label{gamma2} 
\end{equation}
where in parentheses we indicate their decay rates.
The channels (\ref{gamman}) and (\ref{gammap}) are globally indicated as {\it one-nucleon induced decays}, in 
particular one--neutron induced decay (\ref{gamman}) and one--proton induced decay (\ref{gammap}). 
The channel (\ref{gamma2}), referred to as {\it two--nucleon induced decay} and suggested in 
Ref.~\cite{al91}, 
can be interpreted by assuming that the pion from the weak vertex is absorbed by a pair of nucleons 
($np$, $pp$ or $nn$), correlated by the strong interaction. 
Note that the non--mesonic processes can also be mediated by the exchange of mesons more massive 
than the pion. 

The NMWD mode is possible only in nuclei; the Q--value of the elementary reactions (\ref{gamman})--(\ref{gamma2})  
($\sim 175$ MeV) 
is high enough to avoid any Pauli blocking effect, being the outgoing nucleons momenta as high as $\sim 420$ MeV/c 
for the one--nucleon induced process and $\sim 340$ MeV/c for a two--nucleon induced process, if the available 
energy is equally distributed among the final state particles; the final nucleons thus have a great probability to escape from the 
nucleus. Indeed, the NMWD dominates over the MWD for all but the $s$--shell Hypernuclei and only for very light systems ($A\leq 5$) 
the two decay modes are competitive.

The NMWD is a four--fermion, $\Delta S=1$, baryon--baryon weak interaction 
and represents the only 
practical way to obtain information on the weak process $\Lambda N \rightarrow n N$, which represents the first 
extension of the weak, $\Delta S = 0$, $N N \rightarrow N N$ interaction to the strange baryon sector. 
We recall that the short $\Lambda$ lifetime prevents from producing hyperon targets or beams  of suitable intensity and only 
scarce $\Lambda N$ scattering data are presently available.
Moreover, the four--body $\Delta S = 1$ interactions (\ref{gamman}) and (\ref{gammap}) are the best candidate 
to allow the investigation of 
the parity--conserving part of the weak interaction, which is completely masked by the strong component 
in the $N N \rightarrow N N$ reaction \cite{AG02,alberico,outa}. 

The total decay width of a $\Lambda$--Hypernucleus
$\Gamma_{\rm T}({^A_{\Lambda}\mathrm{Z}})$ is given by the sum 
of the mesonic and the non--mesonic decay widths: 
\begin{equation} 
\Gamma_{\rm T} = \Gamma_{\rm M} + \Gamma_{\rm NM}\, , 
\label{gamma} 
\end{equation} 
where the first term can be further expressed as the sum 
of the decay widths for the emission of negative ($\Gamma_{\pi^{-}}$) and 
neutral ($\Gamma_{\pi^{0}}$) pions:
\begin{equation} 
\Gamma_{\rm M} = \Gamma_{\pi^{-}} + \Gamma_{\pi^{0}}\, ,
\label{gamma_m} 
\end{equation} 
and the second term can be written as the sum of the one-nucleon ($\Gamma_{1}$) 
and two--nucleon induced ($\Gamma_{2}$) decay widths:
\begin{equation} 
\Gamma_{\rm NM} = \Gamma_{1} + \Gamma_{2}\, ,
\label{gamma_nm} 
\end{equation} 
with $\Gamma_{1} = \Gamma_{p} + \Gamma_{n}$. 
The total decay width $\Gamma_{\rm T}(^A_{\Lambda}\mathrm{Z})$ is expressed in terms of the 
Hypernuclear lifetime by: 
\begin{equation} 
\Gamma_{\rm T}(^A_{\Lambda}\mathrm{Z}) = {\hbar}/ \tau(^A_{\Lambda}Z)\, . 
\label{tau} 
\end{equation} 

The decay observables which can be directly measured are the Hypernucleus lifetime, 
$\tau_{\Lambda}$, the branching ratios and decay rates for the MWD channels, $\Gamma_{\pi^{-}}$ and 
$\Gamma_{\pi^{0}}$, and the spectra of both MWD and NMWD light decay products (pions and nucleons). 
The total non--mesonic rate can thus be obtained in an indirect way
as $\Gamma_{\rm NM}=\Gamma_{\rm T}-\Gamma_{\rm M}$. 
The partial decay rates for the NMWD channels ($\Gamma_n$, $\Gamma_p$, $\Gamma_{np}$, etc) cannot be directly determined from the 
data due to presence of final state interactions (FSI) for the weak decay nucleons;
these rates are not quantum--mechanical observables \cite{Ba10b}. Each one of the possible elementary non--mesonic decays 
occurs in the nuclear environment, thus subsequent FSI modify the quantum numbers of the weak decay nucleons and
new, secondary nucleons are emitted as well. Daughter nuclei can hardly be detected.

A complete review of the experimental results and theoretical models describing the weak decay of Hypernuclei 
before 1990 can be found in Ref.~\cite{cohen}; more recent review papers on the results obtained from 1990 on   
and on the development of the theoretical interpretations can be found in Ref.~\cite{AG02,alberico,outa,Pa07}.

\subsection{Hypernuclear Lifetimes}
\label{subsec:tau_meas}
Among the experimental observables, the Hypernucleus lifetime $\tau$, or equivalently the total 
decay width $\Gamma_{\rm T}$, is the one which can be measured with the highest accuracy and is 
free from all the distortions and corrections connected to final state interactions of the emitted particles. 
Being an inclusive quantity, for its measurement one has to detect any of the possible products of either mesonic or
non--mesonic decays (typically protons from non--mesonic decays) as a function
of time and then fit the observed distribution with an exponential decay law.
The lifetime  measurement, moreover, is the starting point for obtaining, from the measured MWD branching ratios, 
the $\Gamma_{\rm NM}$ rates as well as the ratio $\Gamma_{n}/\Gamma_{p}$ between the one--neutron 
induced and the one--proton induced NMWD rates. However, one has to note that for the determination of the
the ``experimental'' value of $\Gamma_n/\Gamma_p$ (the same occurs for the individual rates $\Gamma_n$ and $\Gamma_p$)
one needs 1) a measurement of the nucleon emission spectra and 2) a theoretical model of nucleon FSI. 
Indeed, the nucleon spectra turn out to strongly depend on these FSI 
effects: the experimental $\Gamma_{n}/\Gamma_{p}$ ratio must thus be obtained by a deconvolution of the nucleon FSI effects 
contained in the measured spectra.

The weak decay of $\Lambda$--Hypernuclei has been studied since their discovery in 1953; the first 
determinations of $\tau$ date back to the sixties when visualizing techniques were used, by 
exposing, mainly, photographic emulsions and, less frequently, $LH_{2}$ filled bubble chambers to the 
worldwide available $K^{-}$ beams (Argonne ZGS, BNL AGS and Bevatron). In emulsions, Hypernuclei were produced 
as hyperfragments on the emulsion components, like hydrogen, carbon, nitrogen, oxygen, silver and bromine, 
with a strong prevalence of the lighter nuclei; in bubble chambers, the helium was an active target. 
Only events with unambiguously identified hyperfragments were used to determine the lifetime, and both decays at rest 
and in flight were considered. A maximum likelihood method was developed in order to extract $\tau(^A_{\Lambda}Z)$ from 
the measurements of moderation times or flight times of the selected events.  
Meager samples (1-30 events each) were typically collected due to the reduced intensity of the beams and, consequently, big statistical errors 
were obtained, ranging from $15\%$ to more than $100\%$. 
Lifetimes of $s$--shell Hypernuclei were measured for $^{3}_{\Lambda}$H \cite{prem,phillips,bohm,keyes70,keyes73}, 
$^{4}_{\Lambda}$H \cite{prem,kang65,phillips}, $^{4}_{\Lambda}$He \cite{phillips} and $^{5}_{\Lambda}$He \cite{prem,kang65,phillips,bohm}. 
The obtained values are all consistent with the free $\Lambda$ lifetime within errors.

Obvious limitations of the visualizing techniques prevent from performing a complete study of the Hypernucleus decay. 
Concerning lifetime, since the decay timing information cannot be obtained, only indirect determinations of 
$\tau(^A_{\Lambda}Z)$ were available, as discussed before. Moreover, concerning the study of the different decay modes, 
visualizing devices are blind to neutral particles, such as $\gamma$ and neutrons; thus the study was limited to the 
weak decay modes with charged particles in the final state. Finally,  
the apparatuses were not able to count the number of formed Hypernuclei, obstructing the determination of the 
decay branching ratios.

As described in Sec.~\ref{prod}, starting from the eighties, the counter experiments at BNL AGS and at the 12 GeV PS at KEK opened a new era in the 
study of Hypernuclear weak decay, overcoming the difficulties of the visualizing techniques. These experiments 
used the $(K^{-},\pi^{-})$ and $(\pi^{+},K^{+})$ reactions to copiously produce Hypernuclei, thus allowing to 
perform coincidence measurements for the study of weak decay. Actually, the counter experiments dedicated to 
Hypernuclear spectroscopy started in the early seventies, but high intensity $K^{-}/\pi^{+}$ beams and large solid angle 
spectrometers for the high statistics ($10^{3}-10^{5}$ events) production of 
$\Lambda$--Hypernuclei, needed for the 
decay coincidence study, became available only about ten years later. 

For a direct determination of the lifetime it is necessary to measure the delay in the emission of the decay products from the 
ground state of the Hypernucleus. In counter experiments this normally requires the detection of a particle emitted in the decay in 
coincidence with a particle originating from the production reaction, $\pi^{-}$ or $K^{+}$,  which tags 
the formation of the ground state or of a low--lying excited state which is known to de--excite to the ground state by 
electromagnetic decay; in this way it is possible to identify uniquely the decaying system. 

The typical apparatuses operated both at BNL and at KEK and consisted of double spectrometers to measure with high precision the 
momenta of the incident ($K^{-}$ or $\pi^{+}$) and scattered ($\pi^{-}$ or $K^{+}$) mesons and to determine, by simple momentum 
and energy conservation, the mass of the produced Hypernuclear state. Around the reaction target, coincidence detectors were placed 
to detect charged and neutral  decay particle and to measure the decay time; 
for charged particle the identification was performed by specific ionization and range measurements and the energy was determined 
by the range, while for neutrals both the identification and the energy evaluation were performed by time of flight measurements. 
For lifetime determination, the formation time was given by a fast detector located in front of the target and the delay in the emission 
was determined by the difference between decay and formation time. 
Lifetimes of light-- and medium--$A$ Hypernuclei were extracted from the measured delay time distributions for 
$^{4}_{\Lambda}$H \cite{outa95,outa98}, $^{4}_{\Lambda}$He \cite{zeps,outa98,parker}, $^{5}_{\Lambda}$He \cite{szym,kame05}, 
$^{9}_{\Lambda}$Be \cite{szym}, $^{11}_{\Lambda}$B \cite{szym,grace,park}, 
$^{12}_{\Lambda}$C \cite{szym,kame05,grace,park}, $^{27}_{\Lambda}$Al, $^{28}_{\Lambda}$Si  and $_{\Lambda}$Fe \cite{park}, 
with statistical errors ranging from $3\%$ to $20\%$. BNL and KEK values, when available for the same Hypernuclear specie, are consistent within errors. 

\begin{figure} [h]
\begin{center}
\resizebox{0.5\textwidth}{!}{%
  \includegraphics{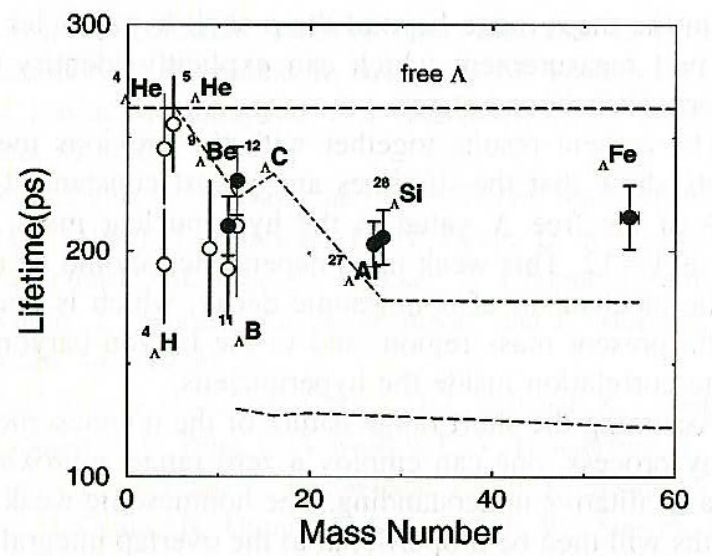}
}
\caption{Mass--dependence of Hypernuclear lifetime, from Ref.~\cite{park}. Full circles: KEK--E307 results \cite{park};
open circles: previous counter experiments \cite{outa98,szym,grace}.
The dot--dashed line shows the calculations by Ref.~\cite{itonaga} and the dashed line the calculations by Ref.~\cite{ramos}.}
\label{fig:decay1}       
\end{center}
\vspace{-4mm}
\end{figure}

Fig.~\ref{fig:decay1}, from Ref.~\cite{park}, shows the mass number dependence of Hypernuclear lifetime for $A<60$ obtained 
by various counter experiments at BNL and KEK (open and full circles), from 1985 to 2000. The data indicate that the lifetime is quite stable 
from light-- to medium--$A$ Hypernuclei and is rather constant above $A=20$, at $\sim$210 ps, which corresponds to $\sim$80$\%$ of the free $\Lambda$ 
lifetime. This smoothly decreasing behaviour is sign of an anticorrelation between the MWD and NMWD modes: the rapid decrease of the MWD, due to Pauli 
blocking effect, seems to be balanced by the NMWD decay mechanism and the asymptotic behaviour of the lifetime indicates the presence of saturation 
properties for the $\Lambda N \rightarrow n N$ weak interaction for increasing $A$ analogous to those of the finite range $NN$ strong interaction. 

In Fig.~\ref{fig:decay1} the dot--dashed line shows 
calculations by Ref.~\cite{itonaga}, where $2\pi/\rho$ and $2\pi/\sigma$ exchange terms were added to the one--pion--exchange potential for the 
NMWD. These calculations reproduce Hypernuclear lifetime quite well, with a slight 
underestimation of the saturation value for $^{28}_{\Lambda}$Si and $^{56}_{\Lambda}$Fe. The 
dashed line represents lifetime calculations by Ref.~\cite{ramos} in which a one--pion--exchange model approach is used, including 
the two--nucleon induced decay modes on strongly correlated $np$ pairs, to evaluate lifetimes for medium and heavy Hypernuclei from 
$^{12}_{\Lambda}$C to $^{208}_{\Lambda}$Pb. The calculated values are short by a $30-40\%$ with respect to the measured values. 
By a detailed study of baryon--baryon short--range correlations and of the $\Lambda$ wave function in Hypernuclei, which was obtained 
by reproducing the data on $s$ and $p$ $\Lambda$ level energies with Woods--Saxon $\Lambda$--nucleus potentials, 
this calculation was then updated in Ref.~\cite{albe2}. It was then 
possible to reproduce the experimental lifetimes from light to heavy Hypernuclei (see Fig.~4 in Ref.~\cite{albe2}).

In the medium mass region, a lifetime measurement was also performed for A$\sim$16 Hypernuclei ($^{~16}_{\Lambda}Z$) \cite{nield} using a 
2.1 GeV/nucleon $^{16}$O beam and a polyethylene target at the LBL Bevalac. The trigger for the Hypernuclear production was given by the 
decay at rest of $K^{+}$ produced together with $K^{-}$ by the associated production reaction, but the Hypernuclear mass spectrum 
could not be reconstructed and the Hypernuclear specie could not be identified. The lifetime was extracted from the recoil distance distribution, 
measured by spark chambers. Only 22 decay events were identified and the lifetime from the 
maximum likelihood fitting of the recoil distance was $\sim$86 ps, much smaller than the value for neighbor $^{12}_{\Lambda}$C and 
$^{28}_{\Lambda}$Si reported in Fig.~\ref{fig:decay1}. The datum on $^{16}$O, however, is very likely to be affected by some experimental 
drawback due to the presence of a very strong instrumental background and, unfortunately, the measurement was not repeated so that no 
reliable value is available for $A=16$. 

More recently, the lifetimes of $^{3}_{\Lambda}$H and $^{4}_{\Lambda}$H were obtained with relativistic ions beams ($^{3}$He, $^{4}$He, 
$^{6}$Li and $^{7}$Li ions with 2.2-5.1 GeV/nucleon energies) and $^{12}$C targets, at JINR \cite{avra}; 
the technique used at LBL was improved by means of a more effective trigger system for the Hypernuclear production, able to identify 
events in which a Hypernucleus of charge $Z$ decays, through $\pi^{-}$ MWD, to a nucleus with charge $Z+1$. The lifetime was determined 
by measuring the Hypernucleus recoil distance distribution with a Ne streamer chamber. Values of 220 ps were reported for $^{4}_{\Lambda}$H 
with errors at 20$\%$ level and of 240 ps for $^{3}_{\Lambda}$H with errors of $\sim$ 60$\%$.

In the heavy mass region, Hypernucleus lifetimes have been measured, between 1985 and 2003, in electroproduction experiments on $^{209}$Bi 
targets \cite{noga}, in low momentum $\bar{p}$--$A$ annihilation at CERN on $^{209}$Bi \cite{bocquet} and $^{238}$U \cite{armstrong} targets and 
in $p$--$A$ collisions at COSY on $^{197}$Au, $^{209}$Bi and $^{238}$U \cite{cassing} targets. 
For heavy Hypernuclei the application of direct timing methods, as adopted for light and medium--$A$ systems, is not feasible due to the huge 
background of the light particles produced. This problem is overcome by detecting heavy fragments from fission processes, induced by the high energy 
release due to the NMWD reaction of the heavy Hypernucleus.  A common feature of both $\bar{p}$--$A$ and $p$--$A$ reactions is that the Hypernucleus 
is  produced with a sizable 
energy, in particular for high--momentum $p$--$A$ collisions and recoils over a distance of some $mm$ before decaying; the recoil shadow 
method \cite{metag}, used in nuclear physics for the measurement of fission isomers, can thus be applied with good sensitivity to evaluate the lifetime. 
The determination, however, is not direct and depends strongly on theoretical models which describe the time evolution of the system during the reaction: 
transport calculations for the initial fast non--equilibrium phase and statistical calculations for the final evaporation phase. These calculations 
give the mass ($A$) and charge ($Z$) distribution of the Hypernuclei which undergo delayed fission together with their individual, $A$-- and 
$Z$-- dependent, velocity distribution in the laboratory frame; a fit of the recoil distance distribution allows to extract the lifetime. 

Following this procedure, from $\bar{p}$--$A$ experiments $\tau(^A_{\Lambda}Z)$ values of $180\pm40$ ps have been obtained for the $^{209}$Bi target 
and $130\pm30$ ps for $^{238}$U target \cite{armstrong}, and from $p$--$A$ experiments values of $130\pm20$ ps for $^{197}$Au target, $161\pm16$ ps 
for $^{209}$Bi target and $138\pm18$ ps for $^{238}$U target were reported \cite{cassing}. It must be noted that, unlike ($K^{-}, \pi^{-}$) and 
($\pi^{+}, K^{+}$) reactions, both $\bar{p}$--$A$ and $p$--$A$ production mechanisms do not allow to identify the decaying system; the $A$ and $Z$ 
values of the produced Hypernuclei are, indeed, obtained from the calculated $(Z,A)$ distributions which span over 35-40 mass units and 8-10 
charge units \cite{cassing}. Considering the overlapping of the 
individual $(A, Z)$ distributions for the three targets used in $p$--$A$ experiments, it is possible to average over them to obtain a mean lifetime 
for Hypernuclei with masses $A\sim 180$-$225$, with a dispersion in charge $\Delta Z \sim 3$ for fixed $A$, of $\tau(^Z_{\Lambda}A) = 145 \pm 11$ ps,
while, by averaging over the results obtained in $\bar{p}$--$A$ experiments on Bi and U, the value $\tau(^Z_{\Lambda}A) = 143 \pm 36$ ps can be 
quoted.

The value from $p$--$A$ collisions, figuring a global $8\%$ error, shows an uncertainty level only a factor two worse than that obtained in recent 
very high statistics ($\pi^{+}, K^{+}$) experiments on $^{12}_{\Lambda}$C at KEK \cite{kame05}, 212$\pm$6~ps. 
Both $\bar{p}$--$A$ and $p$--$A$ averaged values represent about the $55\%$ of the free $\Lambda$ lifetime and suggest a very different behaviour of 
the $\Lambda$ when bound in heavy nuclei with respect to the case of medium--$A$ nuclei. Indeed, if in nuclei with $A=12$-$56$ the lifetime seems to 
saturate at about the $80\%$ of the free particle value, indicating the short--range nature of the weak decay interaction, the presence of a much more 
massive nuclear system seems to produce a stronger interaction of the $\Lambda$ with the core nucleons, as if the 
range of the weak four--body NMWD interaction could increase considerably between $A=56$ and $A=180$ to include large part of the nucleons. 
On the other hand, in $\bar{p}$--$A$ and $p$--$A$ collision experiments, the production of 
$\Lambda$--Hypernuclei is not identified by the detection of a particle which 
undoubtedly signs the formation of a bound strange nuclear system; 
the complexity of systems with 180-225 components leads to suppose the presence of other interaction mechanisms which 
could open for some threshold $A$ value, higher than 60 mass units, and could explain the strong enhancement of the interaction rate for $A>180$. 
It is worth to remind, however, that the lifetime determination is indirect and relies on calculated, model dependent, $A$, $Z$ and velocity 
distributions, which take into account the interaction dynamics during all the development of the process. 
The big difference between the lifetime values obtained with the recoil shadow method and those obtained directly in $(K^{-}, \pi^{-})$ 
and $(\pi^{+}, K^{+})$ reactions suggests that the first method could be affected by sources of systematic errors, possibly connected to the nature 
of the process which causes the fission, that need to be carefully considered; anyway, further experimental work in the
$A\sim 200$ region is necessary, with an explicit tagging of the strange nuclear system formation.

\subsection{Mesonic Weak Decay}
\label{subsec:mwd_meas}
In MWD, light-- and medium--$A$ Hypernuclei are converted to non--strange nuclei through the reactions:
\begin{equation} 
^{A}_{\Lambda}\mathrm{Z}  \rightarrow  ^{A}(\mathrm{Z+1}) + \pi^{-} \quad(\Gamma_{\pi^{-}})
\label{mwd+}
\end{equation}
 \begin{equation}
^{A}_{\Lambda}\mathrm{Z}  \rightarrow  ^{A}\mathrm{Z} + \pi^{0}  \quad(\Gamma_{\pi^{0}})\, ,
\label{mwd0} 
\end{equation} 
corresponding to the elementary reactions (\ref{lfreep}) and (\ref{lfreen}) for bound $\Lambda$; 
the final nuclear states in (\ref{mwd+}) and (\ref{mwd0}) are not necessarily particle stable. 

The theory of Hypernuclear MWD was initiated by Dalitz \cite{dal1,dal2}, 
based on a phenomenological Lagrangian describing the elementary decay 
processes (\ref{lfreep}) and (\ref{lfreen}), and motivated by the observation of MWD reactions in 
the pioneering Hypernuclear physics experiments with photographic emulsions 
that provided means of extracting Hypernuclear ground--state spins and 
parities. The $J^{\pi}$ assignment for $^{3}_{\Lambda}$H, $^{4}_{\Lambda}$H, $^{4}_{\Lambda}$He, $^{8}_{\Lambda}$Li, 
$^{11}_{\Lambda}$B and $^{12}_{\Lambda}$B \cite{dal2,bertrand,bloch,davis63} was done 
by applying the recently established properties of the free $\Lambda$ mesonic decay ($\Delta I=1/2$ rule; prevalence 
of the $s$--wave, parity--violating, spin--non--flip amplitude; $\pi$ angular distribution dependence on $\Lambda$ spin axis) 
to $\Lambda$ bound in nuclear systems; see Ref.~\cite{davis05} for a recent summary. 

Following the development of counter techniques for use in $(K^-,\pi^-)$ 
and $(\pi^+,K^+)$ reactions in the 1970s and 1980s, a considerable body of 
experimental data on $\Gamma_{\pi^{-}}$ and/or $\Gamma_{\pi^{0}}$ is now 
available on light $\Lambda$--Hypernuclei up to ${^{15}_{\Lambda}\mathrm{N}}$: 
${^{4}_{\Lambda}\mathrm{H}}$ \cite{outa95}, ${^{4}_{\Lambda}\mathrm{He}}$ 
\cite{outa98}, 
${^{5}_{\Lambda}\mathrm{He}}$ \cite{szym,kame05,okada,fnd_mwd}, ${^{7}_{\Lambda}\mathrm{Li}}$ \cite{fnd_mwd}, 
${^{11}_{\Lambda}\mathrm{B}}$ \cite{szym,fnd_mwd,saka,noumi,sato05}, 
${^{12}_{\Lambda}\mathrm{C}}$ \cite{szym,saka,noumi,sato05} and 
${^{15}_{\Lambda}\mathrm{N}}$ \cite{fnd_mwd}. 

Comprehensive calculations of the main physical properties of MWD were performed 
during the 1980s and 1990s for $s$--shell \cite{motoba4,motoba2}, 
$p$--shell \cite{motoba2,motoba3,motoba1} and $sd$--shell Hypernuclei 
\cite{motoba2,motoba1}. The basic ingredients of the calculations are the 
Pauli suppression effect, the enhancement of MWD owing to the pion
polarization effect in the nuclear medium \cite{bando,oset}, the sensitive final--state shell--structure dependence
and the resulting charge--dependence of the decay rates. 

An important ingredient of MWD calculations is the pion--nucleus 
potential which generates pion distorted waves that strongly affect 
the magnitude of the pionic decay rates. Indeed, for low--energy pions, 
the pion--nucleus potential has been studied so far through $\pi$--nucleus 
scattering experiments \cite{friedman1} and measurements of $X$--rays from 
pionic atoms \cite{friedman2}; the study of MWD in which a pion is created 
by the decay of a $\Lambda$ hyperon deep inside the nucleus offers important 
opportunities to investigate in--medium pions and to discriminate between 
different off--shell extrapolations inherent in potential models. For this 
reason, MWD continues to be an interesting item of Hypernuclear physics, 
and precise and systematic determinations of $\Gamma_{\pi^{-}}$ and 
$\Gamma_{\pi^{0}}$ are very welcome. 

In the following two subsections $s$--shell and $p$--shell Hypernuclei MWD will be discussed 
separately.

\subsubsection{MWD of $s$--shell Hypernuclei and the $\Lambda$N potential} 
\label{subsubsec:mwd_s}
As discussed before, due to the short $\Lambda$ lifetime, the production of $\Lambda$--Hypernuclei 
and the investigation of their structure is the only practical way to study the $\Lambda N$ interaction. 
To date, the theory of few--body systems is able to calculate directly up to five--body systems starting 
from the elementary two--body $\Lambda N$ interaction; on the other hand, to construct a reliable $\Lambda N$ 
interaction model, the basic requirement is to reproduce the measured binding energy values of all $s$--shell Hypernuclei.

In the determination of a realistic $\Lambda N$ phenomenological potential, a long--standing problem 
was that potentials determined by fits to $B_{\Lambda}(^{3}_{\Lambda}{\rm H})$, $B_{\Lambda}(^{4}_{\Lambda}{\rm H})$ 
and $B_{\Lambda}(^{4}_{\Lambda}{\rm He})$, together with low--energy $\Lambda p$ scattering data, lead to 
overestimate data on $B_{\Lambda}(^{5}_{\Lambda}{\rm He})$. A better agreement with data is obtained 
by explicitly incorporating the $\Lambda$N--$\Sigma$N coupling \cite{nemura}. 

The possible existence of a central repulsion in the hyperon--nucleus mean potential has been discussed by several authors
by starting from the realistic two--body $YN$ interaction: the strength of the long--range attraction of the $YN$ 
interaction is much weaker than that of the $NN$ interaction and it would be almost counterbalanced by the short--range 
repulsion. Consequently, inner repulsion would be present also in the $Y$--nucleus potentials constructed from the 
$YN$ interaction with the folding procedure. The effect would be seen most clearly for very light Hypernuclei ($A=4,5$): the 
hyperon would be pushed outward the nucleus due to this repulsion and consequently the overlap of the hyperon wave function 
with the nucleus would be much reduced. 

The existence of such a repulsive core is established experimentally for 
$^{4}_{\Sigma}$He, where the $\Sigma N \rightarrow \Lambda N$ conversion is suppressed thanks to 
the expulsion of the $\Sigma$ outward the nuclear core due to the inner repulsion. This  mechanism is considered to be responsible 
for the narrow width of the $^{4}_{\Sigma}$He system \cite{sigma}.

Moreover, the experimental evidence for the presence of a central repulsion in the $\Lambda$--nucleus potential can be 
obtained from the Hypernuclear decay observables, which provide significant information on the $\Lambda$ single--particle 
behaviour. In particular, the MWD rate, influenced by Pauli suppression, is sensitive to the overlap between 
the wave functions of the $\Lambda$ and of the nuclear core; this overlap, in turn, reflects the potential shape felt by the 
$\Lambda$ in nuclei. 

A repulsive core in the $\Lambda N$ interaction was introduced in Ref.~\cite{kurihara} to lead to a repulsion in the 
$\alpha$--$\Lambda$ potential: the $\Lambda$ wave function is spread outward the nucleus and the MWD rate is 
enhanced as a consequence of the Pauli suppression relaxation, since on the Hypernuclear surface the local 
nuclear Fermi momentum is smaller than in the nuclear core. The $\alpha$--$\Lambda$ potential is expressed by a 
two--range Gaussian function, indicated as Isle, and yields a 30$\%$ larger MWD rate than the single--gaussian 
$\alpha$--$\Lambda$ potential, indicated as SG, without the repulsive central part. In the calculations, the strength parameters of both 
Isle and SG potentials are determined to reproduce $B_{\Lambda}(^{5}_{\Lambda}{\rm He})$ and only MWD rate measurements 
can distinguish between them. 

\begin{figure} [h]
\begin{center}
\resizebox{0.52\textwidth}{!}{%
 \hspace{-5mm}
  \includegraphics{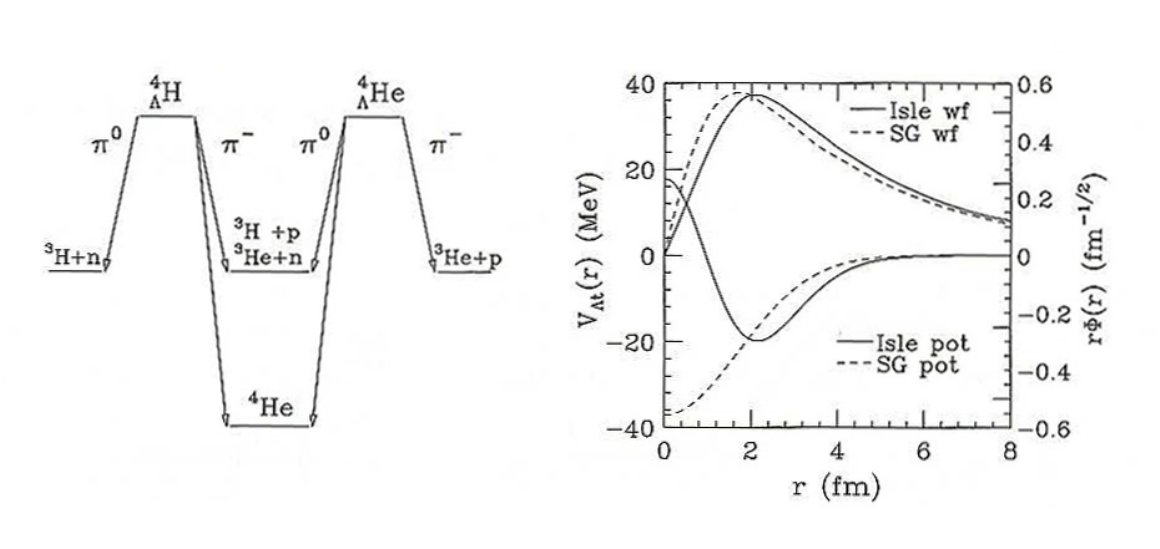}
}
\caption{Left: mesonic weak decay scheme for $^{4}_{\Lambda}$H and $^{4}_{\Lambda}$He. Right: $\Lambda$ radial wave functions
and $\alpha$--$\Lambda$ folding potentials for $^{4}_{\Lambda}$H, calculated by the SG and Isle interactions. 
From Ref.~\cite{kumagai}.}
\label{fig:decay2}       
\end{center}
\vspace{-4mm}
\end{figure}

Other authors too suggested a central repulsion in the $\Lambda$--nucleus potential and evaluated MWD rates for $s$--shell 
Hypernuclei \cite{motoba4,kumagai}. In Fig.~\ref{fig:decay2} the left part shows the decay scheme for 
$^{4}_{\Lambda}$H and $^{4}_{\Lambda}$He; the charge symmetry features of $A=4$ Hypernuclei can be observed by comparing the two mesonic decays. 
In the right part of the figure the $\Lambda$ radial wave functions and the $\alpha$--$\Lambda$ folding potentials for $^{4}_{\Lambda}$H, 
calculated by the SG and Isle interactions, are shown.
Two--body decays are allowed only for reactions producing $^{4}$He in the final state, being the residual nuclei of the other two body channels 
only loosely bound. A $\Lambda$--nucleus potential for $A=4$ with a central repulsion gives a two--body decay rate smaller than that of a 
SG potential, due to the reduction of the overlap of the $\Lambda$ and final nucleon wave functions. This produces an enhancement 
of the three--body decay rates and hence the total MWD rate should not change so much, whereas the ratio between two-- and three--body decays 
is very sensitive to the potential shape and is thus useful to investigate the nature of the short--range part of the $\Lambda N$ interaction. 

From the experimental side, a precise measurement of the Hypernuclear decay rates requires as a pre--requisite a precise measurements of both 
lifetime and branching ratios of reactions (\ref{mwd+}) and (\ref{mwd0}). The E167 experiment at KEK measured the MWD and NMWD rates of 
$^{4}_{\Lambda}$H and $^{4}_{\Lambda}$He  \cite{outa98} produced with the $(K^{-}_{\rm stop}, \pi^{-})$ reaction on a liquid He target, measuring the 
momentum of the outgoing formation $\pi^{-}$ with a large acceptance ($\sim$100 msr) magnetic spectrometer with a resolution of $1$-$2\%$ FWHM. 
$\pi^{0}$, $\pi^{-}$ and $p$ coming from decay reactions 
were detected by a system of NaI(Tl) counters covered by thin plastic scintillators for energy deposition measurement and particle identification. 
The MWD rates obtained by the experiment are summarized in Tables~\ref{tab:1} and \ref{tab:2} together with theoretical calculations from 
\cite{kumagai}. In Table~\ref{tab:2} results of older visualizing techniques experiments \cite{bertrand,bloch} are also reported for 
$^{4}_{\Lambda}$He.
\begin{table}[h]
\begin{center}
\caption{MWD rates of $^{4}_{\Lambda}$H in units of the free $\Lambda$ decay rate ($\Gamma_{\Lambda}$) 
measured by the KEK--E167 experiment \cite{outa98}. The values are compared with theoretical estimates obtained with SG and Isle $\Lambda N$ 
potentials \cite{kumagai}.}
\label{tab:1}       
\begin{tabular}{llll}
\hline\noalign{\smallskip}
Decay Rate & Data & SG & Isle  \\
\noalign{\smallskip}\hline\noalign{\smallskip}
$\Gamma_{\rm T}/\Gamma_{\Lambda}$ & $1.03^{+0.12}_{-0.10}$ & & \\
$\Gamma_{\pi^{0}}/\Gamma_{\Lambda}$ & $0.53 \pm 0.07$ & 0.49 & 0.51 \\
$\Gamma_{\pi^{-}}/\Gamma_{\Lambda}$ & $0.33 \pm 0.05$ & 0.25 & 0.31 \\
$\Gamma_{\pi^{0}}/\Gamma_{\pi^{-}}$ & $1.59 \pm 0.20$ & 2.02 & 1.59 \\
\noalign{\smallskip}\hline
\end{tabular}
\end{center}  
\end{table}
\begin{table}[h]
\begin{center}  
\caption{MWD rates of $^{4}_{\Lambda}$He in units of the free $\Lambda$ decay rate ($\Gamma_{\Lambda}$) 
measured by the KEK--E167 experiment \cite{outa98} and by Refs.~\cite{bertrand,bloch}. The values are compared with theoretical 
estimates obtained with SG and Isle $\Lambda N$ potentials \cite{kumagai}.}
\label{tab:2}       
\begin{tabular}{llll}
\hline\noalign{\smallskip}
Decay Rate & Data & SG & Isle  \\
\noalign{\smallskip}\hline\noalign{\smallskip}
$\Gamma_{\rm T}/\Gamma_{\Lambda}$ & $1.36^{+0.21}_{-0.15}$ & & \\
$\Gamma_{\pi^{0}}/\Gamma_{\Lambda}$ & & 0.16 & 0.19 \\
$\Gamma_{\pi^{-}}/\Gamma_{\Lambda}$ & $1.00^{+ 0.18}_{-0.15}$ & 0.93 & 0.88 \\
$\Gamma_{\pi^{-}\ ^{4}{\rm He}}/\Gamma_{\Lambda}$ & $0.69^{+ 0.12}_{-0.10}$ & 0.71 & 0.61 \\
$\Gamma_{\pi^{-}\ ^{4}{\rm He}}/\Gamma_{\pi^{-}}$ & $0.69 \pm 0.02$ & 0.76 & 0.69 \\
\noalign{\smallskip}\hline
\end{tabular}
\end{center}  
\end{table}

From the comparison between the experimental results and the theoretical calculations of Ref.~\cite{kumagai} it is possible to 
observe that the Isle $\Lambda N$ potential is able to reproduce the data within $\pm 1\sigma$ error, while SG calculations 
figure a disagreement at 2$\sigma$ level for the $\Gamma_{\pi^{0}}/\Gamma_{\pi^{-}}$ ratio of $^{4}_{\Lambda}$H. 
This allows to conclude that the 
existence of a repulsive core in the $\Lambda$--nucleus mean potential seems to be experimentally established.

More recently, the E462 and E508 experiments at KEK measured with very high statistics both $\pi^{-}$ \cite{kame05} and 
$\pi^{0}$ \cite{okada} MWD rates of $^{5}_{\Lambda}$He in $(\pi^{+}, K^{+})$ measurements on $^{6}$Li target with the SKS 
magnetic spectrometer. The results are reported in Table~\ref{tab:3} and compared with SG and Isle based calculations 
\cite{kumagai2,motoba4}.
For  BR($\pi^{-}$) the calculations with the Isle potential are in good agreement with the measured decay
rates, whereas the SG potential gives a significant underestimation \cite{kumagai2}.
For $\Gamma_{\pi^{0}}/\Gamma_{\Lambda}$ the experimental value is located between Isle and SG calculations of Ref.~\cite{motoba4},
but the Isle value reproduces the experimental result within 1$\sigma$, while the SG one differs of more than 2$\sigma$.
Thank to the very small value of the errors on the measured BR's, this fact strongly supports the presence of the central repulsion in
the $\alpha$--$\Lambda$ potential, independently of the considered specific theoretical calculation.
\begin{table}[h]
\begin{center}  
\caption{MWD branching ratios and rates of $^{5}_{\Lambda}$He in units of free $\Lambda$ decay rate ($\Gamma_{\Lambda}$) 
measured by the KEK--E462 and KEK--E508 experiments \cite{kame05,okada}. The values are compared with theoretical 
estimates obtained with SG and Isle $\Lambda N$ potentials for $\Gamma_{\pi^{-}}/\Gamma_{\Lambda}$ 
\cite{kumagai2} and for $\Gamma_{\pi^{0}}/\Gamma_{\Lambda}$ \cite{motoba4} .}
\label{tab:3}       
\begin{tabular}{llll}
\hline\noalign{\smallskip}
Decay Rate & Results & SG & Isle  \\
\noalign{\smallskip}\hline\noalign{\smallskip}
$\Gamma_{\rm T}/\Gamma_{\Lambda}$ & $0.947 \pm 0.038$ & & \\
BR($\pi^{-}$) & $0.359 \pm 0.009$ &  & \\
$\Gamma_{\pi^{-}}/\Gamma_{\Lambda}$ & $0.340 \pm 0.016$ & 0.271 & 0.354 \\
BR(${\pi^{0}}$) & $0.212 \pm 0.008$ &  & \\
$\Gamma_{\pi^{0}}/\Gamma_{\Lambda}$ & $0.201 \pm 0.011$ & 0.177 & 0.215 \\
\noalign{\smallskip}\hline
\end{tabular}
\end{center}  
\end{table}

The last conclusion is representative of the 
deep interplay between particle and nuclear physics that is realized in Hypernuclear 
physics. Indeed, being at the frontier between the two fields, Hypernuclei can be studied exploiting  experimental techniques and 
models typical of both approaches; in particular we can see here how, from measurement of typical nuclear physics observables, like 
decay rates, it is possible to obtain information on the baryon--baryon potentials, a typical subject of particle physics. 


\subsubsection{MWD of $p$--shell Hypernuclei and the
$\pi$ nuclear polarization effect} 
\label{subsubsec:mwd_p}

In Hypernuclei MWD is disfavoured by the Pauli principle, particularly in heavy systems. It is strictly forbidden in normal infinite nuclear matter, 
where the nucleon Fermi momentum is about 270 MeV/c, while in finite nuclei it can occur because of three important effects: 1) in nuclei the hyperon 
has a momentum distribution, being confined in a limited spatial region, that allows larger momenta to be available to the final nucleon; 2) the final 
pion feels an attraction by the nuclear medium, due to the $p$--wave part of the optical $\pi$--nucleus potential, which modifies the pion dispersion 
relation; for a fixed momentum, the pion carries an energy smaller than if it was free and the energy conservation increases the chance of the final 
nucleon to lie above the Fermi surface; indeed, it has been shown that the pion distortion increases the MWD width by more than one order of 
magnitude for very heavy Hypernuclei ($A\sim 200$) with respect to the value obtained without the medium distortion \cite{oset2}; 3) at the nuclear 
surface the local Fermi momentum can be smaller than 270 MeV/c and the Pauli blocking is less effective in forbidding the decay. 

In any case, the MWD rate rapidly decreases as the Hypernucleus mass number increases. 
Table~\ref{tab:4} reports the actual experimental knowledge of both $\Gamma_{\pi^{-}}$ and $\Gamma_{\pi^{0}}$ for $p$--shell Hypernuclei. For the sake 
of completeness, also $^{5}_{\Lambda}\mathrm{He}$, $^{27}_{\Lambda}\mathrm{Al}$ and $^{28}_{\Lambda}\mathrm{Si}$ counter experiments results are 
included. 
A significant part of the available data comes from experiments at KEK \cite{saka,sato05,kame05,okada} and are affected
by relative errors of about 20-25$\%$, which diminish significantly in high statistics measurements \cite{kame05} to about 5$\%$.
\begin{table}[h]
\begin{center}  
\caption{MWD rates $\Gamma_{\pi^{-}}$ and $\Gamma_{\pi^{0}}$ in units of free $\Lambda$ decay rate $\Gamma_{\Lambda}$ for $p$--shell 
$\Lambda$--Hypernuclei obtained in recent counter experiments. $^{5}_{\Lambda}\mathrm{He}$, 
$^{27}_{\Lambda}\mathrm{Al}$ and $^{28}_{\Lambda}\mathrm{Si}$ 
results are also reported for comparison.}
\label{tab:4}       
\begin{tabular}{llll}
\hline\noalign{\smallskip}
Hypernucleus & $\Gamma_{\pi^{-}}/\Gamma_{\Lambda}$ & $\Gamma_{\pi^{0}}/\Gamma_{\Lambda}$ & Ref.  \\
\noalign{\smallskip}\hline\noalign{\smallskip}
$^{5}_{\Lambda}\mathrm{He}$ & 0.44$\pm$0.11 & 0.14$\pm$0.19 & \cite{szym} \\ 
 & 0.340$\pm$0.016 & & \cite{kame05} \\
 & & 0.201$\pm$0.011 & \cite{okada} \\
 & 0.332$\pm$0.069 & & \cite{fnd_mwd} \\
$^{7}_{\Lambda}\mathrm{Li}$ & 0.353$\pm$0.059 &  & \cite{fnd_mwd}  \\ 
$^{9}_{\Lambda}\mathrm{Be}$ & 0.178$\pm$0.050 &  & \cite{fnd_mwd}  \\ 
$^{11}_{\Lambda}\mathrm{B}$ & 0.22$\pm$0.05 &  & \cite{montwill}  \\ 
 & & 0.192$\pm$0.056 & \cite{saka}  \\
 & 0.23$\pm$0.06 & & \cite{noumi}  \\
 & 0.212$\pm$0.036 & & \cite{sato05}  \\
 & 0.249$\pm$0.051 & & \cite{fnd_mwd}  \\
$^{12}_{\Lambda}\mathrm{C}$ & 0.052$^{+0.063}_{-0.035}$ &  & \cite{szym}  \\ 
 & & 0.217$\pm$0.073 & \cite{saka}  \\
 & 0.14$\pm$0.07 & & \cite{noumi}  \\
 & 0.113$\pm$0.014 & & \cite{sato05} \\
$^{15}_{\Lambda}\mathrm{N}$ & 0.108$\pm$0.038 &  & \cite{fnd_mwd}  \\ 
$^{27}_{\Lambda}\mathrm{Al}$ & 0.041$\pm$0.010 &  & \cite{sato05}  \\ 
$^{28}_{\Lambda}\mathrm{Si}$ & 0.046$\pm$0.011 &  & \cite{sato05}  \\ 
\noalign{\smallskip}\hline
\end{tabular}
\end{center}  
\end{table}

Very recently, the FINUDA experiment has performed a systematic study of the charged MWD 
channel of $p$--shell Hypernuclei \cite{fnd_mwd}, $^{7}_{\Lambda}\mathrm{Li}$, 
$^{9}_{\Lambda}\mathrm{Be}$, $^{11}_{\Lambda}\mathrm{B}$  and $^{15}_{\Lambda}\mathrm{N}$, and of $^{5}_{\Lambda}\mathrm{He}$ 
with errors ranging between 15$\%$ (for $A=5$) 
and 35$\%$ (for $A=15$). Very thin target materials were used to stop the low momentum ($\sim$ 127 MeV/c) $K^{-}$'s coming from the 
$\Phi\rightarrow K^{-}K^{+}$ decay; moreover the high transparency of the FINUDA tracker and the very large solid angle 
($\sim 2\pi$ sr) covered by the detector ensemble make the FINUDA apparatus 
suitable to study the formation and the decay of $\Lambda$--Hypernuclei 
by means of high--resolution magnetic spectroscopy of the charged particles 
emitted in the processes. In particular, a $\pi^{-}$ of about 250-280 MeV/c momentum has been required to identify the Hypernucleus bound state production 
on $p$--shell  nuclear targets, in coincidence with a second $\pi^{-}$ of momentum less than 120 MeV/c to detect the MWD.  

It must be stressed that the information available before the FINUDA measurements on 
the charged MWD of light Hypernuclei consisted almost entirely of 
$\Gamma_{\pi^{-}} / \Gamma_{\Lambda}$ and $\Gamma_{\pi^{0}} / \Gamma_{\Lambda}$ values obtained by means of counting 
measurements in coincidence with the Hypernuclear formation $\pi^{-}$ 
detection, with no magnetic analysis of the decay meson; $\pi^{-}$ kinetic 
energy spectra were reported for $^{12}_{\Lambda}\mathrm{C}$ MWD 
only \cite{sato05}. 
The $\pi^{-}$ spectra of Ref.~\cite{fnd_mwd}, on the other hand, allow one to have a more 
careful confirmation of the elementary mechanism that is supposed to underlie 
the decay process, through the determination of the decay rates, as well as to have information on the spin--parity of the 
initial Hypernuclear ground state through the analysis of their energy dependence. In this respect, the study of pion spectra 
from MWD can be regarded as an indirect spectroscopic investigation tool. 

\begin{figure} [h]
\begin{center}
\resizebox{0.5\textwidth}{!}{%
 \hspace{-5mm}
  \includegraphics{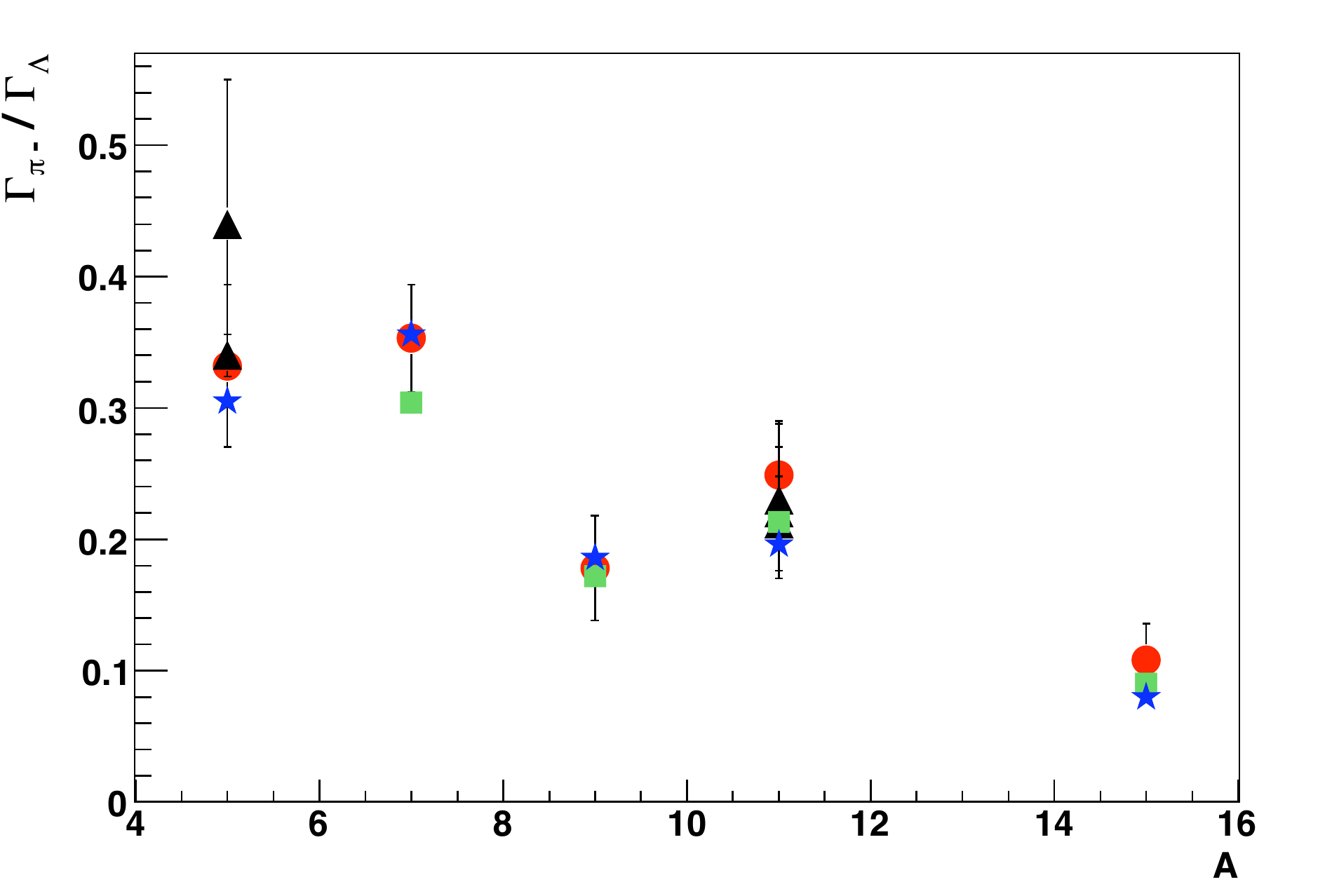}
}
\caption{The $\Gamma_{\pi^{-}} / \Gamma_{\Lambda}$ ratios obtained by the FINUDA experiment (red circles) \cite{fnd_mwd}, 
compared with previous measurements (black triangles) \cite{szym,kame05,montwill,noumi,sato05} and with theoretical calculations 
\cite{motoba3,motoba1} (green squares) and \cite{gal} (blue stars). 
See Ref.~\cite{fnd_mwd} for more details.}
\label{fig:br_mwd}       
\end{center}
\vspace{-4mm}
\end{figure}

In particular, the ratios $\Gamma_{\pi^{-}} / \Gamma_{\Lambda}$ were obtained from the measured 
branching ratios, using available $\Gamma_{\rm T}/\Gamma_{\Lambda}$ values or relying on a linear fit to the 
known values of all measured $\Lambda$--Hypernuclei in the mass range $A=4$-$12$ \cite{sasao}. 
Fig.~\ref{fig:br_mwd} reports the values obtained by the FINUDA experiment (red circles) compared with 
previous experimental data (black triangles) and theoretical calculations (green squares and blue stars) for $^{5}_{\Lambda}\mathrm{He}$, 
$^{7}_{\Lambda}\mathrm{Li}$, $^{9}_{\Lambda}\mathrm{Be}$, $^{11}_{\Lambda}\mathrm{B}$ and $^{15}_{\Lambda}\mathrm{N}$. 

Being characterized by a small Q--value, the MWD mode is strongly affected by the details of both the Hypernucleus 
and the daughter nucleus structure. Indeed, the theoretical calculations of Refs.~\cite{motoba3,motoba1} evaluate the $p$--shell 
Hypernuclei total and partial MWD decay rates by incorporating the pion final state interaction using pion--nuclear distorted waves and by describing 
the structure of the nuclear core with the Cohen--Kurath spectroscopic approach \cite{CK}. The authors found that for $p$--shell 
Hypernuclei the total $\pi^{-}$ decay rate is dominated by $\Lambda_{1s} \rightarrow p_{1p}$ transitions, while only little contributions are 
given by higher energy configurations of the final nuclear system, mainly through $\Lambda_{1s} \rightarrow p_{2s,1d}$ transitions.

Recently, in Ref.~\cite{gal} these calculations were revisited, following the same approach, and a new sum rule was introduced to 
encapsulate the suppressive effect of the Pauli principle on the total and partial $\pi^{-}$ decay rates. 
In Fig.~\ref{fig:br_mwd}, both calculations of Refs.~\cite{motoba1,motoba3} (green squares) and Ref.~\cite{gal} (blue stars) are
reported, for ground state spin--parity $1/2^{+}$ for $^{5}_{\Lambda}\mathrm{He}$ and $^{7}_{\Lambda}\mathrm{Li}$, $5/2^{+}$ for 
$^{11}_{\Lambda}\mathrm{B}$ and $3/2^{+}$ for $^{15}_{\Lambda}\mathrm{N}$.
A good agreement holds among the FINUDA results and previous measurements (black triangles), 
when existing, and among the FINUDA results and the theoretical calculations. The total 
decay rates for the other choice of ground state spin--parity calculated in Refs.~\cite{motoba1,motoba3,gal} ($3/2^{+}$ for 
$^{7}_{\Lambda}\mathrm{Li}$, $7/2^{+}$ for $^{11}_{\Lambda}\mathrm{B}$ and $1/2^{+}$ for $^{15}_{\Lambda}\mathrm{N}$) are 
substantially lower and disagree with the experimentally derived values; an exception is given by $^{15}_{\Lambda}\mathrm{N}$,
for which the calculated values reported in Ref.~\cite{motoba1} does not allow one to fix the ground state spin--parity, as will be discussed in the 
following. 

The agreement between experimental data and calculations supports the correctness of the hypotheses on which the theoretical descriptions are based. 
It clearly indicates the possibility to evaluate typical nuclear physics observables, like the rates of the MWD of Hypernuclei, starting from the knowledge 
of the elementary phenomenological Hamiltonian which describes the decay of the free $\Lambda$; it also gives a confirmation of the presence of the 
distortion of the outgoing $\pi^{-}$ 
wave function due to the polarization effect of the nuclear medium and of the effectiveness of the Cohen-Kurath spectroscopic calculations in describing both 
the Hypernucleus core and the daughter nucleus, allowing to account naturally for the strong final state shell--structure--dependence and 
charge--dependence of the measured decay rates.  

The decay $\pi^{-}$ kinetic energy spectra obtained by the FINUDA experiment
for MWD of $^{7}_{\Lambda}\mathrm{Li}$, $^{9}_{\Lambda}\mathrm{Be}$,
$^{11}_{\Lambda}\mathrm{B}$ and $^{15}_{\Lambda}\mathrm{N}$ \cite{fnd_mwd}
show interesting structures which can be directly
related to the excitation function of the daughter nucleus calculated by Refs.~\cite{motoba1,motoba3,gal}, allowing to determine the
spin--parity configuration of the Hypernucleus ground state.

\begin{figure} [h]
\vspace{-2mm}F
\begin{center}
\resizebox{0.55\textwidth}{!}{%
 \hspace{-5mm}
  \includegraphics{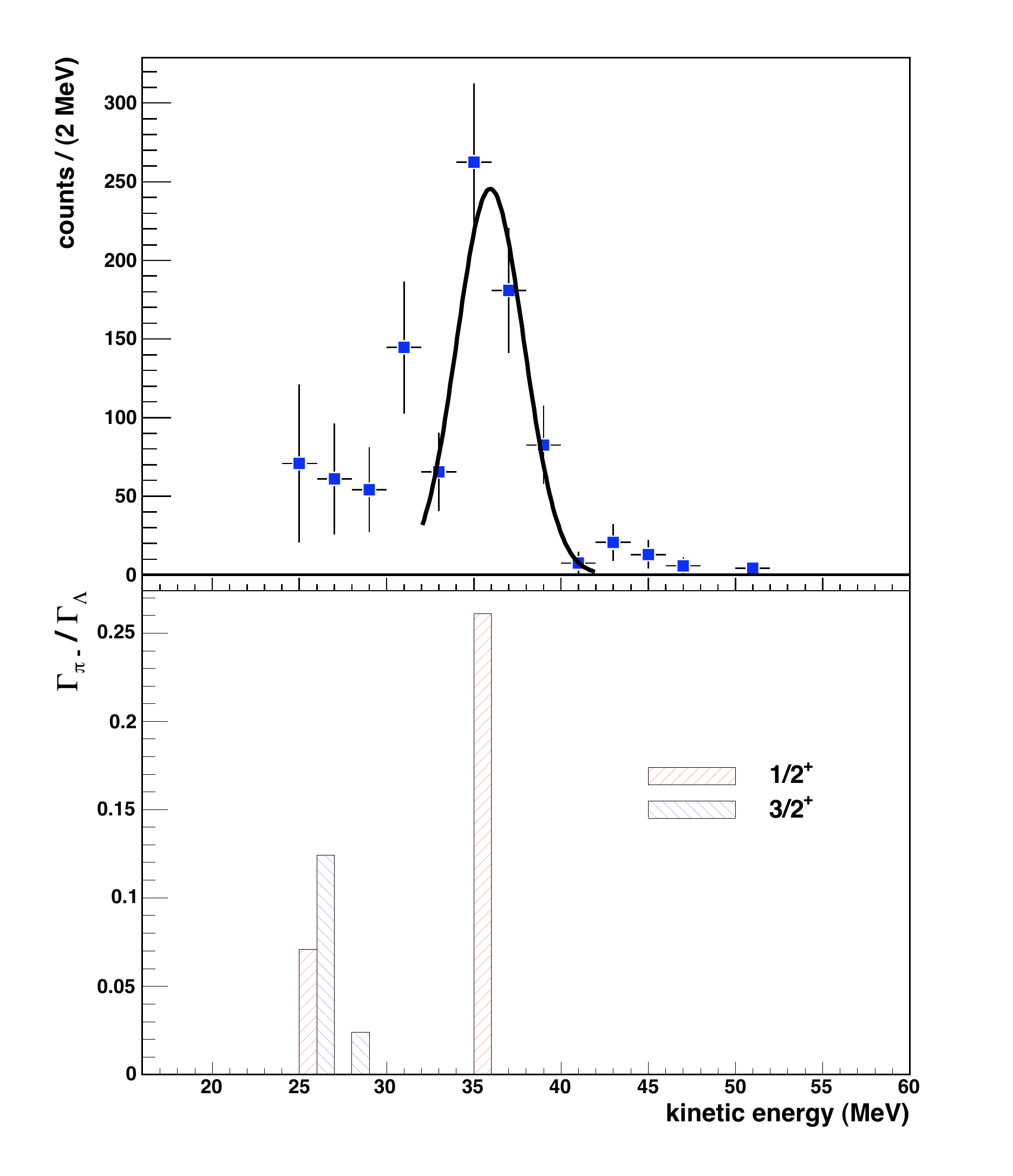}
}
\caption{Top: kinetic energy spectrum of MWD $\pi^{-}$ from $^{7}_{\Lambda}\mathrm{Li}$; the solid line is a Gaussian fit to compare with
theoretical calculation. Bottom: calculated major decay rates to $^{7}$Be final states \cite{gal} for $^{7}_{\Lambda}\mathrm{Li}$ $1/2^{+}$
(red bars) and $3/2^{+}$ (blue bars) ground state spin--parity. From Ref.~\cite{fnd_mwd}.}
\label{fig:br_Li}       
\end{center}
\vspace{-4mm}
\end{figure}

In Fig.~\ref{fig:br_Li} the $\pi^{-}$ spectrum from MWD of $^{7}_{\Lambda}\mathrm{Li}$ is shown 
(upper part) and compared the with calculated decay ratios $\Gamma_{\pi^{-}}/\Gamma_{\Lambda}$ to final $^{7}\mathrm{Be}$ states 
\cite{gal} (lower part). These calculated rates are close to those obtained by Ref.~\cite{motoba1}. The correspondence of 
the structures observed in the experimental spectra with the rates for decays to different excited states of the daughter nucleus, assuming 
a $1/2^{+}$ initial spin--parity state, is clear. The peak structure corresponds to the production of $^{7}\mathrm{Be}$ in its $3/2^{-}$ 
ground state and in its only bound $1/2^{-}$ excited state, at $429$ keV, not resolved due to the FINUDA experimental resolution (4.2 MeV FWHM). 
The part of the spectrum at lower energies is due to three--body decays.
The shape of the spectrum confirms the spin assigned to the Hypernuclear
ground state of $^{7}_{\Lambda}\mathrm{Li}$ \cite{sasao}. Indeed, only
a $1/2^+$ spin--parity for $^{7}_{\Lambda}\mathrm{Li}$ ground state, shown by
red bars, reproduces the fitted peak at $\sim 36$ MeV due to the
$^{7}\mathrm{Be}$ ground state and excited state at 429 keV. A $3/2^+$ spin--parity
for $^{7}_{\Lambda}\mathrm{Li}$ ground state would imply a radically
different spectral shape \cite{motoba1,gal}, as indicated in Fig.~\ref{fig:br_Li}
by the blue bars.

\begin{figure} [h]
\vspace{-1mm} 
\begin{center}
\resizebox{0.51\textwidth}{!}{%
 \hspace{-5mm}
  \includegraphics{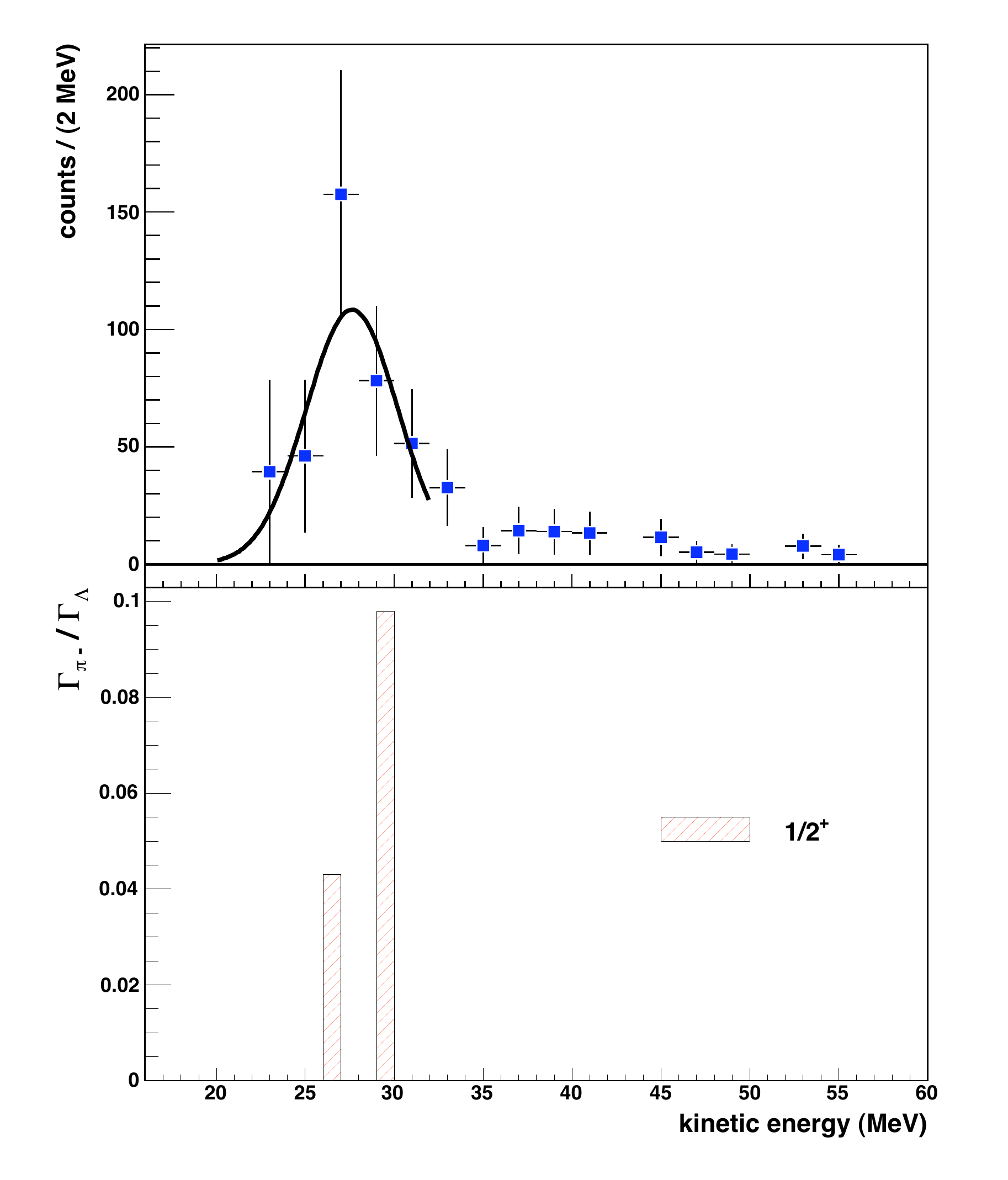}
}
\caption{Top: kinetic energy spectrum of MWD $\pi^{-}$ from $^{9}_{\Lambda}\mathrm{Be}$; the solid line is a Gaussian fit to compare with
the theoretical calculation. Bottom: calculated major decay rates to $^{9}$B final states \cite{gal} for $^{9}_{\Lambda}\mathrm{Be}$
$1/2^{+}$ ground state (red bars). From Ref.~\cite{fnd_mwd}.}
\label{fig:br_Be}       
\end{center}
\vspace{-4mm}
\end{figure}

In Fig.~\ref{fig:br_Be} the spectrum for $^{9}_{\Lambda}\mathrm{Be}$ is shown in the 
upper part and compared with calculated decay ratios $\Gamma_{\pi^{-}}/\Gamma_{\Lambda}$ to $^{9}\mathrm{B}$ final states 
\cite{gal} shown in the lower part. These calculated rates too are close to those predicted by Ref.~\cite{motoba1}. 
In the $^{9}_{\Lambda}\mathrm{Be}$ spectrum the energy resolution does not 
allow a separation between the two components predicted to dominate the 
spectrum \cite{motoba1,gal}, the $3/2^{-}$ $^{9}\mathrm{B}$ ground state and 
the $1/2^{-}$ excited state at $2.75$ MeV. 
The correspondence between the experimental spectrum and the calculated rates 
of decay to different excited states of the daughter nucleus is clear. 
The spectrum is consistent with the interpretation from $(\pi^+,K^+)$ reactions 
\cite{hashim,tamu}, according to which the $^{9}_{\Lambda}\mathrm{Be}$ ground state 
is dominantly a $1s$ $\Lambda$ coupled to the $^{8}\mathrm{Be}(0^{+})$ ground 
state. 

\begin{figure} [h]
\begin{center}
\resizebox{0.50\textwidth}{!}{%
 \hspace{-5mm}
  \includegraphics{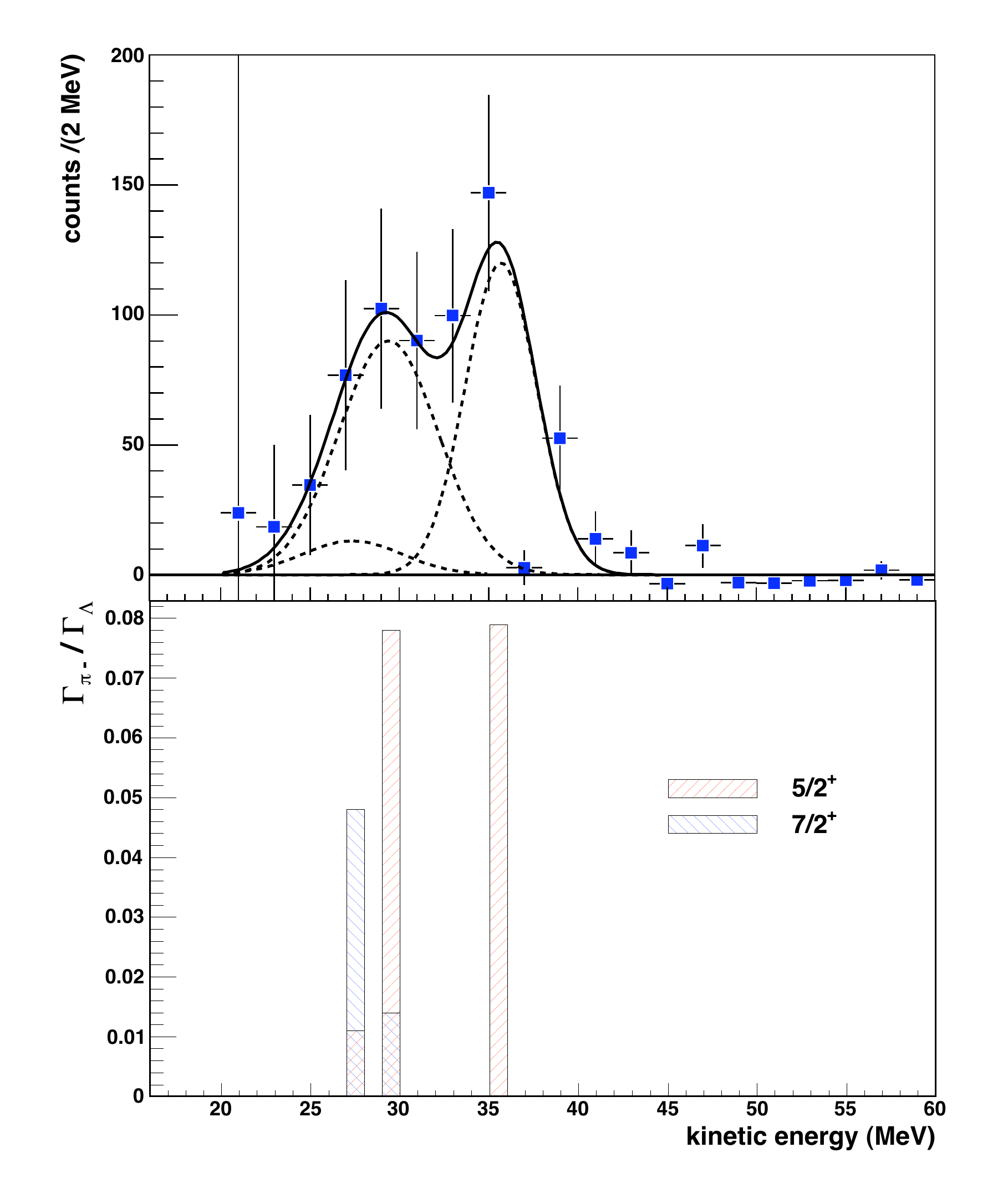}
}
\caption{Top: kinetic energy spectrum of MWD $\pi^{-}$ from $^{11}_{\Lambda}\mathrm{B}$; the solid line is a Gaussian fit to compare with
the theoretical calculation. Bottom: calculated major decay rates to $^{11}$C final states \cite{gal} for $^{11}_{\Lambda}\mathrm{B}$ 
$5/2^{+}$ (red bars) and $7/2^{+}$ (blue bars) ground states. From Ref.~\cite{fnd_mwd}.}
\label{fig:br_B}       
\end{center}
\vspace{-4mm}
\end{figure}

In Fig.~\ref{fig:br_B} the spectrum for $^{11}_{\Lambda}\mathrm{B}$ is shown and compared with calculated decay rates to  
$^{11}\mathrm{C}$ final states \cite{gal}. By assuming a $5/2^{+}$ ground state, it is possible to identify 
two major contributions in the $^{11}_{\Lambda}\mathrm{B}$ spectrum due to a $3/2^{-}$
$^{11}\mathrm{C}$ ground state and its $7/2^{-}$ excited state at 6.478 MeV, both shown by red bars. 
It is clear from the figure that the shape 
of the spectrum is well reproduced by assigning a $5/2^{+}$ spin--parity to 
$^{11}_{\Lambda}\mathrm{B}$ ground state, while by assuming a $7/2^{+}$ ground state, 
the $^{11}\mathrm{C}$ ground state peak is missing and the dominant decay is to the $5/2^{-}$ 
excited state at 8.420 MeV, shown by a blue bar.
A $5/2^{+}$ assignment for $^{11}_{\Lambda}\mathrm{B}$ ground state, 
first made by Zieminska by studying emulsion spectra \cite{ziem}, was 
experimentally confirmed by the KEK measurement \cite{sato05}, comparing 
the derived value of the total $\pi^-$ decay rate with the calculation of Ref.~\cite{motoba2}. 
The FINUDA measurement of the decay spectrum shape provides a 
confirmation of the fact that $J^{\pi}(^{11}_{\Lambda}\mathrm{B}_{\rm g.s.}) = 5/2^{+}$ 
by a different observable. 

\begin{figure} [h]
\begin{center}
\resizebox{0.50\textwidth}{!}{%
 \hspace{-5mm}
  \includegraphics{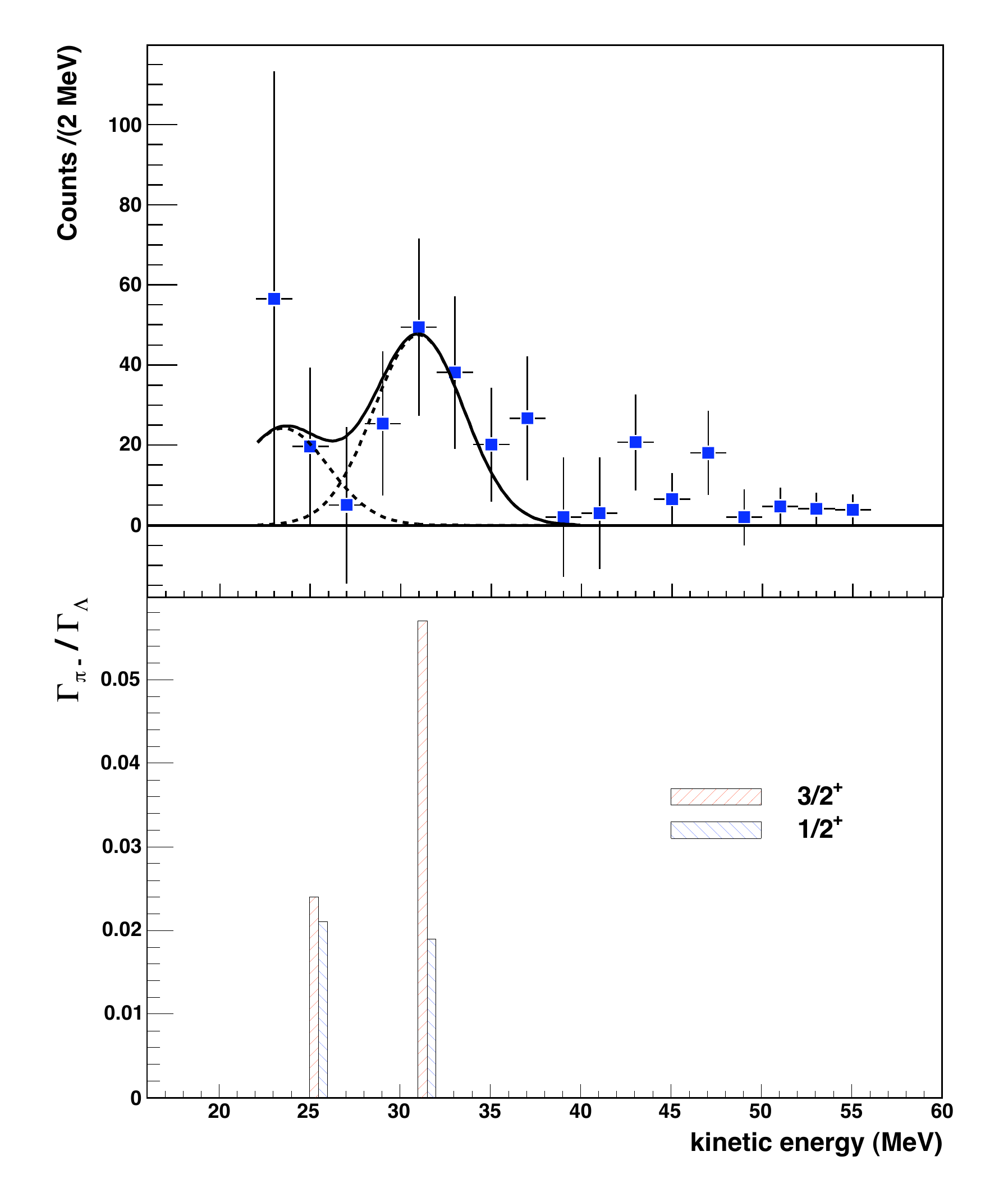}
}
\caption{Top: kinetic energy spectrum of MWD $\pi^{-}$ from $^{15}_{\Lambda}\mathrm{N}$; the solid line is a Gaussian fit to compare with
the theoretical calculation. Bottom: calculated major decay rates to $^{15}$O final states \cite{gal} for 
$^{15}_{\Lambda}\mathrm{N}$ 3/2$^{+}$ (red bars) and 1/2$^{+}$ (blue bars) ground states. From Ref.~\cite{fnd_mwd}.}
\label{fig:br_N}       
\end{center}
\vspace{-4mm}
\end{figure}

In Fig.~\ref{fig:br_N} the spectrum for $^{15}_{\Lambda}\mathrm{N}$ is shown and compared with calculated decay rates to 
$^{15}\mathrm{O}$ final states \cite{gal}. In the experimental spectrum, 
the $^{15}\mathrm{O}$ $1/2^{-}$ ground state contribution stands out clearly, 
along with a hint for a secondary structure separated by about 6 MeV. 
The fit to the lower energy secondary structure is strongly influenced by the 
substantial error affecting the data point at the lowest energy. 
According to Refs.~\cite{motoba3,gal}, this secondary structure derives most 
of its strength from $sd$ states scattered around 6 MeV excitation, while the 
contribution of the $p_{3/2}^{-1}p_{1/2}$ $^{15}\mathrm{O}$ excited state at 6.176 MeV is negligible.
Before the FINUDA measurement, the ground state spin of $^{15}_{\Lambda}\mathrm{N}$ was not determined experimentally. 
The most recent theoretical study of Hypernuclear spin--dependence \cite{boh3} 
predicts $J^{\pi}(^{15}_{\Lambda}\mathrm{N}_{\rm g.s.}) = 3/2^{+}$, setting the 
$1/2^{+}$ excited state of the ground state doublet about 90 keV above the $3/2^{+}$ 
state. The spin ordering, however, could not be determined from the 
$\gamma$--ray de--excitation spectra measured recently on a $^{16}\mathrm{O}$ target at BNL \cite{boh4}. 
The prominence of $^{15}\mathrm{O}_{\rm g.s.}$ in the spectrum of Fig.~\ref{fig:br_N} supports this 
$3/2^{+}$ theoretical assignment. Moreover, as already observed, the total $\pi^-$ decay rate of 
$^{15}_{\Lambda}\mathrm{N}$ agrees with calculations by Refs.~\cite{motoba1,gal} by assuming 
a $3/2^+$ ground state spin--parity assignment. These two calculations 
disagree for a $1/2^+$ spin--parity assignment, and, following the new calculation \cite{gal} for 
$^{15}_{\Lambda}\mathrm{N}$, which corrects the older calculations of Refs.~\cite{motoba3,motoba1}, a 
$1/2^+$ spin--parity is excluded and the assignment $J^{\pi}(^{15}_{\Lambda}\mathrm{N}_{g.s.}) 
={ 3/2}^+$ is made based mainly on the decay rate and the shape of the MWD spectrum. 

It is thus evident that the study of the MWD spectra has strong potentialities for the determination 
of the ground state spin--parity for $s$-- and $p$--shell Hypernuclei; for $sd$--shell strange nuclear systems too, the spectroscopy 
of particles emitted in the decay processes can help in fixing the spectroscopic configuration of the ground state, although 
the MWD branching ratios diminish below 10$\%$ for these Hypernuclei and the errors become consequently very sizable. The MWD spectra analysis has 
thus demonstrated its reliability as light-- and medium--$A$ Hypernuclei spectroscopic tool, complementary to the $\gamma$--ray spectroscopy of 
low--lying excited states \cite{tamu} when the spin ordering cannot be determined; it represents a new version of the old technique based on the 
study of the angular distribution of the MWD $\pi^{-}$, which allowed the ground state spin--parity determination of light Hypernuclei from emulsion 
experiments \cite{dal2,bertrand,bloch,davis63}, based on the known properties of the free $\Lambda$ weak 
decay. 

\begin{figure} [h]
\begin{center}
\resizebox{0.50\textwidth}{!}{%
 \hspace{-5mm}
  \includegraphics{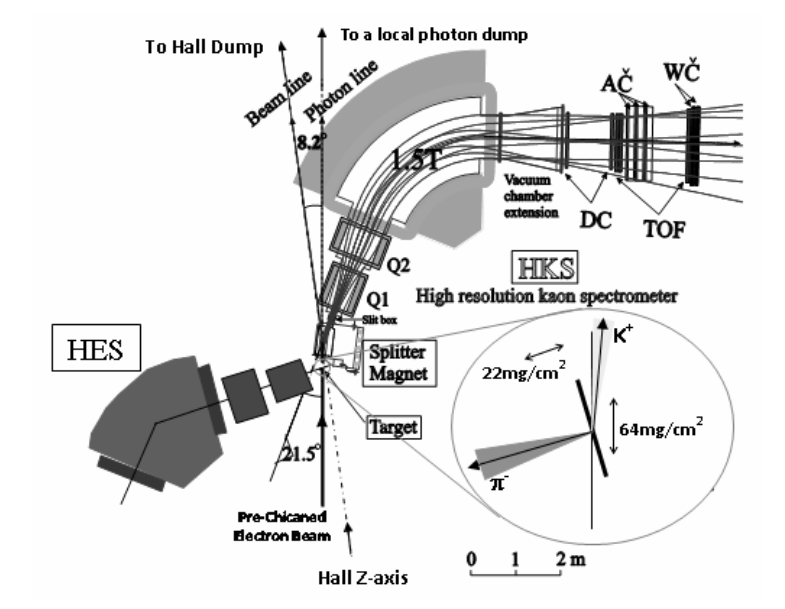}
}
\caption{Schematic view of the experimental layout proposed at JLab Hall A for performing high resolution spectroscopy from MWD of Hyperfragments produced by 
electroproduction. From Ref.~\cite{tang}.}
\label{fig:jlab_mes}       
\end{center}
\vspace{-4mm}
\end{figure}

As a matter of fact, very recently plans for performing high resolution ($\sim$100 keV) spectroscopy of $\pi^{-}$ from MWD of Hyperfragments produced by 
electroproduction at JLab and MAMI-C were put forward \cite{tang}. Fig. \ref{fig:jlab_mes} shows a schematic view of the experimental layout proposed for operation 
at JLab, Hall A. 
However, it is difficult from such experiments to determine the spin--parity of the 
Hypernuclear ground states, since the quantum numbers of the initial state are difficult, if not impossible, to determine. 

\subsection{Non--Mesonic Weak Decay}
\label{subsec:nmwd_gen}

In NMWD, Hypernuclei are converted into non--strange nuclei through the reactions:
\begin{equation}
^{A}_{\Lambda}\mathrm{Z}  \rightarrow  ^{A-2}\mathrm{Z} + n + n\, ,
\label{nmwdn} 
\end{equation} 
\begin{equation} 
^{A}_{\Lambda}\mathrm{Z}  \rightarrow  ^{A-2}(\mathrm{Z-1}) + p + n\, ,  
\label{nmwdp}
\end{equation}
corresponding to the elementary neutron-- and proton--induced reactions (\ref{gamman}) and (\ref{gammap}); 
as for MWD, the final nuclear states in (\ref{nmwdn}) and (\ref{nmwdp}) are not necessarily particle stable and are actually not detectable, so that only inclusive 
measurements can be performed. 

The  possibility of NMWD of $\Lambda$--Hypernuclei was suggested for the first time in 1953 
\cite{ches} and interpreted in terms of the free space $\Lambda \rightarrow N \pi$ decay, where the pion was considered as virtual and then absorbed 
by a bound nucleon, as already mentioned before. 
In the 60's, Block and Dalitz \cite{dalitz} developed a phenomenological model of NMWD, which was recently updated (see for instance 
Ref.~\cite{al_gar}).  
Within this 
approach, some important characteristics of the NMWD of $s$--shell Hypernuclei, mainly the degree of validity of the $\Delta I=1/2$ rule on the 
isospin change, can be reproduced in terms of elementary spin--dependent branching ratios for the processes (\ref{gamman}) and (\ref{gammap}), by 
fitting the available experimental data.

After the first analysis by Block and Dalitz, microscopic models of the $\Lambda N \rightarrow nN$ interaction began to be developed. The first of 
these approaches adopted a one--pion--exchange (OPE) model \cite{Ad67}. This model is based on a $\Delta I=1/2$ $\Lambda N \pi$ vertex, with the 
absorption of the virtual pion by a second nucleon of the nuclear medium. The results of the decay width calculation for the one--nucleon induced 
NMWD were not realistic also because the employed $\Lambda N \pi$ coupling was too small to reproduce the free $\Lambda$ lifetime.

Since the NMWD channel is characterized by a large momentum transfer, the details of the Hypernuclear structure should not have a substantial
influence, thus providing useful information directly on the four--baryon, strange-ness changing, $\Lambda N \rightarrow n N$ weak interaction from  
nuclear physics measurements, analogously to what occurs in the description of the MWD observables starting from the free $\Lambda$ decay
knowledge. The large momentum transfer implies that NMWD probes short--distance interactions and might, therefore, elucidate the
effect of heavy--meson exchange or even the r$\hat{\mathrm{o}}$le of the explicit quark/gluon substructure of baryons in
baryon weak interactions.

In order to improve the OPE model, mesons heavier than the pion were thus introduced as mediators of the $\Lambda N \rightarrow nN$ interaction. McKellar 
and Gibson \cite{mckellar} evaluated the width for a $\Lambda$ in nuclear matter, adding the exchange of the $\rho$--meson and taking into account the $\Lambda N$ 
relative $s$--states only. They used a $\Delta I=1/2$ $\Lambda N \pi$ vertex and made the calculation by using the two possible relative 
signs (being at that time unknown and not fixed by their model) between the pion and the $\rho$ potentials. It is important to note that, for mesons heavier than the pion, 
no experimental indication supports the validity of the $\Delta I=1/2$ isospin rule in $\Lambda N \to nN$ (transitions with $\Delta I=3/2$ are also possible).
Some years later, Nardulli \cite{nardulli} determined the relative sign ($-$) between $\rho$-- and $\pi$-- exchange by implementing the available information from weak 
non--leptonic and radiative decays. Refs.~\cite{mckellar,nardulli} obtained a NMWD width in the ($\rho + \pi$) exchange model smaller than the OPE one. 
In 1986, Dubach {\it et al.} \cite{dubach} extended the OPE to a one--meson--exchange (OME) model, in which the $\pi$, $\rho$, $K$, $K^{*}$, $\omega$ 
and $\eta$ mesons were considered, in a nuclear matter calculation.
Shell model approaches were then considered \cite{Pa97} in terms of OME models including
the mesons of the pseudoscalar and vector octets and, more recently, also
uncorrelated and correlated two--pion--exchange (TPE) were added to the OME potentials \cite{Os01,CGPR07}.  
For most of the OME and OME$+$TPE calculations performed up to date, the $\Delta I=1/2$ rule is assumed as valid,
although this is justified only for the $\Lambda N \pi$ weak vertex. In Sec.~\ref{subsec:nmwd12} we shall discuss in detail
the status of the studies performed on the $\Delta I=1/2$ rule in NMWD.

Another theoretical approach to the short--range part of the $\Lambda N\to nN$ weak interaction is provided by the quark model. Cheung, 
Heddle and Kisslinger \cite{cheung} considered an hybrid quark--hadron approach for the NMWD in which the decay is explained by two separate 
mechanisms with different interaction ranges: the long--range term ($r\geq$0.8 fm) was described by the OPE with the $\Delta I=1/2$ rule, while 
the short--range interaction was described by a six--quark cluster model including both $\Delta I=1/2$ and $\Delta I=3/2$ contributions.
More recently, Inoue {\it et al.} \cite{inoue} calculated the NMWD widths with a direct quark (DQ) model combined with the OPE description.
In the DQ model, the $\Lambda N$ and $NN$ short--range repulsion originates from quark exchange between baryons and gluon exchange between 
quarks. The effective weak Hamiltonian for quarks was obtained from the so--called operator product expansion, which contains perturbative
QCD effects and, by construction, both $\Delta I=1/2$ and $\Delta I=3/2$ transitions. It was found that the DQ 
mechanism gives a significant $\Delta I=3/2$ contribution in the $J=0$ $\Lambda N\to nN$ channel. The approach was then extended to incorporate 
a OME model containing $\pi$, $K$ and $\sigma$ exchange \cite{sasaki}.

A method alternative to the finite nucleus shell model approach makes use of a nuclear matter formalism.
This is a many--body technique, first introduced in Ref.~\cite{oset},
in which the calculation is performed in infinite nuclear matter
and then it is extended to finite nuclei through the local density approximation.
It provides a unified picture of both mesonic and non--mesonic decay channels.
In nuclear matter one has to evaluate the $\Lambda$ self--energy $\Sigma$, which
provides the various mesonic and non--mesonic decay widths through the relation
$\Gamma_i = -2\, {\rm Im}\, \Sigma_i$ ($i=n$, $p$, $np$, etc). The $\Lambda$ self--energy is derived from a diagrammatic approach.

Shell model and quark model based calculations restricted their analyses
to one--nucleon induced NMWD. In the first nuclear matter calculations proposed for the
evaluation of the two--nucleon induced decay rates, a phenomenological approach was used \cite{al91,ramos}.
In these works, the Feynman diagrams contributing to the two--nucleon induced NMWD were not explicitly
evaluated, but an approximate calculation was done by using data on pion absorption
in nuclei. In the work by Ramos at al.~\cite{ramos}, additionally, an argument for the phase space
available for the $2p2h$ configurations was introduced. This
resulted in a non--negligible contribution of the two--nucleon induced rate to the total NMWD rate.
According to a quasi--deuteron approximation, the rate was
assumed to be due to a decay induced by a neutron--proton pair: $\Gamma_2=\Gamma_{np}$.

More recently, a program was started for a microscopic calculation of the one-- and two--nucleon induced
decays \cite{bau2}. A nuclear matter formalism extended to finite Hypernuclei by the local density
approximation is adopted. All isospin channels of the two--nucleon stimulated decay, $nn$--, $np$-- and 
$pp$--induced, are included in this diagrammatic approach. 
In particular, Pauli exchange (ground state correlation) contributions were recently evaluated \cite{bau_garb} (\cite{Ba10b}):
the results showed that Pauli exchange terms and ground state correlations are very important for a detailed
calculation of all the NMWD rates and that,
as expected, the two--nucleon stimulated decay width is dominated by the $np$--induced process.
In Ref.~\cite{BG-PRC}, the microscopic diagrammatic approach was then applied to the calculation of the
nucleon spectra emitted in Hypernuclear NMWD.

For a long period of time, all these theoretical efforts were mainly devoted to the solution of an important question concerning the weak 
decay rates. In fact, the study of the NMWD was characterized by a long--standing disagreement between theoretical estimates and experimental 
determinations for the $\Gamma_{n} / \Gamma_{p}$ ratio between the neutron-- and proton--induced decay widths: this became known as the 
{\it $\Gamma_{n} / \Gamma_{p}$ puzzle.}

It is worth to remind here that, up to a few years ago, all theoretical calculations appeared to strongly underestimate the available data measured in 
several Hypernuclei for the $\Gamma_n/\Gamma_p$ ratio. Concerning measured values, the situation can be summarized by the following inequalities:
\begin{equation}
\left \{ \frac{\Gamma_{n}}{\Gamma_{p}} \right \}^{\rm theory} \ll  \left \{ \frac{\Gamma_{n}}{\Gamma_{p}} \right \}^{\rm exp}, 
\ \ \ \ \ \ 0.5 \lsim  \left \{ \frac{\Gamma_{n}}{\Gamma_{p}} \right \}^{\rm exp} \lsim 2 
\label{puzzle}
\end{equation}
(only for $^4_\Lambda$He the experimental value of this ratio is less than 0.5), although the large experimental error bars did not
allow to reach any definite conclusion. 

Nowadays, it is widely recognized that a solution has been found to the $\Gamma_{n} / \Gamma_{p}$ puzzle. This achievement
was possible mainly by a rapid development of the experiments, which measured a variety of NMWD nucleon emission spectra, and, from the theoretical 
side, thanks to: 1) the inclusion in the $\Lambda N \rightarrow nN$ transition potential of mesons heavier than the pion
(two--meson--exchange was also considered), 2) the description of the short--range $\Lambda N$ and $NN$ correlations in terms of quark degrees of 
freedom, 3) the analysis of the effect of the two--nucleon induced decay mechanism and (especially) 4) the accurate study of the nucleon FSI following 
the elementary one-- and two--nucleon induced decays (\ref{gamman})--(\ref{gamma2}). In Sec.~\ref{subsec:nmwd_exp} we shall 
describe these recent advances in detail.

The experimental observables in NMWD are: the total NMWD width $\Gamma_{\rm NM}$, the spectra of the nucleons emitted in the decay 
(from which one can determine, via some theoretical model, the partial decay rates $\Gamma_n$, $\Gamma_p$, $\Gamma_{np}$, etc) and
the asymmetry of protons from NMWD of polarized Hypernuclei. A  theory which aims to explain the NMWD reaction mechanisms  
must account coherently for the description of the experimental behaviour of all these observables.

In the next Sections, the most important topics related to NMWD are discussed: 
in Sec.~\ref{subsec:nmwd12}, the present status on the $\Delta I =1/2$ isospin rule in NMWD is concerned; 
in Sec.~\ref{subsec:nmwd_Gtot}, measurements and calculations of the total decay are described; 
in Sec.~\ref{subsec:nmwd_exp}, the present status of the experimental and theoretical results on the $\Gamma_{n} / \Gamma_{p}$ ratio is reported, 
with particular attention to the latest results from the FINUDA experiment; 
in Sec.~\ref{subsec:nmwd_2N}, the existence of the two--nucleon induced mechanism is discussed and different methods for the 
determination of its strength are described;
in Sec.~\ref{subsec:nmwd_rare}, recent measurements of branching ratios for two--body rare decays of $s$--shell Hypernuclei are reported;
in Sec.~\ref{subsec:nmwd_asym} the asymmetry in the spatial distribution of protons from 
NMWD of polarized Hypernuclei is discussed. 

\subsection{NMWD and the $\Delta I =1/2$ Rule}
\label{subsec:nmwd12}

The $\Delta I=1/2$ isospin rule is valid to a good degree
of approximation in various non--leptonic strangeness changing processes,
for instance in free $\Lambda$ and $\Sigma$ hyperon and pionic kaon decays.
While this is a well established empirical rule, it is still unknown whether
the large suppression of the $\Delta I=3/2$ transition amplitudes
with respect to the $\Delta I=1/2$ amplitudes holds as
a universal feature of all non--leptonic weak processes.
Also, different mechanisms seem to be responsible for the
above isospin rule in the various hadronic processes.

A possible relevance of $\Delta I=3/2$ terms in the Hypernuclear NMWD
would represent the first evidence for a $\Delta I=1/2$ rule violation
in non--leptonic strangeness changing interactions.
One should note that the $\Lambda N\to nN$ process has an important 
short--range part which is not accessible to those non--leptonic strangeness
changing particle physics processes that respect the $\Delta I=1/2$ rule.
Indeed, while the MWD only involves
$\Lambda \pi N$ vertex, which in free space respects the
$\Delta I=1/2$ rule, the NMWD also calls into play
mesons different from the pion in the $\Lambda$ vertexes and is thus
a complex source of information.
Nowadays, no experimental indication supports nor excludes the validity of the
$\Delta I=1/2$ rule for these couplings with heavy mesons.
Indirect information could thus come from Hypernuclear NMWD.

A possible violation of the $\Delta I=1/2$ rule in the NMWD was studied within a shell model   
framework in Ref.~\cite{Pa98}. In this work, a one--meson--exchange model with hadronic
couplings evaluated in the factorization approximation was adopted.
The conclusion reached by the authors is that only large $\Delta I=3/2$ factorization
terms (of the order of the $\Delta I=1/2$ ones) have a relevant effect on
the decay rates (and also on the asymmetry parameter) for $^{12}_\Lambda$C.   

Tests of the $\Delta I=1/2$ rule in NMWD are customarily
discussed by adopting a model by Block and Dalitz \cite{dalitz}. Such an approach
allows one to easily extract information on the spin--isospin
dependence of the $\Lambda N\to nN$ process directly from data on $s$--shell   
Hypernuclei. The neutron-- and proton--stimulated decay widths of $s$--shell  
Hypernuclei are obtained in terms of a few spin-- and isospin--dependent
rates for the elementary process $\Lambda N\to nN$.
The relationship among the elementary rates is strongly affected
by the isospin change experienced in the NMWD, both
$\Delta I=1/2$ and $\Delta I=3/2$ transitions being in principle possible.

Within the Block--Dalitz approximation, the width $\Gamma_{\rm NM}=
\Gamma_n+\Gamma_p$ of the Hypernucleus $^{A}_{\Lambda}Z$ turns
out to be factorized as follows into a density--dependent factor
and a term incorporating the dynamics of the decay \cite{dalitz}:
\begin{eqnarray}
\label{dalitz1}
\Gamma_{\rm NM}(^{A}_{\Lambda}Z)&=&\bar R(^{A}_{\Lambda}Z) \rho_A \\
&=& \frac{N\bar {R}_n(^{A}_{\Lambda}Z)+Z
\bar {R}_p(^{A}_{\Lambda}Z)}{A}\rho_A~,
\end{eqnarray}
where
\[
\rho_A \equiv \int d{\vec r} \rho_A({\vec r}) \mid
\psi_{\Lambda}({\vec r})\mid^2
\]
is the average nucleon density at the position of the $\Lambda$ baryon,
$\psi_{\Lambda}({\vec r})$ is the $\Lambda$ wave function in the
Hypernucleus and the nuclear density $\rho_A({\vec r})$ is normalized to the
mass number $A=N+Z$. Moreover, $\bar R$ denotes a rate
(per unit nucleon density at the $\Lambda$ position) averaged over spin and
isospin and is given in the second equality in Eq.~(\ref{dalitz1}) in
terms of the spin--averaged rates $\bar R_n$ and $\bar R_p$.

For $s$--shell Hypernuclei the $\Lambda N$ initial
pair is in the $L=0$ relative orbital angular momentum state and the possible
$\Lambda N\rightarrow nN$ transition channels are given in Table~\ref{partialt}
together with their main properties. 
\begin{table}[t]
\begin{center}
\caption{Amplitudes for the $\Lambda N\to nN$ decay in $s$--shell Hypernuclei.
The spectroscopic notation $^{2S+1}L_J$ is used.
$I_f$ is the isospin of the final $NN$ pair. With PC and PV
we denote parity--conserving and parity--violating channels,
respectively.}
\label{partialt}
\begin{tabular}{l c c c} \hline
\mc {1}{c}{Amplitude} &
\mc {1}{c}{Channel} &
\mc {1}{c}{$I_f$} &
\mc {1}{c}{Parity} \\ \hline
$a_p$, $a_n$ & ${^1S_0} \to {^1S_0}$ & 1 & PC  \\
$b_p$, $b_n$ & ${^1S_0} \to {^3P_0}$ & 1 & PV  \\
$c_p$        & ${^3S_1} \to {^3S_1}$ & 0 & PC  \\
$d_p$        & ${^3S_1} \to {^3D_1}$ & 0 & PC  \\
$e_p$        & ${^3S_1} \to {^1P_1}$ & 0 & PV  \\
$f_p$, $f_n$ & ${^3S_1} \to {^3P_1}$ & 1 & PV  \\ \hline
\end{tabular}
\end{center}
\end{table} 

In terms of the amplitudes of Table~\ref{partialt}, the rates $R_{NJ}$ for the spin--singlet
($R_{n0}$, $R_{p0}$) and spin--triplet ($R_{n1}$, $R_{p1}$) elementary $\Lambda N\to nN$ interactions are given by:
\begin{eqnarray}
\label{erre}
R_{n0}&=&|a_n|^2+|b_n|^2~,  \\
R_{p0}&=&|a_p|^2+|b_p|^2~, \nonumber \\
R_{n1}&=&|f_n|^2~, \nonumber \\
R_{p1}&=&|c_p|^2+|d_p|^2+|e_p|^2+|f_p|^2~. \nonumber
\end{eqnarray}
The NMWD widths of $s$--shell Hypernuclei are thus derived in the following form \cite{dalitz}:
\begin{eqnarray}
\label{phen}
\Gamma_{\rm NM}(^3_{\Lambda}{\rm H})&=&
\left(3R_{n0}+R_{n1}+3R_{p0}+R_{p1}\right)\frac{\rho_3}{8}~,\\
\Gamma_{\rm NM}(^4_{\Lambda}{\rm H})&=&
\left(R_{n0}+3R_{n1}+2R_{p0}\right)\frac{\rho_4}{6}~, \\
\Gamma_{\rm NM}(^4_{\Lambda}{\rm He})&=&
\left(2R_{n0}+R_{p0}+3R_{p1}\right)\frac{\rho_4}{6}~, \\
\label{last-phen}
\Gamma_{\rm NM}(^5_{\Lambda}{\rm He})&=&
\left(R_{n0}+3R_{n1}+R_{p0}+3R_{p1}\right)\frac{\rho_5}{8}~.
\end{eqnarray}

The isospin--dependence of the Block--Dalitz rates can be summarized by the following relation:
\begin{equation}
\label{uno2}
\frac{R_{n1}}{R_{p1}}\leq \frac{R_{n0}}{R_{p0}}=2~,
\end{equation}
which holds for pure $\Delta I=1/2$ transitions. Instead, for pure $\Delta I=3/2$ transitions one has:
\begin{equation}
\label{uno3}
\frac{R_{n1}}{R_{p1}}= \frac{R_{n0}}{R_{p0}}=\frac{1}{2}~.
\end{equation}

Various equalities and inequalities among the Hypernuclear decay rates can
be obtained from Eqs.~(\ref{phen})--(\ref{uno2}) in the limit of pure $\Delta I=1/2$ transitions.
The most interesting relation reads:
\begin{equation}
\label{best-tested}
\frac{R_{n0}}{R_{p0}} \equiv
\frac{\Gamma_n(^4_\Lambda {\rm He})}{\Gamma_p(^4_\Lambda {\rm H})}=2~.
\end{equation}
Predictions of this kind turn out to be relevant for an experimental refutation of the $\Delta I =1/2$ rule.

In various works, the Block--Dalitz rates were determined phenomenologically   
by fitting existing data on $s$--shell Hypernuclei for $\Gamma_{\rm NM}$ and $\Gamma_n/\Gamma_p$.
We consider here in some detail the work by Schumacher \cite{schumacher}.
He examined the existing data on NMWD of $s$--shell Hypernuclei to obtain a quantitative estimate of the 
relative strength of the two isospin channels of the $\Lambda$ decay. 
He observed that the NMWD approaches available at that time were able to achieve reasonable agreement 
with the data on the total NMWD rates, by assuming the validity of the $\Delta I=1/2$ rule, but no framework had succeeded in reproducing the measured 
$\Gamma_{n}/\Gamma_{p}$ ratios. To further test the isospin rule for the less understood parity--conserving part of the weak interaction, he 
examined the available data on $\Gamma_{n}/\Gamma_{p}$ for $^{4}_{\Lambda}$He and $^{5}_{\Lambda}$He and the ratio 
$\Gamma_{\rm NM}(^{4}_{\Lambda}{\rm He})/\Gamma_{\rm NM}(^{4}_{\Lambda}{\rm H})$ from Refs.~\cite{dalitz,szym}. 
In the calculations, the $^{1}S_{0}$ $\Lambda N$ initial state was isolated because it leads to isospin $I_f=1$ final states, populated by both 
neutron-- and proton--induced processes, while the $^{3}S_{1}$ $\Lambda N$ initial state leads to $I_f=0$ final states that can be produced only 
by the proton--induced reaction, being the neutron--induced transition forbidden by the Pauli principle. 
From the data, Schumacher determined $R_{n0}/R_{p0} = 0.23 \pm 0.17$, to be compared with the value 2 predicted by pure
$\Delta I=1/2$ $\Lambda N\to nN$ transitions (see Eq.~(\ref{uno2})). This seems to suggest a rather large violation of the 
$\Delta I=1/2$ rule for $s$--shell Hypernuclei NMWD.
Due to the quite large errors affecting the experimental values of Refs.~\cite{dalitz,szym}, ranging from $38\%$ to $59\%$, 
this result, however, cannot be considered as definitive. 

An analysis of more recent experimental data on $s$--shell Hypernuclei was performed in Ref.~\cite{al_gar},
again with the phenomenological Block--Dalitz model. It was concluded that the large error bars in the data do not allow
to draw definite conclusions about the possible violation of the
$\Delta I=1/2$ rule and the spin--dependence of the Block--Dalitz transition rates.
The data turned out to be consistent with the hypothesis of validity of the $\Delta I=1/2$ rule at the level of 60\%. In other words,
the $\Delta I=1/2$ rule could be excluded only at the 40\% confidence level. This result is not changed much if present data are used.

The usual outcome of all investigations is that more precise measurements of $\Gamma_n$ for $^{4}_\Lambda$He
and $\Gamma_p$ for $^{4}_\Lambda$H (see Eq.~(\ref{best-tested})) are needed to reach reliable conclusions on the possible violation of the 
$\Delta I=1/2$ rule.
At present, we known that $\Gamma_{n} (^{4}_{\Lambda}{\rm He})$ is very small \cite{outa98}, compatible with zero within the experimental error, 
while there is no experimental data available for $\Gamma_{p} (^{4}_{\Lambda}{\rm H})$: this prevents 
any further study for the possible violation of the isospin rule in the sector of $s$--shell Hypernuclei with the Block--Dalitz model.
New and more precise experiments are thus necessary. In particular,
for a measurement of the $^{4}_{\Lambda}$H NMWD one needs to tag the formation of the Hypernucleus through the detection of 
a neutral meson, like in ($K^{-}, \pi^{0}$) or ($\pi^{-}, K^{0}$) reactions, and so very intense beams are necessary.
Such beams will be available very soon at the J--PARC accelerator complex, where the NMWD rates of $A=4$ Hypernuclei 
will be determined with improved precision \cite{jparc-4}.

\subsection{NMWD: Total Decay Width}
\label{subsec:nmwd_Gtot}
\begin{figure} [h]
\begin{center}
\resizebox{0.52\textwidth}{!}{%
 \hspace{-5mm}
  \includegraphics{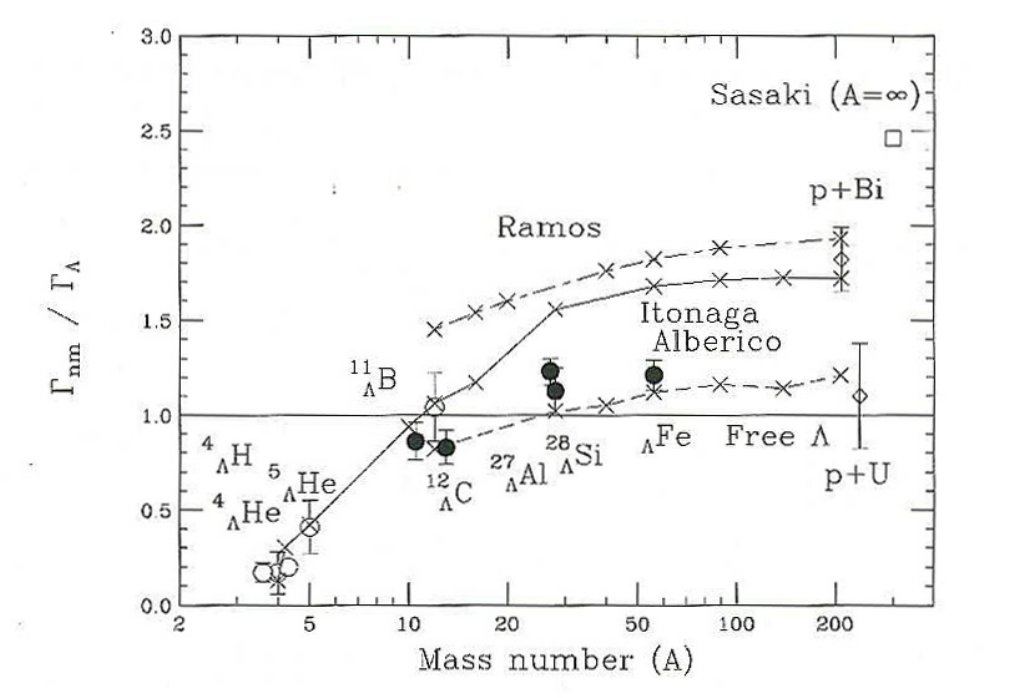}
}
\caption{Mass number dependence of the NMWD rate measured before 2001. Full circles are data from the KEK--E307 experiment \cite{sato05}, open circles 
are previous data 
from counter experiments in which the formation of the Hypernucleus was explicitly identified \cite{szym,outa95,outa98}, open diamonds are data 
obtained at COSY on heavy nuclei, Bi and U,  \cite{ohm,kulessa}. Solid, dot--dashed and dashed lines correspond, respectively, to calculations by 
Refs.~\cite{itonaga2}, \cite{ramos} and \cite{albe2}. The open square shows a result from direct quark exchange by Ref.~\cite{sasaki}
for a $\Lambda$ bound in nuclear matter. From 
Ref.~\cite{sato05}.}
\label{fig:nmwd_rate}       
\end{center}
\vspace{-4mm}
\end{figure}

The total NMWD rate, $\Gamma_{\rm NM}$, can be determined experimentally from the lifetime and the MWD width measurements: 
$\Gamma_{\rm NM} = \Gamma_{T} - \Gamma_{\pi^{-} }- \Gamma_{\pi^{0}}$.  
In Fig.~\ref{fig:nmwd_rate}, taken from Ref.~\cite{sato05}, results obtained in the $4\leq A\leq 238$ mass number range before 2001
are reported.
The full circles are data from the KEK--E307 experiment \cite{sato05}, the open circles are previous data from counter experiments in which the 
formation of the Hypernucleus was 
explicitly identified \cite{szym,outa95,outa98}, the open diamonds are data obtained at COSY on heavy nuclei, Bi and U,  \cite{ohm,kulessa}. 
Solid, dot--dashed and dashed lines correspond, respectively, to calculations by Itonaga {\it et al.} \cite{itonaga2}, Ramos {\it et al.} \cite{ramos} 
and Alberico {\it et al.} \cite{albe2}. The open square shows a result from the direct quark mechanism \cite{sasaki} for a $\Lambda$ in nuclear 
matter. The experimental data up to $A=56$ clearly indicate an increasing trend up to $A\sim 30$, followed by a saturation of  
$\Gamma_{\rm NM}$ around 1.2-1.3 $\Gamma_{\Lambda}$, which mainly affects the Hypernucleus lifetime $A$--dependence. 
In Ref.~\cite{ramos}, the OPE potential was considered with pion renormalization effects in the nuclear medium, while the local density 
approximation was used to study finite Hypernuclei. 
The results for $\Gamma_{\rm NM}$ are sensibly larger than the experimental data. 
Recently, the calculation of Ref.~\cite{ramos} was updated by Ref.~\cite{albe2}, which   
obtained values of $\Gamma_{\rm NM}$ in agreement with data by tuning the Landau--Migdal parameters which fixes the short--range 
part of the $\Lambda N$ and $NN$ interactions and by using more realistic in--medium $\Lambda$ wave functions.

More recently, the KEK high statistics experiments E462 and E508 measured the lifetime, $\Gamma_{\pi^{-}}$ and $\Gamma_{\pi^{0}}$ for 
$^{5}_{\Lambda}$He and $^{12}_{\Lambda}$C with errors reduced to the 4-5$\%$ level \cite{kame05,okada}; the 
resulting values of $\Gamma_{\rm NM}$, in units of the free $\Lambda$ decay rate $\Gamma_{\Lambda}$, are: 
$\Gamma_{\rm NM}(^{5}_{\Lambda}{\rm He})=0.406\pm0.020$ and 
$\Gamma_{\rm NM}(^{12}_{\Lambda}{\rm C})=0.953\pm0.032$, which adhere completely to the trend indicated by the previous data in 
Fig.~\ref{fig:nmwd_rate}.

In the high mass number region, the very low values of lifetime obtained by Ref.~\cite{cassing} 
for $A=180$-$225$, give a very high value of the total NMWD rate, $\Gamma_{\rm NM}/\Gamma_{\Lambda}=1.82\pm0.14$, being 
the MWD mechanism completely negligible in this region. However, we have to remind the reader that,
as discussed in Sec.~\ref{subsec:tau_meas}, this result was obtained from delayed fission experiments with the recoil shadow method, 
by detecting nuclear fragments which are supposed to originate from Hypernuclear fission, induced by NMWD.
This method does not allow the identification of the produced Hypernucleus and requires model--dependent assumptions.

In Table~\ref{tot-nmwd} we summarize the theoretical and experimental determinations of the total NMWD rate
for $^{5}_{\Lambda}$He and $^{12}_{\Lambda}$C. We only list works which also determined the $\Gamma_n/\Gamma_p$ 
ratio. We note that only the theoretical works of Refs.~\cite{Os01,Ba10b} include two--nucleon induced weak decays.
Most of the theoretical predictions agree with data. However, we have to note that with the most recent and precise KEK--E462
and KEK--E508 data one is able to distinguish among the different theoretical approaches. 
Moreover, within a particular approach, one could also determine the weak transition potential which better reproduce
the KEK--E462 and KEK--E508 data. 
\begin{table}[h]
\caption{Theoretical and experimental results for the NMWD rate $\Gamma_{\rm NM}$ in units of the free $\Lambda$ decay rate
$\Gamma_{\Lambda}$.}
\vskip 2mm
\label{tot-nmwd}
\begin{center}
\resizebox{8.7cm}{!} {
\begin{tabular}{lcc} \hline
\mc {1}{c}{Ref. and Model} &
\mc {1}{c}{$^5_{\Lambda}$He} &
\mc {1}{c}{$^{12}_{\Lambda}$C} \\ \hline
Sasaki et al. \cite{sasaki} &0.52 &  \\
$\pi +K+$ DQ               & &  \\
Jido et al. \cite{Os01} & &1.04 \\
$\pi +K+2\pi/\sigma+2\pi +\omega$      & &  \\
Parre\~{n}o and Ramos \cite{parreno} &$0.43$ & $0.73$  \\
$\pi +\rho +K + K^* + \omega +\eta$                        & &  \\
Itonaga et al. \cite{itonaga2}    & 0.42 & 1.06  \\
$\pi + 2\pi/\sigma + 2\pi/\rho+\omega$ & &       \\
Barbero et al. \cite{Ba03} & 0.69 & 1.17  \\
$\pi +\rho +K + K^* + \omega +\eta$ &  &      \\
Bauer and Garbarino \cite{Ba10b}  & & 0.98    \\
$\pi +\rho +K + K^* + \omega +\eta$ &  &      \\ \hline
BNL \cite{szym} &$0.41\pm0.14$    &$1.14\pm0.20$ \\
KEK \cite{noumi} & &$0.89\pm0.18$ \\
KEK \cite{noumi2} &$0.50\pm0.07$ &  \\
KEK--E307 \cite{sato05} & &$0.828\pm 0.056 \pm 0.066$ \\    
KEK--E462 \cite{kame05} & $0.404\pm 0.020$ & \\
KEK--E508 \cite{okada} &  &$0.953\pm 0.032$ \\\hline
\end{tabular}}
\end{center}
\end{table}


\subsection{NMWD: the $\Gamma_{n} / \Gamma_{p}$ puzzle}
\label{subsec:nmwd_exp}

Table~\ref{tab:nmwd_data} summarizes, in chronological order of publication, the results of measurements on NMWD of $A=5$-$56$ Hypernuclei, performed 
from 1991 on: they represent the subsequent steps toward the solution of the $\Gamma_{n}/\Gamma_{p}$ puzzle from the experimental side.
\begin{table}[h] 
\caption{NMWD measurements for $A=5$-$56$ Hypernuclei after 1991, listed in chronological order of publication. In the first column details of the 
experiment are reported, in column two the reference is given and in column three details of the measurement are reported.}
\label{tab:nmwd_data}       
\resizebox{8.7cm}{!} {
\begin{tabular}{lll}
\hline\noalign{\smallskip}
Experiment & Ref. & Measurement  \\
\noalign{\smallskip}\hline\noalign{\smallskip}
BNL AGS, LESB I & \cite{szym} & $^{5}_{\Lambda}$He: p spectrum, $\Gamma_{p}$, $\Gamma_{n}$,  \\
($K^{-}, \pi^{-}$), 800 MeV/c &  & $\Gamma_{\rm NM}$; $^{12}_{\Lambda}$C: $\Gamma_{\rm NM}$, $\Gamma_{n}/\Gamma_{p}$ \\
\noalign{\smallskip}\hline\noalign{\smallskip}
KEK PS E160 &  \cite{noumi} & $^{12}_{\Lambda}$C: p spectrum;   \\
 ($\pi^{+}, K^{+}$), 1.05 GeV/c & &  $^{11}_{\Lambda}$B and $^{12}_{\Lambda}$C: $\Gamma_{p}/\Gamma_{\Lambda}$, \\
 & & $\Gamma_{\rm NM}/\Gamma_{\Lambda}$, $\Gamma_{n}/\Gamma_{p}$ \\
\noalign{\smallskip}\hline\noalign{\smallskip}
KEK PS E278 &  \cite{noumi2} & $^{5}_{\Lambda}$He:  \\
 ($\pi^{+}, K^{+}$), 1.05 GeV/c & &  $\Gamma_{\rm NM}$, $\Gamma_{n}/\Gamma_{p}$ \\
\noalign{\smallskip}\hline\noalign{\smallskip}
KEK PS E307 &  \cite{hashi3} & $^{12}_{\Lambda}$C and $^{28}_{\Lambda}$Si: \\
 ($\pi^{+}, K^{+}$), 1.05 GeV/c & & p spectrum, $\Gamma_{n}/\Gamma_{p}$ \\
 \noalign{\smallskip}\hline\noalign{\smallskip}
KEK PS E369 &  \cite{hjkim} & $^{12}_{\Lambda}$C and $^{89}_{\Lambda}$Y: n spectrum \\
 ($\pi^{+}, K^{+}$), 1.05 GeV/c & & $^{12}_{\Lambda}$C: $\Gamma_{n}/\Gamma_{p}$ \\ 
 \noalign{\smallskip}\hline\noalign{\smallskip}
KEK PS E462, E508 &  \cite{okada2} & $^{5}_{\Lambda}$He and $^{12}_{\Lambda}$C: \\
 ($\pi^{+}, K^{+}$), 1.05 GeV/c & & p and n spectra, $\Gamma_{n}/\Gamma_{p}$ \\ 
 \noalign{\smallskip}\hline\noalign{\smallskip}
KEK PS E307 &  \cite{sato05} &$^{11}_{\Lambda}$B, $^{12}_{\Lambda}$C, $^{27}_{\Lambda}$Al, $^{28}_{\Lambda}$Si, $_{\Lambda}$Fe: \\
 ($\pi^{+}, K^{+}$), 1.05 GeV/c & & p spectrum, $\Gamma_{\rm NM}/\Gamma_{\Lambda}$ \\
 \noalign{\smallskip}\hline\noalign{\smallskip}
 KEK PS E462 &  \cite{kang} & $^{5}_{\Lambda}$He: nn and np spectra,  \\
 ($\pi^{+}, K^{+}$), 1.05 GeV/c & & $\Gamma_{n}/\Gamma_{p}$ \\ 
 \noalign{\smallskip}\hline\noalign{\smallskip}
 KEK PS E508 &  \cite{mjkim} & $^{12}_{\Lambda}$C: nn and np spectra, \\
 ($\pi^{+}, K^{+}$), 1.05 GeV/c & &  $\Gamma_{n}/\Gamma_{p}$ \\ 
 \noalign{\smallskip}\hline\noalign{\smallskip}
KEK PS E462, E508 &  \cite{bhang} & $^{5}_{\Lambda}$He: p and n spectra \\
 ($\pi^{+}, K^{+}$), 1.05 GeV/c & & $^{12}_{\Lambda}$C: p and n spectra, \\ 
  & & $\Gamma_{n}/\Gamma_{p}$ \\ 
 \noalign{\smallskip}\hline\noalign{\smallskip}
BNL AGS, LESB II & \cite{parker} & $^{4}_{\Lambda}$He: p spectra, \\
($K^{-}, \pi^{-}$), 750 MeV/c &  & $\Gamma_{p}$, $\Gamma_{n}$,  $\Gamma_{n}/\Gamma_{p}$ \\
\noalign{\smallskip}\hline\noalign{\smallskip}
LNF DA$\Phi$NE, FINUDA & \cite{npa804} &   $^{5}_{\Lambda}$He,  $^{7}_{\Lambda}$Li, $^{12}_{\Lambda}$C: p spectra \\
($K^{-}_{\rm stop}, \pi^{-}$) & & \\
\noalign{\smallskip}\hline\noalign{\smallskip}
 KEK PS E508 &  \cite{mkim} & $^{12}_{\Lambda}$C: nn and np spectra, \\
 ($\pi^{+}, K^{+}$), 1.05 GeV/c & &  $\Gamma_{n}$, $\Gamma_{p}$,  $\Gamma_{2}$ \\ 
 \noalign{\smallskip}\hline\noalign{\smallskip}
LNF DA$\Phi$NE, FINUDA & \cite{fnd_nmwd} &  $^{5}_{\Lambda}$He,  $^{7}_{\Lambda}$Li, $^{9}_{\Lambda}$Be, $^{11}_{\Lambda}$B, $^{12}_{\Lambda}$C \\
($K^{-}_{\rm stop}, \pi^{-}$)& & $^{15}_{\Lambda}$N, $^{16}_{\Lambda}$O: p spectra, $\Gamma_2$ \\
 \noalign{\smallskip}\hline\noalign{\smallskip}
LNF DA$\Phi$NE, FINUDA & \cite{nmwd_n} &  $^{5}_{\Lambda}$He,  $^{7}_{\Lambda}$Li, $^{9}_{\Lambda}$Be, $^{11}_{\Lambda}$B, $^{12}_{\Lambda}$C \\
($K^{-}_{\rm stop}, \pi^{-}$)& & $^{15}_{\Lambda}$N, $^{16}_{\Lambda}$O: np spectra, $\Gamma_2$ \\
\noalign{\smallskip}\hline\noalign{\smallskip}
\end{tabular}}
\end{table}

As can be seen from the table, the major part of the data have been obtained at KEK on $^{5}_{\Lambda}$He and $^{12}_{\Lambda}$C, which are the best 
known Hypernuclei as for NMWD, and were published from 1995 to 2009 \cite{noumi,hashi3,hjkim,okada2,sato,kang,mjkim,bhang,mkim}. 
BNL experiments explored NMWD of $^{4}_{\Lambda}$He, $^{5}_{\Lambda}$He and $^{12}_{\Lambda}$C during the 90's \cite{szym,parker}. 
Recently, the FINUDA experiment performed a systematic study of the spectra of protons emitted in NMWD of $^{5}_{\Lambda}$He and the complete set of 
$p$--shell Hypernuclei \cite{npa804,fnd_nmwd,nmwd_n}. 

Concerning the $\Gamma_{n}/\Gamma_{p}$ ratio, the first counter experiments studying the NMWD processes detected only protons 
\cite{szym,noumi,noumi2} and determined the $\Gamma_{p}$ value for $^{5}_{\Lambda}$He,  $^{11}_{\Lambda}$B and $^{12}_{\Lambda}$C; the $\Gamma_{n}$ 
values were obtained by subtraction of all the other decay rates. Thus, the results might have been affected by an underestimation of the number
of emitted protons due to the rescattering of the proton inside the residual nucleus, i.e. the nucleon FSI, 
and/or to the existence of two--nucleon induced decay modes (\ref{gamma2}), 
which at that time had been suggested by Ref.~\cite{al91}, but was not yet observed. In this way, missing protons could be considered as neutrons and 
resulted in an artificial increase of the $\Gamma_{n}/\Gamma_{p}$ ratio, contributing, from the experimental side, to the birth of the 
$\Gamma_{n}/\Gamma_{p}$ puzzle. $\Gamma_{n}/\Gamma_{p}=0.93\pm0.55$ for $^{5}_{\Lambda}$He, $1.04^{+0.59}_{-0.48}$ for $^{11}_{\Lambda}$B 
and $1.33^{+1.12}_{-0.81}$ for $^{12}_{\Lambda}$C were reported by Ref.~\cite{szym}, while Ref.~\cite{noumi} obtained 
$\Gamma_{n}/\Gamma_{p}=2.16\pm0.58^{+0.45}_{-0.95}$ for $^{11}_{\Lambda}$B 
and $1.87\pm0.59^{+0.32}_{-1.00}$ for $^{12}_{\Lambda}$C and Ref.~\cite{noumi2} reported $\Gamma_{n}/\Gamma_{p}=1.97\pm0.67$ for 
$^{5}_{\Lambda}$He; the quite big statistical errors, varying from about 25$\%$ to about 50$\%$, were due to the limited 
statistics of the data samples. On the contrary, smaller values of $\Gamma_n/\Gamma_p$, between 0.3 and 0.5, were predicted by
the most realistic theoretical approaches for all these Hypernuclei.

\subsubsection{Single Nucleon Spectra Analyses}
\label{subsec:nmwd_single}

It must be stressed that nucleon rescattering in the residual nucleus strongly modifies the shape of the emitted nucleon distributions. FSI affect not only the 
energy and direction of nucleons, through scattering processes, but also their charge, through charge--exchange processes, and number, through 
knock--out reactions: FSI thus tend to enhance (decrease) the low--energy (high--energy) region in the nucleon energy spectra.

The highest statistics experiment measuring single proton spectra were carried out at KEK--E307 \cite{hashi3,sato}. They measured the number of 
protons emitted in the NMWD of $^{12}_{\Lambda}$C, $^{28}_{\Lambda}$Si and $_{\Lambda}$Fe (a mixture of $^{56}_{\Lambda}$Fe, $^{55}_{\Lambda}$Fe and 
$^{55}_{\Lambda}$Mn), formed in the ($\pi^{+}, K^{+}$) reaction with the SKS spectrometer and 
a coincidence detector, located below and above the reaction target, consisting of timing and veto scintillators, range counters made up of stacks of 
scintillators and drift chambers to detect the charged decay products, to make particle identification (p.id.) between $\pi^{-}$ and p and to measure their energy. 
The proton yield $R^{exp}_{p}$ was measured as a function of the detected proton kinetic energy $E_p$ and normalized to the number of produced 
Hypernuclei, giving $R^{exp}_{p} = Y_{coinc}(E_{p})/Y_{hyp}$, where $Y_{coinc}$ is the number of protons detected in coincidence with the $K^{+}$ 
identifying Hypernucleus production and $Y_{hyp}$ is the number of produced Hypernuclei.  

\begin{figure} [h]
\begin{center}
\resizebox{0.47\textwidth}{!}{%
 \hspace{-5mm}
  \includegraphics{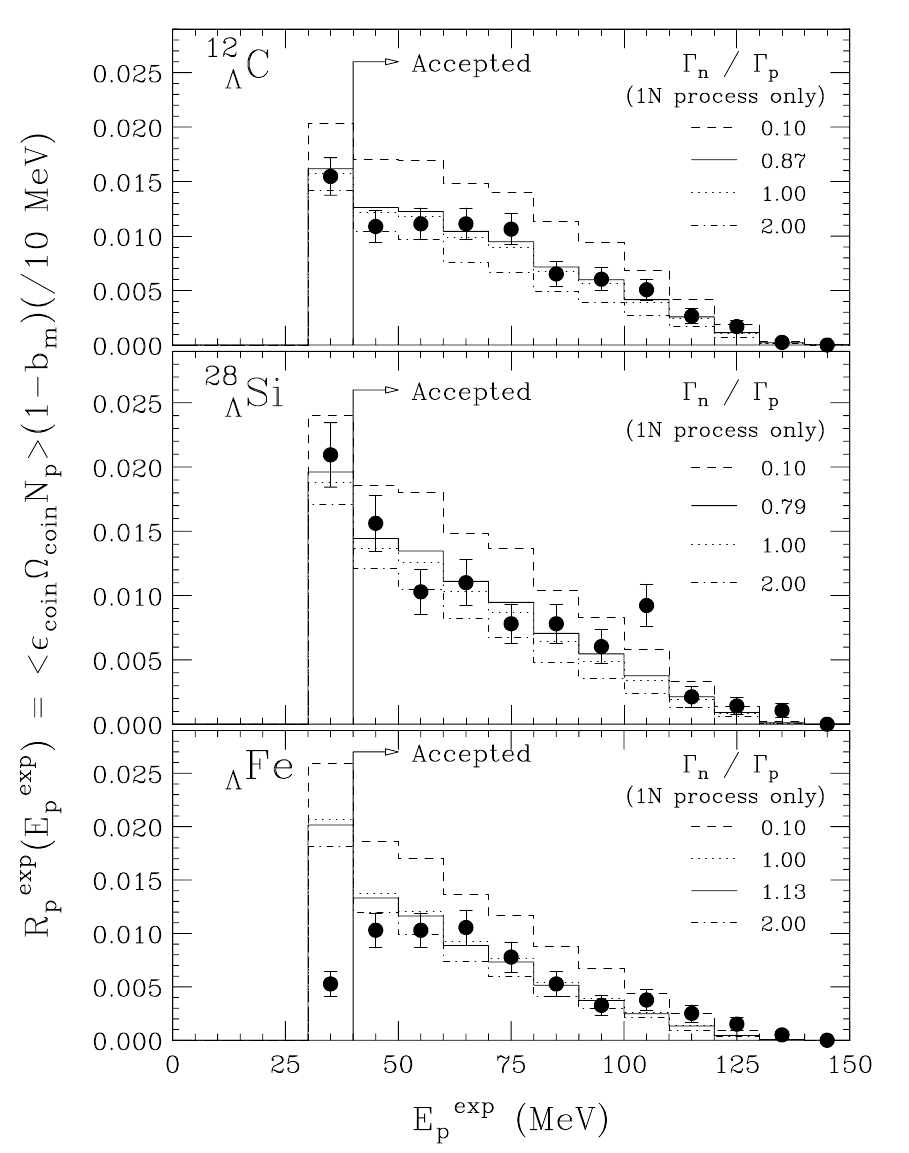}
}
\caption{Protons spectra from NMWD of $^{12}_{\Lambda}$C, $^{28}_{\Lambda}$Si and $_{\Lambda}$Fe as a function of the measured proton kinetic 
energy. The histograms superimposed to the data represent the simulated spectra from Refs.~\cite{ramos1,ramos2} for different values of the 
$\Gamma_{n}/\Gamma_{p}$ ratio; the continuous line histograms refer to the best reproduction of the experimental spectra. From Ref.~\cite{sato05}.}
\label{fig:sato8}       
\end{center}
\vspace{-4mm}
\end{figure}

\begin{figure} [h]
\begin{center}
\resizebox{0.5\textwidth}{!}{%
 \hspace{-5mm}
  \includegraphics{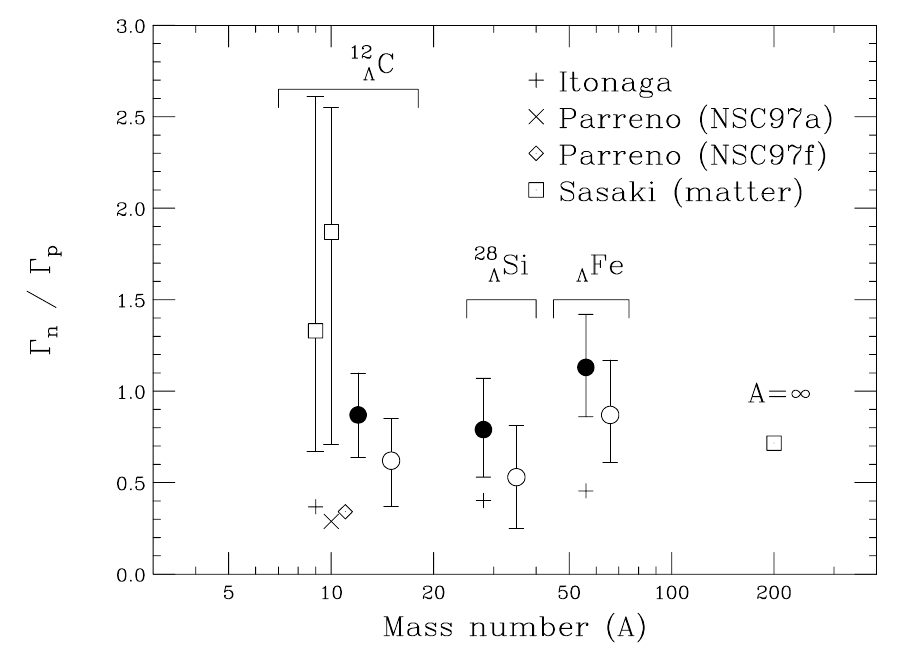}
}
\caption{$\Gamma_{n}/\Gamma_{p}$ values for the Hypernuclei studied by Refs.~\cite{szym,noumi}, open squares, and by Ref.~\cite{sato05}, full 
(with only one--nucleon induced decays) and open (with the inclusion of two--nucleon induced decays) circles. 
Theoretical calculations by Refs.~\cite{itonaga2,parreno,sasaki} are also indicated. From Ref~\cite{sato05}.}
\label{fig:sato11}       
\end{center}
\vspace{-4mm}
\end{figure}

Fig.~\ref{fig:sato8} shows the raw proton spectra, normalized per NMWD and not corrected for the detector acceptance and efficiency, as a 
function of the measured proton energy; a lower cut at about 30 MeV, due to the target thickness, is visible. 
The $R^{exp}_{p}$ spectra were then compared with those from theoretical calculations obtained by describing the  
one--nucleon induced NMWD processes (\ref{gamman}) and (\ref{gammap}) through OPE models, using the local density approximation \cite{ramos1,ramos2} 
and taking into account nucleon FSI effects in the residual nucleus by an intranuclear cascade (INC) calculations and FSI effects in the target 
material by the GEANT simulation \cite{geant}. 
The $\Gamma_{n}/\Gamma_{p}$ values for the three targets were obtained in an indirect way, by a comparison between the calculated 
spectra and the experimental ones. The theoretical 
calculations are reported in Fig.~\ref{fig:sato8} as histograms superimposed to the experimental data, for different values of the 
$\Gamma_{n}/\Gamma_{p}$ ratio. The final values of $\Gamma_{n}/\Gamma_{p}$ are given in the same figure and correspond to the 
continuous line histograms; in Fig.~\ref{fig:sato11} they are reported as a function of the mass number together with the results of Refs.~\cite{szym,noumi} 
and the calculations of Refs.~\cite{itonaga2,parreno,sasaki}. 
Although all available experimental data agree within $1\sigma$ level, the values of Ref.~\cite{sato05} are in the region around 0.5-1.0, whereas the 
previous ones are significantly larger. It turns out that there is no theoretical calculation able to reproduce 
simultaneously the measured values of $\Gamma_{n}/\Gamma_{p}$ and 
$\Gamma_{\rm NM}$. This is the core of the $\Gamma_{n}/\Gamma_{p}$ puzzle. 

The single proton experiments show evident drawbacks in the determination of $\Gamma_{n}/\Gamma_{p}$. In particular, to obtain sufficient statistics, 
thick targets have to be used, in which protons strongly suffer from energy losses and an important deformation of their spectrum is produced. 
Nucleon FSI inside the nucleus reduce the population of the higher energy part of the spectrum and have to be taken into account in the theoretical
calculations with a weight which is not known experimentally.
Moreover, if neutrons are not directly counted, the MWD branching ratio has to be used, that contributes with its uncertainty to the total error. 
Finally, the proton energy threshold of the experimental apparatus produces a loss in 
the low--energy proton region, which is populated not only by one--nucleon induced decays but also by FSI and two--nucleon induced 
processes, giving a systematically higher value for $\Gamma_{n}/\Gamma_{p}$ if these last processes are ignored in the analysis. 

It is thus clear that the measurement of the NMWD neutrons is mandatory for a reliable measurement of $\Gamma_n/\Gamma_p$. 
The detection of neutrons provides cleaner data than for protons: neutral particles, 
indeed, do not undergo energy loss in the target material; moreover, if neutrons are counted, it is no more necessary to use MWD branching ratios in 
the analysis. As for FSI, the effect on neutrons and on protons can be assumed to be similar, due to the charge symmetry of the $NN$ scattering cross 
section, and cancel out, at least at the first order, if one considers the ratio between neutron and proton spectra. 
Neutron spectra from NMWD of $^{12}_{\Lambda}$C and $^{89}_{\Lambda}$Y were measured with high statistics by the KEK--E369 experiment \cite{hjkim}, as 
a by--product of the Hypernuclear spectroscopy measurements. For $^{12}_{\Lambda}$C the yield of neutrons from Ref.~\cite{hjkim} was compared to that 
of protons from Ref.~\cite{sato05}; the $N_{n}/N_{p}$ ratio between the neutron and proton numbers, corrected for acceptance and efficiency, was found to be 
slightly less than 2; this suggests, through the simple relation $N_{n}/N_{p} \sim 2(\Gamma_{n}/\Gamma_{p}) + 1$ that 
holds considering the neutron and proton multiplicities in (\ref{gamman}) and (\ref{gammap}), that $\Gamma_{n}/\Gamma_{p} \sim 0.5$,
improving the agreement with the most realistic theoretical predictions ranging from 0.3 to 0.5. 
The dominance of the proton--induced decay channel over the neutron--induced one was thus shown experimentally for the first time. 

\begin{figure} [h]
\begin{center}
\resizebox{0.5\textwidth}{!}{%
 \hspace{-5mm}
  \includegraphics{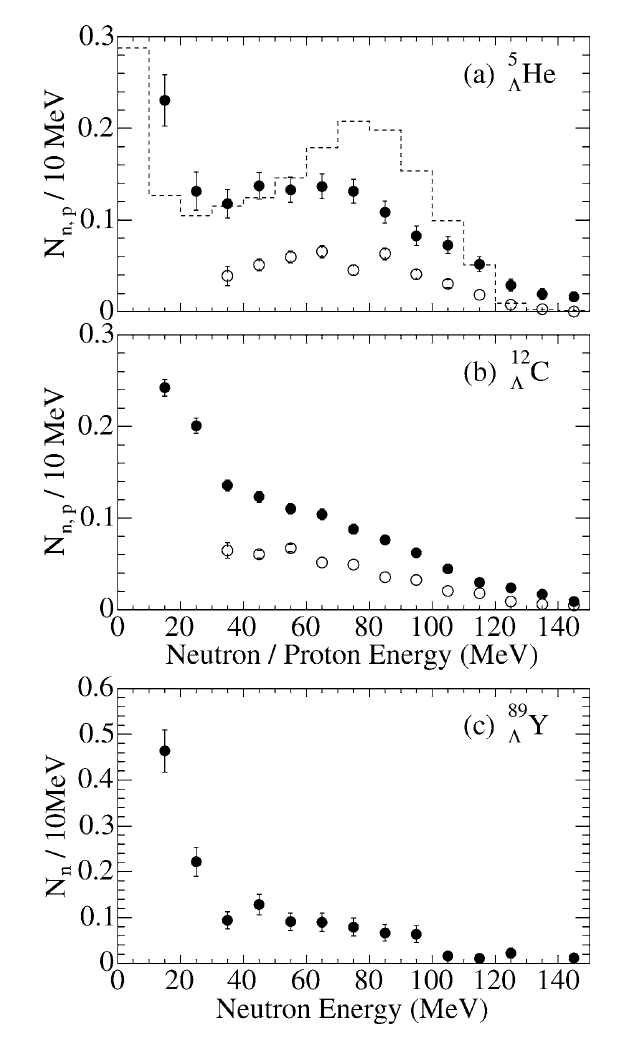}
}
\caption{Neutron (filled circles) and proton (open circles) kinetic energy spectra of (a) $^{5}_{\Lambda}$He and (b) $^{12}_{\Lambda}$C 
\cite{okada}; (c) neutron spectrum for $^{89}_{\Lambda}$Y \cite{hjkim}. The spectra are normalized per NMWD.
The dashed histogram in the upper figure shows the neutron spectrum 
calculated in Ref.~\cite{garb}. From Ref.~\cite{okada2}.}
\label{fig:okada}       
\end{center}
\vspace{-4mm}
\end{figure}

A very high statistics measurement of NMWD proton and neutron spectra was performed by the KEK--E462 experiment for $^{5}_{\Lambda}$He and by the 
KEK--E508 experiment for $^{12}_{\Lambda}$C \cite{okada,okada2}. Protons and neutrons were identified by a decay coincidence system analogous to that of 
Ref.~\cite{sato05}, but figuring a larger angular acceptance and detection efficiency and a clear separation capability between $\pi$/p/d for charged 
particles and $\gamma$/n for neutral particles. The kinetic energy detection threshold was 30 MeV for protons and 15 MeV for neutrons. The number of 
nucleons emitted per NMWD was determined as a function of the measured kinetic energy; the spectra are reported in Fig.~\ref{fig:okada}. It is 
possible to observe that the neutron spectra of both $^{5}_{\Lambda}$He and $^{12}_{\Lambda}$C have a shape similar to those of protons above the 
proton energy threshold. To evaluate the $\Gamma_{n}/\Gamma_{p}$ ratio, a threshold was set at 60 MeV on both nucleons spectra, to reduce the 
contribution from FSI and two--nucleon induced processes; the values of $N_{n}/N_{p}=2.17\pm0.15\pm0.16$ and $N_{n}/N_{p}=2.00\pm0.09\pm0.14$ 
were obtained for $^{5}_{\Lambda}$He and $^{12}_{\Lambda}$C, respectively, leading to $\Gamma_{n}/\Gamma_{p}=
(N_n/N_p-1)/2 \sim0.5$-$0.6$. 

By comparing the proton and neutron spectra from Ref.~\cite{okada} with the neutron spectrum for $^{89}_{\Lambda}$Y from Ref.~\cite{hjkim} 
of Fig.~\ref{fig:okada}(c), it is possible to see the suppression of the high--energy neutron yield with the mass number and the enhancement of 
the low--energy region in the heavy Hypernuclei, which can be naturally interpreted as an effect of the FSI. If one--nucleon induced processes dominate and 
FSI can be neglected, the neutron energy spectrum is expected to show a broadened peak, due to the Fermi motion, at about one half of the reaction 
Q--value: this should be the case of $^{5}_{\Lambda}$He, due to its small mass number. On the other hand, 
the shape of the KEK $^{5}_{\Lambda}$He neutron spectrum of Fig.~\ref{fig:okada} shows a strong increase below 30 MeV, so that FSI and two--nucleon 
induced processes are supposed to give an important contribution even in such a light Hypernucleus. 

The first theoretical analysis of the $^{5}_{\Lambda}$He and $^{12}_{\Lambda}$C nucleon spectra was performed in Refs.~\cite{garb,INC},
where a OME model for the $\Lambda N\to nN$ transition in finite nuclei was incorporated in the INC code of Refs.~\cite{ramos1,ramos2} to take 
into account nucleon FSI.
The two--nucleon stimulated channel was also included, using the phenomenological approach of Ref.~\cite{ramos}. 
This approach predicts $\Gamma_{n}/\Gamma_{p}=0.46$ and $\Gamma_{2}/\Gamma_{1}=0.20$
for $^{5}_{\Lambda}$He and $\Gamma_{n}/\Gamma_{p}=0.34$ and $\Gamma_{2}/\Gamma_{1}=0.25$ for $^{12}_{\Lambda}$C.
Neutron and proton spectra were calculated and compared with the KEK data for $^5_\Lambda$He and $^{12}_\Lambda$C
of Fig.~\ref{fig:okada}, which are normalized per NMWD.
For $^{12}_\Lambda$C, the KEK neutron spectrum could be explained, while the proton spectrum was overestimated, 
although the shape of the experimental distribution was qualitatively reproduced. For $^5_\Lambda$He, the theoretical
neutron spectrum is overlaid on Fig.~\ref{fig:okada} by the dashed histogram and 
looks significantly different from the observed one (both in shape and magnitude), 
which shows a maximum at about Q/2 but levels off in the region from 20 to 80 MeV.
The theoretical spectrum for protons from $^5_\Lambda$He instead has the same shape of the KEK one, but overestimates the data.
A calculation of the ratio between the neutron and proton spectra, $N_n/N_p$, in the KEK experimental conditions was also performed in 
Ref.~\cite{garb}. By applying a 60 MeV kinetic energy threshold for both neutrons and protons, the results are:
$N_n/N_p=1.98$ for $^5_\Lambda$He and $N_n/N_p=1.42$ for $^{12}_\Lambda$C, to be compared with the KEK determinations:
$N_{n}/N_{p}=2.17\pm0.15\pm0.16$ for $^{5}_{\Lambda}$He and $N_{n}/N_{p}=2.00\pm0.09\pm0.14$ for $^{12}_{\Lambda}$C.


Recently, the FINUDA experiment obtained single proton energy spectra from NMWD of $^{5}_{\Lambda}$He, $^{7}_{\Lambda}$Li and $^{12}_{\Lambda}$C 
\cite{npa804}. Exploiting the very good momentum resolution of the magnetic spectrometer (0.6$\%$ FWHM for $\pi^{-}$ of $\sim$270 MeV/c, 2$\%$ FWHM 
for $p$ of $\sim$400 MeV/c), the thinness of the targets ($\sim$0.2 g/cm$^{2}$ for $^{6}$Li and $^{7}$Li, $\sim$0.4 g/cm$^{2}$ for $^{12}$C ) in which 
the low momentum $K^{-}$'s from $\Phi$ decay are stopped and interact at rest, the large angular coverage, $\sim 2\pi$ sr, and the very good p.id. 
capabilities of the detector ensemble, the energy spectra of NMWD protons were measured, in coincidence with a $\pi^{-}$ identifying the formation of 
the ground state of the Hypernuclei, with an energy threshold as low as 15 MeV. 

\begin{figure} [h]
\begin{center}
\vspace{-4mm}
\resizebox{0.50\textwidth}{!}{
  \includegraphics{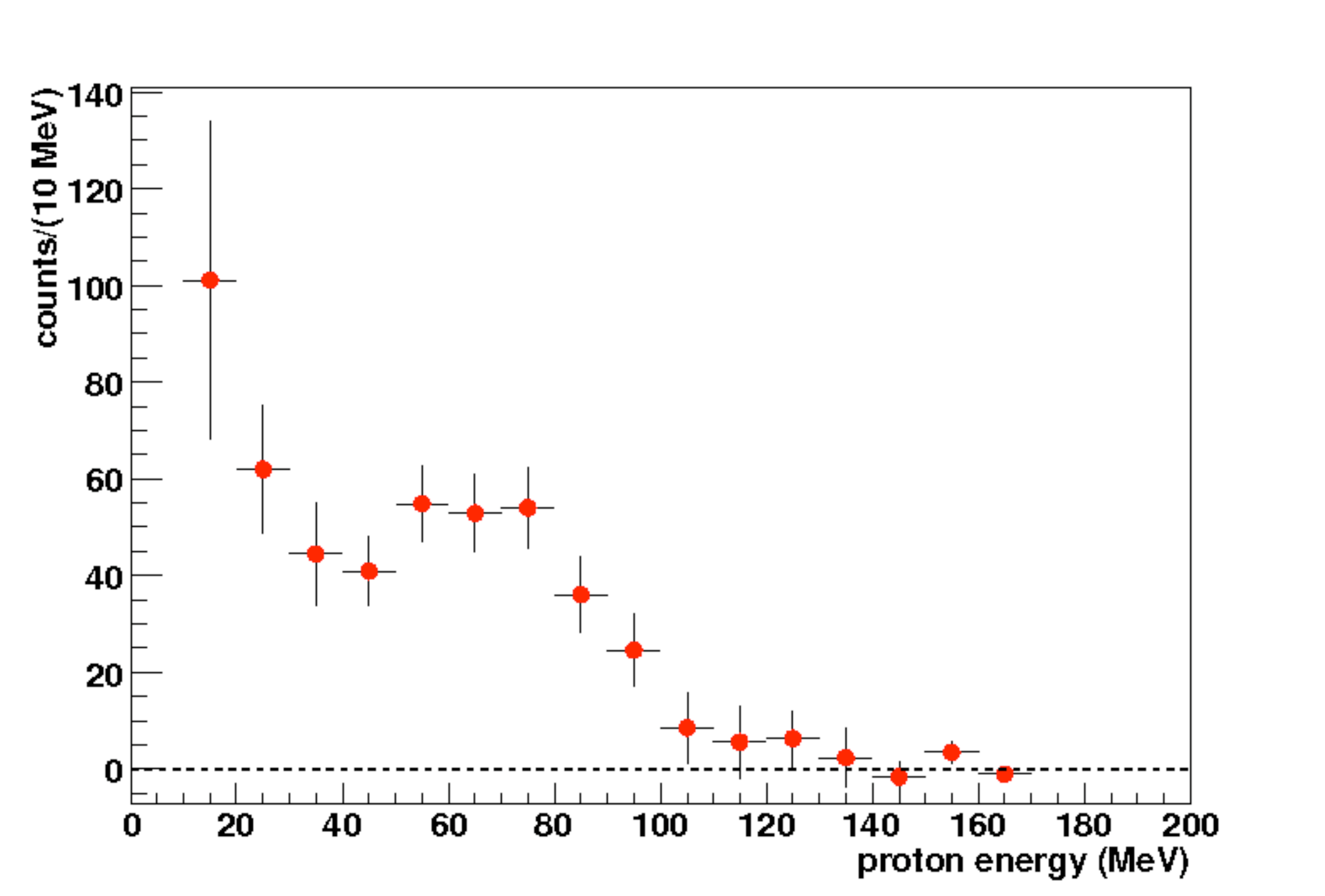}
}
\resizebox{0.50\textwidth}{!}{
  \includegraphics{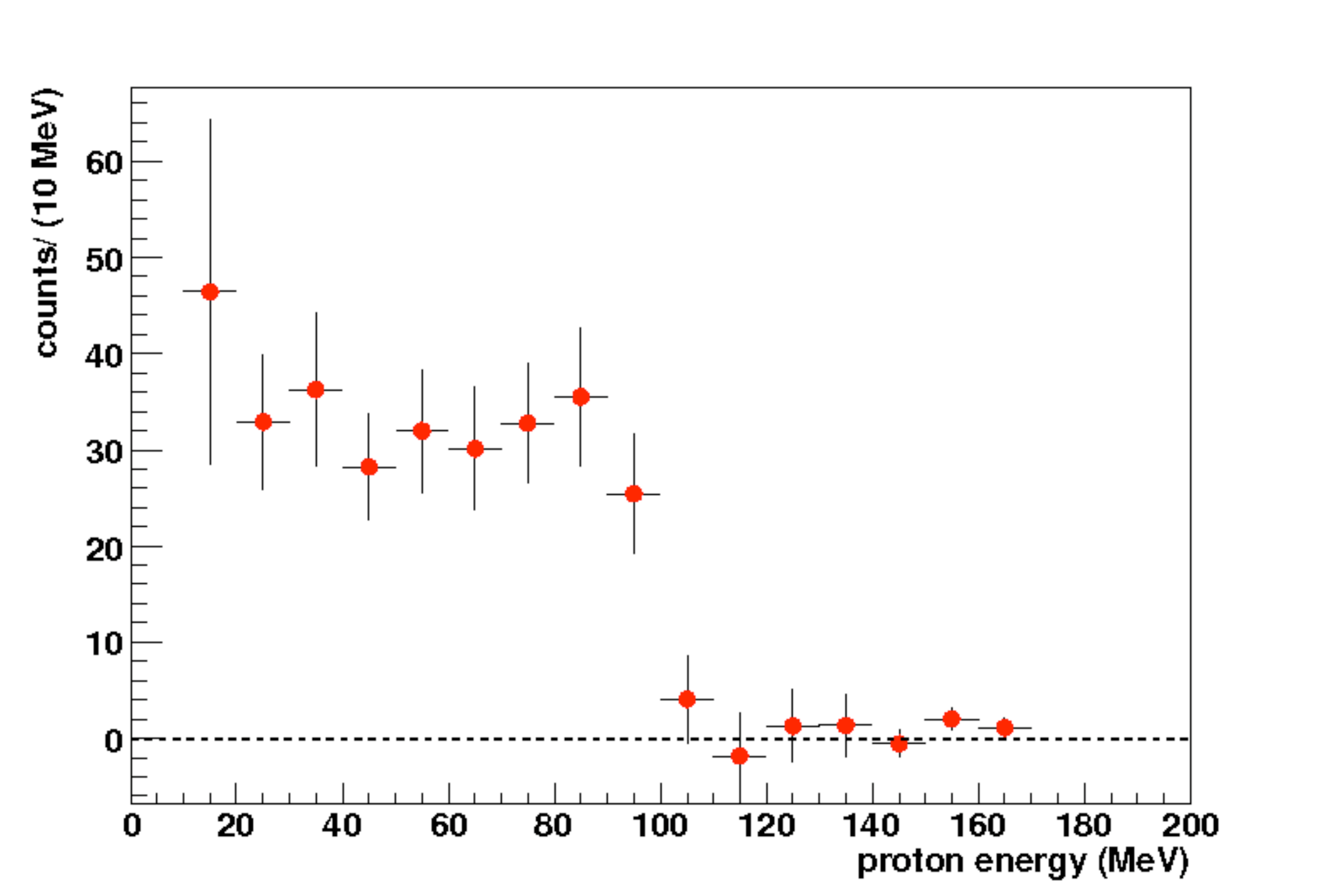}
}
\resizebox{0.50\textwidth}{!}{
  \includegraphics{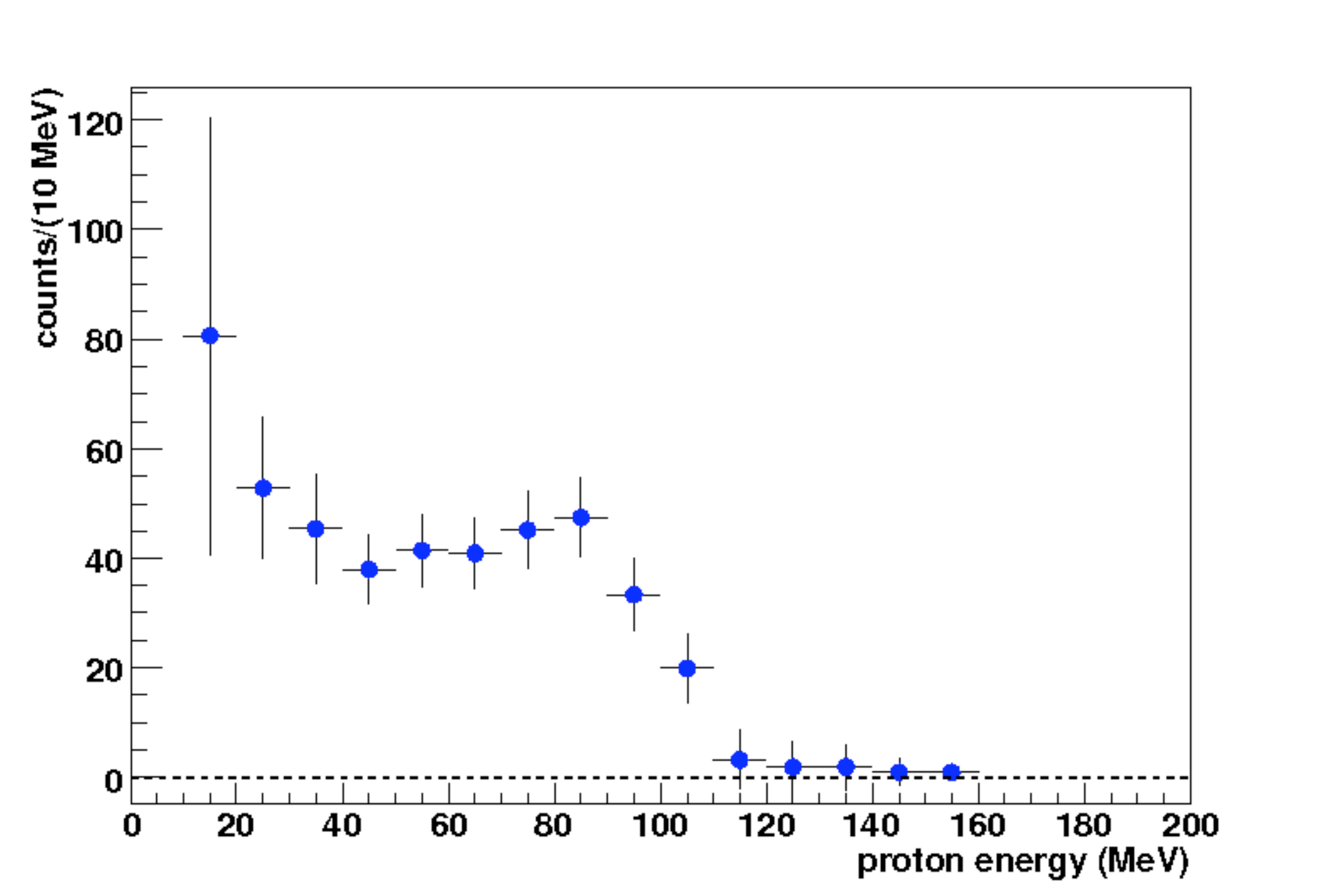}
}
\caption{Proton energy spectrum of $^{5}_\Lambda$He (top), $^{7}_\Lambda$Li (center) and $^{12}_\Lambda$C (bottom) obtained by the FINUDA experiment.
From Ref.~\cite{npa804}.}
\label{fig:sp804}       
\end{center}   
\vspace{-4mm}
\end{figure}
The FINUDA proton spectra from NMWD of $^{5}_\Lambda$He, $^{7}_\Lambda$Li and $^{12}_\Lambda$C  are given in Fig.~\ref{fig:sp804}. They all show 
a similar shape, i.e., a peak around 80 MeV, corresponding to about a half the Q--value for the $\Lambda p\to np$ weak reaction, with a low--energy rise, 
due to the FSI and/or to two--nucleon induced weak decays \cite{AG02,alberico,garb,INC,BGPR}, in spite of the large mass number difference of these 
nuclei. If the low energy rises were predominantly due to FSI effects, one should naturally expect that the broad peak structure at 80 MeV (coming 
from clean $\Lambda p \rightarrow n p$ weak processes broadened by the Fermi motion of nucleons) would be smeared out for the heavier nuclei.  
As for the second effect, as already discussed, if the weak decay Q--value is shared by three nucleons, a low--energy rise may exist even for the very 
light $s$--shell Hypernuclei. 
\begin{figure} [h]
\begin{center}
\vspace{-4mm}
\resizebox{0.50\textwidth}{!}{%
 \hspace{-5mm}
  \includegraphics{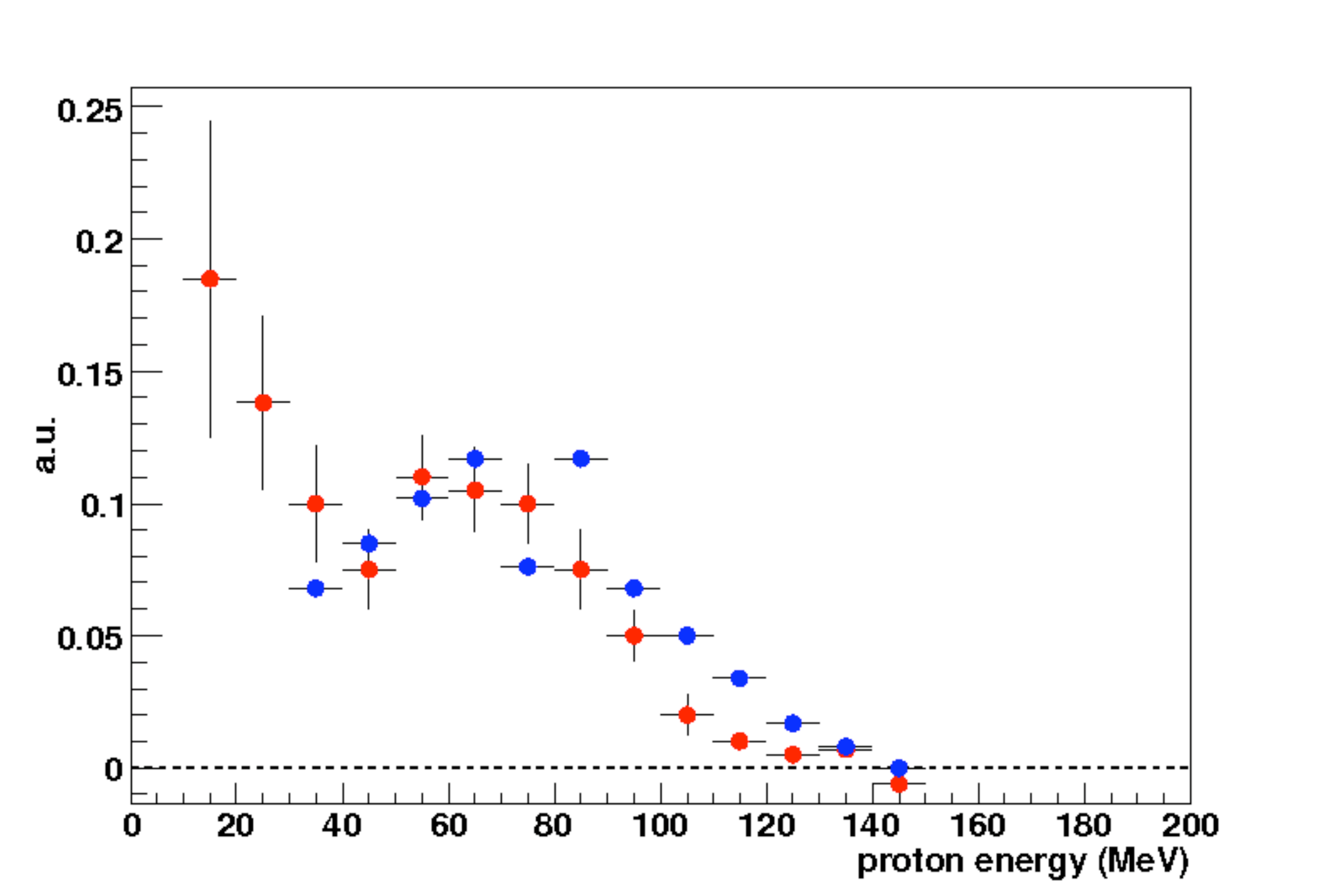}
}
\caption{FINUDA (red dots) \cite{npa804} and KEK (blue dots) \cite{okada} proton spectrum from NMWD of $^{5}_{\Lambda}$He.
The two spectra are normalized to have the same area beyond 35 MeV. From Ref.~\cite{npa804}.}
\label{fig:He5fk}       
\end{center}
\vspace{-4mm}
\end{figure}
The FINUDA data, thus, seem to agree with the hypothesis of a substantial contribution of the two--nucleon induced NMWD. 

Fig.~\ref{fig:He5fk} shows the comparison of the FINUDA spectrum for $^{5}_{\Lambda}$He with the KEK--E462 one \cite{okada}.
The two spectra were normalized beyond 35 MeV (the proton energy threshold of Ref.~\cite{okada}) and  are compatible at a C.L. of $75\%$.
\begin{figure} [h]
\begin{center}
\resizebox{0.50\textwidth}{!}{%
 \hspace{-5mm}
  \includegraphics{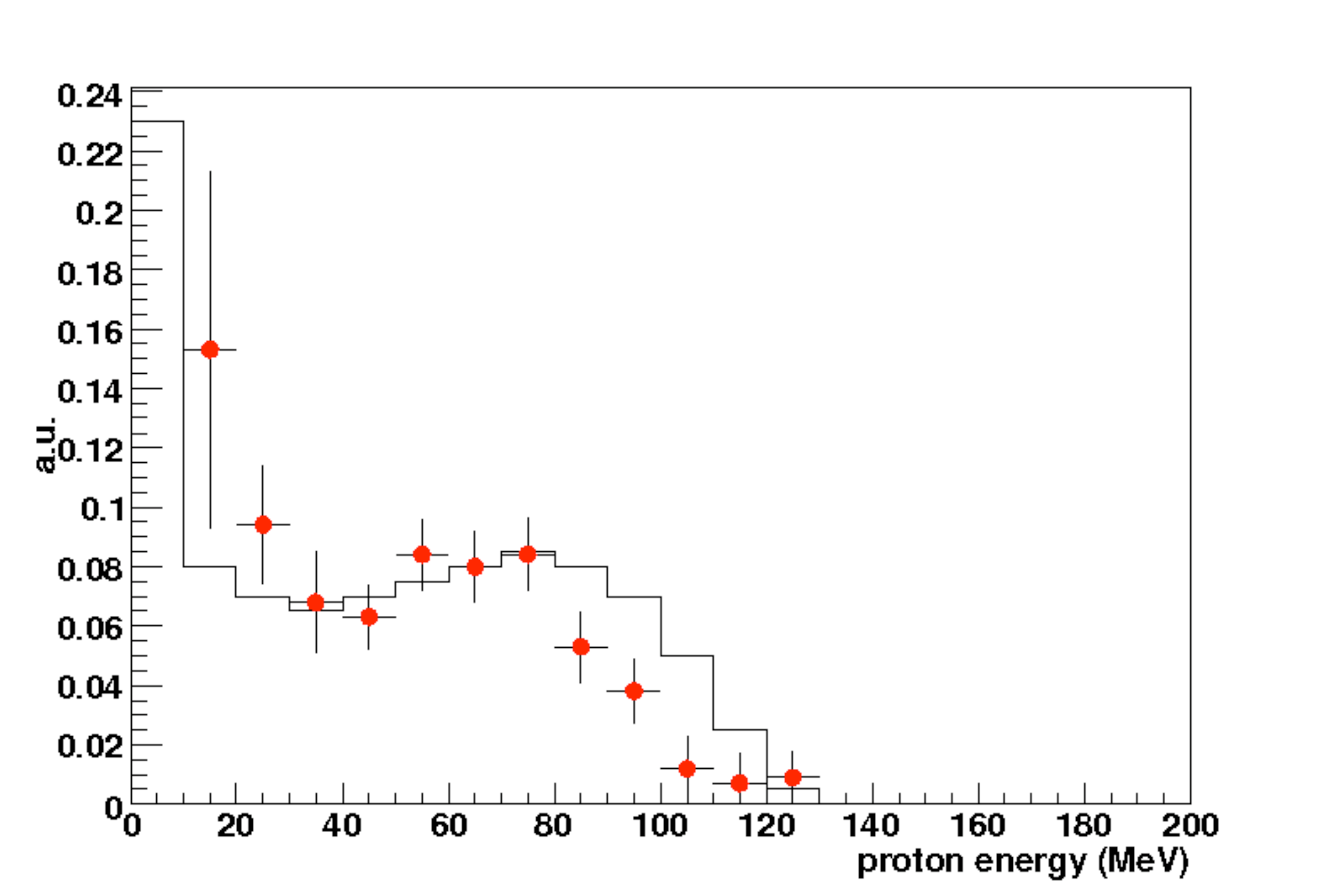}
}
\caption{FINUDA (red dots) \cite{npa804} and theoretical (continuous histogram) \cite{garb} proton spectrum from NMWD of $^{5}_{\Lambda}$He; 
The two spectra are normalized to have the same area beyond 15 MeV. From Ref.~\cite{npa804}.}
\label{fig:He5fth}       
\end{center}
\vspace{-4mm}
\end{figure} 
Fig.~\ref{fig:He5fth} shows the comparison of the FINUDA proton spectrum with the theoretical one calculated by Ref.~\cite{garb},
which considered both two--nucleon induced decays and nucleon FSI.
\begin{figure} [h]
\begin{center}
\resizebox{0.50\textwidth}{!}{%
 \hspace{-5mm}
  \includegraphics{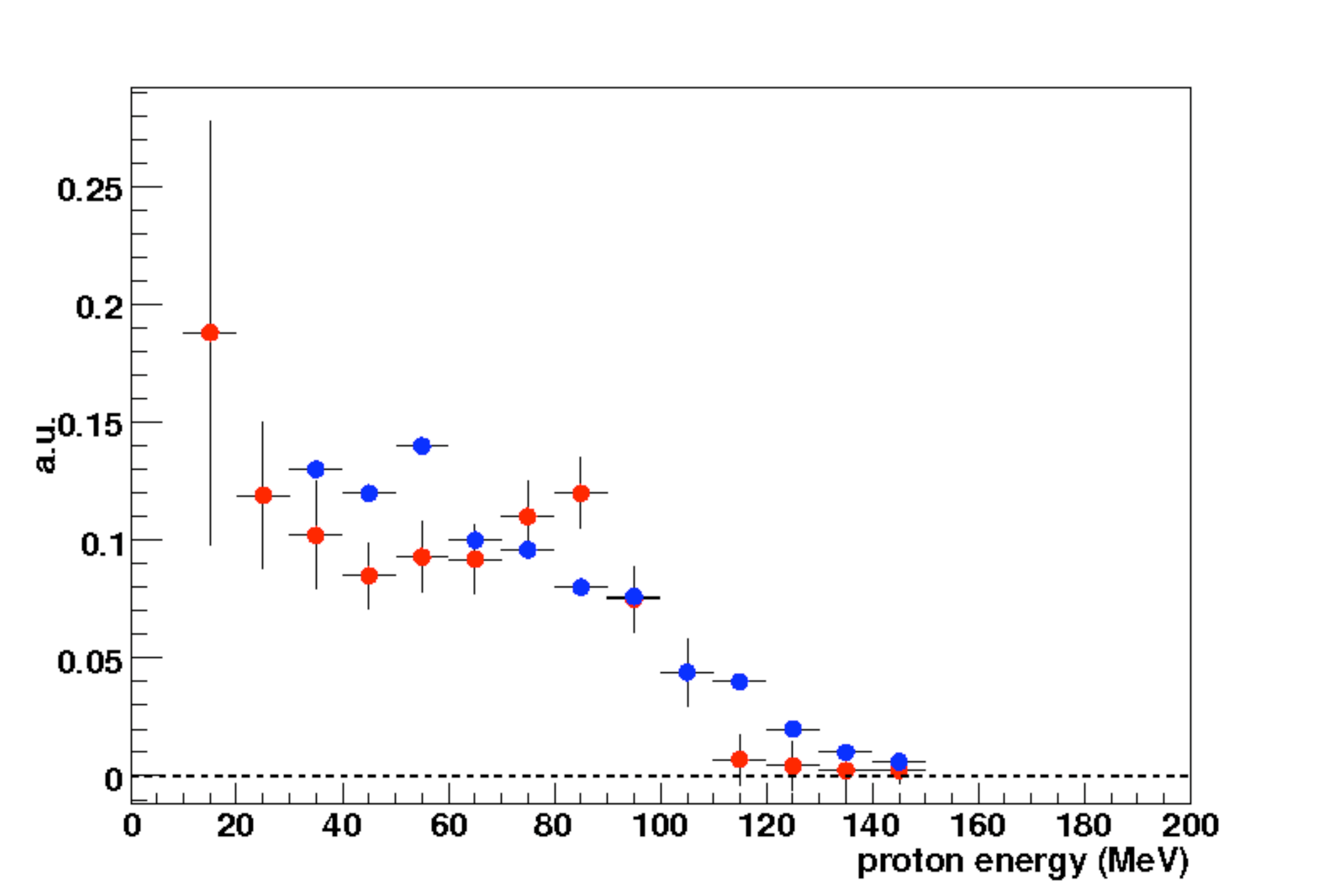}
}
\caption{FINUDA (red dots) and KEK (blue dots) \cite{okada} proton spectra from NMWD of $^{12}_{\Lambda}$C; 
the spectra are normalized to have the same area beyond 35 MeV. From Ref.~\cite{npa804}.}
\label{fig:C12fk}       
\end{center}
\vspace{-4mm}
\end{figure}
The two spectra were normalized to have the same area beyond 15 MeV (the FINUDA proton energy threshold). A compatibility test between the 
two spectra provides a C.L. of $80\%$, indicating that there is a disagreement between the two experiments, and also with theory. 
However, these incompatibilities are not so severe.

The situation for $^{12}_{\Lambda}$C is completely different. Fig.~\ref{fig:C12fk} shows the comparison of the FINUDA and KEK \cite{okada}
proton spectra: the spectra were normalized to the same area beyond 35 MeV.
A compatibility test between the two data sets provides a C.L. of $20\%$. 
\begin{figure} [h]
\begin{center}
\resizebox{0.50\textwidth}{!}{%
 \hspace{-5mm}
  \includegraphics{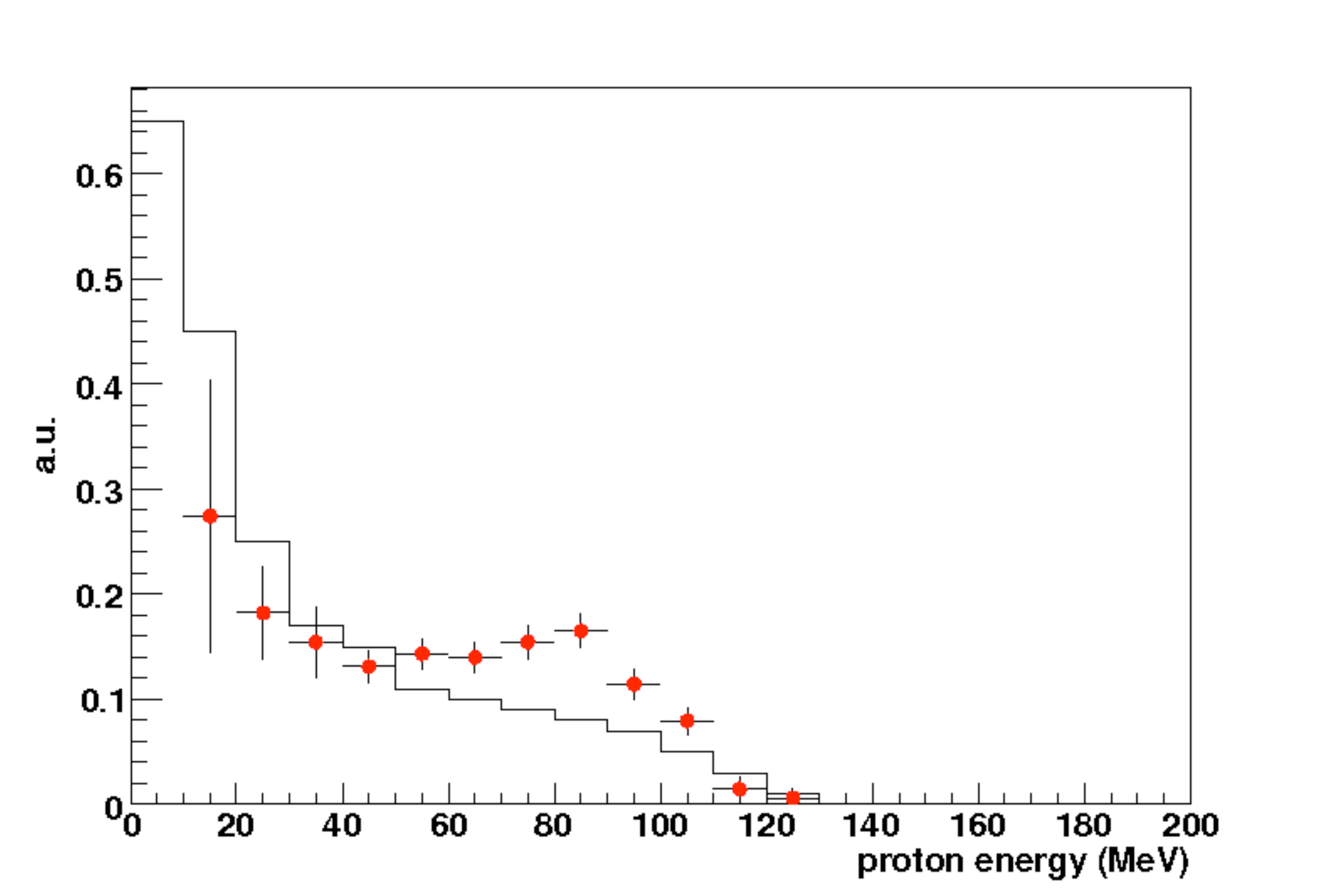}
}
\caption{FINUDA (red dots) and theoretical (continuous histogram) \cite{garb} proton spectra from NMWD of $^{12}_{\Lambda}$C; 
The two spectra are normalized to the same area beyond 15 MeV. From Ref.~\cite{npa804}.}
\label{fig:C12fth}       
\end{center}
\vspace{-4mm}
\end{figure}
Fig.~\ref{fig:C12fth} shows the comparison of the FINUDA proton spectrum with the calculated one by Ref.~\cite{garb}. The two spectra were 
normalized to the same area beyond 15 MeV. The compatibility test between the theoretical and experimental spectra of 
Fig.~\ref{fig:C12fk} provides 
a C.L. of $5\%$. The conclusion is that there is a strong disagreement between the two experiments and with theory. 

Concerning the discrepancy between the FINUDA and KEK sets of data, it can be observed that in Ref.~\cite{okada} the proton energy was measured by a 
combination of time--of--flight and total energy deposit measurements. The energy loss inside the thick targets was corrected on an event--by--event
basis. The energy resolution becomes poorer in the high energy--region, especially above 100 MeV, with the consequence that the KEK spectra could be 
strongly distorted. On the contrary, with FINUDA the proton momenta were measured by means of a magnetic analysis, with an excellent resolution and no 
distortion on the spectra is expected, in particular in the high--energy region. The FINUDA spectrum thus is less biased, even if it has a limited 
statistics. 

\begin{figure*}[ht]
\begin{center}
\vspace{-5mm}
\resizebox{0.9\textwidth}{!}{%
 \hspace{-5mm}
  \includegraphics{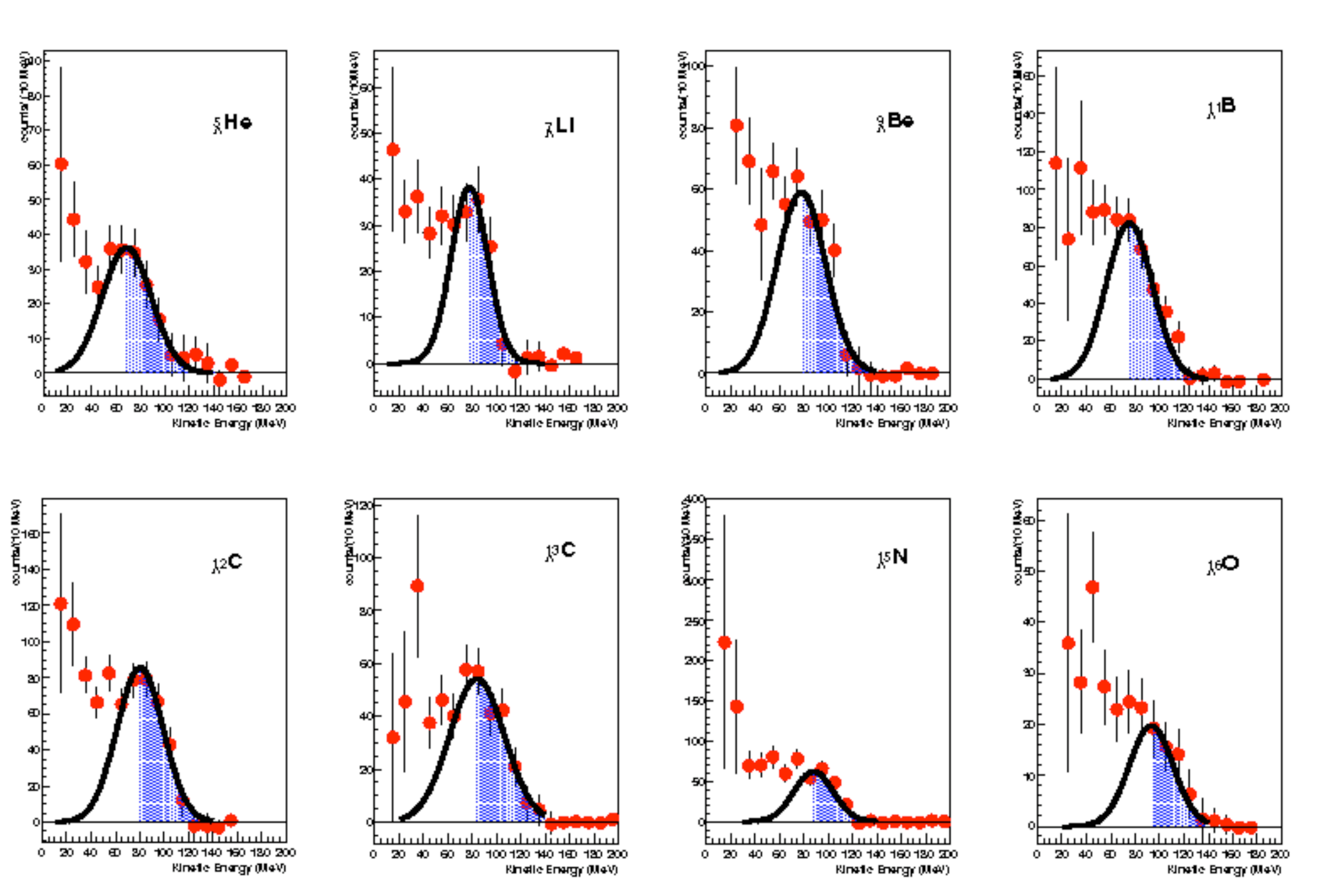}
}
\caption{Proton spectra from the NMWD of ${\mathrm{^{5}_{\Lambda}He}}$, ${\mathrm{^{7}_{\Lambda}Li}}$, ${\mathrm{^{9}_{\Lambda}Be}}$, 
${\mathrm{^{11}_{\Lambda}B}}$,  ${\mathrm{^{12}_{\Lambda}C}}$, 
${\mathrm{^{13}_{\Lambda}C}}$, ${\mathrm{^{15}_{\Lambda}N}}$ 
and ${\mathrm{^{16}_{\Lambda}O}}$ measured by FINUDA. For more details, see Sec.~\ref{subsec:nmwd_2N}. From Ref.~\cite{fnd_nmwd}.}
\label{FINUDA-tot}       
\end{center}
\vspace{-4mm}
\end{figure*}

The FINUDA experiment has then reported the NMWD proton spectra for 
${\mathrm{^{9}_{\Lambda}Be}}$, ${\mathrm{^{11}_{\Lambda}B}}$,  ${\mathrm{^{12}_{\Lambda}C}}$, 
${\mathrm{^{13}_{\Lambda}C}}$, ${\mathrm{^{15}_{\Lambda}N}}$ and  ${\mathrm{^{16}_{\Lambda}O}}$ \cite{fnd_nmwd}; they are given
in Fig.~\ref{FINUDA-tot} together with the previous results for ${\mathrm{^{5}_{\Lambda}He}}$ and ${\mathrm{^{7}_{\Lambda}Li}}$. 
Despite these distributions are affected by considerable errors, in particular 
in the low--energy region, they show a clear trend as a function of $A$ (from 5 to 
16): a peak around 80 MeV is broadened by the Fermi motion of nucleons and more and more blurred as $A$ increases. 
The peak is smeared, on its low--energy side, by a rise that can be ascribed to FSI and two--nucleon induced weak decays. 

The spectra of Fig.~\ref{FINUDA-tot} complete the FINUDA systematic study of the MWD and NMWD of $p$--shell Hypernuclei. In particular, as we shall 
discuss in Sec.~\ref{subsec:nmwd_2N}, these spectra make possible a determination of the two--nucleon induced decay contribution to NMWD in a 
model--indepen-dent way. It must be noted that FINUDA has not determined the value of $\Gamma_p$ from the proton distributions, which are affected 
by unavoidable FSI and two--nucleon induced reaction effects; the extraction of $\Gamma_p$ would have been based mainly on the choice of the 
particular theoretical model adopted to describe the NMWD. The study of the $A$--dependence of the spectral shape, on the other hand, made 
possible to obtain information on the two--nucleon induced decay mechanism by relying only on simple theoretical assumptions, whose validity can 
be verified from the results of the measurements, and on previous experimental data on the one--nucleon induced decay rates.  

\subsubsection{Nucleon--Nucleon Coincidence Spectra Analyses}
\label{subsec:nmwd_coinc}

Simultaneous measurements of NMWD neutron and proton distributions turned out to be very important in clarifying the inconsistencies
of the old experiments, in which only protons were measured and the determined $\Gamma_n/\Gamma_p$ ratios were in clear disagreement with theory.
To extract the value of $\Gamma_{n} / \Gamma_{p}$ from a measured nucleon spectra it is always necessary to correct 
the distributions for nucleon FSI; by measuring the yield ratio $N_{n}/N_{p}$, the problem was 
only partially removed and a correction based on FSI calculations was still needed to determine a reliable value of $\Gamma_n/\Gamma_p$. 
In addition, the results of these single nucleon measurements could be possibly affected by a two--nucleon induced NMWD contribution. 

Indeed, a real step forward in the solution of the $\Gamma_{n} / \Gamma_{p}$ puzzle came from the measurement of the nucleon spectra in
double coincidence and their theoretical analyses. 
Both the $\Lambda n\to nn$ and $\Lambda p\to np$ processes in nuclei are (quasi) two--body decay processes, so that the two NMWD nucleons 
have a clean back--to--back correlation if both of them do not suffer FSI and do not originate from two--nucleon induced effects; the one--nucleon induced 
processes could be thus clearly observed by measuring the yields of $np$ and $nn$ pairs in the back--to--back 
configuration and requiring that the sum of their kinetic energies correspond to the Q--value of the NMWD reaction. With these
angular and energy restrictions, nucleon FSI and two--nucleon induced decay processes should provide small effects and
a measurement of the $N_{nn}/N_{np}$ ratio could directly provide the experimental value of $\Gamma_n/\Gamma_p$.

\begin{figure} [ht]
\begin{center}
\resizebox{0.47\textwidth}{!}{%
 \hspace{-5mm}
  \includegraphics{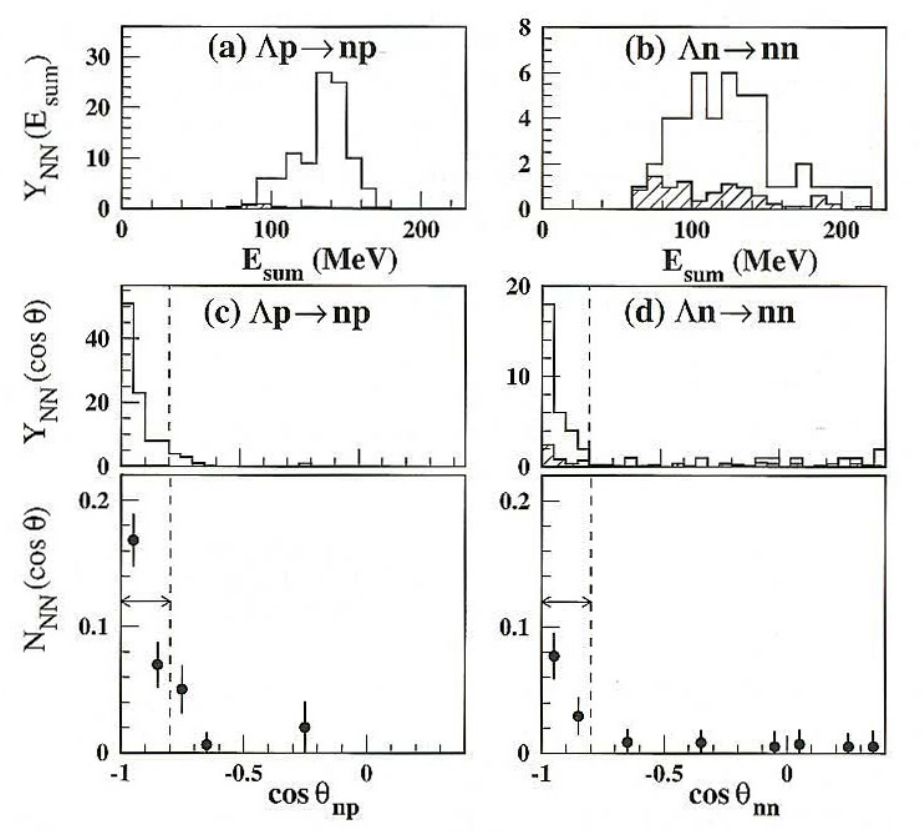}
}
\caption{KEK--E462 nucleon--nucleon coincidence spectra for the NMWD of $^5_\Lambda$He.
(a) and (b): $np$ and $nn$ pair raw yields as a function of the energy sum of the pair nucleons, before efficiency
correction; a 30 MeV threshold was applied to both nucleons. (c) and (d): The upper panel shows the opening angle dependence of the 
$np$ and $nn$ raw yields; the lower panel shows the final angular spectra normalized for NMWD. From Ref.~\cite{kang}.}
\label{fig:kang4}       
\end{center} 
\vspace{-4mm}
\end{figure}

A nucleon--nucleon coincidence measurement for $^{5}_{\Lambda}$He NMWD \cite{kang} was performed at KEK--E462; the 
light target was chosen to reduce the influence of FSI. The upper panels of Fig.~\ref{fig:kang4} show the $np$ and $nn$ pair raw 
yields, $Y_{np}$ and $Y_{nn}$, as a function of the kinetic energy sum of the nucleon pairs, before the 
application of the efficiency correction; only events in which each nucleon of the pair has a kinetic energy $E_N$ above 30 MeV are considered,
while the back--to--back topology requirement is not yet applied. The hatched histograms represent the 
contamination coming from the absorption of the $\pi^{-}$ emitted in the MWD of $^{5}_{\Lambda}$He. In the energy sum spectrum of $np$ pairs, a peak 
located around the Q--value of the NMWD process can be seen; its sharp shape indicates that the FSI effect is not severe and one--nucleon 
induced NMWD gives the main contribution. The $nn$ pair energy distribution is instead broader, due to the limited resolution in the neutron energy 
measurement, and has a bigger background contribution. In the lower panel, the raw yields, $Y_{np}$ and $Y_{nn}$, are 
reported together with the efficiency corrected yields normalized per NMWD, $N_{np}$ 
and $N_{nn}$, as a function of the opening angle between the two nucleons, $\theta_{np}$ and $\theta_{nn}$. A strong 
angular correlation is evident and can be interpreted as a signature of the direct observation of the neutron--
and proton--induced NMWD processes. Similar correlation measurements were then performed for $^{12}_{\Lambda}$C by KEK--E508 \cite{mjkim}.

To determine $\Gamma_{n} / \Gamma_{p}$ from the coincidence measurements, only events with $E_N>30$ MeV and $\cos\, \theta_{NN} < -0.8$ were 
considered. In particular,  taking into account that, in the $N_{nn}/N_{np}$ ratio, factors as the NMWD branching ratio, the FSI effect 
and the solid angle acceptance cancel out in first approximation, it is possible to simply write: 
$\Gamma_{n} / \Gamma_{p} = N_{nn}/N_{np} = (Y_{nn} \epsilon_{p}) / (Y_{np} \epsilon_{n})$, where $\epsilon_{N}$ is the detection efficiency of the 
nucleon $N$, if two--nucleon induced NMWD is neglected. 

In this way, the results obtained by KEK--E462 and KEK--E508 are:
$\Gamma_{n} / \Gamma_{p}$($^{5}_{\Lambda}$He) = 0.45 $\pm$ 0.11 $\pm$0.03 \cite{kang} and
$\Gamma_{n} / \Gamma_{p}$($^{12}_{\Lambda}$C) = 0.51 $\pm$ 0.13 $\pm$0.05 \cite{mjkim}. 
These values were considered in Refs.~\cite{kang,mjkim} to be free from ambiguities due to FSI and two--nucleon induced NMWD, due to the 
particular cuts imposed on the nucleon energy and nucleons angular correlation.

An evaluation of the effects of FSI and two--nucleon induced NMWD in coincidence nucleon spectra was performed in Refs.~\cite{garb,INC},
with a OME model for the $\Lambda N\to nN$ transition in finite nuclei, a phenomenological approach of the two--nucleon stimulated channel
and an INC code to take into account nucleon FSI. In particular, a weak--decay model--independent analysis of KEK--E462 and KEK--E508 
coincidence measurements was performed. A fit of the $N_{nn}/N_{np}$ data (taken with $E_N>30$ MeV and $\cos\, \theta_{NN} < -0.8$) allowed the 
determination of the values of $\Gamma_n/\Gamma_p$ reported in Table~\ref{nn-np}. 
We see that the theoretical fits of data prefer smaller values for $\Gamma_n/\Gamma_p$ than the purely experimental determinations 
of Refs.~\cite{kang,mjkim}, given in the previous paragraph, which neglected nucleon FSI and two--nucleon induced decays. As one may expect, the 
differences are smaller with the fits neglecting the two--nucleon induced channels. This may signal a non--negligible effect of both FSI and 
two--nucleon induced processes in the $N_{nn}/N_{np}$ data (however, we note that theoretical fits and pure data 
agree with each other within $1\sigma$).%
This conclusion was corroborated by a recent analysis \cite{BGPR}, analogous to the one of Ref.~\cite{garb} but performed
entirely within a microscopic, nuclear matter formalism adapted to finite nuclei via the local density approximation.
All the isospin channels for the two--nucleon induced mechanism were taken into account in Ref.~\cite{BGPR}: this also includes the channels
$\Lambda nn\to nnn$ and $\Lambda pp\to npp$ besides the standard mode $\Lambda np\to nnp$ of the phenomenological approach of 
Ref.~\cite{garb}. The results of Ref.~\cite{BGPR} are also reported in Table~\ref{nn-np} and agree with the fits of Ref.~\cite{garb} despite
the difference between the two frameworks.
\begin{table}[h]
\caption{Theoretical determinations of the $\Gamma_n/\Gamma_p$ ratio obtained by fitting the KEK--E462 and KEK--E508 data
on $N_{nn}/N_{np}$. The $1N$ results only include one--nucleon induced NMWD, while the $1N+2N$ ones also include 
two--nucleon induced NMWD.}
\label{nn-np}
\begin{center}
\begin{tabular}{l c c} \hline
\mc {1}{c}{Ref.} &
\mc {1}{c}{$^5_{\Lambda}$He} &
\mc {1}{c}{$^{12}_{\Lambda}$C} \\ \hline
\cite{garb} 
 & $0.40\pm 0.11$ ($1N$) & $0.38\pm 0.14$ ($1N$)\\
 & $0.27\pm 0.11$ ($1N+2N$) & $0.29\pm 0.14$ ($1N+2N$) \\
\cite{BGPR}
 &   & $0.37\pm 0.14$ ($1N$)\\
 &   & $0.34\pm 0.15$ ($1N+2N$) \\\hline
\end{tabular}
\end{center}
\end{table}

In Table~\ref{gngp} we summarize the theoretical and experimental determinations of $\Gamma_n/\Gamma_p$;
the listed values correspond to the theoretical results and experiments of Table~\ref{tot-nmwd} for $\Gamma_{\rm NM}$.
Some calculation can explain the recent KEK--E462 and KEK--E508 data for both $\Gamma_n/\Gamma_p$ and
$\Gamma_{\rm NM}$. It is a widely shared opinion that this fact definitely provides a solution of the $\Gamma_n/\Gamma_p$ 
puzzle. However, we note that, concerning $\Gamma_n/\Gamma_p$, the calculations tend to be more in agreement with the fits of 
Refs.~\cite{garb,BGPR}, given Table~\ref{nn-np}, which do not neglect FSI and two--nucleon induced decays,
than with the experimental determinations by Refs.~\cite{kang,mjkim}, in which it was assumed that $\Gamma_n/\Gamma_p=N_{nn}/N_{np}$
when $E_p>30$ MeV and $\cos\, \theta_{NN}< -0.8$. Together with the discussion of the previous paragraph,
this comparison may signal a non--negligible effect of FSI and two--nucleon induced decays even in the favourable 
conditions of the performed correlation measurements.
\begin{table}[h]
\caption{Theoretical and experimental determinations of the $\Gamma_n/\Gamma_p$ ratio.}
\label{gngp}
\begin{center}
\resizebox{8.7cm}{!} {
\begin{tabular}{l c c} \hline
\mc {1}{c}{Ref. and Model} &
\mc {1}{c}{$^5_{\Lambda}$He} &
\mc {1}{c}{$^{12}_{\Lambda}$C} \\ \hline
Sasaki et al. \cite{sasaki} &0.70 &  \\
{\small $\pi +K+$ DQ}               & & \\
Jido et al. \cite{Os01} & &0.53  \\
$\pi +K+2\pi/\sigma+2\pi +\omega$      & & \\
Parre\~{n}o and Ramos \cite{parreno} &$0.46$ & $0.34$  \\
$\pi +\rho +K + K^* + \omega +\eta$                        & &  \\
Itonaga et al. \cite{itonaga2}    & 0.39 & 0.37  \\
$\pi + 2\pi/\sigma + 2\pi/\rho+\omega$ & &      \\
Barbero et al. \cite{Ba03} & 0.24 & 0.21  \\   
$\pi +\rho +K + K^* + \omega +\eta$ &  &      \\
Bauer and Garbarino \cite{Ba10b}  & & 0.34    \\
$\pi +\rho +K + K^* + \omega +\eta$ &  &      \\ \hline
BNL \cite{szym} &$0.93\pm0.55$
&$1.33^{+1.12}_{-0.81}$  \\
KEK \cite{noumi} & &$1.87^{+0.67}_{-1.16}$  \\
KEK \cite{noumi2} &$1.97\pm0.67$ &  \\
KEK--E307 \cite{sato05} & &$0.87\pm 0.23$  \\
KEK--E462 \cite{kang} & $0.45\pm 0.11\pm 0.03$ &  \\
KEK--E508 \cite{mjkim} & & $0.51\pm 0.13\pm 0.05$  \\\hline
\end{tabular}}
\end{center}
\end{table}

\subsection{Two--Nucleon Induced Process and FSI effect in NMWD}
\label{subsec:nmwd_2N}

In Ref.~\cite{bhang}, KEK--E508 single and coincidence spectra of nucleons from the NMWD of $^{12}_{\Lambda}$C were analyzed theoretically.
One-- and two--nucleon induced NMWD processes were described by phase--space arguments, without resorting to any weak decay model,
while an INC code was used to incorporate nucleon FSI effects. Without the two--nucleon induced mechanism, a quenching of the observed 
nucleon spectra was found with respect to the calculated spectra. This quenching was interpreted as the effect of the two--nucleon induced NMWD, 
which cannot be separated kinematically from FSI--induced decays processes, since, 
in first approximation, these decays share the same phase--space.

\begin{figure} [h]
\begin{center}
\resizebox{0.5\textwidth}{!}{%
 \hspace{-5mm}
  \includegraphics{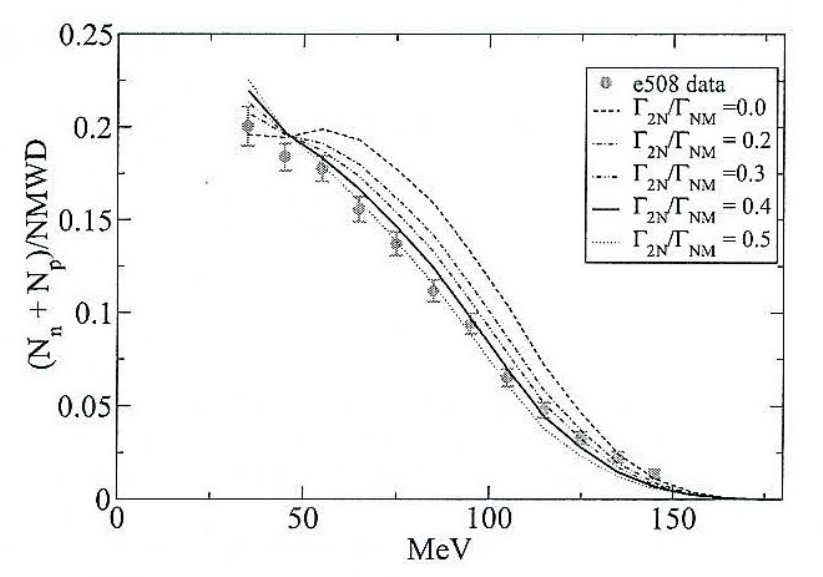}
}
\caption{Kinetic energy dependence of the sum of single proton and neutron spectra per NMWD. Data are from 
KEK--E508 \cite{okada2}. The curves represent calculations of the
energy distribution with different contributions from the two--nucleon induced NMWD. From Ref.~\cite{bhang}.}
\label{fig:bhang4}       
\end{center} 
\vspace{-4mm} 
\end{figure}

It was argued that the value of $\Gamma_{2}$ could be inferred from the measured spectra with the above theoretical description,
by considering this rate as a free parameter.
Fig.~\ref{fig:bhang4} shows the experimental energy dependence of the sum of proton and neutron distributions, normalized per NMWD, from 
Ref.~\cite{bhang}. The experimental spectra was simulated with different values of the $\Gamma_{2}/\Gamma_{\rm NM}$ ratio and an indication of a 
contribution as high as 40$\%$ to the total NMWD width was deduced. 

\begin{figure} [h]
\begin{center}
\resizebox{0.5\textwidth}{!}{%
 \hspace{-5mm}
  \includegraphics{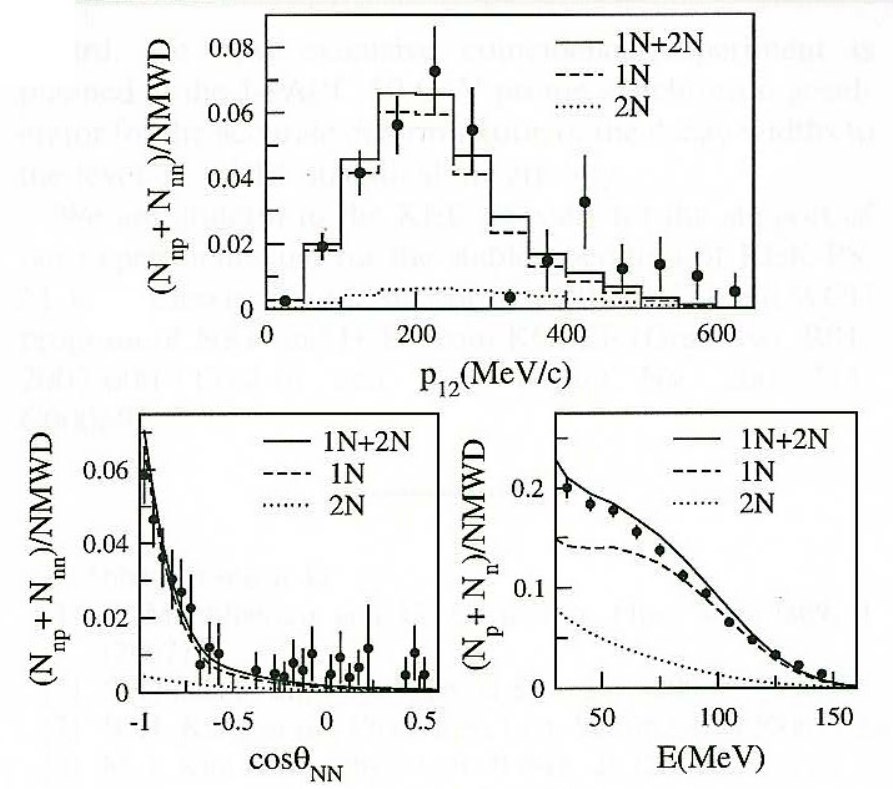}
}
\caption{Upper panel: sum of the $nn$ and $np$ distributions as a function of the pair momentum sum correlation ($p_{12}=|\vec p_1+\vec p_2|$);
Lower panel: sum of the $nn$ and $np$ distributions as a function of the pair opening angle and
kinetic energy dependence of the sum of single $n$ and $p$ spectra. Data are from Ref.~\cite{mjkim,okada2}.
The curves represent the different contributions to the spectra obtained by the theoretical simulation of Ref.~\cite{mkim}.
Normalization is per NMWD.}
\label{fig:mkim4}       
\end{center}
\vspace{-4mm}
\end{figure}

In Ref.~\cite{mkim}, the same authors applied a different implementation of the INC calculations to estimate the FSI effect. 
The strengths of the used $NN$ cross sections were varied to fit the measured inelastic total cross section data for $^{12}$C(p,p') reactions, 
thus eliminating the uncertainties left in the previous INC implementation. The $^{12}_{\Lambda}$C experimental single and double--coincidence nucleon 
spectra were reproduced by using the value $\Gamma_{2}/\Gamma_{\rm NM}$=0.29 $\pm$ 0.13, which is sizably lower than the previous indication. 
The following experimental values of the partial decay rates for $^{12}_{\Lambda}$C (in units of the free $\Lambda$ decay rate $\Gamma_{\Lambda}$) 
were consequently determined by using previous KEK--E508 data on 
$\Gamma_{\rm NM}$ and $\Gamma_n/\Gamma_p$: $\Gamma_{2}$= 0.27 $\pm$ 0.13, $\Gamma_{1}$= 0.68 $\pm$ 0.13, $\Gamma_{n}$= 0.23 $\pm$ 0.08 and 
$\Gamma_{p}$= 0.45 $\pm$ 0.10. Fig.~\ref{fig:mkim4} shows KEK--E508 $^{12}_{\Lambda}$C data together with the results of the theoretical fit
for the momentum sum correlation ($p_{12}=|\vec p_1+\vec p_2|$) of the sum of the $nn$ and $np$ distributions, 
the angular correlation of the sum of the $nn$ and $np$ distributions and the sum of the single $n$ and $p$ kinetic energy distributions.

We also mention that an old BNL experiment, E788, which observed single nucleon spectra from $^4_\Lambda$He NMWD, published only recently 
the final results \cite{parker}. An upper limit for the contribution of the two--nucleon induced mechanism to NMWD was established: 
$\Gamma_2/\Gamma_{\rm NM}\leq 0.24$ (95 \% C.L.). This result agrees with the discussed KEK--E508 determination of Ref.~\cite{mkim}, although 
it is for a lighter Hypernucleus.

The FINUDA experiment adopted a completely different approach to determine $\Gamma_{2}/\Gamma_{\rm NM}$.
Starting from the measured inclusive single proton spectra of ${\mathrm{^{5}_{\Lambda}He}}$, ${\mathrm{^{7}_{\Lambda}Li}}$,  
${\mathrm{^{9}_{\Lambda}Be}}$, ${\mathrm{^{11}_{\Lambda}B}}$, ${\mathrm{^{12}_{\Lambda}C}}$, ${\mathrm{^{13}_{\Lambda}C}}$, 
${\mathrm{^{15}_{\Lambda}N}}$ and  ${\mathrm{^{16}_{\Lambda}O}}$, a technique was introduced to disentangle the contribution coming from the 
two--nucleon induced decays from those of the one--nucleon induced decays and FSI \cite{fnd_nmwd}. 
The systematics over the mass number range $A=5$-$16$, covered by the FINUDA measurements, was exploited and 
only simple assumptions were made in the calculations. Each spectrum was fitted, from 80 MeV proton kinetic energy on, with a Gaussian function to 
determine, by its mean value, the energy corresponding to the maximum of the one--proton induced contribution. The spectrum was then divided in two 
parts, one below the mean value, with area $A_{\rm low}$, and one above the mean value, with area $A_{\rm high}$. It was assumed that the first part 
is populated by one--proton induced decays, two--nucleon induced decays and FSI processes, while the second part has contributions from one--proton 
induced decays and FSI, neglecting the two--nucleon induced decays contribution, which, above 70 MeV, accounts for only 5$\%$ of the total 
two--nucleon induced strength \cite{garb}. Then, the areas could be written as follows:
\begin{equation}
A_{\rm low} = 0.5 N(\Lambda p \rightarrow n p) + N(\Lambda n p \rightarrow n n p) + N^{\rm FSI}_{\rm low}\, ,
\label{alow}
\end{equation}
\begin{equation}
A_{\rm high} = 0.5 N(\Lambda p \rightarrow n p) + N^{\rm FSI}_{\rm high}\, ,
\label{ahigh}
\end{equation}
where $N(\Lambda p \rightarrow n p)$ is the number of protons coming from the one--proton induced decay and $N(\Lambda n p \rightarrow n n p)$ is the 
number of protons coming from the two--nucleon induced decay $\Lambda n p \rightarrow n n p$. Here the approximation $\Gamma_{2} \sim 
\Gamma_{np}$ was made, which is suggested by Ref.~\cite{bau_garb}.
Moreover, $N^{\rm FSI}_{\rm low}$ is the difference between the number of detected protons and the 
number of primary protons, including one-- and two--nucleon induced sources, for the low energy part of the spectrum and 
$N^{\rm FSI}_{\rm high}$ is the corresponding difference for the high energy region. 

\begin{figure} [h]
\begin{center}
\resizebox{0.5\textwidth}{!}{%
 \hspace{-5mm}
  \includegraphics{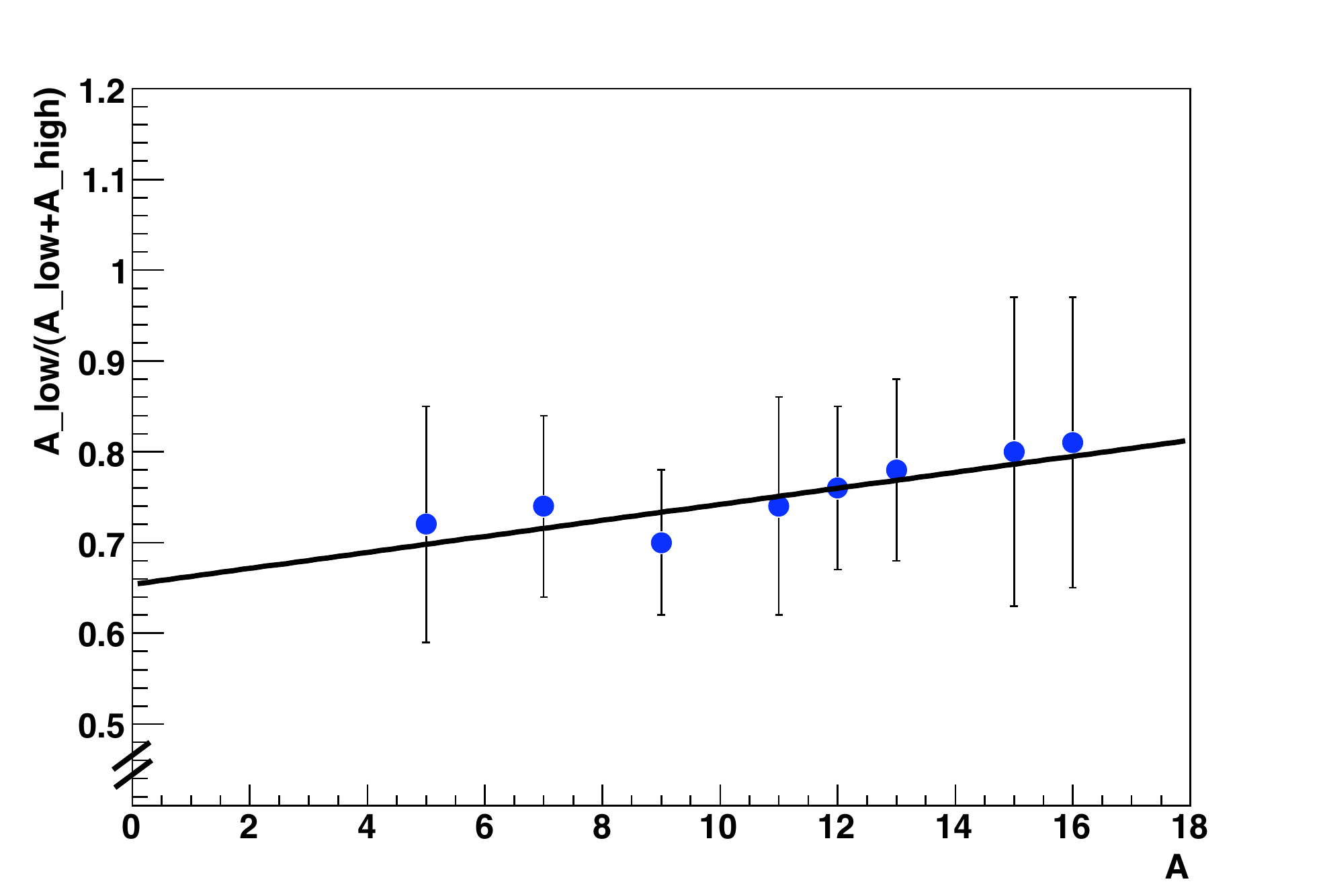}
}
\caption{The ratio $A_{\rm low}/(A_{\rm low}+A_{\rm high})$ as a function of the Hypernuclear mass number. 
The continuous line is a linear fit of the ratio. From Ref.~\cite{fnd_nmwd}.}
\label{fig:G2p}       
\end{center}
\vspace{-4mm}
\end{figure}

For each Hypernucleus, of mass number $A$, the ratio
\begin{equation}
R(A) = \frac{A_{\rm low}(A)}{ A_{\rm low}(A)+A_{\rm high}(A)}
\label{ratio}
\end{equation}
was evaluated. Fig.~\ref{fig:G2p} shows the $A$--dependence of data for $R(A)$ obtained for $^5_\Lambda$He and $p$--shell Hypernuclei. 
Taking into account that, following INC models, nucleon FSI effects can be assumed to be proportional to the number of nucleons 
of the residual nucleus and introducing the hypothesis that the $\Gamma_2/\Gamma_p$ ratio is constant in the $A=5$-$16$ Hypernuclear mass range 
\cite{AG02,alberico}, one can then rewrite Eq.~(\ref{ratio}) as:
\begin{equation}
\label{r-a}
R(A)=\frac{\displaystyle 0.5+\frac{\Gamma_2}{\Gamma_p}}
{\displaystyle 1+\frac{\Gamma_2}{\Gamma_p}}+bA,
\end{equation}
where the $b$ coefficient is the slope of the linear fit of Fig.~\ref{fig:G2p} and expresses the $A$--dependence of the FSI contribution. 
This equation  can be solved for $\Gamma_2/\Gamma_p$, giving:
\begin{equation}
 \frac{\Gamma_{2}}{\Gamma_{p}}=0.43\pm 0.25~.
\label{a6} 
\end{equation}

To determine $\Gamma_{2}$/$\Gamma_{\rm NM}$, the data for $\Gamma_{n}$/$\Gamma_{p}$ obtained by Ref.~\cite{bhang} was used.
The result is:
\begin{equation}
 \frac{\Gamma_{2}}{\Gamma_{\rm NM}}=0.24\pm 0.10~.
\label{a9} 
\end{equation}
This value supports both theoretical predictions \cite{ramos,albe2,BGPR,bau_garb,bau2,bau3,Ba10b} and the latest KEK experimental results of 
Ref.~\cite{mkim}, while it is lower than the KEK value suggested in Ref.~\cite{bhang} by about 2$\sigma$. 

A similar approach was afterwards used by the FINUDA experiment to extract the $\Gamma_{2}$/$\Gamma_{\rm NM}$ ratio for $A=5$-$16$ Hypernuclei,
but using $np$ coincidence spectra \cite{nmwd_n}. Neutrons were detected by the external scintillator barrel of the FINUDA apparatus and their 
kinetic energy was determined by means of time of flight measurements. ($\pi^{-},\, n,\, p$) triple--coincidence events were selected, with a 
$\pi^{-}$ momentum corresponding to the formation of bound Hypernuclear states. 

To deduce the contribution of the two--nucleon induced process, $np$ events with a proton of energy lower than an adequate threshold, 
$E_p^{\rm thr}=\mu-20$ MeV, determined considering the mean value $\mu$ of the Gaussian functions used in the proton inclusive spectra analysis, and a
restriction on the opening angle between the two nucleons, $\cos \theta_{np} \ge -0.8$, were considered for each Hypernucleus. 
In such a way, neutrons from two--nucleon induced processes were selected and could be counted for. With the applied selections,
events due to the one--nucleon induced process would be excluded and only a small contribution of FSI is present. 

\begin{figure} [h]
\begin{center}
\resizebox{0.45\textwidth}{!}{%
 \hspace{-5mm}
  \includegraphics{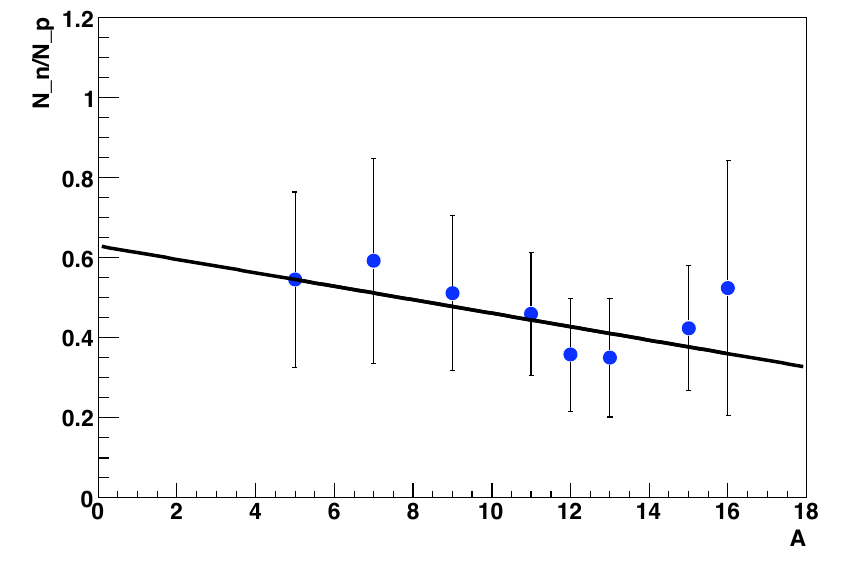}
}
\caption{The experimental and fitted $R_2$ ratio of Eq.~(\protect\ref{R_n}) as a function of the Hypernuclear mass number. 
From Ref.~\cite{nmwd_n}.}
\label{fig:G2n}       
\end{center}
\vspace{-4mm}
\end{figure}

For each nucleus the following ratio was then considered:
\begin{eqnarray}
R_2(A) & = & \frac{N_{np}(E_{p} \leq E_p^{\rm thr}, \cos \theta_{np} \ge -0.8)} {N_{p}(E_{p} > \mu)} \nonumber \\ 
 & = & \frac{0.8 N(\Lambda n p \rightarrow n n p) + FSI_{np}}{0.5 N(\Lambda p \rightarrow n p) + FSI_{p} }\, ,
\label{R_n}
\end{eqnarray}
where $N_{np}(E_{p} \leq E_p^{\rm thr}, \cos \theta_{np} \ge -0.8)$ indicates the number of $np$ events satisfying the condition which selects 
two--nucleon induced candidates and $N_{p}(E_{p} > \mu)$ is the number of protons in the part of the spectrum above the Gaussian mean value of 
Ref.~\cite{fnd_nmwd}; $FSI_{np}$ and $FSI_{n}$ indicate the residual effect of FSI processes on the corresponding samples. The factor 0.8 is due to 
the selections applied in the analysis. Fig.~\ref{fig:G2n} shows the dependence of $R_2$ on the Hypernuclear mass number. 

Also in this case it is possible to apply the same considerations on the $A$--dependence of FSI and $\Gamma_{2}$ used for the study of the 
single proton spectra. 
In Fig.~\ref{fig:G2n} the trend of the ratio (\ref{R_n}) is shown as a function of $A$: a roughly linear behaviour of the above ratio, decreasing with 
$A$, is apparent. This is consistent with the simple hypothesis that the ratio $R_2$ is linearly correlated with the two--nucleon induced rate, which 
is in first approximation independent of $A$ \cite{fnd_nmwd}, with a small contribution of FSI, which, in first approximation, increases 
linearly with $A$. 
By indicating with $\Gamma_{np}$ the width of the $np$--induced decay of the Hypernucleus, it is possible to write:
\begin{equation}
R(A) = \frac{0.8 \Gamma_{np}}{0.5 \Gamma_{p}} + bA\, ,
\label{R_n2}
\end{equation}
where the coefficient $b$ is the slope of the linear fit of Fig.~\ref{fig:G2n} and expresses the $A$--dependence of the FSI contribution. This 
equation can be solved for $\Gamma_{np}/\Gamma_p$, getting:
\begin{equation}
\frac{\Gamma_{np}}{\Gamma_{p}}=0.39\pm 0.16_{\rm stat} \hspace{0.5mm}^{+0.04}_{-0.03\, \rm sys}\, .
\label{an1} 
\end{equation} 

The obtained $\Gamma_{np}$ width can be used to obtain $\Gamma_{2}$ by following the theoretical predictions of Ref.~\cite{bau_garb}; 
moreover, by assuming again that $\Gamma_n/\Gamma_p$ is independent of the Hypernuclear species and
using for this ratio the data obtained by KEK \cite{bhang}, the FINUDA final result is:
\begin{equation}
\frac{\Gamma_{2}}{\Gamma_{\rm NM}}=0.21\pm 0.07_{\rm stat} \hspace{0.5mm}^{+0.03}_{-0.02\, \rm sys}\, ,
\label{an2} 
\end{equation}
where the error is reduced with respect to the one of Ref.~\cite{fnd_nmwd}. 

\begin{figure} [h]
\begin{center}
\resizebox{0.40\textwidth}{!}{%
 \hspace{-5mm}
  \includegraphics{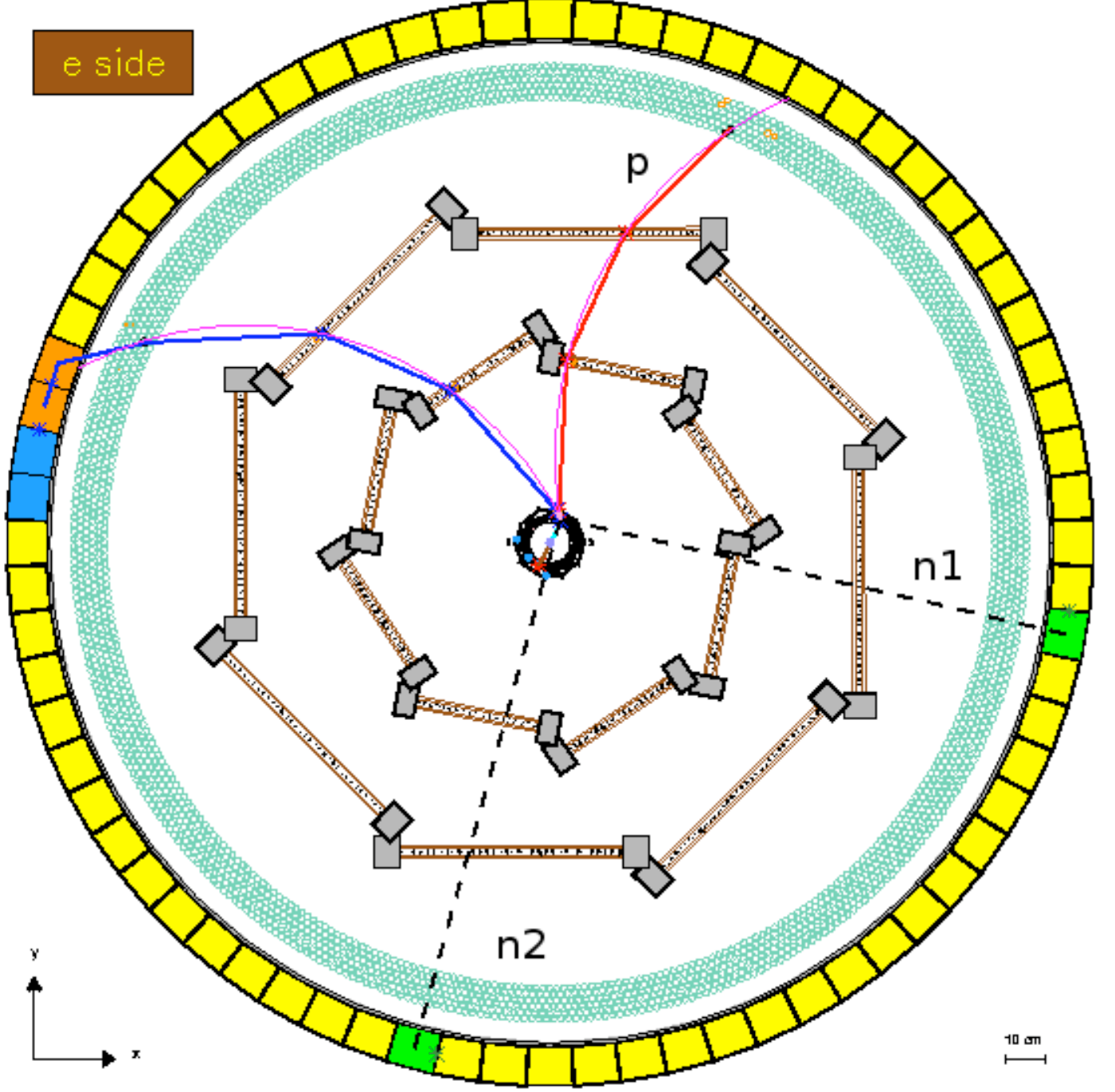}
}
\caption{Pictorial representation of the $\Lambda n p \rightarrow n n p$ event seen by FINUDA on a $^{7}$Li target. The blue track represents 
the trajectory of the $\pi^{-}$ from the hypernucleus formation reaction. From \cite{event_nnp}.}
\label{fig:event}       
\end{center}
\vspace{-4mm}
\end{figure}
Finally,  thanks to the large solid angle coverage of the FINUDA apparatus ($\sim 2 \pi$ srad) and 
to its detection efficiency for neutrons and for protons with low kinetic energy, it has been possible 
to find one event \cite{event_nnp} in which also the second neutron from pn-induced NMWD was 
detected, for which the angles between each neutron and the proton, and between the two neutrons are greater than 
$\pi/2$, which is the maximum angle expected for a simple nucleon-nucleon scattering. 
The projection of this event, occurring on a $^{7}$Li target, onto the $\rho-\phi$ plane of the apparatus is shown in figure \ref{fig:event}. 
The analysis of the momentum of the production $\pi^{-}$ ((276.9$\pm$1.2) MeV/c), indicated by the blue trajectory in figure \ref{fig:event}, of the 
total energy of the detected nucleons 
($T_{nnp} = (172.7 \pm 16.3)$ MeV), together with the determination of the kinetic energy of the missing daughter nucleus 
($T_{miss} = (6.3 \pm 4.3)$ MeV), leads to conclude that the event is a strong candidate for the decay reaction:
\begin{equation}
^{7}_{\Lambda}Li \rightarrow ^{4}He + n + n + p~, 
\label{Li7evt}
\end{equation}
where the high binding energy of the $^{4}He$ ($\sim$28 MeV) allows to perform an exclusive identification of the final state, giving a value for the 
Q of the reaction as high as Q = 167 MeV. Moreover, the topology of the event and 
the particular values of the angles between the three detected nucleons 
($\theta_{n_{2}p} = (154.4\pm2.8)^{\circ}$, $\theta_{n_{1}p} = (102.3\pm2.2)^{\circ}$, $\theta_{n_{2}n_{1}} = (94.8\pm3.6)^{\circ}$)
allow to exclude the possibility of a proton induced decay followed by a 
nucleon-nucleon scattering. 
A final remark is that both $\pi^{-}$ and $p$ are emitted in the same hemisphere (relative angle of (33.4$\pm$3.7)$^{\circ}$), avoiding the possibility that they come 
from a physical event due to K$^{-}$ capture on a correlated $np$ pair, with a $\pi^{-}$ of momentum comprised in the range for ground state 
hypernucleus formation, in coincidence with two random signals from the external TOF system (see Fig.\ref{fig:event}), taken as true neutron events. 
As a matter of fact, $\pi^{-}$ and $p$ 
from a K$^{-}$ capture on a correlated $np$ pair are emitted in opposite emisphere \cite{tamura}. 
All the reconstructed kinematical observables are detailed in Table \ref{tab:nnp}. 
\begin{table}[h]
\vspace{-2mm}
\centering
\caption{Kinematics of the $\Lambda (np) \rightarrow nnp$ event of figure \ref{fig:event}: 
$p_{\pi^{-}}$ indicates the momentum of the hypernucleus formation $\pi^{-}$, $T_{p}$ the $p$ kinetic energy, $T_{n1}$ the kinetic energy 
of one of the neutrons, $T_{n2}$ the kinetic energy of the second one, 
$\theta_{n1n2}$ the angle between the two neutrons, $\theta_{n1p}$ the angle between the first $n$ and the $p$, $\theta_{n2p}$ the 
angle between the second $n$ and the $p$, $\theta_{\pi^{-}p}$ the 
angle between the formation $\pi^{-}$ and the $p$, $p_{miss}$ the missing momentum of the final state, 
and MM the missing mass for the decay of the 
hypernucleus from its ground state. The quoted error is statistical.}
\label{tab:nnp}       
\vspace{2mm}
\begin{tabular}{cc}
\hline\noalign{\smallskip}
$p_{\pi^{-}}$ (MeV/c) & $276.9\pm1.2$ \\ 
$T_{p}$ (MeV) & $51.11\pm0.85$  \\
$T_{n1}$ (MeV) & $110\pm23$  \\
$T_{n2}$ (MeV) & $16.9\pm1.7$  \\
$\theta_{n1n2}$ (deg) & $94.8\pm3.8$  \\
$\theta_{n1p}$ (deg) & $102.2\pm3.4$ \\
$\theta_{n2p}$ (deg) &  $154\pm19$  \\
$\theta_{\pi^{-}p}$ (deg) &  $33.4\pm3.7$   \\
$p_{miss}$ (MeV/c) & $217\pm44$  \\
MM (MeV/c$^{2}$) &$3710\pm23$ \\
\noalign{\smallskip}\hline
\end{tabular}
\end{table}

So, from the kinematic analysis, it is possible to conclude that the event can safely be attributed to the 
exclusive reaction (\ref{Li7evt}) and represents the first direct ("smoking gun") experimental evidence of a two-nucleon induced NMWD. 


Various theoretical approaches were applied to the calculation of $\Gamma_2$ \cite{ramos,albe2,BGPR,bau_garb,bau2,bau3,Ba10b}.
We summarize the theoretical and experimental determinations of $\Gamma_2$ and $\Gamma_2/\Gamma_{\rm NM}$ in Table~\ref{tab-gamma2}
together with information on the Hypernucleus and the isospin channels considered in the calculations and on the
Hypernucleus and the kind of spectra (single or in coincidence) analyzed in the experiments.
All the calculations were performed with a nuclear matter approach and then extended to $^{12}_\Lambda$C via the local density 
approximation. In Refs.~\cite{ramos,albe2}, a phenomenological scheme based on data on pion absorption in nuclei was used
to predict the $np$--induced decay rate $\Gamma_{np}$. Ref.~\cite{albe2} updates the predictions of Ref.~\cite{ramos} by using more realistic 
$\Lambda$ wave functions in Hypernuclei and short--range correlations in the baryon--baryon strong interactions. 
All isospin channels of the two--nucleon stimulated decay, $\Lambda nn\to nnn$, $\Lambda np\to nnp$ and $\Lambda pp\to npp$,
are included in the diagrammatic microscopic approach of Refs.~\cite{bau_garb,bau2,bau3,Ba10b}. 
In particular, Pauli exchange and ground state contributions were also studied in this kind of calculation \cite{bau_garb,Ba10b}. 
Among the predictions of the microscopic approach, only the most updated, from Ref.~\cite{Ba10b}, are reported in Table~\ref{tab-gamma2}.
The microscopic approach also lead to the following predictions for the two--nucleon induced partial rates:
$\Gamma_{np} : \Gamma_{pp} : \Gamma_{nn} = 0.83 : 0.12 : 0.04$;
as expected, the two--nucleon stimulated decay width is dominated by the $np$--induced process.  
\begin{table*}[t]
\caption{Summary of the theoretical and experimental values of $\Gamma_2$ and $\Gamma_{2}/\Gamma_{\rm NM}$.}
\label{tab-gamma2}
\begin{center}
\begin{tabular}{lccl}
\hline
Ref. & $\Gamma_2$ & $\Gamma_{2}/\Gamma_{\rm NM}$ & Notes\\ \hline
\cite{ramos} & 0.23 & 0.16 & $^{12}_\Lambda$C, $\Gamma_2=\Gamma_{np}$\\
\cite{albe2} & 0.16 & 0.16 & $^{12}_\Lambda$C, $\Gamma_2=\Gamma_{np}$\\
\cite{Ba10b} & 0.25 & 0.26 & $^{12}_\Lambda$C, $\Gamma_2=\Gamma_{np}+\Gamma_{nn}+\Gamma_{pp}$\\\hline
BNL--E788 \cite{parker} & & $\leq 0.24$ & $^{4}_\Lambda$He, $n$ and $p$ spectra \\
KEK--E508 \cite{mkim} & $0.27\pm 0.13$ & $0.29\pm 0.13$ &  $^{12}_\Lambda$C, $nn$ and $np$ spectra \\
FINUDA \cite{fnd_nmwd} & & 0.24$\pm$0.10 &  $A=$5-16, $p$ spectra \\
FINUDA \cite{nmwd_n} & & $0.21 \pm 0.07_{\rm stat} \hspace{0.5mm}^{+0.03}_{-0.02\, \rm sys}$ &  $A=$5-16, $np$ spectra \\ \hline
\end{tabular}
\end{center}
\end{table*}

Considering instead the data reported in Table~\ref{tab-gamma2}, it must be noted that the two values obtained by FINUDA are not independent of each 
other, since the proton required in coincidence with a $\pi^{-}$ in Ref.~\cite{fnd_nmwd} and with a ($\pi^{-}$, n) pair in Ref.~\cite{nmwd_n} belongs 
to the same experimental sample. To compare with the KEK--E508 determination \cite{mkim}, the FINUDA value of Ref.~\cite{nmwd_n}, affected by a 
smaller error, can be used: the two results are completely compatible within errors and by evaluating their weighted mean it is  possible to conclude 
that, for $p$--shell $\Lambda$--Hypernuclei, $\Gamma_{2}/\Gamma_{\rm NM}=0.23 \pm 0.06$.


All the predictions of Table~\ref{tab-gamma2} are consistent with the individual KEK and FINUDA experimental determinations of $\Gamma_{2}$ and 
$\Gamma_{2}/\Gamma_{\rm NM}$ (however, only the microscopic calculation of Ref.~\cite{Ba10b} reproduces within $1\sigma$ the weighted mean datum of 
the previous paragraph). One may conclude that the two--nucleon induced NMWD of Hypernuclei is rather well understood at present.

A general conclusion concerning the study of NMWD of Hypernuclei is now in order.
As discussed in Sec.~\ref{subsec:nmwd_gen}, thanks to the large momentum transfer of the NMWD channel, one expects to obtain information 
on the four--baryon, strangeness changing, $\Lambda N \rightarrow n N$ weak interaction from Hypernuclear decay measurements. However, 
the study of NMWD of $p$--shell Hypernuclei clearly indicates that the nuclear effects cannot be disregarded, since 
mechanisms such as FSI and two--nucleon induced decay are not at all inessential but partially mask the elementary four--baryon reaction. 

Therefore, the best way to proceed for a detailed study of the $\Lambda N \rightarrow n N$ reaction seems to be to learn how to parametrize the 
nuclear effects from the systematic study of Hypernuclear NMWD over an appropriate mass number range, as done by FINUDA for $p$--shell Hypernuclei, 
and then to use this information to try to disentangle the four--baryon reaction properties from a very high statistics decay 
measurement on a single nucleus, possibly with low mass number. 

A determination of $\Gamma_n$, $\Gamma_p$ and $\Gamma_2$ for $^{12}_\Lambda$C with a 10\% error level is expected from J--PARC 
\cite{jparc-e18}, from double-- and triple--nucleon coincidence measurements.
This will allow to lower the error on the determination of $\Gamma_2/\Gamma_{\rm NM}$.

\subsection{Rare Two--Body NMWD of $\bf s$--Shell Hypernuclei}
\label{subsec:nmwd_rare}

The decay of the light Hypernuclei $^{4}_{\Lambda}$He and $^{5}_{\Lambda}$He in two--body channels is a rare process because of the large 
momentum transfer and of the possible two--step mechanisms involved. It can occur through the following reactions:
\begin{eqnarray}
\label{rare_1} 
^{4}_{\Lambda}{\rm He} &\rightarrow & d + d\, , \\
\label{rare_2} 
^{4}_{\Lambda}{\rm He} &\rightarrow & p + t\, ,\\
\label{rare_3} 
^{4}_{\Lambda}{\rm He} &\rightarrow & n + ^{3}{\rm He}\, ,\\
\label{rare_4} 
^{5}_{\Lambda}{\rm He} &\rightarrow & d + t\, .
\end{eqnarray}
The existing observations of two--body NMWD of Hypernuclei are very scarce  and date back to bubble 
chamber and emulsion experiments \cite{coremans,bloch2,keyes76}; in particular, reaction (\ref{rare_3}) has not yet been observed. 
The only existing calculation for the $^{4}_{\Lambda}$He two--body NMWD rates was performed long ago
by Ref.~\cite{rayet}, and only one theoretical evaluation for the expected decay rates of $^{5}_{\Lambda}$He exists \cite{thurnauer}. 

Recently, rare two--body NMWD have been studied by means of the FINUDA spectrometer \cite{nmwd_rare}; the large solid angle coverage, 
the good p.id. properties and the good momentum resolution of the experimental apparatus allowed the identification 
of reactions (\ref{rare_1}), (\ref{rare_2}) and (\ref{rare_4}), which are characterized by a back--to--back topology of the outgoing 
particles/clusters and by quite high momenta (p$_{d}$=570 MeV/c for (\ref{rare_1}), p$_{p}$=508 MeV/c for (\ref{rare_2}) and 
p$_{d}$=597 MeV/c for (\ref{rare_4})).

Reactions (\ref{rare_1}), (\ref{rare_2}) and (\ref{rare_4}) have been studied as decays of $^{4}_{\Lambda}$He and $^{5}_{\Lambda}$He 
hyperfragments produced by the interaction of stopped $K^{-}$'s on the various targets used in FINUDA and identified by means of the 
detection of the $\pi^{-}$ emitted in the production reaction; in particular, (\ref{rare_1}) and (\ref{rare_2}) have been looked for on all the nuclear 
targets, while (\ref{rare_4}) has been searched for on $^{6}$Li and $^{7}$Li targets only.

Concerning the $^{4}_{\Lambda}{\rm He} \rightarrow d+ d$ decay, an average yield value 
$Y(^{4}_{\Lambda}{\rm He} \rightarrow d + d) = (2.82\pm0.62) \times10^{-5} /K^{-}_{\rm stop}$ is obtained.
The $^{4}_{\Lambda}{\rm He} \rightarrow p+ t$ decay at rest looks favoured: an average yield of  
$Y(^{4}_{\Lambda}{\rm He} \rightarrow p + t) = (5.42\pm3.43) \times10^{-5} /K^{-}_{\rm stop}$ is obtained. 
As for the $^{5}_{\Lambda}{\rm He} \rightarrow d+ t$ decay, the average yield value over $^{6}$Li and $^{7}$Li nuclei is 
$Y(^{5}_{\Lambda}{\rm He} \rightarrow d + t) = (1.23\pm0.70) \times10^{-4} /K^{-}_{\rm stop}$, corresponding to a branching ratio of  (2.8$\pm$ 1.4) 
$\times$ 10$^{-3}$, in rough agreement with the theoretical expectation of a factor 100 less than the total NMWD branching ratio.

\subsection{Asymmetry in NMWD of Polarized Hypernuclei}
\label{subsec:nmwd_asym}

$\Lambda$--Hypernuclei can also be produced in polarized states \cite{Ba89}.
Thanks to the large momentum transfer involved, the $n(\pi^+,K^+)\Lambda$ reaction was used,
at $p_{\pi}=1.05$ GeV and small $K^+$ laboratory scattering angles
($2^\circ\lsim \theta_{K}\lsim 15^\circ$),
to produce Hypernuclear states with a substantial amount of spin--polarization,
preferentially aligned along the axis normal to the reaction plane \cite{Aj92,Aj00},
the so--called polarization axis.
The origin of Hypernuclear polarization is twofold \cite{Ba89}.
It is known that the distortions (absorptions) of the initial ($\pi^+$)
and final ($K^+$) meson--waves produce a small polarization of the
Hypernuclear orbital angular momentum up to laboratory scattering angles 
$\theta_{K}\sim 15^\circ$
(at larger scattering angles, the orbital polarization increases with a negative sign).
At small but non--zero angles, the main source of polarization is due to
an appreciable spin--flip interaction term in the elementary reaction $\pi^+ n\to \Lambda K^+$,
which interferes with the spin--nonflip amplitude.
In a typical experimental situation with $p_{\pi}=1.05$ GeV and $\theta_{K}\sim 15^\circ$,
the polarization of the hyperon spin in the free $\pi^+ n\to \Lambda K^+$ process is about 0.75.

The distribution of protons produced in one--nucleon induced NMWD of polarized Hypernuclei
shows an angular asymmetry: the difference between the number of protons emitted along the
polarization axis and the number of protons outgoing in the opposite direction determines this
asymmetry. It can be shown that the proton asymmetry is originated by the interference
between the parity--violating and parity--conserving $\Lambda N\to nN$ transition amplitudes
with different values of the isospin of the final $NN$ pair \cite{Ba90}. 
Asymmetry studies are thus expected to provide new constraints on
the strengths and (especially) the relative phases of the decay amplitudes,
i.e., on the dynamics of Hypernuclear decay.

Despite the important progress on the determination
of the NMWD rates (especially the $\Gamma_n/\Gamma_p$ and $\Gamma_2/\Gamma_{\rm NM}$ ratios)
discussed above, another intriguing problem  which was solved only 
recently concerns a strong disagreement between theory and experiment on the weak decay asymmetry.

The intensity of protons emitted in $\Lambda p \to np$ decays
along a direction forming an angle $\theta$
with the polarization axis is given by \cite{Ra92}:
\begin{equation}
\label{int-w}
I(\theta, J)=I_0(J)[1+\mathcal{A}(\theta, J)]\, ,
\end{equation}
$J$ being the total spin of the Hypernucleus and
$I_0$ the (isotropic) intensity for an unpolarized Hypernucleus.
In the shell model weak--coupling scheme, one can write:
\begin{equation}
\mathcal{A}(\theta, J)=P_{\Lambda}(J)\, a_{\Lambda}\, {\rm cos}\, \theta\, ,
\end{equation}
where $P_{\Lambda}$ is the polarization of the $\Lambda$ spin and
$a_{\Lambda}$ the \emph{intrinsic} $\Lambda$ asymmetry parameter, which is
a characteristic of the elementary process $\Lambda p\to np$.

Nucleon FSI acting after the NMWD modify the weak decay intensity (\ref{int-w}): this means that the intrinsic asymmetry $a_\Lambda$
is not an observable. The \emph{observable} asymmetry, $a^{\rm M}_\Lambda(J)$, which is expected to depend
on the considered Hypernucleus, is therefore derived from the observable proton intensity, $I^{\rm M}$.
Assuming for $I^{\rm M}$ the same $\theta$--dependence as in the weak decay 
intensity $I$, one obtains the observable asymmetry as:
\begin{equation}
\label{asym-exp}
a^{\rm M}_\Lambda(J)=\frac{1}{P_\Lambda(J)} \frac{I^{\rm M}(0^{\circ},J)- 
I^{\rm M}(180^{\circ},J)}{I^{\rm M}(0^{\circ},J)+I^{\rm M}(180^{\circ},J)}\, .
\end{equation}
One expects an attenuation of the asymmetry due to FSI, $|a^{\rm M}_\Lambda(J)|<|a_\Lambda|$.

Eq.~(\ref{asym-exp}) is actually the relation used to determine experimentally $a^{\rm M}_\Lambda(J)$.
These measurements generally suffer from large uncertainties, principally due to
limited statistics and to the poor knowledge of the hyperon--spin polarization.
For $^5_\Lambda$He, $P_\Lambda$ was measured \cite{Aj98} by observing 
the asymmetric emission of negative pions in its mesonic decay and by assuming that
the pion asymmetry in the mesonic decay of $^5_\Lambda$He coincides with the
value measured in the free $\Lambda \to \pi^- p$ decay. Unfortunately, the small branching ratio
and asymmetry for the mesonic decay of $p$--shell Hypernuclei makes a measurement
of the $\Lambda$--spin polarization very difficult for these systems. Therefore,
in the proton asymmetry measurements for $^{11}_\Lambda$B and $^{12}_\Lambda$C \cite{Aj00,Mar06},
$P_\Lambda$ was evaluated theoretically by adopting the distorted wave impulse approximation
of Ref.~\cite{Mot94}. Such kind of calculation requires a delicate analysis of
1) the polarization of the Hypernuclear states directly produced in the   
production reaction and 2) the depolarization
effects due to strong and electromagnetic transitions of the populated excited states,
which take place before the weak decay.

In Table~\ref{other-res} we compare the theoretical and experimental 
determinations of the decay asymmetries for $^{5}_\Lambda$He and $^{12}_\Lambda$C.
While inexplicable inconsistencies appeared between the first KEK asymmetry experiments of
Refs.~\cite{Aj92,Aj00}, the recent and more accurate KEK data \cite{Mar06} reported
in Table~\ref{other-res} favor small $a^{\rm M}_\Lambda$ values, compatible
with a vanishing value, for both $^5_\Lambda$He and $^{12}_\Lambda$C. 

On the contrary, theoretical 
models based on OME potentials and/or direct quark (DQ) mechanisms \cite{sasaki,parreno,It03,Ba05} predicted rather large and 
negative $a_\Lambda$ values. It must be noted that, on the contrary, the mentioned models 
were able to account fairly well for the weak decay rates measured for $s$--
and $p$--shell Hypernuclei. As can be seen from Table~\ref{other-res}, not even the inclusion of FSI 
in the theoretical analysis with OME potentials can explain the experimental data. To illustrate this point, 
we quote the result of Ref.~\cite{Al05} for a proton kinetic energy threshold 
of 30 MeV (the data of Table~\ref{other-res} were obtained for such an energy cut).
\begin{table*}
\begin{center}
\caption{Theoretical and experimental determinations of
the asymmetry parameters.}
\label{other-res}
\begin{tabular}{l c c} \hline
\mc {1}{c}{Ref. and Model} &
\mc {1}{c}{$^5_\Lambda {{\rm H}}{\rm e}$} &
\mc {1}{c}{$^{12}_\Lambda {{\rm C}}$} \\ \hline
Sasaki {\it et al.} \cite{sasaki} &   & \\
$\pi+K+{\rm DQ}$                & $-0.68$     \\
Parre\~no and Ramos \cite{parreno} &    & \\
OME: $\pi+\rho+K+K^*+\omega+\eta$    & $-0.68$ & $-0.73$    \\
Itonaga {\it et al.} \cite{It03}           &    & \\
$\pi+K+\omega+2\pi/\rho+2\pi/\sigma$   & $-0.33$ &  \\
Barbero {\it et al.} \cite{Ba05}           &    & \\
OME: $\pi+\rho+K+K^*+\omega+\eta$            & $-0.54$ & $-0.53$   \\
Alberico {\it et al.} \cite{Al05}  &    & \\
OME + FSI      & $-0.46$  & $-0.37$  \\
Chumillas {\it et al.} \cite{CGPR07}  &    & \\
${\rm OME}+2\pi+2\pi/\sigma$  & $+0.041$  & $-0.207$ \\
${\rm OME}+2\pi+2\pi/\sigma$ + FSI & $+0.028$  &  $-0.126$   \\
Itonaga {\it et al.} \cite{It08}      &    & \\
$\pi+K+\omega+2\pi/\rho+2\pi/\sigma+\rho\pi/a_1+\sigma\pi/a_1$     &
$+0.083$ &  $+0.045$ \\ \hline
   KEK--E462  \cite{Mar06}   & $0.07\pm0.08^{+0.08}_{-0.00}$  & \\
   KEK--E508  \cite{Mar06}    &                & $-0.16\pm0.28^{+0.18}_{-0.00}$  \\
\hline
\end{tabular}
\end{center}
\end{table*}

At this point we must note that an effective field theory approach to Hypernuclear
decay \cite{Pa04}, in which the weak two--body transition is based on tree--level
pion-- and kaon--exchange and leading--order contact interactions,  
suggested a dominating spin-- and isospin--independent contact term.
Such central term turns out to be particularly important if one wants to fit the small
and positive value of the intrinsic asymmetry for $^{5}_\Lambda$He
indicated by KEK--E462. In a calculation scheme based on a OME model,
this result can be interpreted dynamically as the need for the introduction
of a meson--exchange contribution in the scalar--isoscalar channel.
Prompted by the work of Ref.~\cite{Pa04}, 
models based on OME and/or DQ mechanisms \cite{Sas05,Ba06} were supplemented
with the exchange of the scalar--isoscalar $\sigma$--meson. 
Despite the phenomenological character of these works (the unknown $\sigma$ weak couplings are fixed
to fit NMWD data for $^5_\Lambda$He and $^{12}_\Lambda$C), they showed the importance of $\sigma$--exchange
in the NMWD.

Then, an investigation was performed on the effects of a chirally motivated
two--pion--exchange mechanism on the NMWD
rates and asymmetries for $s$-- and $p$--shell Hypernuclei \cite{CGPR07}. The
uncorrelated ($2\pi$) and correlated (in the isoscalar channel, $2\pi/\sigma$) two--pion--exchange
weak potentials were adopted from Ref.~\cite{Os01} and added to the exchange of the pseudoscalar
and vector mesons, $\pi$, $\rho$, $K$, $K^*$, $\omega$ and $\eta$, of the standard
OME potentials. These scalar--isoscalar contributions are based on a
chiral unitary model which describes $\pi\pi$
scattering data in the scalar sector up to around 1 GeV, where all the coupling
constants are determined from chiral meson--meson and meson--baryon
Lagrangians by imposing SU(3) symmetry.   
It was found that the two--pion--exchange mechanism modifies moderately
the partial decay rates $\Gamma_n$ and $\Gamma_p$ but has a tremendous
influence on the asymmetry parameter, due to the change of sign of the central,
spin-- and isospin--independent amplitudes \cite{CGPR07}:
the obtained asymmetry values agreed with data for the first time.
It must to be noted that the main mechanism which permitted to achieve this result
is the uncorrelated two--pion--exchange.
The resulting values of the asymmetry parameter are presented in 
Table~\ref{other-res} for the case where no FSI are accounted for
and also when a proton kinetic energy threshold of 30 MeV is imposed in the calculation with
FSI. 

From Table~\ref{other-res} we see that a more recent and different approach
\cite{It08}, whose main characteristic is the use of correlated meson--pair exchange, 
proved that the exchange of the axial--vector $a_1$--meson is also relevant in asymmetry calculations.
We note however that the one--meson--exchange plus two--pion--exchange model of 
Ref.~\cite{CGPR07} tur-ned out to be able to reproduce satisfactorily
the total and partial NMWD rates as well as the asymmetries, for both $^5_\Lambda$He and 
$^{12}_\Lambda$C, within a minimal framework, i.e., without invoking  
exotic decay mechanisms, nor a violation of the $\Delta I=1/2$ isospin rule.

We can thus conclude that with the solution of the asymmetry puzzle a
deeper theoretical understanding of the Hypernuclear decay mechanisms
is now reached.

\section{Mesonless  Multinucleon Absorption of Stopped K$^{-}$}
\label{kns}
\subsection{Recent experimental data}

\begin{figure} [h]
\begin{center}
\resizebox{0.5\textwidth}{!}{%
\hspace{-5mm}
  \includegraphics{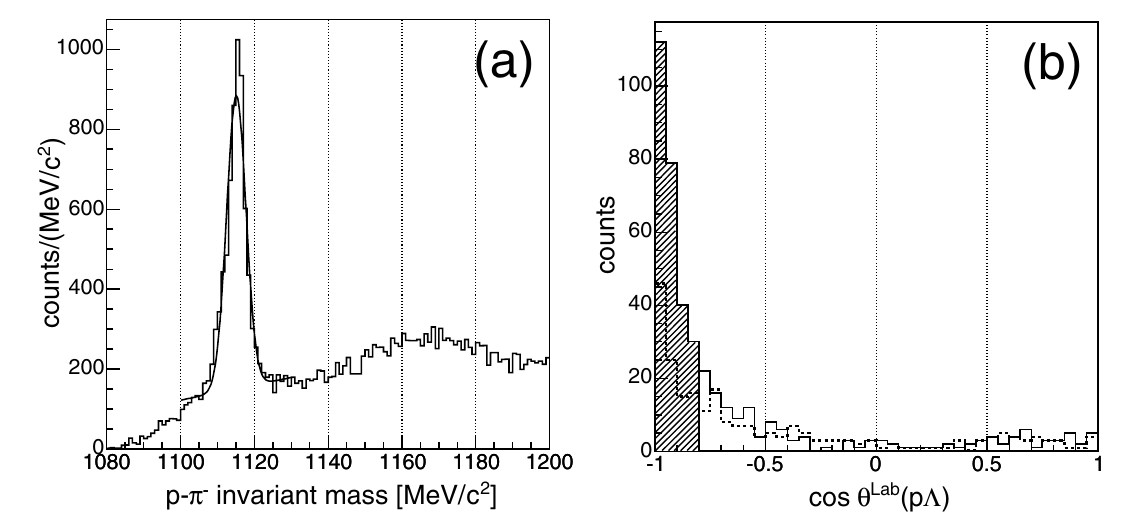}
}
\caption{(a) $\pi^{-}$-p invariant mass distribution; the continuous line is a fit with a Gaussian function plus a linear background. 
(b) Opening angle distribution between a $\Lambda$ and a proton: solid line $^{6}$Li, $^{7}$Li and $^{12}$C, dashed line $^{27}$Al, $^{51}$V. 
The shaded area is selected as back--to--back events ($\cos\, \theta^{\rm lab} <-0.8$). From Ref.~\cite{prl94}.} 
\label{fig:figu2}       
\end{center}
\end{figure}

Data on reactions of nuclear absorption of stopped K$^{-}$ with emission of $\Lambda$, nucleons or light nuclei (deuterons, tritons) were very scarce up to a few years ago. 
The bulk of the data was coming from bubble chamber experiments, which mainly aimed at assessing the K$^{-}$ capture rates of multipionic final states \cite{katz}. 
Thanks to the clever design of the FINUDA spectrometer, described in Sec.~\ref{prod},  it was possible to identify in a clean way the $\Lambda$ 
hyperons emitted following the interaction of stopped K$^{-}$ in nuclei ($^{6}$Li, $^{7}$Li, $^{12}$C, $^{27}$Al, $^{51}$V) \cite{prl94}. 
They can be identified by reconstructing the invariant mass of a proton and a negative pion, as shown in Fig.~\ref{fig:figu2}(a). 
The peak position agrees with the known $\Lambda$ mass, and the width of the peak is as narrow as 6 MeV/c$^{2}$ FWHM. 

When a K$^{-}$ interacts with two protons, one expects that a hyperon--nucleon pair ($\Lambda$+p, $\Sigma^{0}$+p or $\Sigma^{+}$+n) is emitted 
in opposite directions, ignoring a final state interaction inside the nucleus. 
The angular correlation between a $\Lambda$ and a proton from the same point in the target (Fig.~\ref{fig:figu2}(b)) clearly indicates 
the existence of this kind of reaction. Even for heavy nuclei such as $^{27}$Al and $^{51}$V similar correlations were observed, which might suggest that the 
absorption would take place at the surface of a nucleus. 
In the following steps of the analysis  $\Lambda$-p pairs emitted in opposite directions ($\cos\, \theta^{\rm lab} <-0.8$) only from the 
light nuclear targets ($^{6}$Li, $^{7}$Li and $^{12}$C) were used.  Since the back--to--back angular correlation between a $\Lambda$ and a proton is so clear, it is 
naturally expected that the two particles are emitted from a K$^{-}$pp intermediate system. The angular correlation is smeared out due to the Fermi 
motions of the two protons at the surface of the nucleus by which the K$^{-}$ is absorbed after cascading down the atomic orbits by emitting X--rays. 
\begin{figure} [h]
\begin{center}
\resizebox{0.5\textwidth}{!}{%
  \includegraphics{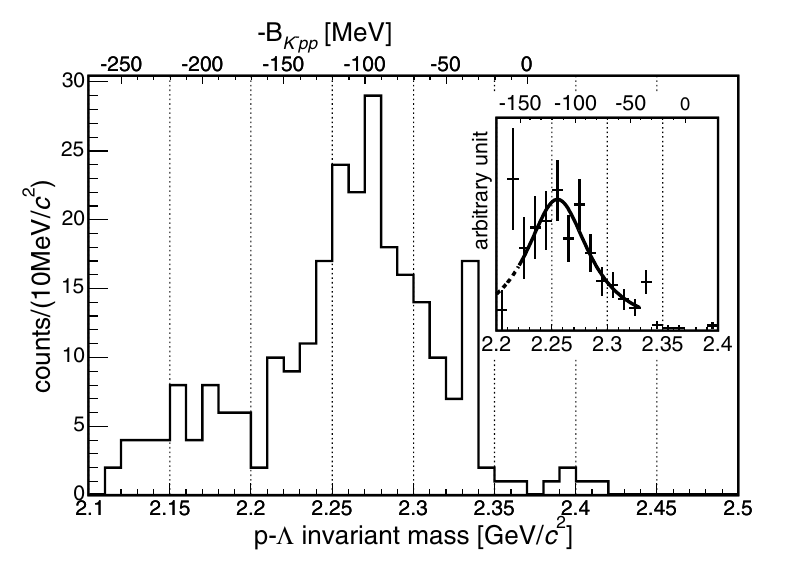}
}
\caption{$\Lambda$-p invariant mass distribution for events with $\cos \theta^{\rm lab} <-0.8$ from light targets before the acceptance correction. 
The inset shows the result after the acceptance correction: the continuous line is a fit to the 2.22$\div$2.33 GeV/c$^{2}$ region. From Ref.~\cite{prl94}.}
\label{fig:figu3}       
\end{center}
\vspace{-4mm}
\end{figure}
If the reaction process were simply a two--nucleon absorption process, the mass of the system should be close to the sum of a kaon and two proton 
masses, namely 2.370 MeV/c$^{2}$. The initial motion of the two protons does not affect the invariant mass distribution.
The invariant mass distribution of the $\Lambda$-p pairs is shown in Fig.~\ref{fig:figu3}. 
A significant  mass decrease of the K$^{-}$pp system with respect to its expected mass is observed. 
It can be interpreted as a bound state composed of a kaon and two protons, hereafter abbreviated as K$^{-}$pp. 
In the inset of Fig.~\ref{fig:figu3}, the acceptance corrected invariant mass distribution is shown. 
Since the trigger and detection acceptance are monotonically increasing functions of the invariant mass in this mass region, the peak further shifts to a lower mass side.
By fitting the acceptance corrected peak with a Lorentzian function (folded with a Gaussian with $\sigma$=4 MeV/c$^{2}$, corresponding to the detector resolution 
estimated with a Monte Carlo simulation) in the region of 2.22-2.33 GeV/c$^{2}$ values for the  binding energy and width of such an hypothetical 
K$^{-}$pp bound state of 115$^{+6}_{-5}$(stat)$^{+3}_{-4}$(syst) MeV and 67$^{+14}_{-11}$(stat)$^{+2}_{-3}$(syst) MeV were obtained. 
\begin{figure} [h]
\begin{center}
\resizebox{0.5\textwidth}{!}{%
 \includegraphics{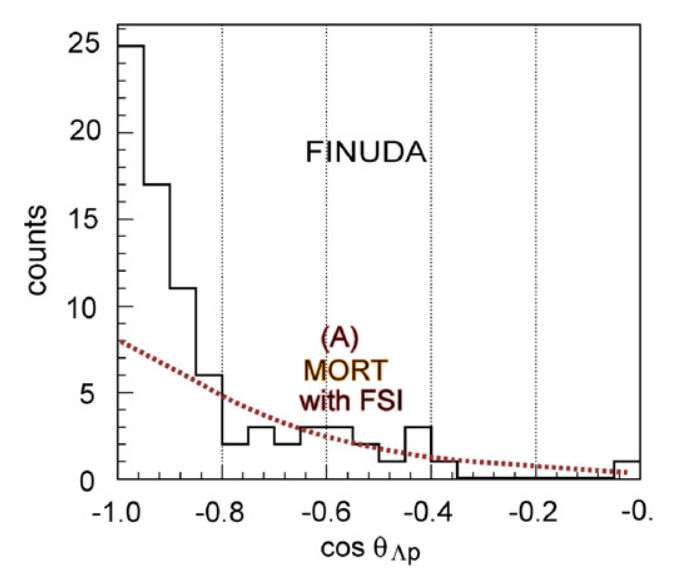} 
}
\caption{Opening angle distribution between $\Lambda$ and p. Histogram: from experiment; dotted line: from FSI mechanism \cite{magas}. 
From Ref.~\cite{yama_db}.}
\label{fig:figu4}       
\end{center}
\vspace{-4mm}
\end{figure}

An alternative, conventional explanation for the above observed bump was put forward. 
In Ref.~\cite{magas} the bump in the $\Lambda$-p invariant mass spectrum was explained as an artifact of the angular cuts applied to the flat spectrum 
of invariant mass 
of $\Lambda$-p events resulting from genuine back--to--back pairs formed in simple K$^{-}$-(np) interactions in the target nuclei that suffered a FSI. 
Even though such a mechanism cannot be completely excluded, other arguments are contradicting it. A first one is the shape of the angular correlation in the approach 
of Ref.~\cite{magas}, showing a strong disagreement \cite{yama_db} with the steepness measured experimentally, as shown in fig.~\ref{fig:figu4}. 
A second one is a possible inadequacy of the FSI calculations leading to an overestimation of these interactions. 
As a matter of fact, as shown in Fig.~\ref{fig:C12fth}, calculations following a similar approach \cite{garb} 
failed to reproduce the spectrum of protons from NMWD of $^{12}_{\Lambda}$C, recently measured with a good precision \cite{npa804}.
A rough estimate on the yield of K$^{-}$(pp)$\to \Lambda$+p provided 0.1$\%$ per stopped K$^{-}$, that is the same order of magnitude of the 
production of bound $\Lambda$--Hypernuclei. 
\begin{figure} [h]
\begin{center}
\resizebox{0.45\textwidth}{!}{%
 \includegraphics{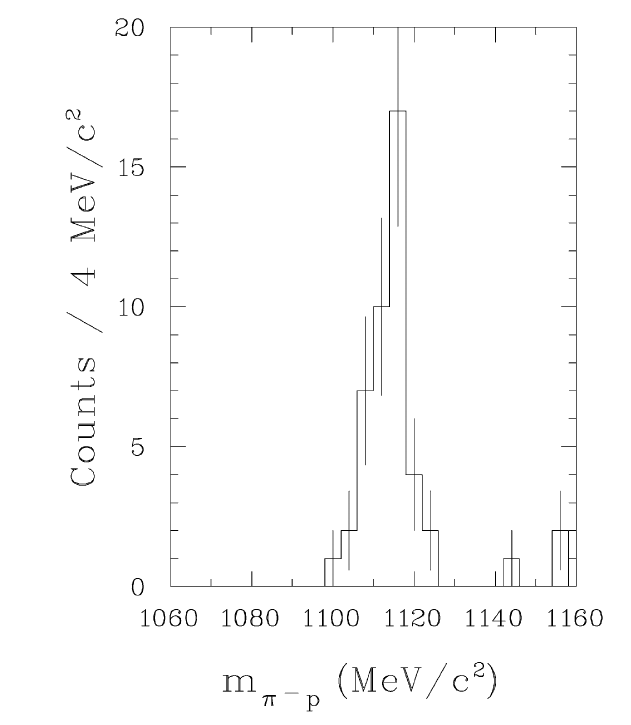} 
}
\caption{Invariant mass distribution of $\pi^{-}$ p pairs, in coincidence with deuterons, for $^{6}$Li. From Ref.~\cite{Lambdad}.}
\label{fig:figu5}       
\end{center}
\vspace{-4mm}
\end{figure}

Stimulated by these unexpected findings, the FINUDA Collaboration searched for correlated $\Lambda$-d pairs emitted following the absorption of  stopped K$^{-}$ in some 
selected targets ($^{6}$Li, $^{12}$C) \cite{Lambdad}. The measurement took advantage from both the excellent identification of $\Lambda$ hyperons by 
the invariant mass measurements 
and of deuterons by three independent measurements of dE/dx and time of flight.  Fig.~\ref{fig:figu5} shows the invariant mass distribution of $\pi^{-}$-p pairs emitted in 
coincidence with deuterons from $^{6}$Li targets. A well defined and narrow peak, centered at the right mass of the $\Lambda$ and nearly background free is evident.
\begin{figure} [h]
\begin{center}
\resizebox{0.5\textwidth}{!}{%
\includegraphics{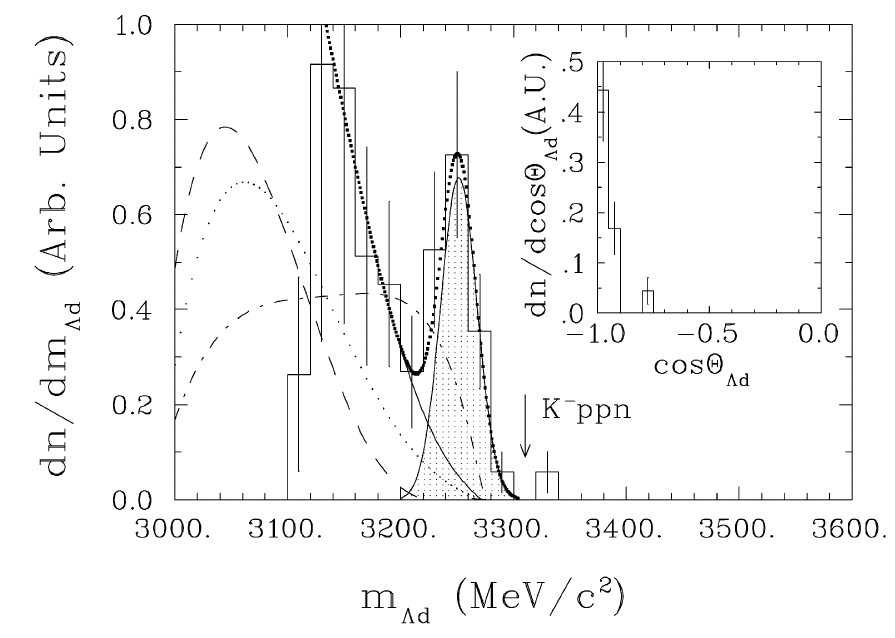} 
}
\caption{Invariant mass distribution of $\Lambda$d pairs from the $^{6}$Li(K$^{-}_{\rm stop}$, $\Lambda$d)3N reaction. From Ref.~\cite{Lambdad}.}
\label{fig:figu6}       
\end{center}
\vspace{-4mm}
\end{figure}
Fig.~\ref{fig:figu6} shows the Invariant  Mass distribution for the $\Lambda$-d events selected with the above criteria. A hint for a peak at a mass of 3251$\pm$6 MeV appears. 
By selecting the events contained in this peak, they appear strongly back--to--back correlated, as shown by the inset of Fig.~\ref{fig:figu6}. They 
could be interpreted as the decay of  a K$^{-}$ppn bound state with a binding energy of 58$\pm$6 MeV and a width of 36.6$\pm$14.1 MeV. 
This object could be formed thanks to the predominant ($\alpha$+d) substructure of $^{6}$Li, already observed in other K$^{-}$ induced reactions 
\cite{p_spectra}.  
\begin{figure} [h]
\begin{center}
\resizebox{0.5\textwidth}{!}{%
  \includegraphics{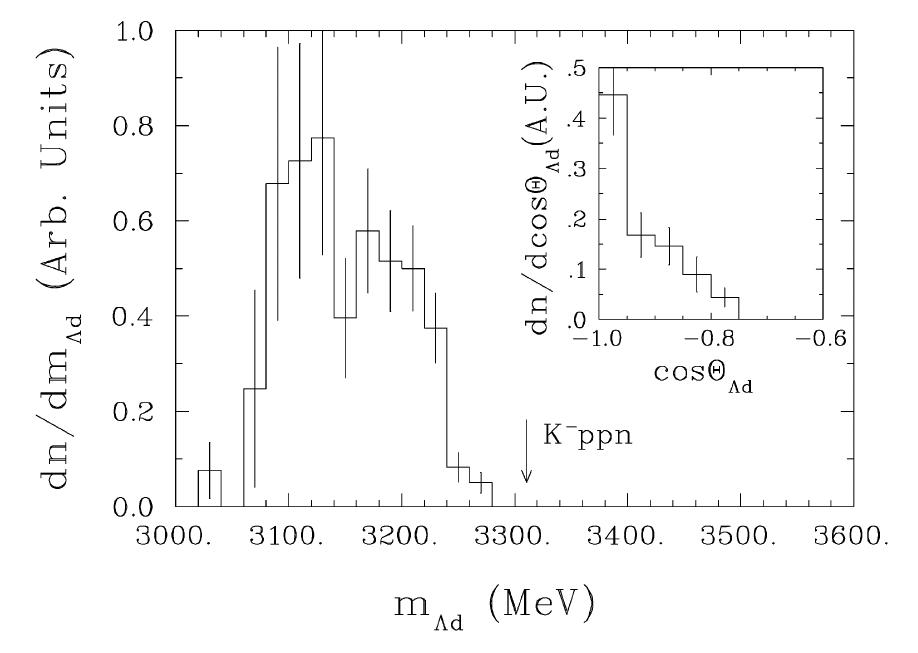} 
}
\caption{Invariant mass distribution of $\Lambda$d pairs from the $^{12}$C(K$^{-}_{\rm stop}$, $\Lambda$d)A' reaction. From Ref.~\cite{Lambdad}.}
\label{fig:figu7}       
\end{center}
\vspace{-4mm}
\end{figure}
As a matter of fact, the analogous distributions relative to events obtained with the $^{12}$C target did not present  similar features, as shown 
by Fig.~\ref{fig:figu7}. There is no 
evidence for a peak in the invariant mass distribution and the angular correlation of the events in the invariant mass range corresponding to the 
peak observed in $^{6}$Li is quite broad. 
The peak observed for $^{6}$Li was interpreted in Ref.~\cite{magas2} in terms of K$^{-}$ absorption from three nucleons, leaving the remaining ones 
as spectators. The back--to--back correlation reported in the experiment of Ref.~\cite{Lambdad} is reproduced too, not so well other observables. 
Such an explanation looks realistic, supporting also analogous measurements performed in Ref.~\cite{suzuki}. 
The yield for the production of the above peak was reported as (4.4$\pm$1.4)$\cdot$10$^{-3}$/ stopped K$^{-}$, quite similar to that for the 
production of the K$^{-}$pp system from light nuclei.

A further effort was then done by the FINUDA Collaboration for the search of correlated $\Lambda$-triton pairs emitted following the absorption 
of stopped K$^{-}$ in $^{6}$Li, $^{7}$Li and $^{9}$Be. The identification of tritons was very hard, since tracks due to these particles were 
completely blurred in the huge amount of other charged particles (pions, protons, deuterons). 
However a very satisfactory separation was achieved, thanks to a clever use of the dE/dx and time of flight information from the different layers 
of detectors \cite{Lambdat}. 
\begin{figure} [h]
\begin{center}
\resizebox{0.5\textwidth}{!}{%
  \includegraphics{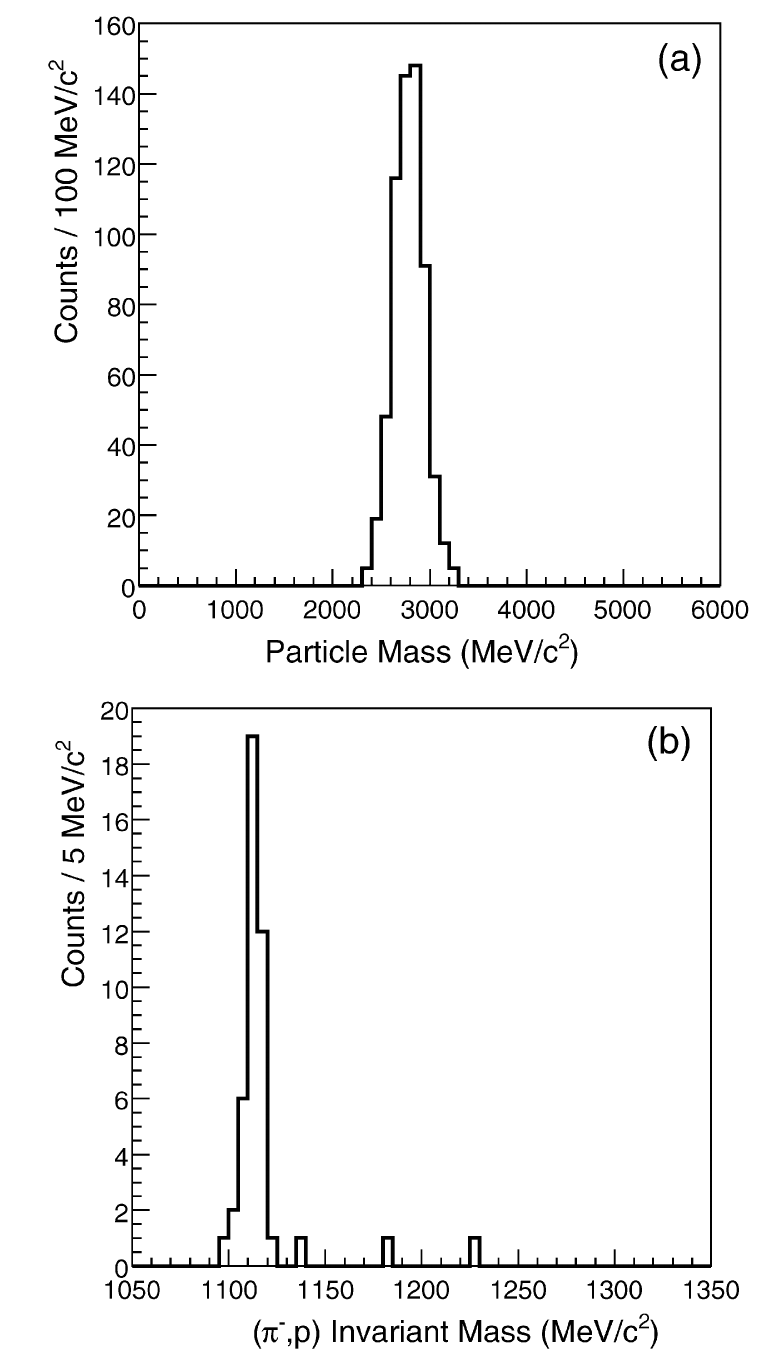} 
}
\caption{(a) Triton mass identification. (b) Invariant mass distribution of $\pi^{-}$p pairs in coincidence with tritons. From Ref.~\cite{Lambdat}.}
\label{fig:figu8}       
\end{center}
\vspace{-4mm}
\end{figure}
The triton mass identification by conditioned dE/dx is shown in fig.~\ref{fig:figu8}a. It was measured that it contains only 3$\%$ of other particles. 
As for the previous investigations the $\Lambda$ hyperons were finally identified by the value of the $\pi$-p invariant mass. Fig.~\ref{fig:figu8}b 
shows the invariant mass of $\pi$-p pairs detected in coincidence with tritons. The distribution of background events is flat outside the peak thus 
allowing for estimation of the number of events inside the peak itself. Such background events constitute 0.63$\pm$0.67 out of 40 $\Lambda$ events. 
The $\Lambda$-triton correlated events show a very nice 
back--to--back correlation, given in Fig.~\ref{fig:figu10}. This distribution is inconsistent with that predicted by a four--body phase space simulation 
of the reaction K$^{-}_{\rm stop}$A $\to \Lambda$tNA$^{\prime}$, the simplest that could mimic the observed events. 
Detailed simulations of other possible reaction chains that could mask the events attributed to mesonless $\Lambda$t emission showed that their 
importance was negligible. 
\begin{figure} [h]
\begin{center}
\resizebox{0.40\textwidth}{!}{%
  \includegraphics{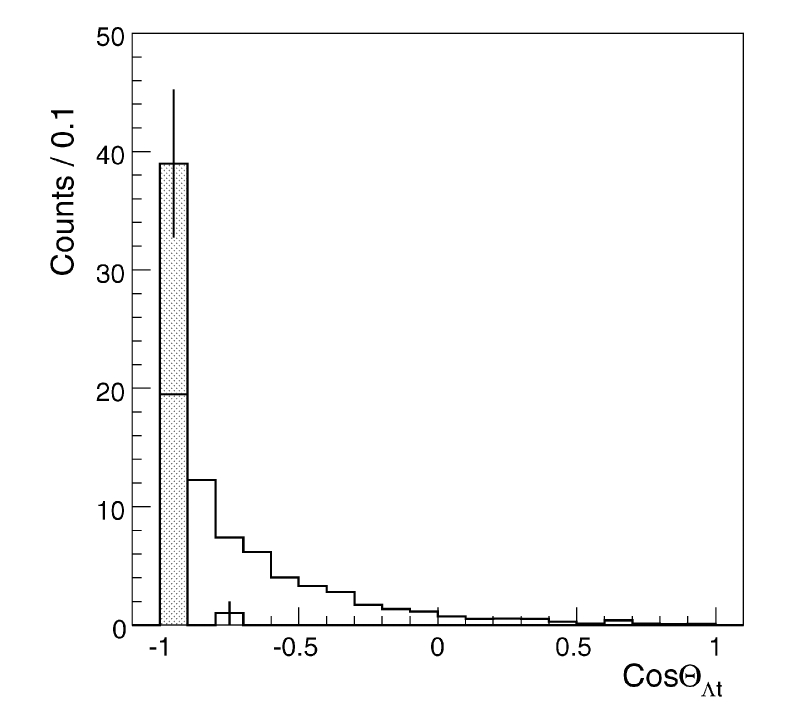} 
}
\caption{Opening angle distribution between $\Lambda$ and t pairs ($\cos\, \Theta_{\Lambda t}$). Filled histogram: experimental data;  open histogram: 
phase space 
distribution for the $^{6}$Li (K$^{-}_{\rm stop}$, $\Lambda$t)A reaction. From Ref.~\cite{Lambdat}.}
\label{fig:figu10}       
\end{center}
\vspace{-4mm}
\end{figure}
Due to the scarce number of events, it was not possible to obtain statistically reasonable distributions for the invariant mass of the $\Lambda$-t correlated events. 
The yield for the production of the $\Lambda$-t correlated events averaged over the 3 stopping targets is 
(1.01$\pm$0.18$_{\rm stat}$ \hspace{0.1mm}$^{+0.17}_{-0.10\, \rm syst}$)$\cdot$10$^{-3}$/K$^{-}_{\rm stop}$. We note that it is the same, within the errors,
of that measured for the production 
of $\Lambda$-p correlated events, and somehow lower than that for the production of $\Lambda$-d correlated events from $^{6}$Li. 
The similarity of the production rates for mesonless $\Lambda$--few (1,2,3) nucleons pairs induced by stopped K$^{-}$ raises doubts on the 
interpretation, 
put forward in Ref.~\cite{prl94} for $\Lambda$-p, that they are due to the decay, at rest, of a K$^{-}$--(1,2,3) nucleon(s) bound state. 
One should admit that these hypothetical states are produced with similar strengths, and this circumstance seems difficult to explain with simple models, 
given the inherent dynamical differences in the possible structure of these objects.
To our knowledge there are no models, at present, developed in order to explain the above production rates of mesonless absorption of stopped K$^{-}$ by 
nuclear clusters in light nuclei.

\subsection{The saga of the AntiKaonic Nuclear Clusters}
\label{saga}

\begin{figure} [h]
\begin{center}
\resizebox{0.50\textwidth}{!}{%
  \includegraphics{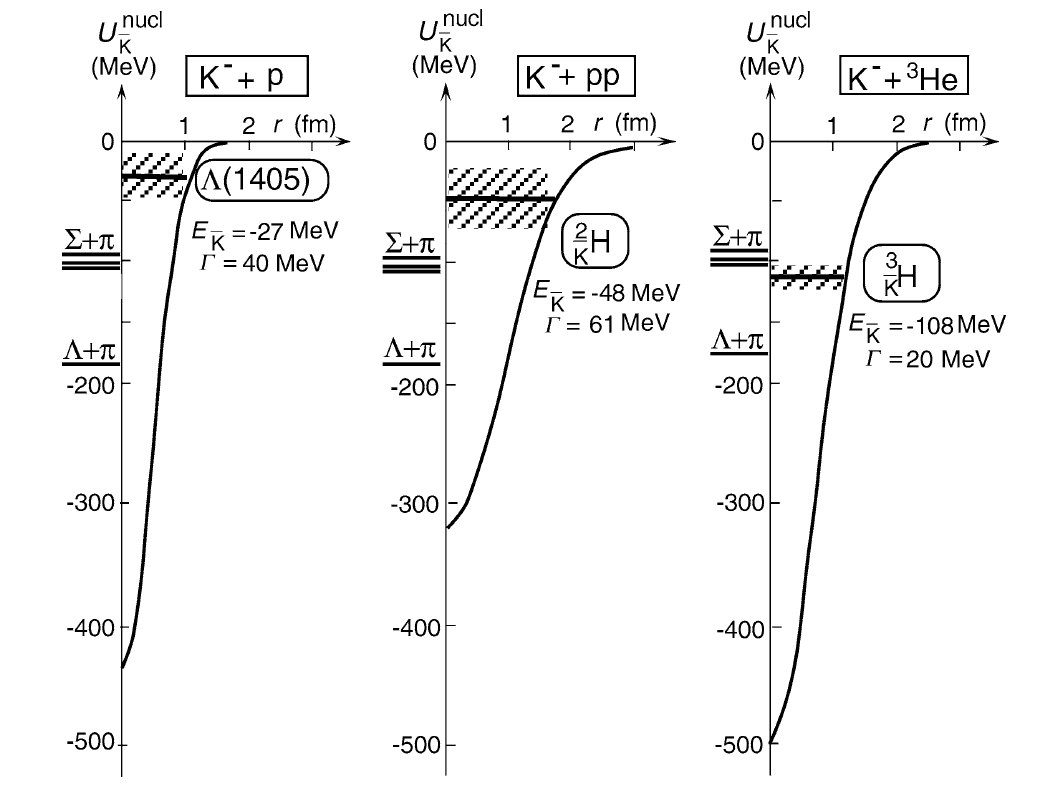} 
}
\caption{Calculated $\bar{K}$N and $\bar{K}$--nucleus potentials and bound levels: $\Lambda$(1405), $^{2}_{\bar{K}}$H and $^{3}_{\bar{K}}$H for K$^{-}$p, 
K$^{-}$pp and K$^{-}$ppn systems, respectively. From Ref.~\cite{plb535}.}
\label{fig:figu11}       
\end{center}
\vspace{-4mm}
\end{figure}

The above mentioned possible K$^{-}$--(few nucleons) bound systems were named Anti Kaonic Nuclear Clusters (AKNC) or Deeply Bound anti(Kaon) States (DBKS). 
The first speculation about the possible existence of AKNC was put forward in Ref.~\cite{wycech}, based on the observation that the driving $\bar{K}$N interaction in the 
isospin I=0 channel is strongly attractive near threshold. A large binding energy B of about 100 MeV was found, but with a similarly large value of the width $\Gamma$. 
The theme received a strong boost by the prediction from Ref.~\cite{plb535} of the possible existence of narrow discrete AKNC in few--body nuclear 
systems. 
The $\bar{K}$--nucleus potential was derived from a phenomenological  $\bar{K}$N potential accounting for several observables, with particular emphasis to the 
r$\hat{\mathrm{o}}$le of the $\Lambda$(1405), assumed to be a bound (K$^{-}$p) system. The predicted binding energies B for $\bar{K}$--few nucleon systems were 
quite large (from 50 to more than 100 MeV), but the distinctive feature was the narrowness ($\Gamma$ of 20-30 MeV). 
It was due to the circumstance that, due to the high value of B, the main decay channel K$^{-}$p (I=0)$\to \Sigma \pi$ is energetically forbidden, and the decay to 
$\Lambda \pi$ is suppressed by the isospin selection rules. 
Fig.~\ref{fig:figu11} shows the predicted binding energies and widths for the K$^{-}$p ($\Lambda$(1405)), K$^{-}$pp and K$^{-}$ppn 
AKNC. 

A dynamical approach, allowing for the polarization of the nucleus by the strong $\bar{K}$--nucleus potential has been proposed in Ref.~\cite{mares}: according 
to this calculation, the depth of the potential is density dependent and AKNC are predicted with B of 100-200 MeV and narrow enough widths of 50 MeV, but only 
for relatively  heavy nuclei. According to this approach, they could not be observed in the case of light targets, since in this case they should be too broad. 
The K$^{-}$pp bound system was studied in Ref.~\cite{shevchenko} by a coupled--channel Faddeev calculation obtaining a B of 55-70 MeV and a quite 
large width, of the order of 95-100 MeV. 

Ref. \cite{weise} examined the possible existence of AKNC by using energy dependent $\bar{K}$N interactions derived with the $s$--wave coupled--channels amplitudes 
involving the $\Lambda$(1405) and resulting from chiral SU(3) dynamics, plus $p$--wave amplitudes dominated by the $\Sigma$(1385): it was concluded that AKNC can 
possibly exist, with B from 60 to 100 MeV, but with decay widths of similar magnitude. 
                                                 
The theoretical speculation from Ref.~\cite{plb535} received a strong boost from the claimed discovery of an AKNC with narrow width by KEK--PS E471 
in a missing mass experiment on the reaction $^{4}$He (K$^{-}_{\rm stop}$, p/n) X, afterwards withdrawn by the same Group (exp. KEK--PS 
E549) as due to an 
experimental artifact \cite{iwasaki}. 

The experiments on correlated $\Lambda$-p and $\Lambda$-d events described in the previous paragraph were interpreted following 
the hypothesis of the existence of the AKNCs.  For the A=2 AKNC (K$^{-}$pp) the width is close to that predicted, but B is twice. 
More difficult is the interpretation of the invariant mass of $\Lambda$-d correlated events in terms of formation of an A=3 AKNC (K$^{-}$ppn). 
The width is again compatible with that predicted, but the B is a half.  The mass of this state would then be larger than the $\Sigma^{+} \pi$ threshold, with a consequently 
large width.

\begin{figure} [h]
\begin{center}
\resizebox{0.4\textwidth}{!}{%
  \includegraphics{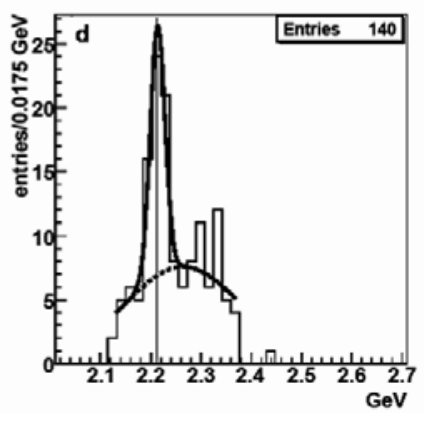} 
}
\caption{Best fit to the $\Lambda$p invariant mass distribution for events observed following the annihilation of $\bar{\mathrm{p}}$'s 
at rest on $^{4}$He after suitable kinematical cuts. From Ref.~\cite{bende2}.}
\label{fig:figu12}       
\end{center}
\vspace{-4mm}
\end{figure}
Due to the large interest and theoretical effort, several experiments that took data with a $\Lambda$ and a proton in the final states re--analysed their data looking for a 
confirmation of  K$^{-}$pp bound state. 
Events containing a $\Lambda$ and a p following the annihilation of $\bar{p}$'s on $^{4}$He nuclei measured by the OBELIX spectrometer at LEAR--CERN were analyzed 
selecting annihilations of $\bar{p}$'s on 3 or 4 nucleons, and using some ad--hoc kinematics cuts \cite{bende,bende2}; the result is shown in 
Fig.~\ref{fig:figu12}. 
A quite nice peak appears at a mass of 2223.2$\pm$3.2$_{\rm stat}$$\pm$1.2$_{\rm syst}$ MeV/c$^{2}$, and a width of less than 33.9$\pm$6.2 MeV/c$^{2}$. 
The binding energy should then be of 151.0$\pm$3.2$_{\rm stat}$$\pm$1.2$_{\rm syst}$ MeV. The statistical significance of the peak over the 
experimental background is 4.71 $\sigma$. 

More recently, the data collected by the DISTO experiment on the exclusive 
pp$\to$p$\Lambda$K$^{+}$ reaction at 2.85 GeV were also analyzed to search for a K$^{-}$pp 
bound system \cite{prl_yama}. The observed spectra of the K$^{+}$ missing mass (p$\Lambda$ 
invariant mass) with high transverse momenta of p and K$^{+}$ revealed a broad distinct peak 
with a mass of 2265$\pm$2$_{\rm stat}$ $\pm$5$_{\rm syst}$ MeV/c$^{2}$ and a width of 
118$\pm$8$_{\rm stat}$$\pm$10$_{\rm syst}$ MeV. Following this experiment, the binding energy of 
the K$^{+}$p system should be 105 MeV. 
\begin{table}[h]
\begin{center}
\caption{Comparison between theory and experiment for the binding energies, B, and widths, $\Gamma$, of the K$^{-}$pp system or 
$^{2}_{\bar{K}}$H.}
\label{tab:knuc}       
\begin{tabular}{ccccc}
\hline
 &  & B (MeV) & $\Gamma$ (MeV) & Ref. \\ 
 \hline
$^{2}_{\bar{K}}$H & K$^{-}$ at rest & -115$\pm$9 & 67$\pm$15 & \cite{prl94} \\
\hline
$^{2}_{\bar{K}}$H & $\bar{p}$ at rest & -151.0$\pm$4.4 & $\leq$33.9 & \cite{bende} \\
\hline
$^{2}_{\bar{K}}$H & pp $\to$ pK$^{+} \Lambda$& -105$\pm$2 & $\leq$118 & \cite{prl_yama} \\
\hline
$^{2}_{\bar{K}}$H & theory & -48 & 61 & \cite{aka_yama} \\
\hline
\end{tabular}
\end{center}  
\end{table}
Table \ref{tab:knuc} reports the binding energies and widths reported, as well as the theoretical prediction, showing in a compact way the present contradictions. 
We hope that new data from the dedicated experiments E15 and E27 at J--PARC and FOPI at GSI could soon clarify the situation.

\section{Conclusions and Perspectives}
\label{concl}

Strangeness Nuclear Physics has demonstrated to be able to provide a definite and precise answer to at least two problems for strangeness $-1$
systems; they are related to the fundamental issues of the $\Lambda N$ strong interaction and the four--baryon, strangeness changing, weak 
interaction. Various experiments have shown that the above tasks are fully achievable. 

Regarding the structure of $\Lambda$--Hypernuclei, in addition to the determination of the hyperon single--particle energy level schemes for heavy 
systems, quite limited up to recent times, by missing mass (magnetic spectrometers) measurements, and the study of the small spin--doublet and 
$\Lambda$ spin--orbit energy splittings performed by $\gamma$--ray measurements in coincidence, a program for the determination of other relevant 
observables (spin, parity and lifetime of hypernuclear levels, hypernuclear
deformation and electric and magnetic moments of hyperons in hypernuclei) should be started. To achieve this purpose, coincidence $\gamma$--rays 
spectroscopy or the recently assessed spectroscopy of  charged pions from mesonic weak decay were shown to be suitable tools. 
A relevant improvement should come from the fine spectroscopy of $\pi^{0}$ from mesonic decay with a resolution of the order of the MeV. 
The experiment is hard, but in the capabilities of the present state--of--art technology. $\gamma$--ray spectroscopy experiments, interpreted within
a shell model, has already allowed to extract important information on the spin--dependence of the $\Lambda N$ effective interaction.

Improved approaches to the structure of single and double strangeness Hypernuclei, possibly establishing solid links with analogous 
calculations in dense stars, will be important. They should incorporate important effects such as the $\Lambda N$--$\Sigma N$ strong coupling 
and, for $S=-2$ Hypernuclei, the coupling among the $\Xi$N, $\Lambda \Lambda$ and $\Sigma \Sigma$ states.
Of paramount importance will be the experimental verification of the possible existence of 
$\Xi$--Hypernuclei, with the related possibility of extracting information on the $\Xi$--nucleus potential and then on
the $\Xi N$ interaction. Studies on $\Lambda \Lambda$--Hypernuclei will allow to determine the $\Lambda \Lambda$ interaction, which is 
poorly known at present.

Particular attention should be devoted to the search for neutron--rich $\Lambda$--Hypernuclei, which are especially important in nuclear 
astrophysics. They provide information on the glue--like r$\hat{\mathrm{o}}$le of the Hyperon, on the $\Sigma$ hyperon admixture in 
$\Lambda$--Hypernuclei and on the neutron drip properties, thus they may play a r$\hat{\mathrm{o}}$le (like normal nuclei) in neutron capture nucleosynthesis.
Other impurity nuclear physics issues such as the deformation properties and the collective motion in Hypernuclei should be clarified by
experiments and new calculations.

Concerning the in--medium properties of the hyperon, the $\Lambda$ magnetic moment is an important observable which is sensitive to
the spin and angular momentum structure of the many--body system, to the spin--dependent part of the $\Lambda N$ effective interaction
and to the $\Lambda N$--$\Sigma N$ coupling. This magnetic moment can be indirectly measured from the transition probability for 
$\Lambda$ spin--flip transitions. In heavy--ion Hypernuclear production experiments such as HypHI, direct measurements
of the $\Lambda$ magnetic moment should be possible by observing the spin precession in a strong magnetic field.


Concerning the weak decay of hypernuclei, convincing evidences have been
achieved for a solution of the long--standing puzzles on the 
$\Gamma_n/\Gamma_p$ ratio between the $\Lambda n\to nn$ and 
$\Lambda p\to np$ decay rates and on the asymmetry parameter in the 
non--mesonic weak decay of polarized hypernuclei. On the one hand, this 
has been possible thanks to experimental and theoretical analyses of 
nucleon--coincidence emission spectra. On the other hand, theoretical 
approaches considering a one--meson--exchange weak potential supplemented 
by a two--pion--exchange mechanism could reproduce all the decay data for 
$^5_\Lambda$He and $^{12}_\Lambda$C. The questions which will be relevant 
for future developments concern
the $\Delta I=1/2$ isospin rule and the possible importance of mechanisms introduced by the $\Delta(1232)$ resonance
in the non--mesonic weak decay of $\Lambda$--Hypernuclei. 

Possible violations of the $\Delta I=1/2$ rule should be studied
for the non--mesonic weak decay of $s$--shell $\Lambda$--Hypernuclei.
A theoretical approach could establish the link with the
elementary $\Lambda N\to nN$ decay amplitudes and rates of the
approach by Bloch and Dalitz, at the same time
demonstrating the degree of reliability of this phenomenological model which is normally adopted
to interpret the experimental data. A study on four--and five--body Hypernuclei will play an important
r$\hat{\mathrm{o}}$le for establishing the detailed
spin--isospin dependence of the $\Lambda N\to nN$ process.

A microscopic calculation of the non--mesonic weak decay emission spectra for $\Lambda$--Hypernuclei is very important,
since it provides a fully quantum--mechanical approach in which a unified treatment of complex
many--body effects such as final state interactions, ground state correlations and ground state normalization is considered.
The nucleon spectra from the non--mesonic Hypernuclear decay are the real observables from which one extracts,
via some theoretical scheme, the experimental values of the partial decay rates. In particular, the contributions of many--body terms
introduced by the $\Delta(1232)$ resonance are expected to play a r$\hat{\mathrm{o}}$le in a detailed calculation of the
non--mesonic decay rates and (especially) the nucleon emission spectra, thus possibly explaining the present incompatibilities
between experiment and theory on the nucleon emission spectra. These contributions should be included in new calculations. 

It will be also important to start a systematic investigation of rare weak decays of $\Lambda \Lambda$--Hypernuclei
such as $\Lambda \Lambda\to \Lambda n$, $\Lambda \Lambda\to \Sigma^- p$,
$\Lambda \Lambda\to \Sigma^0 n$ ($\Delta S=1$) and $\Lambda \Lambda\to n n$ ($\Delta S=2$).
A reliable calculation of the rates for these $\Lambda$--induced $\Lambda$ decay reactions is 
missing and no experimental evidence of such processes is available at present.

Like the majority of items related to experiments with nuclei, the firm and reliable determination of the physical quantities describing the
$\Lambda N$ interaction in the strong and weak sectors requires a systematic series of measurements that may be performed at dedicated laboratories, 
with a dedicated program and a strict collaboration with the theoreticians active in the field.

The situation is rather good worldwide in order to achieve this goal. The J--PARC complex of accelerators at Tokai (Japan) will be the main 
pillar for Strangeness Nuclear Physics in the next decade. More than ten experiments on Strangeness Nuclear Physics were approved with high priority but 
the unfortunate event of the terrifying earthquake in the area delayed the activity by at least one year. 
Experiments will be performed on the $\gamma$--ray spectroscopy of light to heavy $\Lambda$--Hypernuclei, the production of neutron--rich 
$\Lambda$--Hypernuclei, the $\Lambda$ magnetic moment in nuclei and, concerning strangeness $-2$ systems, on 
the $\gamma$--ray spectroscopy of $\Lambda \Lambda$--Hypernuclei and the reaction spectroscopy of $\Xi$--Hypernuclei.
In addition, new coincidence measurements of the non--mesonic weak decay of $A=4$ and $^{12}_\Lambda$C Hypernuclei are planned.
The first J--PARC results are hopefully expected for the forthcoming year. 

Another laboratory which is expected to provide a good wealth of experimental information on high--resolution spectroscopy of $\Lambda$--Hypernuclei 
is TJNAF, after the on--going program of upgrade in energy of the electron accelerator complex. 

New laboratories that will soon enter into the game of Strangeness Nuclear Physics are MAMI--C in Mainz and GSI in Darmstadt. At MAMI--C, the complex 
of high--resolution spectrometers needed to perform Hypernuclear spectroscopy with the ($e,e' K^{+}$) reaction was recently commissioned. 
The results will complement those from TJNAF. At GSI, a pilot experiment on the production of Hypernuclei by heavy--ions (HyPHI), with special 
emphasis on the identification of neutron--rich Hypernuclei and their mesonic decay, has very recently completed the first data taking run.

The experimental teams engaged at the different laboratories are in very close contact and it is natural to expect that the above mentioned program of 
coordinated measurements will be agreed among them. The degree of integration of the experimental teams with the theoreticians has a long tradition, 
initiated by R.H.~Dalitz, and is very good.

In the long term perspective, the FAIR--GSI complex and the powerful PANDA detector will play a major r$\hat{\mathrm{o}}$le in Strangeness Nuclear 
Physics. The $\bar{p}$ source, unique in the world, will allow to produce abundantly $\Lambda \Lambda$--Hypernuclei by $\bar{p}$ annihilation into 
$\Xi \bar{\Xi}$ on nuclei, with subsequent interaction of the $\bar{\Xi}$ with a proton in the same nucleus, providing two low--momentum $\Lambda$'s 
that may stick to other nucleons. These $\Lambda \Lambda$--Hypernuclei should be produced in excited states and the measurement of their energies 
by $\gamma$--ray coincidence would allow the determination of the $\Lambda \Lambda$ strong interaction parameters, an important result in Hadronic 
Physics. If successful, this method should be extended to the production of $\Lambda \Lambda 
\Lambda$--Hypernuclei, by using the $\Omega^{-} \Omega^{+}$ production from $\bar{p}$ annihilation in nuclei.


Finally, the middle--term strategy for the Anti Kaonic Nuclear Clusters is the unambiguous assessment of their existence and of their main features 
by some dedicated experiments already approved and ready to run. In positive case, the long--term strategies are far reaching and could involve 
several laboratories worldwide and the construction of new dedicated detectors.

%

\begin{thebibliography}{300}
%
%


\bibitem{yukawa}
H.~Yukawa,  
{\it Proc. Phys.-Math. Soc. Jpn.} {\bf 17} (1935) 48.

\bibitem{SI08}
For a recent collection of reviews on nuclear physics with strangeness
see the Special Issue on {\em Recent advances in Strangeness
Nuclear Physics}, Nucl. Phys. {\bf A 804} (2008) 1--348.

\bibitem{SI08b}
See also the
proceedings of the {\em  10th International Conference on
Hypernuclear and Strange Particle Physics} (Hyp--X),
Nucl. Phys. {\bf A 835} (2010) 1--470.

\bibitem{tamu}
O.~Hashimoto and H.~Tamura, 
{\it Prog. Part. Nucl. Phys.} {\bf 57} (2006) 564.

\bibitem{vare07}
T.~Bressani, in  {\it Strangeness and Spin in Fundamental Physics}, Proc. of the International School of Physics 
 ``E. Fermi'' Course CLXVII, edited by M. Anselmino, T. Bressani, A. Feliciello and  P. Ratcliffe, (IOS Press, Amsterdam, 2008), p. 3.

\bibitem{AG02}
W.M.~Alberico and G.~Garbarino, {\it Phys. Rep.} {\bf 369} (2002) 1.

\bibitem{Sch10}
J.Schaffner-Bielich, 
{\it Nucl Phys.} {\bf A 835} (2010) 279.

\bibitem{Mi07}
D.J.~Millener, {\it Lect. Notes Phys.} {\bf 724} (2007) 31.

\bibitem{Ko04}
M.~Kohno, Y.~Fujiwara, Y.~Watanabe, K.~Ogata, M.~Kawai,
{\it Prog. Theor. Phys.} {\bf 112} (2004) 895;
{\it Phys. Rev.} {\bf C 74} (2006) 064613.

\bibitem{Da08}
J.~Dabrowski and J.~Rozynek,
{\it Phys. Rev.} {\bf C 78} (2008) 037601.

\bibitem{Ba99}
S.~Bart {\it et al.,}
{\it Phys. Rev. Lett.} {\bf 83} (1999) 5238.

\bibitem{No02}
H.~Noumi {\it et al.},
{\it Phys. Rev. Lett.} {\bf 89} (2002) 072301; 
Erratum-ibid. {\bf 90} (2003) 049902. 

\bibitem{Sa04}
P.K.~Saha {\it et al.},
{\it Phys. Rev.} {\bf C 70} (2004) 044613. 

\bibitem{Ya10}
Y.~Yamamoto, T.~ and T.A.~Rijken,
{\it Prog. Theor. Phys. Suppl.} {\bf 185} (2010) 72. 

\bibitem{Ri10b}
Th.A.~Rijken, M.M.~Nagels and Y.~Yamamoto,
{\it Prog. Theor. Phys. Suppl.} {\bf A 185} (2010) 14.

\bibitem{Fu07}
Y.~Fujiwara, Y.~Suzuki and C.~Nakamoto,
{\it Prog. Part. Nucl. Phys.} {\bf 58} (2007) 439.

\bibitem{Fu98} 
T.~Fukuda {\it et al.,} 
{\it Phys. Rev.} {\bf C 58} (1998) 1306;
P.~Khaustov {\it et al.,}
{\it Phys. Rev.} {\bf C 61} (2000) 054603. 

\bibitem{E05}
K.~Aoki {\it et al.}, "Spectroscopic study of $\Xi$--Hypernucleus, $^{12}_{\Xi}$Be, via the 
$^{12}$C(K$^{-}$, K$^{+}$) 
reaction", Proposal for J-PARK 50 GeV Proton Synchrotron, 1 (available at http:$\//$j-parc.jp/NuclPart$\/$Proposal$\_$e.html). 

\bibitem{Nak10}
K.~Nakazawa and H.~Takahashi,
{\it Prog. Theor. Phys. Suppl.} {\bf 185} (2010) 335.

\bibitem{Tak01}
H.~Takahashi {\it et al.,} 
{\it Phys. Rev. Lett.} {\bf 87} (2001) 212502.

\bibitem{Sc10}
H.J.~Schulze, {\it Nucl. Phys.} {\bf A 835} (2010) 19.

\bibitem{Su10}
K.~Sumiyoski, K.~Nakamoto, C.~Ishizuka, A.~Ohnishi, S.~Yamada and H.~Suzuki,
{\it Nucl. Phys.} {\bf A 835} (2010) 295.

\bibitem{kang} B.H.~Kang {\it et al.,}  {\it Phys. Rev. Lett.} {\bf 96} (2006) 062301.

\bibitem{mjkim} M.J.~Kim {\it et al.,}  {\it Phys. Lett.} {\bf B 641} (2006) 28. 

\bibitem{mkim} M.~Kim {\it et al.,}  {\it Phys. Rev. Lett} {\bf 103} (2009) 182502. 

\bibitem{fnd_nmwd} 
M.~Agnello {\it et al.} (FINUDA Collaboration) and G. Garbarino, {\it Phys. Lett.} {\bf B 685}(2010) 247.

\bibitem{nmwd_n} 
M.~Agnello {\it et al.,} (FINUDA Collaboration) and G. Garbarino {\it Phys. Lett.} {\bf B 701} (2011) 556.

\bibitem{BGPR} 
E.~Bauer, G.~Garbarino, A. Parre\~{n}o and A. Ramos,
{\it Nucl. Phys.} {\bf A 836} (2010) 199.

\bibitem{BG-PRC} 
E.~Bauer and G.~Garbarino,
{\it Phys. Lett.} {\bf B 698} (2011) 306. 

\bibitem{Wa07}
T.~Watanabe {\it et al.,}
{\it Eur. Phys. J.} {\bf A 33} (2007) 265.	

\bibitem{It01} 
K.~Itonaga, T.~Ueda and T.~Motoba, 
{\em Nucl. Phys.} {\bf A 691} (2001) 197c.

\bibitem{Pa02} 
A. Parre\~{n}o, A. Ramos and C. Bennhold,
{\em Phys. Rev.} {\bf C 65} (2002) 015205.


\bibitem{Ta06}
T.~Takatsuka, S.~Nishizaki, Y.~Yamamoto and R.~Tamagaki,
{\it Prog. Theor. Phys.} {\bf 115} (2006) 355.

\bibitem{danysz}
M.~Danysz and J.~Pniewsky, 
{\it Phil. Mag.} {\bf 44} (1953) 348;

\bibitem{fesh}
H.~Feshbach and K.~Kerman, 
{\it Preludes in Theoretical Physics}, edited by A.~De-Shalit,
H.~Feshbach and L.~Van Hove (North Holland, Amsterdam) (1966) 260-278.

\bibitem{bona}
G.C.~Bonazzola {\it et al.}, 
{\it Phys. Lett.} {\bf B 53} (1974) 297.

\bibitem{moto_ito_yama}
T.~Motoba, K.~Itonaga and Y.~Yamamoto, 
{\it Progr. Theor. Phys. Suppl.} {\bf 185} (2010), 197.

\bibitem{moto_bydz}
T.~Motoba, P.~Bydzovsky, M.~Sotona, K.~Itonaga and Y.~Yamamoto, 
{\it Progr. Theor. Phys. Suppl.} {\bf 185} (2010), 224.

\bibitem{tamura}
T.~Tamura {\it et al.,} 
{\it Progr. Theor. Phys. Suppl.} {\bf 117} (1989) 1.

\bibitem{haya}
R.S.~Hayano and T.~Yamazaki, 
{\it Perspectives of Meson Science}, T.~Yamazaki, K.~Nakai and K.~Nagamine Eds.,
(North Holland, Amsterdam) (1992) 493-517.

\bibitem{charpa}
G.~Charpak {\it et al.}, 
{\it Nucl. Instr. Meth.} {\bf 62} (1968) 262.

\bibitem{faess}
M.~Faessler {\it et al.}, 
{\it Phys. Lett.} {\bf B 46} (1973) 468.

\bibitem{bonanim}
G.C.~Bonazzola {\it et al.}, 
{\it Nucl. Instr. Meth.} {\bf 123} (1975) 269.

\bibitem{bambe}
A.~Bamberger {\it et al.}, 
{\it Phys. Lett.} {\bf B 36} (1971) 412.

\bibitem{bruck} 
W.~Bruckner {\it et al.}, 
{\it Phys. Lett.} {\bf B 55} (1975) 107.

\bibitem{bruck78}
W.~Bruckner {\it et al.}, 
{\it Phys. Lett.} {\bf B 79} (1978) 157.

\bibitem{millener}
D.J.~Millener, 
{\it LAMPF Workshop on $(\pi, K)$ Physics}, Edited by B.F.~Gibson, W.R.~Gibbs and 
M.B.~Johnson (AIP, New York) 1991, 128-140.

\bibitem{chrien}
R.E.~Chrien, 
{\it Nucl. Phys.} {\bf A 691} (2001) 501c.

\bibitem{hot}
H.~Hotchi {\it et al.}, 
{\it Phys. Rev.} {\bf C 64} (2001) 044302.   

\bibitem{Lnf}  
T.~Bressani, 
{\it Proc. of Workshop for Physics and detectors for DA$\Phi$NE}, Frascati April 9-12
1991, Edited by G.~Pancheri (Laboratori Nazionali di Frascati) 1991, 475-485.

\bibitem{finrep}
The FINUDA Collaboration, M~Agnello {\it et al.}, 
FINUDA Technical Report, LNF Internal Report LNF - 95/024 (IR) 1995.

\bibitem{agnello}
M.~Agnello {\it et al.}, 
{\it Phys. Lett.} {\bf B 622} (2005) 35.

\bibitem{iodice}
M.~Iodice {\it et al.}, 
{\it Phys. Rev. Lett.} {\bf 99} (2007) 052501.

\bibitem{bocquet} J.P.~Bocquet {\it et al.,} {\it Phys. Lett.} {\bf B192} (1987) 312.

\bibitem{armstrong} T.A.~Armstrong {\it et al.,} {\it Phys. Rev.} {\bf C47} (1993) 1957.

\bibitem{cassing} W.~Cassing {\it et al.,} {\it Eur. Phys. J.} {\bf A16} (2003) 549.

\bibitem{nield} K.J.~Nield {\it et al.,} {\it Phys. Rev.} {\bf C13} (1976) 1263.

\bibitem{avra} S.~Avramenko {\it et al.,} {\it Nucl. Phys.} {\bf A547} (1992) 95c.

\bibitem{Ab09}
B.I.~Abelev {\it et al.,} 
{\it Phys. Rev.} {\bf C 79} (2009) 034909.

\bibitem{Ad05}
J.~Adams {\it et al.,} 
{\it Nucl. Phys.} {\bf A 757} (2005) 102. 


\bibitem{Ab10}
B.I.~Abelev {\it et al.,} 
{\it Science} {\bf 328} (2010) 58.

\bibitem{Ch10}
J.H.~Chen,
{\it Nucl. Phys.} {\bf A 830} (2009) 761;
{\it Nucl. Phys.} {\bf A 835} (2010) 117.


\bibitem{Ko05}
V.~Koch, A.~Majumder and J.~Randrup, 
{\it Phys. Rev. Lett.} {\bf 95} (2005) 182301.

\bibitem{Zh10}
S.~Zhang, J.H.~Chen, H.~Crawford, D.~Keane, Y.G.~Ma and Z.B.~Xu,
{\it Phys. Lett.} {\bf B 684} (2010) 224.


\bibitem{Ri89}
P.M.M.~Maessen, T.A.~Rijken, J.J. de Swart,
{\it Phys. Rev.} {\bf C 40} (1989) 2226.

\bibitem{Ri99}
T.A.~Rijken, V.G.J.~Stoks and Y.~Yamamoto,
{\it Phys. Rev.} {\bf C 59} (1999) 21.

\bibitem{Ri06}
T.A.~Rijken,
{\it Phys. Rev.} {\bf C 73} (2006) 044007;
T.A.~Rijken and Y.~Yamamoto,
{\it Phys. Rev.} {\bf C 73} (2006) 044008.

\bibitem{Ri10}  
Th.A.~Rijken, M.M.~Nagels and Y.~Yamamoto, 
{\it Nucl. Phys.} {\bf A 835} (2010) 160.

\bibitem{Ju89}
B.~Holzenkamp, K.~Holinde, J.~Speth,
{\it Nucl. Phys.} {\bf A 500} (1989) 485.

\bibitem{Ju94}
A.~Reuber, K.~Holinde and J.~Speth,
{\it Nucl. Phys.} {\bf A 570} (1994) 543. 

\bibitem{Ju05}
J.~Haidenbauer and U.G.~Mei{\ss}ner,
{\it Phys. Rev.} {\bf C 72} (2005) 044005.

\bibitem{Po06}
H.~Polinder, J.~Haidenbauer and U.G.~Mei{\ss}ner,
{\it Nucl. Phys.} {\bf A 779} (2006) 244.

\bibitem{Po07} 
H.~Polinder, J.~Haidenbauer and U.G.~Mei{\ss}ner,
{\it Phys. Lett.} {\bf B 653} (2007) 29;          
J.~Haidenbauer and U.G.~Mei{\ss}ner,
{\it Phys. Lett.} {\bf B 684} (2010) 275.         

\bibitem{Sa06}
K.~Sasaki, E.~Oset and M.J.~Vicente Vacas, 
{\it Phys. Rev.} {\bf C 74} (2006) 064002.

\bibitem{Fu96}
Y.~Fujiwara, C.~Nakamoto and Y.~Suzuki, 
{\it Phys. Rev.} {\bf C 54} (1996) 2180.

\bibitem{Fu96b}
Y.~Fujiwara, C.~Nakamoto and Y.~Suzuki, 
{\it Phys. Rev. Lett.} {\bf 76} (1996) 2242.

\bibitem{Be07}
S.R.~Beane {\it et al.,} 
{\it Nucl. Phys.} {\bf A 794} (2007) 62.

\bibitem{Ne09}
H.~Nemura, N.~Ishii, S.~Aoki and T.~Hatsuda,
{\it Phys. Lett.} {\bf B 673} (2009) 136.

\bibitem{In10}
T.~Inoue {\it et al.,}
{\it Prog. Theor. Phys.} {\bf 124} (2010) 591. 

\bibitem{Ha10}   
O.~Hashimoto {\it et al.,}
{\it Nucl Phys.} {\bf A 835} (2010) 121.


%

\bibitem{Mo94}
K.~Itonaga, T.~Motoba, O.~Richter and M.~Sotona, 
{\it Phys. Rev.} {\bf C 49} (1994) 1045.

\bibitem{Mi01}
D.J.~Millener, 
{\it Nucl. Phys.} {\bf A 691} (2001) 93.

\bibitem{Hi00}
E.~Hiyama, M.~Kamimura, T.~Motoba, T.~Yamada and Y.~Yamamoto
{\it Phys. Rev. Lett.} {\bf 85} (2000) 270.

\bibitem{Mi88}
D.J.~Millener, C.B.~Dover and A.~Gal, 
{\it Phys. Rev.} {\bf C 38} (1988) 2700. 


\bibitem{fnd_spec}
M.~Agnello {\it et al.,} {\it Phys. Lett.} {\bf B  698} (2011) 219.  


\bibitem{cusanno}
F.~Cusanno {\it et al.,} {\it Phys. Rev. Lett.} {\bf103} (2009) 202501.  

\bibitem{cieply}
A.~Ciepl\'{y} E.~Friedman, A.~Gal, V.~Krej\u{c}i\u{r}\'{i}k, {\it Phys. Lett.} {\bf B698} (2011) 226.

\bibitem{Hi10b}
E.~Hiyama, M.~Kamimura, Y.~Yamamoto, T.~Motoba and T.A.~Rijken, 
{\it Prog. Theor. Phys. Suppl.} {\bf 185} (2010) 106.

\bibitem{Ko00}
M.~Kohno, Y.~Fujiwara, T.~Fujita, C.~Nakamoto and Y.~Suzuki, 
{\it Nucl. Phys.} {\bf A 674} (2000) 229.

\bibitem{Sc06}
B.J.~Schaefer, M.~Wagner, J.~Wambach, T.T.S.~Kuo and G.E.~Brown, 
{\it Phys. Rev.} {\bf C 73} (2006) 011001.

\bibitem{Dap08}
H.~Dapo, B.J.~Schaefer and J.~Wambach, 
{\it Eur. Phys. J.} {\bf A 36} (2008) 101.

\bibitem{Ak00}
Y.~Akaishi T.~Harada, S.~Shinmura and K.S.~Myint,
{\it Phys. Rev. Lett.} {\bf 84} (2000) 3539.

\bibitem{Hi02}
E.~Hiyama, M.~Kamimura, T.~Motoba and T.~Yamada, 
{\it Phys. Rev.} {\bf C 65} (2002) 011301.

\bibitem{Ta00}
H.~Tamura {\it et al.,}
{\it Phys. Rev. Lett.} {\bf 84} (2000) 5963.

\bibitem{Aj01}
S.~Ajimura {\it et al.,} 
{\it Phys. Rev. Lett.} {\bf 86} (2001) 4255.

\bibitem{Ak02}
H.~Akikawa {\it et al.,}
{\it Phys. Rev. Lett.} {\bf 88} (2002) 082501.

\bibitem{Uk04} 
M.~Ukai {\it et al.,}
{\it Phys. Rev. Lett.} {\bf 93} (2004) 232501.

\bibitem{Ta10}
H.~Tamura {\it et al.,} 
{\it Nucl Phys.} {\bf A 835} (2010) 3.

\bibitem{Mi10}
D.J.~Millener, 
{\it Nucl. Phys.} {\bf A 835} (2010) 11.

\bibitem{boh4} 
M.~Ukai, {\it et al.,} {\it Phys. Rev.} {\bf C 77} (2008) 054315. 

\bibitem{Ga71}
A.~Gal, J.M.~Soper and R.H.~Dalitz, {\it Ann. Phys.} (N.Y.) {\bf 63} (1971) 53;
ibid {\bf 72} (1972) 445; ibid {\bf 113} (1978) 79.

\bibitem{Da78} 
R.H.~Dalitz and A.~Gal, 
{\it Ann. Phys.} (N.Y.) {\bf 116} (1978) 167.

\bibitem{Mi85}
D.J.~Millener, A.~Gal, C.B.~Dover and R.H.~Dalitz
{\it Phys. Rev.} {\bf C 31} (1985) 499. 

\bibitem{Au83}
E.H.~Auerbach, A.J.~Baltz, C.B.~Dover, A.~Gal, S.H.~Kahana, L.~Ludeking and D.J.~Millener, 
{\it Ann. Phys.} (N.Y.) {\bf 148} (1983) 381.


\bibitem{Ko02}
H.~Kohri {\it et al.,}
{\it Phys. Rev.} {\bf C 65} (2002) 034607. 

\bibitem{tamu_hypx}
H.~Tamura {\it et al.,}
{\it Nucl. Phys.} {\bf A 835} (2010) 3. 

\bibitem{Mi10c}
D.J.~Millener and A.~Gal, 
{\it Phys. Lett.} {\bf B 701} (2011) 342. 

\bibitem{Ha08}
D.~Halderson, 
{\it Phys. Rev.} {\bf C 77} (2008) 034304.


\bibitem{Mi10b}
D.J.~Millener, 
{\it J. Phys. Conf. Ser.} {\bf 312} (2011) 022005.

\bibitem{Ba85}
T.~Motoba, H.~Band\={o}, K.~Ikeda and T.~Yamada, 
{\it Prog. Theor. Phys. Suppl.} {\bf 81} (1985) 42.


\bibitem{Ya90}
Y.~Yamamoto and H.~Band\={o},
{\it Prog. Theor. Phys.} {\bf 83} (1990) 254.




\bibitem{Ya94}
Y.~Yamamoto, T.~Motoba, H.~Himeno, K.~Ikeda and S.~Nagata,
{\it Prog. Theor. Phys. Suppl.} {\bf 117} (1994) 361.


\bibitem{hiyama}
H.~Hiyama, T.~Motoba, T.A. Rijken and Y.~Yamamoto,
{\it Progr. Theor. Phys. Suppl.} {\bf 185} (2010) 1.

\bibitem{nemura} 
H.~Nemura, Y.~Akaishi and Y.~Suzuki,
{\it Phys. Rev. Lett.} {\bf 89} (2002) 142504.

\bibitem{Hi99}
E.~Hiyama, M.~Kamimura, K.~Miyazaki and T.~Motoba,
{\it Phys. Rev.} {\bf C 59} (1999) 2351. 

\bibitem{Hi06}
E.~Hiyama, Y.~Yamamoto, T.A.~Rijken and T.~Motoba, 
{\it Phys. Rev.} {\bf C 74} (2006) 054312.

\bibitem{Ts98}
K.~Tsushima, K.~Saito, J.~Haidenbauer, A.~W.~Thomas,
{\it Nucl. Phys.} {\bf A 630} (1998) 691.

\bibitem{Gu08}
P.A.M.~Guichon, A.~W.~Thomas, K.~Tsushima,
{\it Nucl. Phys.} {\bf A 814} (2008) 66.  

\bibitem{Sh02}
H. Shen and H.~Toki,
{\it Nucl. Phys.} {\bf A 707} (2002) 469.


\bibitem{Fu04}
Y.~Fujiwara, M.~Kohno, K.~Miyagawa and Y.~Suzuki, 
{\it Phys. Rev.} {\bf C 70} (2004) 047002.

\bibitem{Fi09}
P.~Finelli, N.~Kaiser, D.~Vretenar and W.~Weise, 
{\it Nucl. Phys.} {\bf A 831} (2009) 163.

\bibitem{Mo83}
T.~Motoba, H.~Band\={o}, K.~Ikeda, 
{\it Prog. Theor. Phys.} {\bf 70} (1983) 189. 

\bibitem{Ta01}
K.~Tanida {\it et al.,} 
{\it Phys. Rev. Lett.} {\bf 86} (2001) 1982.

\bibitem{Ta02}
H.~Tamura, 
{\it Eur. J. Phys.} {\bf A 13} (2002) 181.

\bibitem{Ya11}
J.M.~Yao, Z.P.~Li, K.~Hagino, M.~Thi Win, Y.~Zhang, and J.~Meng, 
e-Print: arXiv:1104.3200 [nucl-th].

\bibitem{Hi97}
E.~Hiyama, M.~Kamimura, T.~Motoba, T.~Yamada and Y.~Yamamoto,
{\it Prog. Theor. Phys.} {\bf 97} (1997) 881;
{\it Phys. Rev. Lett.} {\bf 85} (2000) 270.

\bibitem{Wi08}
M.~Thi Win and K.~Hagino, 
{\it Phys. Rev.} {\bf C 78} (2008) 054311. 

\bibitem{Zh07}
X.R.~Zhou, H.J.~Schulze, H.~Sagawa, C.Xu~Wu and En-Guang~Zhao, 
{\it Phys. Rev.} {\bf C 76} (2007) 034312. 

\bibitem{Sc10b}
H.J.~Schulze, M.~Thi Win, K.~Hagino, and H.~Sagawa, 
{\it Prog. Theor. Phys.} {\bf 123} (2010) 569. 

\bibitem{Wi11}
M.~Thi Win, K.~Hagino and T.~Koibe,
{\it Phys. Rev.} {\bf C 83} (2011) 014301.






\bibitem{Ma95}
L.~Majling, 
{\it Nucl. Phys.} {\bf A 585} (1995) 211c.

\bibitem{Vr98}
D.~Vretenar, W.~Poschl, G.A.~Lalazissis and P.~Ring, 
{\it Phys. Rev.} {\bf C 57} (1998) 1060.

\bibitem{Zh08}
X.R.~Zhou, A.~Polls, H.J.~Schulze and I.~Vidana, 
{\it Phys. Rev.} {\bf C 78} (2008) 054306.

\bibitem{dalitz_levi}
R.H.~Dalitz, R~Levi Setti, {\it Nuovo Cimento} {\bf 30} (1963) 489.

\bibitem{Um09}
A.~Umeya and T.~Harada,
{\it Phys. Rev.} {\bf C 79} (2009) 024315.

\bibitem{Um11}
A.~Umeya and T.~Harada,
{\it Phys. Rev.} {\bf C 83} (2011) 034310.


\bibitem{Ak08}
Y.~Akaishi and K.S.~Myint, 
{\it Few Body Sys. Suppl.} {\bf 12} (2008) 277.


\bibitem{Hi96}
E.~Hiyama, M.~Kamimura, T.~Motoba, T.~Yamada and Y.~Yamamoto, 
{\it Phys. Rev.} {\bf C 53} (1996) 2075.

\bibitem{Hi09} 
E.~Hiyama, Y.~Yamamoto, T.~Motoba, M.~Kamimura, 
{\it Phys. Rev.} {\bf C 80} (2009) 054321.

\bibitem{Yu03}
T.Yu.~Tretyakova and D.E.~Lanskoy,
{\it Phys. At. Nucl.} {\bf 66} (2003) 1651.

\bibitem{Ha09}
T.~Harada, A.~Umeya and Y.~Hirabayashi, 
{\it Phys. Rev.} {\bf C 79} (2009) 014603.

\bibitem{Ku96}   
K.~Kubota {\it et al.,} 
{\it Nucl. Phys.} {\bf A 602} (1996) 327. 

\bibitem{tretyak01}
T.Yu.~Tretyakova and D.E.~Lanskoy, {\it Nucl. Phys.}  {\bf A691} (2001) 51c.

\bibitem{Sa05}
P.K.~Saha {\it et al.,} 
{\it Phys. Rev. Lett.} {\bf 94} (2005) 052502. 
\bibitem{nrich1}
M.~Agnello {\it et al.}, {\it Phys. Lett.} {\bf B640} (2006) 145.

\bibitem{hadron11}   
M.~Agnello {\it et al.,} {\it Phys. Rev. Lett.} {\bf 108} (2012) 042501.


\bibitem{P10}
A.~Sakaguchi {\it et al.}, {\em Production of Neutron--Rich $\Lambda$--Hypernuclei with the double exchange}, 
Proposal for J-PARK 50 GeV Proton Synchrotron, 1 (2006) (available at http:$\//$j-parc.jp/NuclPart$\/$Proposal$\_$e.html).

\bibitem{hyphi}
T.R.~Saito {\it et al.}, {\it Int. J. Mod. Phys.} {\bf E 19} (2010) 2656.

\bibitem{Na10}
K.~Nakamura {\it et al.} (Particle Data Group), 
{\it J. Phys.} {\bf G 37} (2010) 075021.

\bibitem{Ga91}
A.O.~Gattone, M.~Chiapparini and E.D.~Izquierdo, 
{\it Phys. Rev.} {\bf C 44} (1991) 548.

\bibitem{Co92}
J.~Cohen and J.V.~Noble, 
{\it Phys. Rev.} {\bf C 46} (1992) 801.

\bibitem{Do95}
C.B.~Dover, H.~Feshbach and A.~Gal, 
{\it Phys. Rev.} {\bf C 51} (1995) 541.

\bibitem{Sa97}
K.~Saito, M.~Oka and T.~Suzuki, 
{\it Nucl. Phys.} {\bf A 625} (1997) 95.

\bibitem{al91}
W.~M.~Alberico, A.~De Pace, M.~Ericson and A.~Molinari,
{\it Phys. Lett.} {\bf B 256} (1991) 134.

\bibitem{alberico} 
W.~M.~Alberico and G.~Garbarino, in {\it Hadron Physics} (IOS Press, Amsterdam, 2005). 
Proc. of the International School of Physics ``E. Fermi'' Course CLVIII, 
edited by T. Bressani, A. Filippi and  U. Wiedner, p. 125.                 

\bibitem{outa}
 H.~Outa, in  {\it Hadron Physics} (IOS Press, Amsterdam, 2005). Proc. of the International School of Physics 
 ``E. Fermi'' Course CLVIII, edited by T. Bressani, A. Filippi and  U. Wiedner, p. 219.
 
\bibitem{Ba10b}
A.~Bauer and G.~Garbarino,
{\it Phys. Rev.} {\bf C 81} (2010) 064315.

\bibitem{cohen} J.~Cohen, {\it Progr. Part. Nucl. Phys.} {\bf 25} (1990) 139.

\bibitem{Pa07}  
A. Parre\~no, 
{\it Lect. Notes Phys.} {\bf 724} (2007) 141.

\bibitem{prem} R.J.~Prem {\it et al.,} {\it Phys. Rev.} {\bf 136} (1964) B1803.
              
\bibitem{phillips} R.E.~Phillips {\it et al.,} {\it Phys. Rev.} {\bf 180} (1969) 1307.

\bibitem{bohm} G.~Bohm {\it et al.,} {\it Nucl. Phys.} {\bf B16} (1970) 46.

\bibitem{keyes70} G.~Keyes {\it et al.,} {\it Phys. Rev.} {\bf D1} (1970) 66.

\bibitem{keyes73} G.~Keyes {\it et al.,} {\it Nucl. Phys.} {\bf B67} (1973) 269.

\bibitem{kang65} Y.W.~Kang {\it et al.,} {\it Phys. Rev.} {\bf 139} (1965) B401.

\bibitem{outa95} H.~Outa {\it et al.,} {\it Nucl. Phys.} {\bf A585} (1995) 109c.

\bibitem{outa98} H.~Outa {\it et al.,} {\it Nucl. Phys.} {\bf A639} (1998) 251c.

\bibitem{zeps} V.J.~Zeps, {\it Nucl. Phys.} {\bf A639} (1998) 261c.

\bibitem{parker} J.D.~Parker {\it et al.,} {\it Phys. Rev.} {\bf C76} (2007) 035501.

\bibitem{szym} J.J.~Szymanski {\it et al.,} {\it Phys. Rev.} {\bf C43} (1991) 849.

\bibitem{kame05} S.~Kameoka {\it et al.,} {\it Nucl. Phys.} {\bf A754} (2005) 173c.

\bibitem{grace} R.~Grace {\it et al.,} {\it Phys. Rev. Lett.} {\bf 55} (1985) 1055.

\bibitem{park} H.~Park {\it et al.,} {\it Phys. Rev.} {\bf C61} (2000) 054004.

\bibitem{itonaga}K.~Itonaga {\it et al.,} {\it Nucl. Phys.} {\bf A639} (1998) 329c.

\bibitem{ramos} A.~Ramos, E.~Oset and L.L~Salcedo {\it Phys. Rev.} {\bf C50} (1994) 2314.

\bibitem{albe2} W.M.~Alberico, A.~De Pace, G.~Garbarino and A.~Ramos, {\it Phys. Rev.} {\bf C 61} (2000) 044314.

\bibitem{noga} V.I.~Noga {\it et al.,} {\it Sov. J. Nucl. Phys.} {\bf 43} (1986) 856.


\bibitem{metag} V.~Metag {\it et al.,} {\it Nucl. Inst. Meth.} {\bf 114} (1974) 445.






\bibitem{dal1} 
R.H.~Dalitz, {\it Phys. Rev.} {\bf 112} (1958) 605. \\ 
R.H.~Dalitz, L.~Liu, {\it Phys. Rev.} {\bf 116} (1959) 1312. 

\bibitem{dal2} 
R.H.~Dalitz, {\it Nucl. Phys.} {\bf 41} (1963) 78. \\ 
D.~Ziemi\'{n}ska, {\it Nucl. Phys.} {\bf A242} (1975) 461; 
D.~Ziemi\'{n}ska, R.H.~Dalitz, {\it Nucl. Phys.} {\bf A 238} (1975) 453. \\ 
D.~Kie{\l}czewska, D.~Ziemi\'{n}ska, R.H.~Dalitz, {\it Nucl. Phys.} 
{\bf A 333} (1980) 367. 

\bibitem{bertrand} D.~Bertrand {\it et al.,}, {\it Nucl. Phys.} {\bf B 16} (1970) 77.

\bibitem{bloch} M.M.~Bloch {\it al.,}, {\it Proceedings of the International Conference on 
Hyperfragments, St. Cergue}, CERN64-1 (1964), 63. 

\bibitem{davis63} D.H.~Davis {\it et al.,},  {\it Nucl. Phys.} {\bf 41} (1963) 73.

\bibitem{davis05} D.H.~Davis, {\it Nucl. Phys.} {\bf A 754} (2005) 3c. 

\bibitem{okada} S.~Okada {\it et al.,} {\it Nucl. Phys.} {\bf A754} (2005) 178c. 

\bibitem{fnd_mwd} M.~Agnello {\it et al.,}, {\it Phys. Lett.} {\bf B681} (2009) 139.

\bibitem{saka} A.~Sakaguchi, {\it et al.,} {\it Phys. Rev.} {\bf C 43} (1991) 73. 

\bibitem{noumi} H.~Noumi, {\it et al.,} {\it Phys. Rev.} {\bf C 52} (1995) 2936. 

\bibitem{sato05} Y.~Sato, {\it et al.,} {\it Phys. Rev.} {\bf C 71} (2005) 025203. 


\bibitem{motoba4}  
T.~Motoba, H.~Band\={o}, T.~Fukuda, J.~\v{Z}ofka, {\it Nucl. Phys.} 
{\bf A 534} (1991) 597. 

\bibitem{motoba2} 
H.~Band\={o}, T.~Motoba, J.~\v{Z}ofka, in {\it Perspectives of Meson Science}, 
Eds. T. Yamazaki, K. Nakai, K. Nagamine (North Holland, Amsterdam, 1992) 
pp. 571-660. 

\bibitem{motoba3} 
T.~Motoba, K.~Itonaga, H.~Band\={o}, {\it Nucl. Phys.} {\bf A 489} (1988) 683. 

\bibitem{motoba1} 
T.~Motoba, K.~Itonaga, {\it Progr. Theor. Phys. Suppl.} {\bf 117} (1994) 477. 

\bibitem{bando} 
H.~Band\={o}, H.~Takaki, {\it Progr. Theor. Phys. Suppl.} {\bf 72} (1984) 
109; {\it Phys. Lett.} {\bf B 150} (1985) 409. 

\bibitem{oset} 
E.~Oset, L.L.~Salcedo, {\it Nucl. Phys.} {\bf A 443} (1985) 704; 
E.~Oset, P. Fern\'andez de C\'ordoba, L.L.~Salcedo, R.~Brockmann, 
{\it Phys. Rep.} {\bf 188} (1990) 79. 

\bibitem{friedman1} E.~Friedman, {\it et al.,} {\it Phys. Rev.} {\bf C 72} 
(2005) 034609, and references to earlier works therein. 

\bibitem{friedman2} E.~Friedman, A.~Gal, {\it Phys. Rep.} {\bf 452} (2007) 89. 

\bibitem{sigma} R.S.~Hayano {\it et al.,} {\it Phys. Lett.} {\bf B231} (1989) 355. \\
                             H.~Outa {\it et al.,}  {\it Progr. Theor. Phys. Suppl.} {\bf 117} (1994) 177. \\
                             T.~Nagae {\it et al.,} {\it Phys. Rev. Lett.} {\bf 80} (1998) 1605. \\ 
                             T.~Harada {\it et al.,} {\it Nucl. Phys.} {\bf A507} (1990) 715.
                             
\bibitem{kurihara} Y.~Kuriraha  {\it et al.,} {\it Progr. Theor. Phys.} {\bf 71} (1984) 561; 
{\it Phys. Rev.} {\bf C31} (1985) 971.

\bibitem{kumagai} I.~Kumagai-Fuse {\it et al.,} {\it Phys. Lett.} {\bf B345} (1995) 386.     

\bibitem{kumagai2} I.~Kumagai-Fuse, private communication.

\bibitem{oset2} E.~Oset, P.~Fernandez de Cordoba, J.~Nieves, A.~Ramos, L.L.~Salcedo, 
{\it Progr. Theor. Phys. Suppl.} {\bf 117} (1994) 461. 
           
\bibitem{montwill} A.~Montwill {\it et al.,} {\it Nucl. Phys.} {\bf A234} (1974) 413. 

\bibitem{sasao} 
J.~Sasao, {\it et al.,} {\it Phys. Lett.} {\bf B 579} (2004) 258. 

\bibitem{gal} 
A.~Gal, {\it Nucl. Phys.} {\bf A 828} (2009) 72. 

\bibitem{CK} 
S.~Cohen, D.~Kurath,  {\it Nucl. Phys.} {\bf 73} (1965) 1;  {\it Nucl. Phys.} {\bf A101} (1967) 1. 

\bibitem{hashim} 
O.~Hashimoto, {\it et al.,} {\it Nucl. Phys.} {\bf A 639} (1998) 93c.

\bibitem{ziem} 
D.~Ziemi\'{n}ska, {\it Nucl. Phys.} {\bf A 242} (1975) 461. 

\bibitem{boh3}  
D.J.~Millener, {\it Nucl. Phys.} {\bf A 804} (2008) 84. 

\bibitem{tang} 
L.~Tang, {\it Int. Jou. Mod. Phys.} {\bf E 19} (2010) 2638.  

\bibitem{ches} 
W.~Cheston and H.~Primakoff,  {\it Phys. Rev.} {\bf 92} (1953) 1537.  

\bibitem{dalitz}
M.~M.~Block and R.~H.~Dalitz,  {\it Phys. Rev. Lett.} {\bf 11} (1963) 96.

\bibitem{al_gar} W.M.~Alberico and G.~Garbarino,  {\it Phys. Lett.} {\bf B 486} (2000) 362. 

\bibitem{Ad67}
J.~B.~Adams,
{\it Phys. Rev.} {\bf 156} (1967) 1611.

\bibitem{mckellar} B.J.H.~McKellar and B.F.~Gibson, {\it Phys. Rev.} {\bf C 30} (1984) 322.

\bibitem{nardulli} G.~Nardulli, {\it Phys. Rev.} {\bf C 38} (1988) 832.

\bibitem{dubach} J.F.~Dubach {\it et al.,} {\it Nucl. Phys.} {\bf A 450} (1986) 71c.

\bibitem{Pa97}
A. Parre\~{n}o, A. Ramos and C. Bennhold,
{\it Phys. Rev.} {\bf C 56} (1997) 339; 
A. Parre\~{n}o and A. Ramos,
{\it Phys. Rev.} {\bf C 65} (2002) 015204.

\bibitem{Os01}
D.~Jido, E.~Oset and J.E.~Palomar,
{\em Nucl. Phys.} {\bf A 694} (2001) 525.

\bibitem{CGPR07}
C.~Chumillas, G.~Garbarino, A.~Parre\~{n}o and A.~Ramos,
{\em Phys. Lett.} {\bf B 657} (2007) 180.

\bibitem{cheung} 
C.~Y.~Cheung, D.~P.~Heddle and L.~S.~Kisslinger  {\it Phys. Rev.} {\bf C 27} (1983) 335;
D.~P.~Heddle and L.~S.~Kisslinger {\it Phys. Rev.} {\bf C 33} (1986) 608.

\bibitem{inoue} T.~Inoue {\it et al.,} {\it Nucl. Phys.} {\bf A 633} (1998) 312.

\bibitem{sasaki} 
K.~Sasaki {\it et al.,} {\it Nucl. Phys.} {\bf A 669} (2000) 331; 
{\bf  678} (2000) 455.

\bibitem{bau2}  
E.~Bauer and F.~Krmpotic, {\it Nucl. Phys.}  {\bf A 739} (2004) 109.

\bibitem{bau_garb}  
E.~Bauer and G.~Garbarino, {\it Nucl. Phys.} {\bf A 828} (2009) 29.

\bibitem{Pa98}
A. Parre\~{n}o, A. Ramos, C. Bennhold and K. Maltman, 
{\it Phys. Lett.} {\bf B 435} (1998) 1.

\bibitem{schumacher} R.A.~Schumacher, {\it Nucl. Phys.} {\bf A 547} (1992) 143c.


\bibitem{jparc-4}
S. Ajimura {\it et al.}, {\em Exclusive study on the $\Lambda N$ weak interaction in
$A=4$ $\Lambda$--Hypernuclei}, Letter of intent for an experiment (E22) at J--PARC (2007).


\bibitem{ohm}
H.~Ohm {\it et al.,} {\it Phys. Rev.} {\bf C 55} (1997) 3062.

\bibitem{kulessa}
P.~Kulessa {\it et al.,} {\it Nucl. Phys} {\bf A639} (1998) 283c.

\bibitem{itonaga2} K.~Itonaga  {\it et al.,} {\it Phys. Rev.} {\bf C 65} (2002) 034617.

\bibitem{parreno} A.~Parreno and A.~Ramos, {\it  Phys. Rev.} {\bf C 65} (2001) 015204. 
 
\bibitem{Ba03}
C. Barbero, C. De Conti, A. P. Gale\~ao and F. Krmpoti\'c,
Nucl. Phys. {\bf A 726} (2003) 267.

\bibitem{noumi2} H.~Noumi  {\it et al.,} {\it Proc. IV Int. Sym. on Weak and Electrom. Int. in Nuclei}, H.~Ejiri, T.~Kishimoto and T.~Sato Eds. 
                                World Scientific 1995, p.550.  

\bibitem{hashi3} O.~Hashimoto {\it et al.,}  {\it Phys. Rev. Lett.} {\bf 88} (2002) 042503.

\bibitem{hjkim} H.J.~Kim {\it et al.,}  {\it Phys. Rev.} {\bf C 68} (2003) 065201. 



\bibitem{okada2} S.~Okada  {\it et al.,} {\it Phys. Lett.} {\bf B 597} (2004) 249. 

\bibitem{bhang} H.~Bhang {\it et al.,}  {\it Eur. Phys. Jou.} {\bf A 33} (2007) 259.


\bibitem{npa804} M.~Agnello {\it et al.,}  {\it Nucl. Phys.} {\bf A 804} (2008) 151.




\bibitem{ramos1} A.~Ramos {\it et al.,}  {\it Phys. Rev.} {\bf C 55} (1997) 735. 
 
\bibitem{ramos2} A.~Ramos {\it et al.,}  {\it Phys. Rev.} {\bf C 66} (2002) 039903. 

\bibitem{geant} CERN Program Library Entry W 5013, {\it GEANT}.

\bibitem{garb} 
G.~Garbarino, A.~Parreno and A.~Ramos, {\it Phys. Rev.} {\bf C 69} (2004) 054603.

\bibitem{INC}
G.~Garbarino, A.~Parre\~{n}o and A.~Ramos,
{\it Phys. Rev. Lett.} {\bf 91} (2003) 112501.

\bibitem{bau3}  E.~Bauer, {\it Nucl. Phys.} {\bf A 818} (2009) 174.

%
\bibitem{event_nnp}
M.~Agnello {\it et al.,}  "Hypernuclear weak decay studies with FINUDA", 
accepted for publication on {\it Nucl. Phys.} {\bf A}, DOI: 10.1016/j.nuclphysa/2012.01.024.




\bibitem{jparc-e18}
H. Bhang {\it et al.,}, {\em Coincidence measurement of the weak decay of
$^{12}_\Lambda$C and the three--body weak interaction process},
Letter of intent for an experiment (E18) at J--PARC (2006).

\bibitem{coremans} G.~Coremans {\it et al.,} {\it Nucl. Phys.} {\bf B 16} (1970) 209. 

\bibitem{bloch2} M.M.~Bloch {\it et al.,} {\it Proceedings of the 1960 Annual International Conference on 
High Energy Physics at Rochester}  (New York, 1964), 419. 

\bibitem{keyes76} G.Keyes {\it et al.,}  {\it Nuovo Cim.}  {\bf 31 A } (1976) 401. 

\bibitem{rayet} M.~Rayet, {\it Nuovo Cim.}  {\bf 52 B} (1966) 238.

\bibitem{thurnauer} P.G.~Thurnauer, {\it Nuovo Cim.}  {\bf 26 } (1962) 869.  

\bibitem{nmwd_rare} M.~Agnello {\it et al.,} {\it Nucl. Phys.} {\bf A 835} (2010) 439.

\bibitem{Ba89} 
H.~Band$\overline{\rm o}$, T.~Motoba M.~Sotona and J.~\v{Z}ofka,
{\em Phys. Rev.} {\bf C 39} (1989) 587;
T.~Kishimoto, H.~Ejiri and H.~Band$\overline{\rm o}$, {\em Phys. Lett.} {\bf B 232} (1989) 24.

\bibitem{Aj92} 
S.~Ajimura {\sl et al.}, {\em Phys. Lett.} {\bf B 282} (1992) 293.

\bibitem{Aj00}
S.~Ajimura {\sl et al.}, {\em Phys. Rev. Lett.} {\bf 84} (2000) 4052.

\bibitem{Ba90}
H.~Band\={o}, T.~Motoba and J.~\v{Z}ofka,
{\it Int. J. Mod. Phys.} {\bf A 5} (1990) 4021.

\bibitem{Ra92} A. 
A.~Ramos, E.~van Meijgaard, C.~Bennhold and B.K.~Jennings,
{\em Nucl. Phys.} {\bf A 544} (1992) 703.

\bibitem{Aj98} 
S.~Ajimura {\sl et al.}, {\em Phys. Rev. Lett.} {\bf 80} (1998) 3471.

\bibitem{Mar06}
T.~Maruta {\it et al.},
Nucl. Phys. {\bf A 754} (2005) 168c;
Eur. Phys. J. {\bf A 33} (2007) 255.

\bibitem{Mot94} 
T.~Motoba and K.~Itonaga, {\em Nucl. Phys.} {\bf A 577} (1994) 293c.

\bibitem{It03}
K.~Itonaga, T.~Motoba and T.~Ueda, in: K.~Maeda, H.~Tamura, S.N.~Nakamura, O.~Hashimoto 
(Eds.), {\em Electrophotoproduction of Strangeness on Nucleons and Nuclei}, Sendai03, 
World Scientific, 2004, p.397.

\bibitem{Ba05}
C.~Barbero, A.P.~Gale\~ao and F.~Krmpoti\'c,
{\em Phys. Rev.} {\bf C 72} (2005) 035210.


\bibitem{Al05}
W.M.~Alberico, G.~Garbarino, A.~Parre\~{n}o and A.~Ramos, 
{\em Phys. Rev. Lett.} {\bf 94} (2005) 082501.

\bibitem{Pa04}
A.~Parre\~{n}o, C.~Bennhold and B.R.~Holstein, 
{\em Phys. Rev.} {\bf C 70} (2004) 051601(R).

\bibitem{Sas05}
K.~Sasaki, M.~Izaki and M.~Oka, 
{\em Phys. Rev.} {\bf C 71} (2005) 035502.


\bibitem{Ba06}
C.~Barbero and A.~Mariano, 
{\em Phys. Rev.} {\bf C 73} (2006) 024309.

\bibitem{It08}
K.~Itonaga, T.~Motoba, T.~Ueda, and Th.A.~Rijken,
{\em Phys. Rev.} {\bf C 77} (2008) 044605.

\bibitem{katz}
P.A.~Katz {\it et al.}, {\it Phys. Rev.} {\bf D1} (1970), 1267.

\bibitem{prl94}
M.~Agnello {\it et al.}, {\it Phys. Rev. Lett.} {\bf 94} (2005) 212303.

\bibitem{magas}
V.K.~Magas {\it et al.,} {\it Phys. Rev.} {\bf C74} (2006) 025206. 

\bibitem{yama_db}
Y.~Yamazaki, Y.~Akaishi, {\it Nucl. Phys.} {\bf A 792} (2007) 229.

\bibitem{Lambdad}
M.~Agnello {\it et al.}, {\it Phys. Lett.} {\bf B 654} (2007) 80.

\bibitem{p_spectra}
M.~Agnello {\it et al.}, {\it Nucl. Phys.} {\bf A 775} (2006) 35.

\bibitem{magas2}
V.K.~Magas, E.~Oset, A.~Ramos, {\it Phys. Rev.}  {\bf C7} (2008) 065210.

\bibitem{suzuki}
T.~Suzuki {\it et al.,} {\it Phys. Rev.} {\bf C 76} (2007) 068202.

\bibitem{Lambdat}
M.~Agnello {\it et al.,} {\it Phys. Lett.} {\bf B 669} (2008) 229.

\bibitem{wycech}
S.~Wycech, {\it Nucl. Phys.} {\bf A 450} (1986) 399c.

\bibitem{plb535}
T.~Yamazaki, Y.~Akaishi, {\it Phys. Lett.} {\bf B 535} (2002) 70.

\bibitem{mares}
J.~Mares, E.~Friedman, A.~Gal, {\it Nucl. Phys.} {\bf A770} (2006) 84.

\bibitem{shevchenko}
N.V.~Shevchenko, A.~Gal, J.~Mares, {\it Phys. Rev. Lett.} {\bf 98} (2007) 082301. 

\bibitem{weise}
W.~Weise, R.~Hartle, {\it Nucl. Phys.} {\bf A 804} (2008) 173.

\bibitem{iwasaki}
M.~Iwasaki {\it et al.,} {\it Nucl Phys.} {\bf A 804} (2008), 186.

\bibitem{bende}
G.~Bendiscioli {\it et al.,} {\it Nucl. Phys.} {\bf A 789} (2007), 222.

\bibitem{bende2}
G.~Bendiscioli {\it et al.,} {\it Eur. Phys. J.} {\bf A 40} (2009) 11.

\bibitem{prl_yama}
T.~Yamazaki {\it et al.,} {\it Phys. Rev. Lett.} {\bf 104} (2010) 132501.


\bibitem{aka_yama}
Y.~Akaishi, T.~Yamazaki {\it Phys. Rev.} {\bf C 65} (2002) 0444005.


\end{thebibliography}
%

\end{document}